%% file: Draft.tex
\documentclass[11pt, oneside]{article} 
\pdfoutput=1
\usepackage{jheppub}
\usepackage{amsmath,amssymb,amsfonts,amsthm}
\usepackage{color}
\usepackage{booktabs}
\definecolor{darkblue}{rgb}{0.1,0.1,.7}
\usepackage{bbm}
\usepackage{enumitem}
\usepackage[normalem]{ulem}
\usepackage{graphicx}
\usepackage{latexsym}
\usepackage[labelformat=simple]{subcaption}
\usepackage{amscd}
\usepackage[all,cmtip]{xy}
\usepackage{mathrsfs}
\usepackage{bbold}
\usepackage{slashed}
\usepackage[margin=8pt,font=small,labelfont=bf]{caption}
\usepackage{simplewick}
\usepackage{changepage}
\usepackage[]{algorithm2e}
\usepackage{booktabs,multirow}
\usepackage[OT2,OT1]{fontenc}
\usepackage{braket}
\usepackage{tikz}
\usepackage{sidecap}
\usepackage{pgfplots}
\usepackage[framemethod=tikz]{mdframed}
\definecolor{shadecolor}{gray}{0.95}
\pgfplotsset{compat=1.10}
\usepgfplotslibrary{fillbetween}
\usetikzlibrary{patterns}

\newcommand*\pFqskip{8mu}
\catcode`,\active
\newcommand*\pFq{\begingroup
  \catcode`\,\active
  \def ,{\mskip\pFqskip\relax}%
  \dopFq
}
\catcode`\,12

\def\dopFq#1#2#3#4#5{%
  \ifnum#1=2\relax
    \ifnum#2=1\relax
      {}_{2}F_{1}\!\left(#3;#4;#5\right)%
    \else
      {}_{#1}F_{#2}\!\left(\genfrac..{0pt}{}{#3}{#4};#5\right)%
    \fi
  \else
    {}_{#1}F_{#2}\!\left(\genfrac..{0pt}{}{#3}{#4};#5\right)%
  \fi
  \endgroup
}

\newcommand*\regpFqskip{8mu}
\catcode`,\active
\newcommand*\regpFq{\begingroup
	\catcode`\,\active
	\def ,{\mskip\regpFqskip\relax}%
	\doregpFq
}
\catcode`\,12
\def\doregpFq#1#2#3#4#5{%
	{}_{#1}\tilde{F}_{#2}\left(\genfrac..{0pt}{}{#3}{#4};#5\right)%
	\endgroup
}

\newcommand{\abs}[1]{\left\lvert#1\right\rvert}

\newcommand{\assign}{:=}

\newcommand{\bb}[2]{ b^{#1}_{#2}}
\newcommand{\hyperF}[4]{{}_2F_1\left(#1,#2;#3;#4\right)}

\newcommand{\RNum}[1]{\uppercase\expandafter{\romannumeral #1\relax}}

\newcommand{\phih}{\hat{\phi}}

\theoremstyle{remark}

\newmuskip\pFqmuskip

\input{topmatter} 

\makeatletter
\def\@fpheader{\ }
\makeatother

\title{
QFT as a set of ODEs
}
\author[a,b]{Manuel Loparco,}
\author[a]{Gr\'{e}goire Mathys,}
\author[a]{Joao Penedones,}
\author[c,d]{Jiaxin Qiao,}
\author[a,e]{Xiang Zhao}
\affiliation[a]{Fields and Strings Laboratory, Institute of Physics, \\ 
École Polytechnique Fédéral de Lausanne (EPFL), \\
Route de la Sorge, CH-1015 Lausanne, Switzerland}
\affiliation[b]{Istituto Nazionale di Fisica Nucleare, Sezione di Torino, and\\ 
Department of Physics, 
University of Turin,\\ 
	Via P. Giuria 1, 10125, Turin, Italy}
\affiliation[c]{Laboratory for Theoretical Fundamental Physics, Institute of Physics, \\ 
École Polytechnique Fédérale de Lausanne (EPFL), \\ 
Route de la Sorge, CH-1015 Lausanne, Switzerland}
\affiliation[d]{Kavli Institute for the Physics and Mathematics of the Universe (WPI), \\ 
The University of Tokyo Institutes for Advanced Study, The University of Tokyo, \\ 
Kashiwa, Chiba 277-8583, Japan}
\affiliation[e]{Institut de Physique Théorique, \\ 
Université Paris-Saclay, CEA, CNRS, \\ 
91191, Gif-sur-Yvette, France}
\vspace{1cm}
\abstract{
Correlation functions of local operators in Quantum Field Theory (QFT) on hyperbolic space can be fully characterized by the set of QFT data $\{\Delta_i, C_{ijk},b_{j}^{\hat{\mathcal{O}}}\}$. These are the scaling dimensions of boundary operators $\Delta_i$, the boundary Operator Product Expansion (OPE) coefficients $C_{ijk}$ and the Boundary Operator Expansion (BOE)  coefficients $b_{j}^{\hat{\mathcal{O}}}$ that characterize how each bulk operator $\hat{\mathcal{O}}$ can be expanded in terms of boundary operators $\OO_j$. 
For simplicity, we focus on two dimensional QFTs and
derive a universal set of first order Ordinary Differential Equations (ODEs) that encode the variation of the QFT data under an infinitesimal change of a bulk relevant coupling.
In principle, our ODEs can be used to follow a Renormalization Group (RG) flow starting from a solvable QFT into a strongly coupled phase and to the flat space limit.
}

\begin{document}
\maketitle

\section{Introduction and main idea}
Quantum Field Theory (QFT) is the central paradigm of modern theoretical physics.
Despite its ubiquity, computing physical observables in strongly coupled QFTs from first principles remains a significant challenge. Most non-perturbative approaches introduce a UV cutoff (like the lattice spacing) and extrapolate the regulated theory
to the continuum limit.
The Conformal Bootstrap is a distinctive approach that works directly in the continuum \cite{Ferrara:1973yt, Polyakov:1974gs, Rattazzi:2008pe, Poland:2018epd}. By requiring conformal invariance, convergent operator product expansions, and unitarity, this method has achieved remarkable precision in determining critical exponents in systems like the 3D Ising model \cite{El-Showk:2014dwa, Simmons-Duffin:2015qma, Chang:2024whx}.
Unfortunately, this method is limited to Conformal Field Theories (CFTs).
The non-perturbative  S-matrix Bootstrap applies analogous consistency conditions to scattering amplitudes in non-conformal theories \cite{Kruczenski:2022lot}, but so far this approach is limited to 2 to 2 scattering amplitudes.\footnote{The notable exception being \cite{Guerrieri:2024ckc}, whose approach only works for massless particles in 1+1 dimensions.}

In this paper, we propose a new non-perturbative approach to QFT directly in the continuum. The main idea is to study QFT on hyperbolic space of radius $R$ and determine evolution equations for physical observables as $R$ is varied.
Our approach was inspired by \cite{Behan:2017mwi}, who studied evolution equations for exactly marginal deformations of 1D CFTs, and \cite{Hollands:2023txn}, who discussed the OPE in general QFTs.

$D$-dimensional hyperbolic space is a maximally symmetric space with isometry group $SO(D,1)$, which is isomorphic to the conformal group in $D-1$ dimensions.
The Lorentzian version of hyperbolic space is Anti-de Sitter (AdS) spacetime and for this reason we will use AdS to abbreviate hyperbolic space.
If one quantizes QFT in AdS using spatial slices of constant global time then the energy spectrum is discrete as expected for a box of finite size $\sim R$. 
In fact, there is a one-to-one map between the QFT states in this quantization and local boundary operators \cite{Paulos:2016fap}. 
Under this map, the energy of the bulk state (in units of the AdS radius $R$) is equal to the scaling dimension of the boundary operator. 
Correlation functions of boundary operators obey all the Conformal Field Theory (CFT) axioms, except  the existence of a conserved stress-tensor on the boundary. Therefore, they are fully determined by the CFT data $\{\Delta_i, C_{ijk} \}$, where $\Delta_i$ stands for the scaling dimensions and $C_{ijk}$ stands for the Operator Product Expansion (OPE) coefficients. 

Using the Boundary Operator Expansion (BOE),
\begin{equation}
    \hat{\mathcal{O}}(z,x)=\sum_i b^{\hat{\mathcal{O}}}_i z^{\Delta_i} \left[\mathcal{O}_i(x) + \dots \right]\, ,
\end{equation}
any bulk operator can be written as a sum of boundary operators (the $\dots$ stand for boundary operators that are conformal descendants).
Therefore, the QFT data $\{\Delta_i, C_{ijk}, b^{\hat{\mathcal{O}}}_i \}$ 
completely determines all correlators
 involving  local bulk and boundary operators.
This QFT data obeys many constraints that generalize the conformal bootstrap equations \cite{Levine:2023ywq, Levine:2024wqn, Meineri:2023mps}. 


Generically, QFTs in AdS form continuous families. 
Given a bulk local relevant scalar operator $\hat{\Phi}$, we can define a one-parameter family of theories writing their  action as
\begin{align}
    S(\lambda+\delta \lambda) = S(\lambda) + \delta\lambda \int_{AdS} dX \,\hat{\Phi}(X)\,.
    \label{eq:bulk action}
\end{align}
We shall work in units where the radius of AdS is $R=1$. This means that $\lambda$ is a dimensionless coupling, i.e. the dimensionful coupling in units of the AdS radius. Since the integral in \eqref{eq:bulk action} preserves the AdS isometries, there is a well-defined set of QFT data $\{\Delta_i,C_{ijk},b^{\hat{\mathcal{O}}}_i\}$ that continuously varies with $\lambda$.
Using perturbation theory once and for all,\footnote{Here, ``once and for all" means that we express the integrand in (\ref{eq:bulk action 2}) in terms of conformal blocks, so that all the theory dependence is encoded in the QFT data. Moreover, our bulk operators obey $\langle \hat{\Phi}\rangle=0$. This can always be achieved by redefining $\hat{\Phi} \to \hat{\Phi} - \langle\hat{\Phi}\rangle$.}
\begin{align}
\langle\dots\rangle_{\lambda+\delta\lambda} 
= \frac{
\langle\dots e^{-\delta\lambda \int_{AdS} dX \,\hat{\Phi}(X)} \rangle_{\lambda}}
{\langle e^{-\delta\lambda \int_{AdS} dX \,\hat{\Phi}(X)}\rangle_{\lambda}}
=\langle\dots\rangle_\lambda -\delta\lambda \int_{AdS} dX \,\langle \hat{\Phi}(X) \dots\rangle_\lambda \,,
    \label{eq:bulk action 2}
\end{align}
we can express the infinitesimal variation of the QFT data in terms of the QFT data itself. 
This  leads to first order Ordinary Differential Equations (ODEs) that determine the evolution of the QFT data with the coupling $\lambda$.

Although the logic is clear and valid for general spacetime dimension, in this paper we
focus on two dimensional QFTs defined on AdS$_2$, so that the conformal boundary is one dimensional. This means that all  boundary operators are scalars and we can avoid the technical difficulties of dealing with spinning operators. 
Then, the flow equations are\footnote{We do not assume parity symmetry, therefore $C_{lij} \neq C_{lji}$ (see section \ref{sec:OPE on the boundary}).}
\begin{align}
    \frac{d\Delta_i}{d\lambda}&=\sum_{l}b^{\hat\Phi}_l C_{iil} \II^{(\alpha_i)}(\Delta_l)
    \label{eq:dimension flow}
    \\
    \frac{db^{\hat{\Phi}}_i}{d\lambda}
    &=\sum_{j,l} b^{\hat\Phi}_l b^{\hat\Phi}_j \frac{C_{l ji}+C_{lij}}{2} \JJ_{\D_i}^{(\a_{ij})}(\D_l,\D_j)
    \label{eq:BOE flow}
    \\
    \spl{
    \frac{dC_{ijk}}{d\lambda}
    &=\sum_{m,l} \left(
    b^{\hat\Phi}_l C_{ilm}C_{mjk} \KK^{(\a_{im})}_{\D_i\D_j\D_k}(\D_l,\D_m)
    +b^{\hat\Phi}_l C_{lim}C_{mjk} \KK^{(\a_{im})}_{\D_i\D_k\D_j}(\D_l,\D_m)
    \right)
    \\
    &\quad\qquad+\quad (ijk)\rightarrow(jki)\quad+\quad(ijk)\rightarrow(kij)\, ,
    \label{eq:OPE flow}
    }
\end{align}
where Latin indices run over the set of primary boundary operators.
The functions $\II, \JJ, \KK$ are kinematical (\emph{i.e.} theory independent). They
are regulated integrals of (local) conformal blocks, and their explicit expressions are given in section \ref{sec:integrals of blocks}. The $\alpha$'s are auxiliary parameters introduced to improve the convergence of the sums. 
For simplicity, we assumed 
that $\hat\F$ is the only bulk relevant operator. We did not present here the evolution equations for the BOE coefficients $b^{\hat{\mathcal{O}}}_i$ of other bulk operators. We shall discuss those in section \ref{subsec:renormbulk}.

\begin{figure}[t]
    \centering 
    \includegraphics[width=15cm]{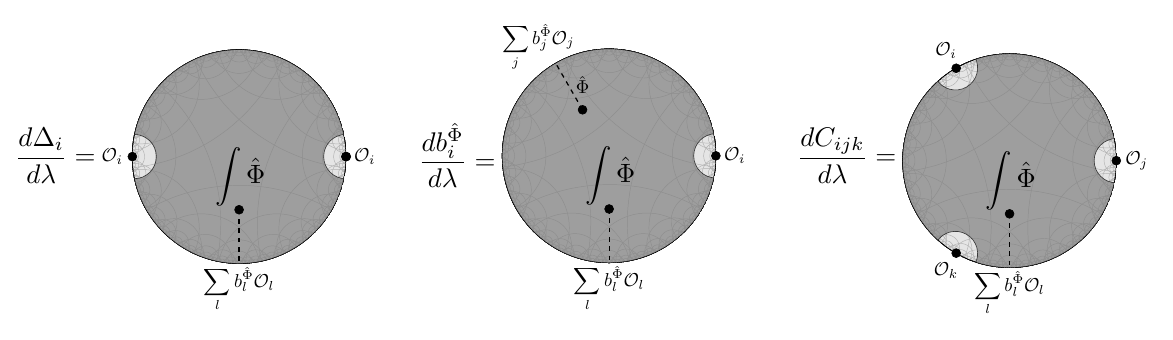}
    \caption{Pictorial derivation of the flow equations for QFT data.}
    \label{fig: flow}
\end{figure}

The careful derivation of the flow equations will be presented in section \ref{sec:derivation of flow eqn}. However, it is easy to understand their structure from figure \ref{fig: flow}.
For example, to derive the first equation, we consider the infinitesimal variation (when $\lambda\to \lambda +\delta \lambda$) of the boundary two-point function $\langle \mathcal{O}_i(\tau) \mathcal{O}_i(0)\rangle_\lambda=|\tau|^{-2\Delta_i(\lambda)}$. This is expressed in terms of the bulk-boundary-boundary (B$\partial \partial$) three-point function $\langle \hat\F \mathcal{O}_i \mathcal{O}_i\rangle_\lambda$, which can be expanded as $\sum_l b^{\hat\F}_l C_{iil}  \times [{\rm kinematical\ block}]_l$ using the BOE and the boundary OPE. Similar reasoning applied to figure \ref{fig: flow}, explains the QFT data dependence of the other flow equations.

The flow equations (\ref{eq:dimension flow}--\ref{eq:OPE flow})  are non-perturbative and make sense in the continuum. 
In principle, they can be used to
follow an RG flow starting from a solvable theory (\emph{i.e.} with known QFT data)
into a strongly coupled regime. 
In practice, the QFT data is infinite and one will have to devise a numerical truncation scheme to solve the flow equations approximately. We discuss some preliminary ideas in section \ref{discussion numerics}.

\paragraph{Summary of assumptions.}
Throughout this work, we consider a local quantum field theory in AdS$_2$ with conformal boundary conditions. We assume the standard convergence properties of the boundary CFT OPE, as well as convergence of the boundary operator expansion for bulk operators in AdS. We further assume that the spectrum of scaling dimensions is discrete and non-degenerate. In the presence of additional symmetries, this assumption may be refined by decomposing the spectrum into symmetry sectors. We also assume the absence of bulk UV divergences in the flow equations \eqref{eq:dimension flow}--\eqref{eq:OPE flow}; this assumption is relaxed in the discussion section. Finally, we assume unitarity, which is used implicitly in establishing the convergence of the local block expansions.


\paragraph{Structure of the paper} In section \ref{sec:BasicsofQFT}, we setup notation and conventions, review basics of one-dimensional CFTs and derive the necessary conformal blocks. In section \ref{sec:derivation of flow eqn}, we carefully derive the three flow equations taking care of various divergences. We also explain how we can interchange sums over primary and descendant operators with integrals over AdS$_2$ in practice. Section \ref{sec:test of flow eqn} is devoted to free massive scalar theory checks of each flow equation in different cases. In section \ref{sec: discussion}, we comment on several open research directions and subtleties that we encountered. Several appendices give further technical details. 

\section{\texorpdfstring{Basics of QFT in AdS$_2$}{Basics of QFT in AdS2}}\label{sec:BasicsofQFT}
In this section we set up the notation and conventions used in this paper. We first briefly review Euclidean AdS$_2$ geometry, the OPE in one-dimensional CFT and the boundary operator expansion (BOE) for QFT in AdS$_2$. Then we derive the conformal blocks that will be used extensively in the rest of this paper. 

Throughout this work, we will use several standard notations. In particular, we define $\Delta_{ij}\equiv \D_i-\D_j$ together with $\D_{ijk}\equiv \D_i+\D_j-\D_k$. We also use $x_{ij}\equiv x_i-x_j$ and $\tau_{ij}\equiv \tau_i-\tau_j$. We will use $\hat\OO$ for bulk operators and $\OO$ for boundary operators.

\subsection{Geometry}
Two-dimensional hyperbolic space, for brevity AdS$_2$, can be defined through its embedding in a three-dimensional Minkowski space. We will denote points in the embedding space that belong to the hyperboloid as $X^A=(X^0,X^1,X^2)\in\mathbb{R}^{1,2}$, satisfying\footnote{As specified in the introduction, in most of this paper we set $R=1$.}. 
\begin{equation}
    -(X^0)^2+(X^1)^2+(X^2)^2=-R^2\,, \qquad X^0>0\,,
    \label{eq:hyperboloid}
\end{equation}
where the restriction $X^0>0$ fixes us on the upper branch of the hyperboloid.

Points on the conformal boundary are embedding space lightrays, denoted as $P^A\in\mathbb{R}^{1,2}$, which  satisfy 
\begin{equation}
    -(P^0)^2+(P^1)^2+(P^2)^2=0\,, \qquad P^0>0\,,
\end{equation}
with the identification $P^A\sim\lambda P^A$.
The usefulness of the embedding space formalism stands in the fact that the group of isometries preserving one branch of the hyperboloid, $SO^+(1,2)$ or $PSL(2,\mathbb{R})$, is realized linearly. Consequently, all invariant cross-ratios that we will define throughout this work can be constructed through 
contractions of vectors $P^A$ and $X^A$ with the embedding metric $\eta_{AB}=\text{diag}(-1,1,1)$ or the Levi-Civita tensor $\epsilon^{ABC}$.


\subsubsection{Poincaré half plane}
It will be useful to introduce the local system of coordinates known as the Poincar\'e patch
\begin{equation}
\begin{aligned}
    X^0&=\frac{1+z^2+\tau^2}{2z}\,,&  \qquad X^1&=\frac{\tau}{z}\,,&\qquad X^2&=\frac{1-z^2-\tau^2}{2z}\,,\\
    P^0&=\frac{1+\tau^2}{2}\,,\qquad & P^1&=\tau\,,&\qquad  P^2&=\frac{1-\tau^2}{2}\,.
\end{aligned}
\end{equation}
The metric of AdS$_2$ in these coordinates reads
\begin{equation}
	\begin{split}
	 ds^2=\frac{d\tau^2+dz^2}{z^2}\,,\qquad\qquad  z>0\,.\label{eq:AdSPoicare}
	\end{split}
\end{equation}
The isometry group of AdS$_2$ is $O^+(1,2)$ which contains two disconnected pieces
\begin{equation}
	\begin{split}
		O^+(1,2)=SO^+(1,2)\sqcup \mathcal{P}SO^+(1,2),
	\end{split}
\end{equation}
where $\mathcal{P}$ is the parity transformation
\begin{equation}\label{def:parity}
	\begin{split}
		\mathcal{P}:\ (X^0,X^1,X^2) \mapsto (X^0,-X^1,X^2)\quad
		\text{or equivalently}\quad (\t,z) \mapsto  (-\t, z)\,.
	\end{split}
\end{equation}
The connected subgroup $SO^+(1,2)$ is isomorphic to $PSL(2;\mathbb{R})$, which is realized by
\begin{equation}
	\begin{split}
		w=\tau+i z
		\quad\mapsto\quad w'=\frac{aw+b}{cw+d},\qquad\text{with}\qquad \left(\begin{matrix}
			a & b \\
			c & d \\
		\end{matrix}\right)\in SL(2;\mathbb{R}),
	\end{split}\label{eq:Sl2action}
\end{equation}
such that $ad-bc=1$. Notice that the coordinate $z'$ remains positive after an $SL(2;\mathbb{R})$ transformation
\begin{equation}
	z' = \frac{(a d - b c) z}{c^2 z^2+(c \tau +d)^2} = \frac{z}{c^2 z^2+(c \tau +d)^2}\, .
\end{equation}


\subsubsection{Poincaré disk}

AdS$_{2}$ can be nicely represented as a unit disk. Geometrically this can be achieved by doing a stereographic projection from the point $X^A=(-1,0,0)$. More precisely, we introduce coordinates
\begin{equation}
    x=\frac{X^1}{1+X^0}\,,\qquad
       y=\frac{X^2}{1+X^0}\,.
\end{equation}
For points obeying \eqref{eq:hyperboloid}, we have $x^2+y^2<1$. The unit circle $x^2+y^2=1$ represents the conformal boundary of AdS$_2$.
The metric is then given by
\begin{equation}
    ds^2=\frac{4(dx^2+dy^2)}{(1-x^2-y^2)^2}\,.
\end{equation}
Geodesics are given by the intersection of the hyperboloid \eqref{eq:hyperboloid} with timelike planes through the origin, which can be defined by
$X^0=r(X^1 \cos\theta + X^2 \sin \theta)$ with $r>1$. These are circles that intersect the boundary of the Poincaré disk at $90^{\circ}$ angles. More precisely,
\begin{equation}
    (x-r \cos\theta)^2 +(y-r \sin \theta)^2=r^2-1\,.
\end{equation}

\subsection{Operator expansions}
\subsubsection{OPE on the boundary}
\label{sec:OPE on the boundary}
For QFTs in AdS$_2$ that preserve the AdS isometries, the boundary is a 1D conformal theory. We normalize the boundary operators such that their two-point functions are given by
\begin{equation}\label{def:bdbdtwopt}
	\begin{split}
		\braket{\mathcal{O}_i(\tau_1)\mathcal{O}_j(\tau_2)}=\frac{\delta_{ij}}{\abs{\tau_1-\tau_2}^{2\Delta_i}}
	\end{split}\,.
\end{equation}
The three-point functions are given by 
\begin{equation}\label{def:bdbdbdthreept}
\braket{\mathcal{O}_i(\tau_1)\mathcal{O}_j(\tau_2)\mathcal{O}_k(\tau_3) }=\begin{cases}
	\frac{C_{ijk}}{\abs{\tau_1-\tau_2}^{\Delta_{ijk}}\abs{\tau_1-\tau_3}^{\Delta_{ikj}}\abs{\tau_2-\tau_3}^{\Delta_{jki}}}, & \quad [ijk], \\
	\frac{C_{jik}}{\abs{\tau_1-\tau_2}^{\Delta_{ijk}}\abs{\tau_1-\tau_3}^{\Delta_{ikj}}\abs{\tau_2-\tau_3}^{\Delta_{jki}}}, & \quad [jik]. \\
\end{cases}
\end{equation}
with $C_{ijk}$ the three-point function coefficients and where we defined $\Delta_{ijk}\equiv \Delta_i + \Delta_j-\Delta_k$. $[ijk]$ is the cyclic ordering of the operators with $\tau_1<\tau_2<\tau_3$, or $\tau_2<\tau_3<\tau_1$, or $\tau_3<\tau_1<\tau_2$. The dependence of the three-point function on cyclic orderings is a special difference between 1D and higher D.
The two  cyclic orderings are distinguished by the $SO^+(1,2)$ invariant
\begin{equation}
    \frac{2 \epsilon(P_1,P_2,P_3) }{\sqrt{(-2P_1\cdot P_2)(-2P_1\cdot P_3)(-2P_2\cdot P_3) }} = \pm 1\,,
\end{equation}
where $\epsilon$ stands for the three-dimensional Levi-Civita tensor
\be
\e(P_1, P_2, P_3)\assign
\e_{ABC}P_1^A P_2^B P_3^C\,,
\qquad
\e_{012}=1\,,
\ee
with $\lbrace A,B,C\rbrace \in \lbrace 0,1,2\rbrace$. The boundary operators obey the usual operator product expansion (OPE). For $\tau_1<\tau_2$, the OPE of a pair of neighboring operators takes the form
\begin{equation}\label{ope1d on tau2}
	\mathcal{O}_i(\tau_1)\mathcal{O}_j(\tau_2)
	=\sum_{k}\frac{C_{ijk}}{\abs{\tau_{12}}^{\Delta_{ijk}}}
	\sum_{n=0}^{\infty}\frac{(\Delta_{kij})_n}{n!\,(2\Delta_k)_n}\,
	(\tau_{12})^n \,\partial^n\mathcal{O}_k(\tau_2)\,,
\end{equation}
where $(a)_n$ denotes the Pochhammer symbol.

An equivalent way to represent the OPE is through an integral kernel\footnote{Similar integral representations were discussed in \cite{Czech:2016xec} for timelike-separated boundary operators. In general dimensions, the integral is taken over the diamond region $
\{x_3\in\mathbb{R}^{1,d-1}\mid x_1<x_3<x_2\}$, where “$<$” denotes causal ordering in Minkowski space. In one dimension, the diamond region reduces to an interval, and Wick rotation to Euclidean signature then leads to \eqref{OPE:integralrepr}. See appendix \ref{app:OPE and BOE blocks} for more details.}
\begin{equation}\label{OPE:integralrepr}
	\begin{split}
		\mathcal{O}_i(\tau_1)\mathcal{O}_j(\tau_2)
		&=\sum_{k}C_{ijk}\int_{\tau_1}^{\tau_2}d\tau_3\,
		\mathcal{B}_{\D_i \D_j \D_k}(\t_1,\t_2,\t_3)\,
		\mathcal{O}_k(\tau_3)\,, \\
	\end{split}
\end{equation}
where the kernel $ \mathcal{B}_{\D_i \D_j \D_k}(\t_1,\t_2,\t_3)$ is essentially a three-point function between two boundary operators and the shadow transform of a third boundary operator, 
\begin{align}
\mathcal{B}_{\D_i \D_j \D_k}(\t_1,\t_2,\t_3)
		 &:=\frac{n_{ijk}}
	 {\abs{\tau_{12}}^{\Delta_i+\Delta_j+\Delta_k-1}\,
 	\abs{\tau_{13}}^{1-\Delta_{kji}}\,
 	\abs{\tau_{23}}^{1-\Delta_{kij}}}\,, \label{eq:Bkernel}\\
 n_{ijk}
 &=\frac{\Gamma(2\Delta_k)}{\Gamma(\Delta_{kij})\,\Gamma(\Delta_{kji})}\,,
\end{align}
where the normalization is fixed in appendix \ref{app:OPE and BOE blocks}.

One may perform the OPE between one of the three operator pairs: $\mathcal{O}_i$ and $\mathcal{O}_j$, $\mathcal{O}_j$ and $\mathcal{O}_k$, or $\mathcal{O}_k$ and $\mathcal{O}_i$, when evaluating the three-point function \eqref{def:bdbdbdthreept}. Consistency of the result implies that the OPE coefficients are invariant under cyclic permutations:
\begin{equation}
	C_{ijk}=C_{jki}=C_{kij}\,.
\end{equation}
This is natural, since a 1D CFT can be regarded as a theory defined on the circle.

If parity invariance is not assumed, in general there is no transformation property of $C_{ijk}$ under odd permutation of the indices, because there is no continuous way to bring one operator across another on a line. Therefore, specializing to three-point functions one finds there are two independent OPE coefficients $C_{ijk}$ and $C_{jik}$.\footnote{See \cite[sec.3.1]{Behan:2017mwi} for a more detailed discussion.} Imposing parity invariance relates the two OPE coefficients through \cite[App.K.1]{Homrich:2019cbt}
\begin{equation}
    C_{jik} = (-1)^{\s_i + \s_j + \s_k}C_{ijk}\, ,\qquad (\text{parity invariance}),
\end{equation}
where $\s_i$ denotes the parity of operator $\OO_i$, which is 0 for even parity and 1 for odd parity. In the following we will not assume parity invariance. 

In a unitary theory in equal-time quantization, we can choose an operator basis such that  $[\mathcal{O}_i(\tau)]^\dagger=\mathcal{O}_i(-\tau)$. Then the odd permutation of indices conjugates the OPE coefficient:
\begin{equation}
	\begin{split}
		C_{ijk}=C_{jik}^*\, ,\qquad (\text{unitarity})\,.
	\end{split}
\end{equation}

\subsubsection{Boundary operator expansion of bulk operators}

In general, a bulk operator $\hat{\mathcal{O}}(\tau,z)$ can be expanded in terms of boundary operators $\mathcal{O}_i(\tau)$ via the so-called \emph{boundary operator expansion} (BOE) \cite{Paulos:2016fap,Levine:2023ywq}:
\begin{equation}\label{eq:BOEdef}
	\hat{\mathcal{O}}(\tau,z)
	=\sum\limits_{i}\bb{\hat{\mathcal{O}}}{i}\sum\limits_{n=0}^{\infty}
	a_n(\Delta_i;\tau,z,\tau_*)\,\partial_{\tau_*}^n\mathcal{O}_i(\tau_*)\,.
\end{equation}
Here the index $i$ runs over boundary primaries $\OO_i$, while the sum over $n$ incorporates the contributions of their descendants. The reference point $\tau_*$ is arbitrary, provided that no other bulk or boundary operator lies closer to $(\tau_*,0)$ than $(\tau,z)$, i.e.\ inside the disk
\begin{equation}
	\Bigl\{(\tau',z')\;\big|\;(\tau'-\tau_*)^2+z'^2\leqslant(\tau-\tau_*)^2+z^2\Bigr\}\,.
\end{equation}
Conventionally, one normalizes $a_0=z^{\Delta_i}$, and $\bb{\hat{\mathcal{O}}}{i}$ is referred to as the BOE coefficient. The remaining $a_n$ are then completely determined by the AdS$_2$ isometries. In particular, choosing $\tau_*=\tau$ yields
\begin{equation}\label{eq:BOE coefficient}
	a_{2n+1}=0,\qquad\quad 
	a_{2n}(\Delta_i;\tau,z,\tau)
	=\frac{(-1)^n}{n!\,4^n\,(\Delta_i+\tfrac{1}{2})_{n}}\,z^{\Delta_i+2n}\,,
\end{equation}
where we used translation invariance to conclude that $a_{n}(\Delta_i;\tau,z,\tau)$ does not depend on $\tau$, dilatation invariance (or dimensional analysis) to write $a_{n}(\Delta_i;\tau,z,\tau)=\alpha_n(\Delta_i)z^{\Delta_i+n}$, and invariance under special conformal transformations to recursively determine the remaining coefficients $\alpha_n$.
Thus the BOE takes the universal form
\begin{equation}\label{eq:BOE}
	\hat{\mathcal{O}}(\tau,z)
	=\sum\limits_{i}\bb{\hat{\mathcal{O}}}{i}\sum\limits_{n=0}^{\infty}
	\frac{(-1)^n}{n!\,4^n\,(\Delta_i+\tfrac{1}{2})_{n}}\,z^{\Delta_i+2n}\,
	\partial_\tau^{2n}\mathcal{O}_i(\tau)\,.
\end{equation}
However, as mentioned above, this expansion is valid only when no other operators stay closer to $(\tau,0)$ than $(\tau,z)$. For generic $\tau_*$, the coefficients $a_n$ are obtained by rearranging the Taylor expansion of \eqref{eq:BOE}:
\begin{equation}
	\begin{split}
		\hat{\mathcal{O}}(\tau,z)
		=\sum\limits_{i}\bb{\hat{\mathcal{O}}}{i}\sum\limits_{n=0}^{\infty}\sum_{m=0}^{\infty}\frac{(-1)^n}{4^n\,n! (\Delta_i+\tfrac{1}{2})_{n}}z^{\Delta_i+2n}\frac{(\tau-\tau_*)^m}{m!}\partial_{\tau_*}^{2n+m}
		\mathcal{O}_i(\tau_*)\,.
        \label{eq:BOEtaustar}
	\end{split}
\end{equation}
Note that this expansion can be rewritten with the help of the differential operator $\mathcal{D}$, which is defined as 
\begin{equation}
\mathcal{D}\left(\Delta_i;\tau,z,\tau'\right)\equiv\sum\limits_{n=0}^{\infty}\sum_{m=0}^{\infty}\frac{(-1)^n}{4^n\,n! (\Delta_i+\tfrac{1}{2})_{n}}z^{\Delta_i+2n}\frac{(\tau-\tau')^m}{m!}\partial_{\tau'}^{2n+m}\,.
\label{eq:Doperator}
    \end{equation}
    such that \eqref{eq:BOEtaustar} can be conveniently rewritten as 
    \begin{equation}
    \label{BOE:general}
        \hat{\mathcal{O}}(\tau,z) = \sum_i\bb{\hat{\mathcal{O}}}{i}\mathcal{D}\left(\Delta_i;\tau,z,\tau_*\right)\mathcal{O}_i(\tau_*)\, .
    \end{equation}
Also, the general form of the coefficients $a_n$ appearing in \eqref{eq:BOEdef} is
\begin{equation}
	\begin{split}
		a_n(\Delta_i;\tau,z,\tau_*)=\sum\limits_{k=0}^{[n/2]}\frac{(-1)^k\,z^{\Delta_i+2k}\,(\tau-\tau_*)^{n-2k}}{4^k\,k!\,(n-2k)!\, (\Delta_i+\tfrac{1}{2})_{k}}\, .
	\end{split}
\end{equation}
The operator-dependent content of the BOE is entirely captured by the coefficients $\bb{\hat{\OO}}{i}$ and by the spectrum of boundary operators $\OO_i(\tau)$ included in the sum. By contrast, the descendant coefficients are universal.

The coefficients can be checked, or determined, by matching to the bulk--boundary two-point function, which is fixed by the AdS$_2$ isometries up to a single normalization:\footnote{An alternative way to obtain \eqref{eq:BOE} is to start from \eqref{def:bkbdtwopt} and expand $\braket{\hat{\mathcal{O}}(\tau_1,z_1)\mathcal{O}_i(\tau_2)}$ as a power series in $z_1$. Matching the coefficients order by order in $z_1$ with the BOE then yields \eqref{eq:BOE}.}
\begin{equation}\label{def:bkbdtwopt}
	\braket{\hat{\mathcal{O}}(\tau_1,z_1)\mathcal{O}_i(\tau_2)}
	=\bb{\hat{\mathcal{O}}}{i}\left(\frac{z_1}{(\tau_1-\tau_2)^2+z_1^2}\right)^{\Delta_i}\,.
\end{equation}

\subsection{Conformal blocks\label{subsec:Confblock}}
There are four types of conformal blocks to be considered in this paper. Their corresponding correlation functions and notations are summarized as follows: 
\al{
\renewcommand{\arraystretch}{1.4} 
\begin{array}{l c||c |c|c}
    \multicolumn{2}{c||}{\text{Correlation function\hspace{0.1cm}}\rule{0pt}{2.5ex}} & \hspace{0.1cm}
   \text{Conformal block} \hspace{0.1cm}& \hspace{0.1cm}
    \text{Definition} \hspace{0.1cm}& \hspace{0.1cm}
    \text{Cross ratios} \hspace{0.1cm}\\ \hline
    BB & \langle \hat\OO_1 \hat\OO_2 \rangle 
        & g_\Delta(\eta) 
        & \eqref{def:bkbk block} 
        & \eqref{def:eta} \\
    B\del\del & \langle \hat\OO \OO_i \OO_j \rangle 
        & G_{\Delta_l}^{\Delta_i \Delta_j}(\chi) 
        & \eqref{def:bkbdbd block} 
        & \eqref{def:chi} \\
    BB\del & \langle \hat\OO_1 \hat\OO_2 \OO_i \rangle
        & R_{\Delta_l \Delta_j}^{\Delta_i}(\upsilon,\zeta) 
        & \eqref{def:bkbkbd block} 
        & \eqref{def:upsilon and zeta} \\
    B\del\del\del & \langle \hat\OO \OO_i \OO_j \OO_k \rangle
        & K_{\Delta_l \Delta_m}^{\Delta_i \Delta_j \Delta_k} (\mu,\omega) 
        & \eqref{def:bkbdbdbd_sblock_normal} 
        & \eqref{def:mu and omega}
\end{array}
\label{tab:conformal blocks}
} 
where the scaling dimensions of external boundary operators appear as superscripts and those of the exchanged operators appear as subscripts. We have also used the abbreviation ``$B$'' and ``$\del$'' to denote ``bulk'' and ``boundary'', respectively.

As usual, these conformal blocks can be computed by applying combinations of BOEs and (boundary) OPEs to the corresponding generic correlation functions, and separating out the theory-independent part. This gives series expansions for conformal blocks, and for simple cases the series can be resummed into a closed form. There is, however, another way to write an integral representation that repackages the sum over descendants by using the so-called ``OPE blocks'' \cite{Czech:2016xec}. It turns out that these integral representations of conformal blocks are more suitable for computing integrals of the blocks in AdS space, especially for the $B\del\del\del$ case. In the following we discuss these conformal blocks in order.

\subsubsection{\texorpdfstring{Bulk-bulk ($BB$) conformal block}{Bulk-bulk (BB) conformal block}}
To compute the conformal block for the bulk-bulk two-point function it is convenient to use the $SL(2;\mathbb{R})$ transformation that sends the bulk points to the configuration\footnote{Note that the configuration for $\h>1$ can be reached by the inversion transformation $\h\to1/\h$.} 
\begin{equation}
(\t_1,z_1)=(0,\eta)\, ,\qquad 
(\t_2,z_2)=(0,1)\, ,\qquad
0<\eta<1\,.\label{eq:BulkBulkConfig}
\end{equation}
Here $\h$ is an $SL(2;\mathbb{R})$ invariant cross ratio, which can be easily seen by considering the scalar product in the embedding space
\be
-2 X_1 \cdot X_2 = \frac{z_1^2 + z_2^2 + \t_{12}^2}{z_1 z_2}\equiv\frac{1+\h^2}{\h}\,,
\label{def:eta}
\ee
where $\t_{ij}\equiv\t_i-\t_j$. For $0<\h<1$ we can write
\begin{equation}
\eta = -X_1\cdot X_2 -\sqrt{(X_1\cdot X_2)^2-1}\, .
\end{equation}

To obtain the conformal block, we use the BOE \eqref{eq:BOE} once and then the bulk-boundary two-point function \eqref{def:bkbdtwopt} and explicitly resum the outcome. In the configuration \eqref{eq:BulkBulkConfig}, this yields (see appendix \ref{app:BKBK block})
\begin{equation}
\braket{\hat{\mathcal{O}}_1(\tau_1,z_1)\hat{\mathcal{O}}_2(\tau_2,z_2)}=
\braket{\hat{\mathcal{O}}_1(0,\eta)\hat{\mathcal{O}}_2(0,1)}
=\sum\limits_{i}\bb{\hat{\mathcal{O}}_1}{i}\,\bb{\hat{\mathcal{O}}_2}{i}\ g_{\Delta_i}(\eta)\, .
\end{equation}
with the conformal block 
\begin{equation}
g_{\Delta}(\eta)=\eta^\Delta\,\hyperF{\Delta}{\frac{1}{2}}{\Delta+\frac{1}{2}}{\eta^2}\, .
\label{def:bkbk block}
\end{equation}

\subsubsection{\texorpdfstring{Bulk-boundary-boundary ($B\del\del$) conformal block}{Bulk-boundary-boundary (Bbb) conformal block}}\label{subsec:bkbdbdCB}

Using $SO^+(1,2)$ invariance, the bulk-boundary-boundary correlator $\braket{\hat{\mathcal{O}}(\tau,z)\mathcal{O}_i(\tau_1)\mathcal{O}_j(\tau_2)}$ can be fixed to
\begin{equation}\label{bkbdbd:generalform}
	\braket{\hat{\mathcal{O}}(\tau,z)\mathcal{O}_i(\tau_1)\mathcal{O}_j(\tau_2)}
	=\frac{1}{\abs{\tau_1-\tau_2}^{\Delta_i+\Delta_j}}
	\left(\frac{(\tau-\tau_1)^2+z^2}{(\tau-\tau_2)^2+z^2}\right)^{\tfrac{\Delta_{ji}}{2}}
	\mathcal{G}^{\hat{\mathcal{O}}}_{ij}(\rho)\,,
\end{equation}
where $\Delta_{ji}\equiv\Delta_j-\Delta_i$, and $\rho$ is an $SO^+(1,2)$-invariant defined by
\begin{equation}\label{def:rho}
	\begin{split}
		\rho&:=\frac{2\,\epsilon(X,P_1,P_2)}{\sqrt{(-2X\cdot P_1)(-2X\cdot P_2)(-2P_1\cdot P_2)}} \\
		&=\text{sgn}(\tau_1-\tau_2)\frac{(\tau-\tau_1)(\tau-\tau_2)+z^2}{\sqrt{\left((\tau-\tau_1)^2+z^2\right)\left((\tau-\tau_2)^2+z^2\right)}}\,.
	\end{split}
\end{equation}
Note that for $(\tau_1,\tau_2)=(0,+\infty)$, we have the simplification $\rho=\tfrac{\tau}{\sqrt{\tau^2+z^2}}$.

\begin{figure}
\centering
	\includegraphics[width=0.48\textwidth]{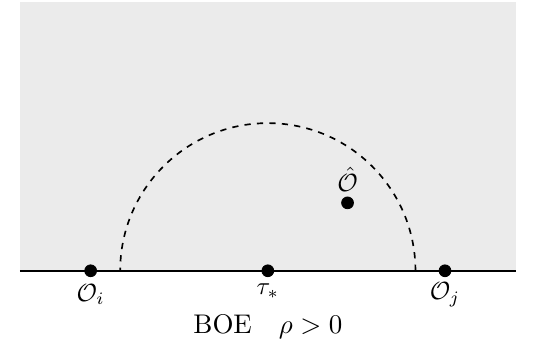}
    \includegraphics[width=0.48\textwidth]{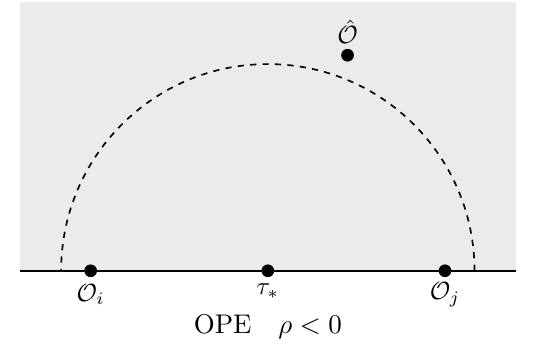}
	\caption{In the radial quantization picture, the Poincar\'e half plane is foliated with semicircles. In configurations like the one on the left, corresponding to $\rho>0$, we must first use the BOE of $\hat{\mathcal{O}}$ and then the OPE with the two boundary operators. In cases like the picture on the right, with $\rho<0$, we must first use the OPE of $\mathcal{O}_i$ and $\mathcal{O}_j$ and then the BOE of $\hat{\mathcal{O}}$.}
        






%
%


	\label{fig:bkbdbd_OPE_vs_BOE}
\end{figure}

Using the BOE of $\hat{\mathcal{O}}$, or equivalently the OPE of $\mathcal{O}_i\mathcal{O}_j$, the function $\mathcal{G}^{\hat{\mathcal{O}}}_{ij}$ can be decomposed into conformal blocks $G_{\D_l}^{\D_i\D_j}$.  
The appropriate expansion depends on the geometric configuration.  
The guiding principle is the radial quantization picture: one should identify a foliation of AdS$_2$ whose time direction corresponds to a Killing vector of the space.  
Practically, this means finding a boundary point $(\tau_*,0)$ such that a semicircle centered at $(\tau_*,0)$ separates the bulk point $(\tau,z)$ from both boundary points (see Figure~\ref{fig:bkbdbd_OPE_vs_BOE}). 
After expanding $\hat{\mathcal{O}}(\tau,z)$ into boundary operators $\mathcal{O}_l(\tau_*,0)$, it becomes evident whether to use $C_{ijl}$ or $C_{jil}$.

It turns out that when $\rho<0$, the relevant OPE is $\mathcal{O}_i\mathcal{O}_j$, with coefficient $C_{ijl}$;  
while for $\rho>0$, it is $\mathcal{O}_j\mathcal{O}_i$, with coefficient $C_{jil}$.  
The resulting decomposition is
\begin{equation}\label{bkbdbd:exp}
	\mathcal{G}^{\hat{\mathcal{O}}}_{ij}(\rho)=
	\begin{cases}
		\sum\limits_{l}\bb{\hat{\mathcal{O}}}{l}C_{ijl}
		G_{\D_l}^{\D_i\D_j}(\chi)\,, & \quad \text{for }\rho<0\,, \\[6pt]
		\sum\limits_{l}\bb{\hat{\mathcal{O}}}{l}C_{jil}
		G_{\D_l}^{\D_i\D_j}(\chi)\,, & \quad \text{for }\rho>0\,,
	\end{cases}
\end{equation}
where $\chi$ is a cross-ratio variable defined by
\begin{equation}\label{def:chi}
	\chi
	:=1-\rho^2 = \frac{-2 P_1\cdot P_2}{(-2 P_1\cdot X)(-2 P_2\cdot X)}
	=\frac{(\tau_{1}-\tau_{2})^2z^2}
	{[(\tau-\tau_1)^2+z^2][(\tau-\tau_2)^2+z^2]}\,.
\end{equation}
The conformal block $G_{\D_l}^{\D_i\D_j}(\chi)$ in \eqref{bkbdbd:exp} is given by
\begin{equation}\label{def:bkbdbd block}
	G_{\D_l}^{\D_i\D_j}(\chi)
	=\chi^{\tfrac{\Delta_l}{2}}
	\hyperF{\tfrac{\Delta_{lij}}{2}}{\tfrac{\Delta_{lji}}{2}}
	{\Delta_l+\tfrac{1}{2}}{\chi}\,,
\end{equation}
where $\D_{ijk}\equiv\D_i+\D_j-\D_k$.  
It can be obtained either from the OPE/BOE construction \cite{Lauria:2020emq} or by solving the quadratic Casimir equation \cite{Meineri:2023mps}. See appendix~\ref{app:BKBDBD block} for the derivation.

Before turning to other correlators, let us comment on the boundary where $\rho$ changes  sign (the location of $\rho=0$ is equivalent to $\chi=1$). In this case, the bulk point and the two boundary points lie on the same geodesic in AdS$_2$, which is a semicircle centered at the boundary point $(\tfrac{\tau_1+\tau_2}{2},0)$. Although nothing singular happens physically at these configurations, the conformal block expansion \eqref{bkbdbd:exp} fails to converge there. Indeed, when expansion \eqref{bkbdbd:exp} is organized in increasing order of the exchanged scaling dimensions $\Delta_l$, it converges exponentially fast for $\chi<1$ but diverges as a power law at $\chi=1$.

\subsubsection{\texorpdfstring{Bulk-bulk-boundary ($BB\del$) conformal block}{Bulk-bulk-boundary (BBb) conformal block}}\label{subsec:bkbkbdCB}

\begin{figure}[ht]
\centering
\begin{minipage}{0.55\textwidth}
  	\includegraphics[]{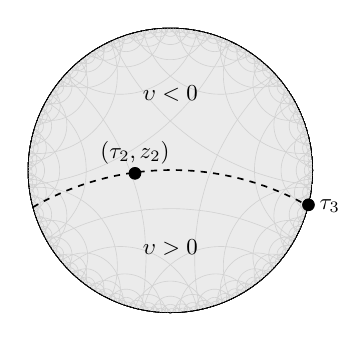}
\end{minipage}\hfill
\begin{minipage}{0.45\textwidth}
  \caption{Sign of $\upsilon$ as a function of $(\tau_1,z_1)$.  
		The dashed line denotes the geodesic connecting the bulk point $(\tau_2,z_2)$ to the boundary point $(\tau_3,0)$.  
		\label{fig:v}}
\end{minipage}
\end{figure}

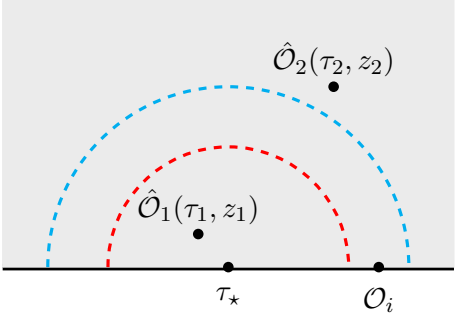
\begin{figure}
	\centering
    \begin{minipage}{0.55\textwidth}
	    \begin{tikzpicture}[scale=2]

\definecolor{light-gray}{gray}{0.92}
\draw[line width=2.5] (0,0) -- (3,0);
\filldraw[light-gray](0,0)--(3,0)--(3,1.8)--(0,1.8)--(0,0);
\fill (1.5,0) circle (1pt) node[below=4pt] {$\tau_\star$};
\fill (2.5,0) circle (1pt) node[below=4pt] {$\mathcal{O}_i$};

\fill (1.3,0.22) circle (1pt) node[above] {$\hat{\mathcal{O}}_1(\tau_1,z_1)$};
\fill (2.2,1.2) circle (1pt) node[above] {$\hat{\mathcal{O}}_2(\tau_2,z_2)$};
\draw[dashed, red,line width=1.2] (0.7,0) arc (180:0:0.8);
\draw[dashed, cyan,line width=1.2] (0.3,0) arc (180:0:1.2);


\end{tikzpicture}
\end{minipage}\hfill
\begin{minipage}{0.45\textwidth}
\caption{In this configuration, we can expand $\hat{\mathcal{O}}_1$ into $\mathcal{O}_l$ and $\hat{\mathcal{O}}_2$ into $\mathcal{O}_j$ via the BOE \eqref{eq:BOE}, with $\mathcal{O}_l$  inside the red semi-circle and $\mathcal{O}_j$  outside the blue semi-circle. Thus, we get the cyclic ordering corresponding to  $C_{jli}$. \label{fig:bkbkbd_OPE_vs_BOE}}
\end{minipage}
\end{figure}

For bulk--bulk--boundary correlators, $SO^+(1,2)$ invariance fixes the general form
\begin{equation}
	\braket{\hat{\mathcal{O}}_{1}(\t_1,z_1)\hat{\mathcal{O}}_{2}(\t_2,z_2)\mathcal{O}_i(\t_3)}
	=
	\left(\frac{z_2}{(\tau_2-\tau_3)^2+z_2^2}\right)^{\Delta_i}
	\mathcal{G}^{\hat{\mathcal{O}}_1\hat{\mathcal{O}}_2}_{i}(\upsilon,\zeta)\,,
	\label{bkbkbd:generalform}
\end{equation}
where the independent $SO^+(1,2)$ invariants are
\begin{equation}\label{def:upsilon and zeta}
	\begin{split}
		\upsilon&:=-\frac{\epsilon(X_1,X_2,P_3)}{X_1\cdot P_3}=\frac{\tau_{12}\tau_{13}\tau_{23}+z_1^2\tau_{23}+z_2^2\tau_{31}}{z_2(z_1^2+\tau_{13}^2)}\,, \\
		\zeta&:=\frac{X_2\cdot P_3}{X_1\cdot P_3}=\frac{z_1(z_2^2+\tau_{23}^2)}{z_2(z_1^2+\tau_{13}^2)}\, , \\
	\end{split}
\end{equation}
with $\tau_{ij}\equiv\tau_i-\tau_j$.
Their geometric meaning can be understood from the following special configuration:
\begin{equation}\label{eq:operator config Bkbkbd}
	\begin{split}
		(\tau_1,z_1)=(\upsilon,\zeta)\,,\quad\quad 
		(\tau_2,z_2)=(0,1)\,,\quad\quad 
		\tau_3=+\infty\,.
	\end{split}
\end{equation}
For a generic configuration, the regions with positive and negative $\upsilon$ are shown in figure~\ref{fig:v}.
The analysis of the conformal block expansion closely parallels that of the $B\partial\partial$ correlator.  
After expanding $\hat{\mathcal{O}}_1$ into $\mathcal{O}_l$ and $\hat{\mathcal{O}}_2$ into $\mathcal{O}_j$ via the BOE \eqref{eq:BOE}, one must determine whether to use $C_{lji}$ or $C_{jli}$.  
The guiding principle is again the radial quantization picture: one searches for a boundary point $(\tau_*,0)$ such that a semicircle centered at $(\tau_*,0)$ separates the bulk point $(\tau_1,z_1)$ from the other points, and another semicircle separates $(\tau_2,z_2)$ from the rest (see figure~\ref{fig:bkbkbd_OPE_vs_BOE}). 

The result is
\begin{equation}\label{bkbkbd:exp}
	\begin{split}
		\mathcal{G}^{\hat{\mathcal{O}}_1\hat{\mathcal{O}}_2}_{i}(\upsilon,\zeta)=
		\begin{cases}
			\sum\limits_{l,j}\bb{\hat{\mathcal{O}}_1}{l}\bb{\hat{\mathcal{O}}_2}{j}C_{lji} R_{\D_l \D_j}^{\D_i}(\upsilon,\z)\,, & \quad \text{for }\upsilon<0\,, \\[4pt]
			\sum\limits_{l,j}\bb{\hat{\mathcal{O}}_1}{l}\bb{\hat{\mathcal{O}}_2}{j}C_{jli} R_{\D_l \D_j}^{\D_i}(\upsilon,\z)\,, & \quad \text{for }\upsilon>0\,. \\
		\end{cases}
	\end{split}
\end{equation}
The corresponding conformal block is\footnote{Notice that, despite the presence of the factor $\Gamma(1-\frac{\Delta_{lij}}{2})$, the block does not diverge when $\Delta_{lij}=2m$ with $m\in\mathbb{N}_0$. That is because \begin{equation}
    \ _3\tilde F_2\left(\begin{matrix}
        a & b & 1-m\\
        & c & 1-m-n 
    \end{matrix};z\right)\propto\frac{1}{\Gamma(-m-n)}\,.
\end{equation}}
\begin{align}\label{def:bkbkbd block}
		R^{\Delta_i}_{\Delta_l\Delta_j}(\upsilon, \zeta)&=\frac{\zeta^{\Delta_l}}{(\zeta^2+\upsilon^2)^{\frac{\Delta_{lji}}{2}}}
		\sum_{n=0}^{\infty}\frac{\left(\frac{\Delta_{lji}}{2}\right)_n\Gamma(1-\frac{\Delta_{lij}}{2})\Gamma(\Delta_j+\frac{1}{2})}{n!\,\left(\Delta_l+\frac{1}{2}\right)_n}\left(\frac{-\zeta^2}{\zeta^2+\upsilon^2}\right)^{n}\\
		&\qquad\quad \times
		\regpFq{3}{2}{\frac{\Delta_{lji}+1}{2},\frac{\Delta_{lji}}{2}+n,1-\frac{\Delta_{lij}}{2}}{\Delta_j+\frac{1}{2},1-\frac{\Delta_{lij}}{2}-n}{-\frac{1}{\zeta^2+\upsilon^2}}.\nonumber
\end{align}
A detailed derivation is given in appendix~\ref{app:BKBKBD block}.

In addition, the $BB\del$ block can also be written in terms of $B\del\del$ blocks. This is achieved by writing the correlator  $\braket{\hat{\mathcal{O}}_{1}(\t_1,z_1)\hat{\mathcal{O}}_{2}(\t_2,z_2)\mathcal{O}_i(\t_3)}$ in two different ways. First, we use the decomposition \eqref{bkbkbd:generalform} together with the expansion  \eqref{bkbkbd:exp}, and compare it with \eqref{bkbkbd:generalform} combined with the BOE \eqref{BOE:general} (assuming $\upsilon>0$)
\be
\left( \frac{z_2}{\tau_{23}^2 + z_2^2} \right)^{\Delta_i}
\sum_{l,j}
b^{\hat\OO_1}_l 
b^{\hat\OO_2}_j  C_{jli} R_{\D_l\D_j}^{\D_i}(\upsilon,\z)
= \sum_j b^{\hat\OO_2}_j \DD(\D_j;\t_2,z_2,\t_2')\<\hat\OO_1(\t_1,z_1)\OO_j(\t'_2)\OO_i(\t_3)\>\,,
\ee
where $\DD$ is the compact form of the differential operators, given in \eqref{eq:Doperator}. Then we use \eqref{bkbdbd:generalform} and \eqref{bkbdbd:exp} on the right-hand side, choose coordinates such that $\r_{2'3}<0$ and factor out the QFT data to get (suppressing the arguments of $\DD$)
\be
\left( \frac{z_2}{\tau_{23}^2 + z_2^2} \right)^{\Delta_i}
R_{\D_l,\D_j}^{\D_i}(\upsilon,\z)
=\DD \left[\frac{1}{|\t_2'-\t_3|^{\D_i+\D_j}}
\left|\frac{(\t_1-\t'_2)^2+z_1^2}{(\t_1-\t_3)^2+z_1^2}\right|^{\D_{ij}}
G_{\D_l}^{\D_j,\D_i}(\chi(\t_1,z_1,\t'_2,\t_3))\right]\,.
\label{eq:Bkbkbd blk from bkbdbd blk}
\ee

We conclude with three remarks: 
first, for configurations where $\upsilon=0$, generically neither of the expansions in \eqref{bkbkbd:exp} converges.  
These correspond to the case where the three points lie on the same geodesic, analogous to the $B\partial\partial$ case.  
Second, the conformal block itself depends on $\upsilon$ only through $\upsilon^2$; the sign of $\upsilon$ merely determines whether one uses $C_{lji}$ or $C_{jli}$.  
Third, the series in \eqref{def:bkbkbd block} converges for all $\zeta^2,\upsilon^2\in(0,\infty)$.  
Although one can express the $R$ block appearing in \eqref{def:bkbkbd block} as a single Appell $F_4$ function (defined by a double series), its primal domain of analyticity does not cover the entire AdS$_2$.

\subsubsection{\texorpdfstring{Bulk-boundary-boundary-boundary ($B\partial\partial\partial$) conformal block}{Bulk-boundary-boundary-boundary (Bbbb) conformal block}}\label{subsec:bkbdbdbdCB}

Lastly, we turn to the bulk-boundary-boundary-boundary four-point function.  
By $SO^+(1,2)$ invariance, it takes the general form
\begin{equation}\label{bkbdbdbd:generalform}
	\begin{split}
		\braket{\hat{\mathcal{O}}(\tau,z)\mathcal{O}_i(\tau_1)\mathcal{O}_j(\tau_2)\mathcal{O}_k(\tau_3)}
		=\frac{\mathcal{G}^{\hat{\mathcal{O}}}_{ijk}(\mu,\omega)}
		{|\tau_{12}|^{\Delta_{ijk}}|\tau_{23}|^{\Delta_{jki}}|\tau_{31}|^{\Delta_{kij}}}\,.
	\end{split}
\end{equation}
The two independent $SO^+(1,2)$ invariants are defined as\footnote{In this case, we can also reuse the cross ratios \eqref{def:chi}. Since there are three boundary points and one bulk-point, there are three different cross ratios $\chi_{12}$, $\chi_{23}$, $\chi_{13}$ where the subscripts label the boundary points. The cross ratios in \eqref{def:mu and omega} are related to $\chi_{ij}$'s through 
\be
\omega^2=\frac{\chi_{13}\chi_{23}}{\chi_{12}}\,,\qquad\quad 
\mu=\frac{\chi_{12}-\chi_{13}+\chi_{23}}{2\chi_{12}}\,.
\ee
The $\chi_{ij}$'s are not all independent. In AdS$_2$, embedding space vectors have three components and thus  $X^{[A}P_1^BP_2^CP_3^{D]}=0$. Taking the square of this identity leads to
\begin{equation*}
    \chi_{12}^2+\chi_{13}^2+\chi_{23}^2-2\left(\chi_{12}\chi_{13}+\chi_{12}\chi_{23}+\chi_{13}\chi_{23}\right)+4\chi_{12}\chi_{23}\chi_{13}=0\, .
    \label{eq: cross ratio relation for B∂∂∂}
\end{equation*}
}
\begin{equation}\label{def:mu and omega}
	\begin{split}
		\mu
		&:=-\frac{\epsilon(X,P_1,P_3)(P_2\cdot P_3)}{\epsilon(P_1,P_2,P_3)(X\cdot P_3)}
		=\frac{\tau_{23}}{\tau_{21}}\frac{\left((\tau-\tau_1)(\tau-\tau_3)+z^2\right)}{(\tau-\tau_3)^2+z^2}\,, \\
		\omega
		&:=\frac{-2\epsilon(P_1,P_2,P_3)}{(-2X\cdot P_3)(-2P_1\cdot P_2)}
		=\frac{\tau_{23}}{\tau_{21}}\frac{z(\tau_1-\tau_3)}{(\tau-\tau_3)^2+z^2}\,.
	\end{split}
\end{equation}
When the boundary points follow the cyclic ordering $[123]$, one can always find an $SO^+(1,2)$ transformation that brings the configuration to the canonical form
\begin{equation}\label{config:4pt}
	(\tau,z)=(\mu,\omega)\,,\qquad 
	\tau_1=0\,,\qquad 
	\tau_2=1\,,\qquad 
	\tau_3=\infty\,.
\end{equation}

\begin{figure}
\centering
   \begin{tikzpicture}[x=0.75pt,y=0.75pt,yscale=-0.9,xscale=0.9]

\draw    (30,70) -- (60,70) ;
\draw    (266,69.33) -- (296,69.33) ;
\draw    (80,70) -- (110,100) ;
\draw    (150,100) -- (181,130) ;
\draw    (150,100) -- (180,70) ;
\draw    (80,130) -- (110,100) ;
\draw    (316,69.33) -- (346,99.33) ;
\draw    (346,139.33) -- (376,169.33) ;
\draw    (346,99.33) -- (376,69.33) ;
\draw    (316,169.33) -- (346,139.33) ;
\draw    (470,70) -- (500,70) ;
\draw    (520,70) -- (550,100) ;
\draw    (550,100) -- (599.5,169.5) ;
\draw    (550,140) -- (562.8,123.4) ;
\draw    (520,170) -- (550,140) ;
\draw    (566.6,117.2) -- (600.3,70.5) ;
\draw    (110,100) -- (150,100) ;
\draw    (346,99.33) -- (346,139.33) ;
\draw    (550,100) -- (550,140) ;

\draw (10,60) node [anchor=north west][inner sep=0.75pt]    {$\hat{\OO }$};
\draw (244,60) node [anchor=north west][inner sep=0.75pt]    {$\hat{\OO }$};
\draw (61,65) node [anchor=north west][inner sep=0.75pt]    {$\mathcal{O}_l$};
\draw (297,65) node [anchor=north west][inner sep=0.75pt]    {$\mathcal{O}_l$};
\draw (448,60) node [anchor=north west][inner sep=0.75pt]    {$\hat{\OO }$};
\draw (500,65) node [anchor=north west][inner sep=0.75pt]    {$\mathcal{O}_l$};
\draw (601.5,65) node [anchor=north west][inner sep=0.75pt]    {$\mathcal{O}_j$};
\draw (601.5,165) node [anchor=north west][inner sep=0.75pt]    {$\mathcal{O}_k$};
\draw (500,165) node [anchor=north west][inner sep=0.75pt]    {$\mathcal{O}_i$};
\draw (377,65) node [anchor=north west][inner sep=0.75pt]    {$\mathcal{O}_j$};
\draw (377,165) node [anchor=north west][inner sep=0.75pt]    {$\mathcal{O}_k$};
\draw (298,165) node [anchor=north west][inner sep=0.75pt]    {$\mathcal{O}_i$};
\draw (182.4,65) node [anchor=north west][inner sep=0.75pt]    {$\mathcal{O}_{j}$};
\draw (183.4,127) node [anchor=north west][inner sep=0.75pt]    {$\mathcal{O}_{k}$};
\draw (61,127) node [anchor=north west][inner sep=0.75pt]    {$\mathcal{O}_{i}$};
\draw (96.67,190) node [anchor=north west][inner sep=0.75pt]   [align=center] {$s$-channel};
\draw (315,190) node [anchor=north west][inner sep=0.75pt]   [align=center] {$t$-channel};
\draw (525,190) node [anchor=north west][inner sep=0.75pt]   [align=center] {$u$-channel};

\end{tikzpicture}
\caption{Conformal block decomposition of $\braket{\hat{\mathcal{O}}(\tau,z)\mathcal{O}_i(\tau_1)\mathcal{O}_j(\tau_2)\mathcal{O}_k(\tau_3)}$, where $\hat{\OO}(\tau,z)$ is expanded into $\mathcal{O}_l$ and the boundary four-point function admits three possible OPE channels.}
	\label{fig:bkbdrybdrybdryope2}
\end{figure}
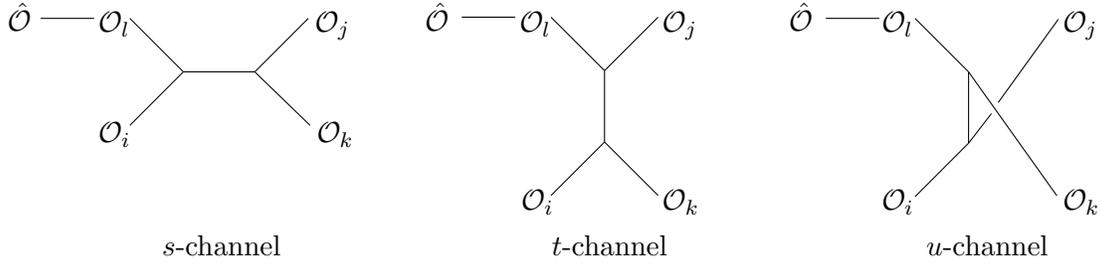

The conformal block expansion of $\mathcal{G}^{\hat{\mathcal{O}}}_{ijk}(\mu,\omega)$ can be obtained by applying the BOE once and the boundary OPE twice on the left-hand side of \eqref{bkbdbdbd:generalform}.  
In this case there are three possible OPE channels, corresponding to the different pairings of the boundary operators:
\begin{equation*}
	\begin{array}{ll}
		s\text{-channel:} & (\mathcal{O}_l\times\mathcal{O}_i)(\mathcal{O}_j\times\mathcal{O}_k)\,,\\[2pt]
		t\text{-channel:} & (\mathcal{O}_l\times\mathcal{O}_j)(\mathcal{O}_i\times\mathcal{O}_k)\,,\\[2pt]
		u\text{-channel:} & (\mathcal{O}_l\times\mathcal{O}_k)(\mathcal{O}_i\times\mathcal{O}_j)\,,
	\end{array}
\end{equation*}
with $\mathcal{O}_l$  the operator appearing in the BOE of $\hat{\mathcal{O}}(\tau,z)$. This is illustrated in figure \ref{fig:bkbdrybdrybdryope2}.

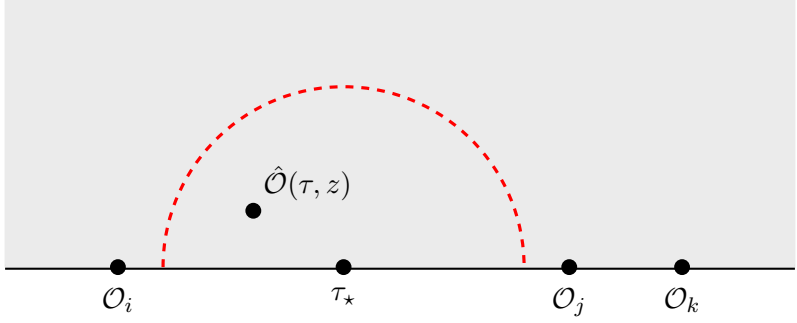
\begin{figure}
	\centering
	\begin{tikzpicture}[scale=3]

\draw[line width=2] (0,0) -- (3.5,0);
\definecolor{light-gray}{gray}{0.92}
\filldraw[light-gray](0,0)--(3.5,0)--(3.5,1.2)--(0,1.2)--(0,0);
\fill (1.5,0) circle (1pt) node[below=4pt] {$\tau_\star$};
\fill (0.5,0) circle (1pt) node[below=4pt] {$\mathcal{O}_i$};
\fill (2.5,0) circle (1pt) node[below=4pt] {$\mathcal{O}_j$};
\fill (3,0) circle (1pt) node[below=4pt] {$\mathcal{O}_k$};

\fill (1.1,0.25) circle (1pt) node[above right] {$\hat{\mathcal{O}}(\tau,z)$};
\draw[dashed, red,line width=1.2] (0.7,0) arc (180:0:0.8);

\end{tikzpicture}
	\caption{In this configuration, we can use the BOE to expand $\hat{\mathcal{O}}(\tau,z)$ in terms of boundary operators $\mathcal{O}_l$ inside the red semi-circle.
    This leads to a boundary four-point function with cyclic ordering $[iljk]$ which can be expanded using the  $s-$channel or the  $t-$channel OPE. This corresponds to region $\Omega_{st}$ in figure \ref{fig:no_stu}.
    }
	\label{fig:bkbdbdbd_OPE_vs_BOE}
\end{figure}

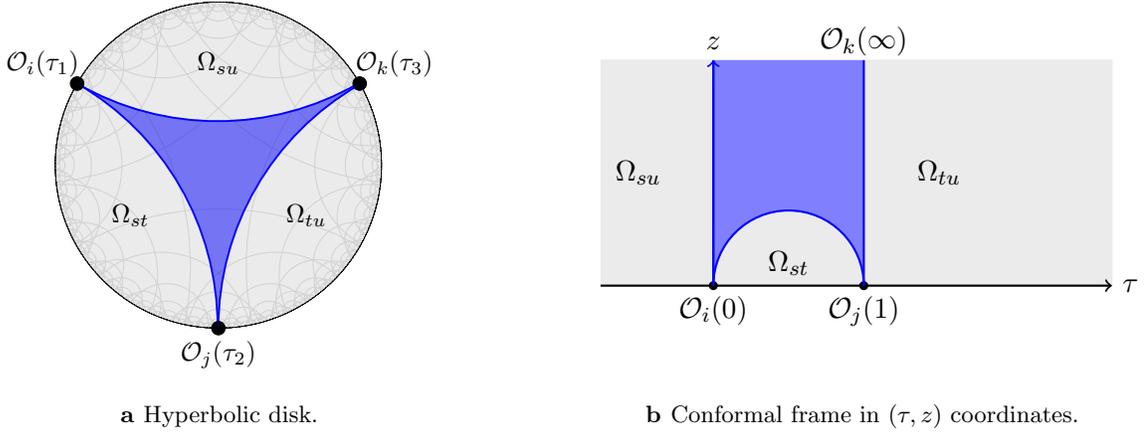
\begin{figure}[htbp]
  \centering
  \begin{subfigure}[b]{0.45\textwidth}
    \centering
    \includegraphics[scale=0.9]{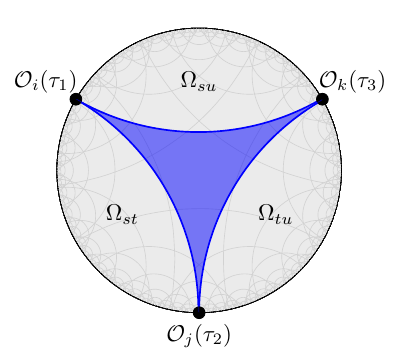}
    \caption{Hyperbolic disk.}
    \label{fig:no_stu_a}
  \end{subfigure}
  \hfill
  \begin{subfigure}[b]{0.45\textwidth}
    \centering
    \begin{tikzpicture}[scale=1]
  
    \definecolor{light-gray}{gray}{0.92}
    \filldraw[light-gray](-1.5,0)--(5.3,0)--(5.3,3)--(-1.5,3);
    \draw[thick,->] (-1.5,0) -- (5.3,0) node[right] {$\tau$};
    \fill[blue!50] (0,0) arc (180:0:1) -- (2,3) -- (0,3) -- cycle;
    \draw[thick,blue,->] (0,0) -- (0,3) node[above,black] {$z$};

    \draw[white] (0,0) -- (0,-1.3) ;
  
    \filldraw (0,0) circle (1.5pt) node[below] {$\mathcal{O}_i(0)$};
    \filldraw (2,0) circle (1.5pt) node[below] {$\mathcal{O}_j(1)$};
    \node at (2,3.25) {$\mathcal{O}_k(\infty)$};

    \draw[thick,blue] (0,0) arc (180:0:1);
    \draw[thick,blue] (2,0) -- (2,3);
    \draw[thick,blue] (0,0) -- (0,3);

    \node at (1,0.3) {$\Omega_{st}$};
    \node at (3,1.5) {$\Omega_{tu}$};
    \node at (-1,1.5) {$\Omega_{su}$};
  
    \end{tikzpicture}
    \caption{Conformal frame in $(\tau,z)$ coordinates.}
    \label{fig:no_stu_b}
  \end{subfigure}
  \caption{AdS$_2$ partitioned by the geodesics connecting the boundary insertion points. The $s$- and $t$-channel conformal block expansions converge in the region $\Omega_{st}$; analogous statements hold for $\Omega_{su}$ and $\Omega_{tu}$. The BOE of the bulk operator is not valid in the blue region.}
	\label{fig:no_stu}
\end{figure}

Similarly to the previous cases, for the BOE to be valid there must exist a boundary point $(\tau_*,0)$ and a semicircle centered at $(\tau_*,0)$ that separates the bulk point from all other insertions (in this case, the three boundary points).  
An illustrative configuration is shown in figure~\ref{fig:bkbdbdbd_OPE_vs_BOE}.  
However, this condition is not satisfied for all $B\partial\partial\partial$ configurations: it fails within the closed region shown in figure~\ref{fig:no_stu}. This region corresponds to the geodesic triangle bounded by the three geodesics connecting the boundary insertion points.\footnote{We thank Miguel Paulos for pointing out this issue.} As we will show later in this paper, this difficulty is resolved by introducing the concept of \emph{local blocks}. With this refinement, the union of the domains of convergence covers the entire AdS space.

In the following we focus on the $s$-channel for concreteness as the conformal blocks in the other two channels can be obtained in a completely analogous manner.

Consider the configuration shown in figure~\ref{fig:bkbdbdbd_OPE_vs_BOE}, where the bulk point lies inside the half-disk bounded by the geodesic connecting $(\tau_1,0)$ and $(\tau_2,0)$.  
In this case, we first apply the integral representation of the OPE to $\mathcal{O}_j\mathcal{O}_k$ (see eq.~\eqref{OPE:integralrepr}), thereby reducing the correlator to a sum of integrated $B\partial\partial$ correlators:
\begin{equation}\label{bkbdbdbd:reduction_bkbdbd}
	\begin{split}
		&\braket{\hat{\mathcal{O}}(\tau,z)\mathcal{O}_i(\tau_1)\mathcal{O}_j(\tau_2)\mathcal{O}_k(\tau_3)} \\
		&\quad\quad \quad  \quad \quad =\sum_{m}C_{jkm}
		\int_{\tau_2}^{\tau_3}\!d\tau_4\,
		\mathcal{B}_{\Delta_j\Delta_k\Delta_m}(\tau_2,\tau_3,\tau_4)\,
		\braket{\hat{\mathcal{O}}(\tau,z)\mathcal{O}_i(\tau_1)\mathcal{O}_m(\tau_4)}\,.
	\end{split}
\end{equation}
Each $B\partial\partial$ correlator can then be decomposed as discussed in section~\ref{subsec:bkbdbdCB}.  
For the configuration in figure~\ref{fig:bkbdbdbd_OPE_vs_BOE}, one can verify that $\rho_{14}>0$ for all $\tau_4\in[\tau_2,\tau_3]$ (here $\rho_{14}$ is defined by \eqref{def:rho} using $P_4$ instead of $P_2$). 
Hence, the conformal block expansion of the $B\partial\partial$ correlator is of the second type in equation \eqref{bkbdbd:exp}.  
Combining all ingredients, we obtain the $s$-channel conformal block expansion of the $B\partial\partial\partial$ correlator:
\begin{equation}\label{bkbdbdbd:sexp}
		\mathcal{G}^{\hat{\mathcal{O}}}_{ijk}(\mu,\omega)
		=\sum_{m,l}
		b^{\hat{\mathcal{O}}}_{l}\,
		C_{jkm}\,C_{ilm}\,
		K^{\Delta_i\Delta_j\Delta_k}_{\Delta_l\Delta_m}(\mu,\omega)\,,
\end{equation}
which converges in the region $\mu^2+\omega^2<\mu$. The $s$-channel conformal block is given by the following integral:
\begin{equation}\label{def:bkbdbdbd_sblock_normal}
	\begin{split}
		K^{\Delta_i\Delta_j\Delta_k}_{\Delta_l\Delta_m}(\mu,\omega)
		&:=\frac{\Gamma(2\Delta_m)}
		{\Gamma(\Delta_{mjk})\,\Gamma(\Delta_{mkj})}
		\int_{1}^{\infty}\!dt\,
		\frac{(t-1)^{\Delta_{mkj}-1}}{t^{\Delta_i+\Delta_m}}
		\left(\frac{\mu^2+\omega^2}{(\mu-t)^2+\omega^2}\right)^{\tfrac{\Delta_{mi}}{2}} \\
		&\quad\times
		G_{\Delta_l}^{\Delta_i\Delta_m}\!\left(
		\frac{\omega^2t^2}{(\mu^2+\omega^2)((\mu-t)^2+\omega^2)}
		\right)\,,
	\end{split}
\end{equation}
where $G_{\Delta_l}^{\Delta_i\Delta_m}$ is the $B\partial\partial$ conformal block defined in \eqref{def:bkbdbd block}.  
We have adopted the canonical configuration \eqref{config:4pt} to simplify the expression.

The $s$-channel expansion is also convergent in the region $\mu<0$.  
In this domain, the conformal block takes the same form as in \eqref{def:bkbdbdbd_sblock_normal}.  
However, since $\rho_{14}<0$ in eq.~\eqref{bkbdbdbd:reduction_bkbdbd}, the relevant OPE coefficient is $C_{lim}$ instead of $C_{ilm}$.  
Thus we obtain
\begin{equation}\label{bkbdbdbd:sexp2}
		\mathcal{G}^{\hat{\mathcal{O}}}_{ijk}(\mu,\omega)
		=\sum_{m,l}b^{\hat{\mathcal{O}}}_{l}\,C_{jkm}\,C_{lim}\,
		K^{\Delta_i\Delta_j\Delta_k}_{\Delta_l\Delta_m}(\mu,\omega)\, ,\qquad
		\qquad(\mu<0)\,.
\end{equation}

For the $t$- and $u$-channel expansions, it is unnecessary to repeat the entire analysis.  
One simply redefines the $SO^+(1,2)$ invariants by cyclic permutations of the boundary coordinates, leading to the substitutions
\begin{align}
		\mu&\rightarrow\frac{\mu(\mu-1)+\omega^2}{\mu^2+\omega^2}\,,\qquad \qquad 
		\omega\rightarrow\frac{\omega}{\mu^2+\omega^2}\, ,
		&&\text{($t$-channel)}\,,\\
		\mu&\rightarrow\frac{1-\mu}{(\mu-1)^2+\omega^2}\,,\qquad\qquad 
		\omega\rightarrow\frac{\omega}{(\mu-1)^2+\omega^2}\, ,&&
		\text{($u$-channel)}\,.
\end{align}
The operator indices $(i,j,k)$ are permuted accordingly.  
Collecting all results, we summarize the conformal block expansions as
\begin{equation}\label{bkbdbdbd:stu_summary}
	\begin{split}
		\mathcal{G}^{\hat{\mathcal{O}}}_{ijk}(\mu,\omega)
		=\begin{cases}
			\displaystyle\sum\limits_{m,l}
			b^{\hat{\mathcal{O}}}_{l}\,C_{jkm}\,C_{ilm}\,
			K^{\Delta_i\Delta_j\Delta_k}_{\Delta_l\Delta_m}(\mu,\omega)
			& \Omega_{st},\quad \hspace{0.2cm} s\text{-channel}, \\[8pt]
			\displaystyle\sum\limits_{m,l}
			b^{\hat{\mathcal{O}}}_{l}\,C_{jkm}\,C_{lim}\,
			K^{\Delta_i\Delta_j\Delta_k}_{\Delta_l\Delta_m}(\mu,\omega)
			& \Omega_{su},\quad\ s\text{-channel}, \\[8pt]
			\displaystyle\sum\limits_{m,l}
			b^{\hat{\mathcal{O}}}_{l}\,C_{kim}\,C_{jlm}\,
			K^{\Delta_j\Delta_k\Delta_i}_{\Delta_l\Delta_m}\!\left(
				\tfrac{\mu(\mu-1)+\omega^2}{\mu^2+\omega^2},
				\tfrac{\omega}{\mu^2+\omega^2}
			\right)
			& \Omega_{tu},\quad\ t\text{-channel}, \\[8pt]
			\displaystyle\sum\limits_{m,l}
			b^{\hat{\mathcal{O}}}_{l}\,C_{kim}\,C_{ljm}\,
			K^{\Delta_j\Delta_k\Delta_i}_{\Delta_l\Delta_m}\!\left(
				\tfrac{\mu(\mu-1)+\omega^2}{\mu^2+\omega^2},
				\tfrac{\omega}{\mu^2+\omega^2}
			\right)
			& \Omega_{st},\quad \hspace{0.15cm} t\text{-channel}, \\[8pt]
			\displaystyle\sum\limits_{m,l}
			b^{\hat{\mathcal{O}}}_{l}\,C_{ijm}\,C_{klm}\,
			K^{\Delta_k\Delta_i\Delta_j}_{\Delta_l\Delta_m}\!\left(
				\tfrac{1-\mu}{(\mu-1)^2+\omega^2},
				\tfrac{\omega}{(\mu-1)^2+\omega^2}
			\right)
			& \Omega_{su},\quad u\text{-channel}, \\[8pt]
			\displaystyle\sum\limits_{m,l}
			b^{\hat{\mathcal{O}}}_{l}\,C_{ijm}\,C_{lkm}\,
			K^{\Delta_k\Delta_i\Delta_j}_{\Delta_l\Delta_m}\!\left(
				\tfrac{1-\mu}{(\mu-1)^2+\omega^2},
				\tfrac{\omega}{(\mu-1)^2+\omega^2}
			\right)
			& \Omega_{tu},\quad u\text{-channel}.
		\end{cases}
	\end{split}
\end{equation}
Here we have assumed that the boundary operators follow the cyclic ordering $[ijk]$. In equation \eqref{bkbdbdbd:stu_summary}, the second column keeps track of the region in which we use a given expansion, and they are defined by 
\begin{equation}
    \begin{split}
        \Omega_{st}&:=\left\{(\mu \in \mathbb{R},\omega>0)\,\big|\,\mu^2+\omega^2<\mu \right\}, \\
        \Omega_{su}&:=\left\{(\mu,\omega)\,\big|\,\mu<0\right\}, \\
        \Omega_{tu}&:=\left\{(\mu,\omega)\,\big|\,\mu>1\right\},
    \end{split}
\end{equation}
where the subscripts label the OPE channels that converge in each domain. They correspond to the three disconnected regions in figure~\ref{fig:no_stu_b}. Note that in each region, two OPE channels simultaneously converge, and their expansions are mutually consistent.

The remaining portion of the $(\mu,\omega)$-domain corresponds to the blue region shown in figure~\ref{fig:no_stu}.  
As discussed above, the standard conformal block expansion fails to converge there.  
In the next section, we will remedy this issue by employing the local block expansion.

\section{Derivation of the flow equations}
\label{sec:derivation of flow eqn}

The goal of this section is to derive the evolution equations for the QFT data $\{\Delta_i,\ b^{\hat{\Phi}}_{i},\ C_{ijk}\}$ under a deformation of the theory triggered by a single relevant bulk operator $\hat{\Phi}$ such that the action becomes
\begin{equation*}
	S_{\lambda+\delta\lambda}=S_\lambda +  \delta\lambda\int_{\text{AdS}_2}d^2x\sqrt{g}\,\hat{\Phi}\, .
\end{equation*}
This will result in the flow equations (\ref{eq:dimension flow}--\ref{eq:OPE flow}).  For simplicity, we will assume in this section that the boundary operators are all irrelevant (having scaling dimensions above 1). We will discuss marginal and relevant boundary operators in section \ref{sec:relevant and marginal boundary op}.

The main idea of the derivation is illustrated in figure~\ref{fig: flow}. In practice, two technical issues arise:
\begin{itemize}
	\item \textbf{Divergences:} integrals of conformal blocks over the full AdS space are generically divergent;
	\item \textbf{Swappability:} the sums over exchanged operators do not necessarily commute with these integrals.
\end{itemize}

The divergence issue can be resolved by, amongst other methods, introducing a cutoff and renormalizing the operators, as is standard in QFT. The swappability issue is more subtle and requires choosing a more suitable basis for expanding correlation functions.

To separate these problems, we first present a derivation using the standard conformal blocks (referred to as \emph{normal blocks} in this paper) in section~\ref{sec:FlowEqNormal}. There, we treat operator renormalization carefully, while formally assuming that sums and integrals commute. Most steps of the derivation do not rely on the explicit form of the blocks and therefore remain valid when switching to a different basis.

We then resolve the swappability issue by introducing the so-called \emph{local blocks} in section~\ref{subsec:local block}. This is based on the results of \cite{Loparco:2025aag}, which build on \cite{Levine:2023ywq,Meineri:2023mps,Levine:2024wqn}. Some properties of the coefficients that appear in the flow equations will be discussed in sections~\ref{subsec:poles} and \ref{sec:integrals of blocks}.

\subsection{Flow equations using the normal blocks}\label{sec:FlowEqNormal}
\subsubsection{Equation I: scaling dimensions and operator renormalization}\label{subsec:eq1}

Consider the two-point functions of boundary operators with finite coupling $\lambda$. We assume that at $\lambda$ all the operators are properly renormalized such that the two-point functions are diagonalized:
\begin{equation}\label{bdrybdrytwopt:normalization}
	\begin{split}
		\braket{\mathcal{O}_i(\tau_1)\mathcal{O}_j(\tau_2)}_\lambda=\frac{\delta_{ij}}{\abs{\tau_1-\tau_2}^{2\Delta_i(\lambda)}}\, .
	\end{split}
\end{equation}
For simplicity, we assume that there is no degeneracy in the spectrum of boundary primary operators.

Let us now change the coupling infinitesimally as  $\lambda\rightarrow\lambda+\delta\lambda$. The two-point function then changes as
\begin{equation}\label{deformation:2pt}
	\begin{split}
		\braket{\mathcal{O}^{\text{(bare)}}_i(\tau_1)\mathcal{O}^{\text{(bare)}}_j(\tau_2)}_{\lambda+\delta\lambda}=\braket{\mathcal{O}_i(\tau_1)\mathcal{O}_j(\tau_2)}_\lambda-\delta\lambda\int\frac{d\tau dz}{z^2}\braket{\mathcal{O}_i(\tau_1)\mathcal{O}_j(\tau_2)\hat{\Phi}(\tau,z)}_\lambda\,.
	\end{split}
\end{equation}
Here we ignore terms that are higher order in $\delta\lambda$ because they do not contribute to our flow equations. 
Note that the operators on the left-hand side of \eqref{deformation:2pt} are not yet the renormalized ones as written in \eqref{bdrybdrytwopt:normalization}, instead they are \emph{bare} operators  that do not vary with $\l$. We will see later that proper field redefinition (field-strength renormalization) is needed to obtain \eqref{bdrybdrytwopt:normalization} at coupling $\lambda+\delta\lambda$.

\begin{figure}
	\centering
	\includegraphics[width=0.48\textwidth]{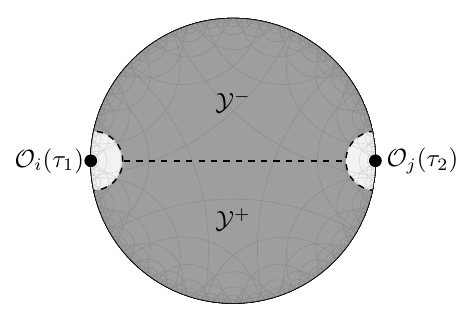}
    \includegraphics[width=0.48\textwidth]{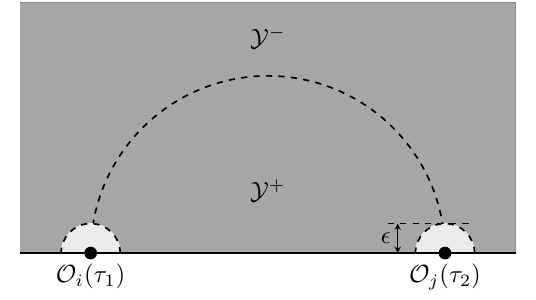}
	\caption{Domain of integration in \eqref{def:Iij}. We remove the circular regions close to where $(\tau,z)$ hits one of the boundary operators at positions $\tau_{1,2}$. We further split the region into $\mathcal{Y}^+$ and $\mathcal{Y}^-$ in \ref{def:Y}.} 
	\label{fig:RegionsAdS2ToIntegrate}
\end{figure}

The integral in \eqref{deformation:2pt} is divergent in general. The divergences arise from the regime where the bulk point $(\tau,z)$ is close to one of the boundary points $\tau_{1,2}$. Therefore, we must introduce a proper renormalization scheme to remove this divergence. In this paper we use the ``hard-cutoff scheme", by which we mean that the bulk-integration domain is the whole AdS$_2$ minus the half-disks with radius $\epsilon$ centered at $\tau_{1,2}$ as shown on figure \ref{fig:RegionsAdS2ToIntegrate}. Thus it is convenient to define
\begin{equation}\label{def:Iij}
	\begin{split}
		I_{ij}^{\hat{\Phi}}(\tau_1,\tau_2,\epsilon):=\int_{(\tau-\tau_k)^2+z^2\geqslant\epsilon^2}\frac{d\tau dz}{z^2}\braket{\hat{\Phi}(\tau,z)\mathcal{O}_i(\tau_1)\mathcal{O}_j(\tau_2)}_\lambda\, .
	\end{split}
\end{equation}

The introduction of the hard cutoff does not preserve the whole $SO^+(1,2)$ symmetry, but it preserves the translation and the dilatation (up to an overall scaling). Consequently, $I_{ij}^{\hat{\Phi}}(\tau_1,\tau_2,\epsilon)$ must be of the form
\begin{equation}
	\begin{split}
		I_{ij}^{\hat{\Phi}}(\tau_1,\tau_2,\epsilon)=\frac{1}{\abs{\tau_{12}}^{\Delta_i+\Delta_j}}F_{ij}\left(\tfrac{\tau_1-\tau_2}{\epsilon}\right)\, .
	\end{split}
\end{equation}
To see the structure of $F_{ij}$, we recall the results of $B\partial\partial$ correlators from section~\ref{subsec:bkbdbdCB}:
\begin{equation}\label{Gbkbdbd:generalform}
		\braket{\hat{\Phi}(\tau,z)\mathcal{O}_i(\tau_1)\mathcal{O}_j(\tau_2)}_\lambda=\frac{1}{\abs{\tau_1-\tau_2}^{\Delta_i+\Delta_j}}\left(\frac{(\tau-\tau_1)^2+z^2}{(\tau-\tau_2)^2+z^2}\right)^{\frac{\Delta_j-\Delta_i}{2}}\mathcal{G}^{\hat{\Phi}}_{ij}(\rho)\,, 
\end{equation}
with the cross-ratio $\rho$ defined as 
\begin{equation}
		\rho\assign\text{sgn}(\tau_1-\tau_2)\frac{(\tau-\tau_1)(\tau-\tau_2)+z^2}{\sqrt{\left((\tau-\tau_1)^2+z^2\right)\left((\tau-\tau_2)^2+z^2\right)}}\,. 
\end{equation}
Moreover, $\mathcal{G}^{\hat{\Phi}}_{ij}(\rho)$ has expansions
\begin{equation}\label{Gbkbdbd:OPE}
		\mathcal{G}^{\hat{\Phi}}_{ij}(\rho)=\begin{cases}
        \sum\limits_{l}b^{\hat{\Phi}}_{l}C_{ijl}G_{\Delta_l}^{\Delta_i,\Delta_j}(\chi), & \quad \text{for }\rho<0 \,, \\\sum\limits_{l}b^{\hat{\Phi}}_{l}C_{ilj}G_{\Delta_l}^{\Delta_i,\Delta_j}(\chi), & \quad \text{for }\rho>0 \,. \\
		\end{cases}
        \end{equation}
Note that the $\rho$ dependence is entirely contained in the choice of operator ordering in $C_{ijk}$. The conformal block $G_{\Delta_l}^{\Delta_i,\Delta_j}$ depends on $\rho$ through the cross-ratio
\begin{equation}
    \chi\equiv 1-\rho^2=\frac{(\tau_1-\tau_2)^2z^2}{[(\tau-\tau_1)^2+z^2][(\tau-\tau_2)^2+z^2]}\,,
\end{equation}
and is given by
\begin{equation}
		G_{\Delta_l}^{\Delta_i,\Delta_j}(\chi)=\chi^{\frac{\Delta_l}{2}}\hyperF{\tfrac{\Delta_{lij}}{2}}{\tfrac{\Delta_{lji}}{2}}{\Delta_l+\tfrac{1}{2}}{\chi}\,.
\end{equation}
Plugging \eqref{Gbkbdbd:generalform} and \eqref{Gbkbdbd:OPE} into \eqref{def:Iij}, \emph{assuming that} the integration and the summation can be freely exchanged, we obtain
\begin{equation}
	\begin{split}
		I_{ij}^{\hat{\Phi}}(\tau_1,\tau_2,\epsilon)=\sum_{l}b^{\hat{\Phi}}_{l}\left[C_{ilj}\mathcal{Y}^{\Delta_l;+}_{\Delta_i,\Delta_j}(\tau_1,\tau_2,\epsilon)+C_{ijl}\mathcal{Y}^{\Delta_l;-}_{\Delta_i,\Delta_j}(\tau_1,\tau_2,\epsilon)\right]\, ,\label{eq:DefIij}
	\end{split}
\end{equation}
where $\mathcal{Y}^{+}$ and $\mathcal{Y}^{-}$ are the integrated blocks in different regimes in \eqref{Gbkbdbd:OPE} (see figure \ref{fig:RegionsAdS2ToIntegrate}):
\begin{equation}\label{def:Y}
	\begin{split}
		\mathcal{Y}^{\Delta_l;\pm}_{\Delta_i,\Delta_j}(\tau_1,\tau_2,\epsilon):=\int_{\substack{\text{sgn}(\rho)=\pm \\ (\tau-\tau_i)^2+z^2\geqslant\epsilon^2}}\frac{d\tau dz}{z^2} \frac{1}{\abs{\tau_{12}}^{\Delta_i+\Delta_j}}\left(\frac{(\tau-\tau_1)^2+z^2}{(\tau-\tau_2)^2+z^2}\right)^{\frac{\Delta_j-\Delta_i}{2}} G_{\Delta_l}^{\Delta_i,\Delta_j}(\chi)\,.
	\end{split}
\end{equation}
We leave the calculation details of $\mathcal{Y}^{\pm}$ to appendix~\ref{app:subsec_IntegralY}, and quote the result 
\begin{equation}\label{Y:result}
	\begin{split}
		\mathcal{Y}^{\Delta_l;\pm}_{\Delta_i,\Delta_j}(\tau_1,\tau_2,\epsilon)&=\frac{1}{\abs{\tau_{12}}^{\Delta_i+\Delta_j}}\sum_{p=0}^{\infty}\frac{a_p^{\pm}(\Delta_{ij},\Delta_l)}{\Delta_{ij}-p}\abs{\frac{\tau_{21}}{\epsilon}}^{\Delta_{ij}}\left(\frac{\epsilon}{\tau_{21}}\right)^p +(\Delta_i\leftrightarrow \Delta_j)\, .
	\end{split}
\end{equation}
The explicit forms of the coefficients $a_p^{\pm}(\Delta_{ij},\Delta_l)$ are not all needed. For this work, we will only need the following properties (which are also derived in appendix~\ref{app:subsec_IntegralY}):
\begin{equation}\label{apm:prop1}
		a^{\pm}_0(0,\Delta_l)=\frac{\sqrt{\pi}\Gamma(\Delta_l+\tfrac{1}{2})}{(\Delta_l-1)\Gamma(\tfrac{\Delta_l}{2})\Gamma(\tfrac{\Delta_l}{2}+1)},\qquad\qquad   \frac{\partial a^{\pm}_0(\Delta_{ij},\Delta_l)}{\partial\Delta_{ij}}\bigg{|}_{\Delta_{ij}=0}=0\,, 
\end{equation}
together with
\begin{align}
		a^{\pm}_p(p,\Delta_l)&=0\, ,&&\text{for }p =1,2,3,\ldots\,,\label{apm:prop2} \\
        a^{+}_p(\Delta_{ij},\Delta_l)&=(-1)^p\,a^{-}_p(\Delta_{ij},\Delta_l),&&\text{for all }p\,. \label{apm:prop3} 
\end{align}
Then, taking into account the BOE and OPE coefficients $\bb{\hat{\Phi}}{l}$ and $C_{ijl}$ in \eqref{eq:DefIij}, the total contribution from the boundary operator $\mathcal{O}_l$ to $I_{ij}^{\hat{\Phi}}(\tau_1,\tau_2,\epsilon)$ is
\begin{equation}\label{IntBlock:Iij}
		\begin{split}
		    \left[I_{ij}^{\hat{\Phi}}(\tau_1,\tau_2,\epsilon)\right]_{l}&=\frac{b^{\hat{\Phi}}_{l}\epsilon^{-\Delta_{ij}}}{\abs{\tau_{12}}^{2\Delta_j}}\sum_{p=0}^{\infty}\frac{C_{jil}a_p^{+}(\Delta_{ij},\Delta_l)+C_{ijl}a_p^{-}(\Delta_{ij},\Delta_l)}{\Delta_{ij}-p}\left(\frac{\epsilon}{\tau_{21}}\right)^p +(\Delta_i\leftrightarrow\Delta_j)\, .
		\end{split}
\end{equation}

We see that, for $\Delta_{ij}\neq0$, the regularized two-point function has power-law divergences $\epsilon^{\Delta_{ij}+p}$ or $\epsilon^{\Delta_{ji}+p}$ as $\epsilon\rightarrow0$. While for $\Delta_{ij}=0$, we will see that it has logarithmic divergence $\log\epsilon$.
These divergences can be removed by renormalizing the boundary operators:
\begin{equation}\label{eq:oprenom}
	\begin{split}
		\mathcal{O}^{\text{(bare)}}_i(\tau)=\sum_{j}\sum_{p=0}^{\infty}Z_{ij;p}\partial^p\mathcal{O}_j^\text{(ren)}(\tau)\,,
	\end{split}
\end{equation} 
where the coefficients $Z_{ij;p}$ are the field-strength renormalization constants,\footnote{We adopt the same terminology as \cite{Peskin:1995ev}. There, the term \emph{field-strength renormalization} refers to the field redefinition that normalizes the momentum-space two-point function at the mass shell $p^2=m^2$. Here, we always normalize the position-space two-point function, but the idea is essentially the same.} which can be written as
\begin{equation}\label{eq:Zansatz}
		Z_{ij;p}=\delta_{ij}\delta_{p,0}+\delta Z_{ij;p}\,, 
\end{equation}
where $\delta Z=O(\delta\lambda)$ but crucially also contain a $p=0$ piece. The coefficients in \eqref{eq:Zansatz} are fixed by the normalization conditions of the boundary two-point function
\begin{equation}
	\begin{split}
		\braket{\mathcal{O}_i^\text{(ren)}(\tau_1)\mathcal{O}_j^\text{(ren)}(\tau_2)}_{\lambda+\delta\lambda}=\frac{\delta_{ij}}{\abs{\tau_1-\tau_2}^{2\Delta_i+2\delta\Delta_i}}.
	\end{split}
\end{equation}

Let us first look at the coefficients with $i\neq j$. We study $\braket{\OO_i(\tau_1)\OO_j(\tau_2)}_{\lambda + \delta\lambda}$ as in \eqref{deformation:2pt}, using the definition \eqref{def:Iij}, the parametrization \eqref{eq:DefIij}, and \eqref{Y:result}, and comparing this with \eqref{eq:oprenom} in the same two-point function.  At leading order, we have the following matching condition:
\begin{equation}\label{matching:bdrybdry}
	\begin{split}
		&-\delta\lambda\frac{b^{\hat{\Phi}}_{l}\epsilon^{-\Delta_{ij}}}{\abs{\tau_{12}}^{2\Delta_j}}\sum_{p=0}^{\infty}\frac{C_{jil}a_p^{+}(\Delta_{ij},\Delta_l)+C_{ijl}a_p^{-}(\Delta_{ij},\Delta_l)}{\Delta_{ij}-p}\left(\frac{\epsilon}{\tau_{21}}\right)^p +(\Delta_i\leftrightarrow\Delta_j)\\
		&\qquad\qquad \quad =\sum_{p=0}^{\infty}\Big[\delta Z_{ij;p}\partial^p_{\tau_1}\frac{1}{\abs{\tau_{12}}^{2\Delta_j}}+\delta Z_{ji;p}\partial^p_{\tau_2}\frac{1}{\abs{\tau_{12}}^{2\Delta_i}}\Big]\,. \\
	\end{split}
\end{equation}
By matching the coefficients order by order in $\tau_1-\tau_2$, we get
\begin{equation}\label{eq:zij}
		\delta Z_{ij;p}=-\frac{\delta\lambda\,\epsilon^{p-\Delta_{ij}}}{\left(2\Delta_j\right)_p}\sum_{l} b^{\hat{\Phi}}_{l}\frac{C_{jil}a_p^{+}(\Delta_{ij},\Delta_l)+C_{ijl}a_p^{-}(\Delta_{ij},\Delta_l)}{\Delta_{ij}-p}\,, \qquad \text{for } \quad \,i\neq j\, ,
\end{equation}
and $\delta Z_{ji;p} = (-1)^p(\delta Z_{ij;p} \text{   with    }\Delta_i\leftrightarrow\Delta_j)$.
The consistency between the two expressions follows from property \eqref{apm:prop3} of $a^{\pm}_p$. 

From \eqref{eq:zij} we see that the only relevant field-strength renormalization constants $\delta Z_{ij;p}$ are the ones with $p\leqslant\Delta_{ij}$, since the others vanish in the limit $\epsilon\rightarrow0$.  Equation \eqref{eq:zij} also implies that $\OO_i$ does not mix with heavier operators (i.e.\,$\D_{ji}>0$)  since $\d Z_{ij;p}$ vanishes for all $p$ in this case. Furthermore, for $p\geqslant1$, it follows from property \eqref{apm:prop2} (and the smoothness of $a^{\pm}_p$ in $\Delta_{ij}$) that $\delta Z_{ij;p}$ does not blow up when $\Delta_{ij}=p$.

We can now look at the coefficients when the two boundary operators are the same, which is the case $i=j$. Block by block, this case can be treated as \eqref{Y:result} in the limit $\Delta_i\rightarrow\Delta_j$. Here we only keep the $p=0$ term since the higher-order terms vanish in the limit $\epsilon\rightarrow0$:
\begin{equation}
	\begin{split}
		\left[I_{ii}(\tau_1,\tau_2,\epsilon)\right]_{l}&=\frac{2b^{\hat{\Phi}}_{l}C_{iil}}{\abs{\tau_{12}}^{2\Delta_i}} \lim\limits_{\Delta_j\rightarrow\Delta_i}\left(\frac{a_0(\Delta_{ij},\Delta_l)}{\Delta_{ij}}\abs{\frac{\tau_{21}}{\epsilon}}^{\Delta_{ij}}+i\leftrightarrow j\right) \\
		&=\frac{4b^{\hat{\Phi}}_{l}C_{iil}}{\abs{\tau_{12}}^{2\Delta_i}} a_0(0,\Delta_l)\log\abs{\frac{\tau_{21}}{\epsilon}}. 
        \label{eq:Iiiepslog}
	\end{split}
\end{equation}
Here we have used the properties of $a^{\pm}_0$ in \eqref{apm:prop1}, and the shorthand $a_0\equiv a^{\pm}_0$.

For the two-point function $\braket{\mathcal{O}_i(\tau_1)\mathcal{O}_i(\tau_2)}$, we use the Ansatz \eqref{eq:oprenom}, keeping only the $p=0$ term. The matching condition at order $O(\delta\lambda)$ becomes
\begin{equation}\label{matching:bdrybdry_nonid}
	\begin{split}
		-\delta\lambda\sum\limits_{l}\left[I_{ii}(\tau_1,\tau_2,\epsilon)\right]_{l} =\frac{2\delta Z_{ii;0}}{\abs{\tau_1-\tau_2}^{2\Delta_i}}-\frac{2\delta\Delta_i}{\abs{\tau_1-\tau_2}^{2\Delta_i}}\log\abs{\tau_1-\tau_2}, \\
	\end{split}
\end{equation}
which leads to 
\begin{equation}\label{zii:result}
	\begin{split}
		\delta Z_{ii;0}&=2\delta\lambda\sum\limits_{l}b^{\hat{\Phi}}_{l}\,C_{iil}\,a_0(0,\Delta_l)\,\log{\epsilon}\,, \\
		\delta\Delta_i&=2\delta\lambda\sum\limits_{l}b^{\hat{\Phi}}_{l}\,C_{iil}\,a_0(0,\Delta_l)\,.
	\end{split}
\end{equation}
The second equation gives us the flow equation for the scaling dimensions:
\begin{equation}\label{eq:flow_scaling_dim_normal}
	\begin{split}
		\frac{d\Delta_i}{d\lambda}=2\sum\limits_{l}b^{\hat{\Phi}}_{l}\,C_{iil}\,a_0(0,\Delta_l)\,,
	\end{split}
\end{equation}
where the explicit expression of $a_0(0,\Delta_l)$ for the normal conformal block is given in \eqref{apm:prop1}.

\subsubsection{Equation II: BOE coefficients}\label{subsec:eq2}

To derive the flow equation for the bulk-to-boundary OPE (BOE) coefficients $b^{\hat{\Phi}}_{i}$, the starting point is the bulk-boundary two-point function
\begin{equation}\label{bulkbdrytwopoint:normalization}
	\braket{\hat{\Phi}(\tau_1, z_1)\mathcal{O}_i(\tau_2)}_{\lambda}
	= b^{\hat{\Phi}}_{i}(\lambda) \left( \frac{z_1}{(\tau_1 - \tau_2)^2 + z_1^2} \right)^{\Delta_i(\lambda)}\,.
\end{equation}

In the following, we restrict our attention to the case where the bulk operator is the deformation operator $\hat{\Phi}$ itself for two reasons.  
First, to obtain a closed system of ODEs which includes the flow equation for scaling dimensions, it suffices to include only $b^{\hat{\Phi}}_l$ as the other bulk operators do not enter equations \,(\ref{eq:dimension flow}--\ref{eq:OPE flow}).  
Second, extending the analysis to all bulk operators would inevitably introduce bulk UV divergences (see below) once the bulk operator becomes heavy.\footnote{We assume that the bulk theory admits a UV completion by a CFT. By “heavy’’ we mean that the UV scaling dimension of the bulk operator is large.}  
In many QFT examples, restricting to the deformation operator alone may avoid such UV divergences.  
We will comment on  general bulk operators in section~\ref{sec: discussion}.

Under an infinitesimal change of the coupling $\lambda \to \lambda + \delta\lambda$, the deformed two-point function becomes
\begin{equation}\label{deformation:bulkbdry}
	\begin{split}
		\braket{\hat{\Phi}^{(\text{bare})}(\tau_1, z_1)\mathcal{O}^{\text{(bare)}}_i(\tau_2)}_{\lambda + \delta\lambda}
		= \braket{\hat{\Phi}(\tau_1, z_1)\mathcal{O}_i(\tau_2)}_{\lambda} - \delta\lambda \int \frac{d\tau\, dz}{z^2} \braket{\hat{\Phi}(\tau, z) \hat{\Phi}(\tau_1, z_1) \mathcal{O}_i(\tau_2)}_\lambda\,,
	\end{split}
\end{equation}
where the operators on the left-hand side are bare operators, similar to \eqref{deformation:2pt}. This expression involves the integral of a bulk-bulk-boundary correlator, which may develop two types of UV divergences:
\begin{enumerate}
	\item A bulk UV divergence when $(\tau, z) \to (\tau_1, z_1)$,
	\item A bulk-boundary UV divergence when $(\tau, z) \to (\tau_2, 0)$.
\end{enumerate}
The bulk UV divergence is controlled by the OPE of the UV CFT in the bulk\footnote{This is the form of the OPE when scalars are exchanged. We neglect the contributions of spinning operators to avoid clutter, but in general they can appear.}:
\begin{equation}
	\hat{\Phi}(\tau, z)\hat{\Phi}(\tau_1, z_1) 
	= \sum_{\hat\OO\in\hat\F\times\hat\F} C^\text{UV}_{\hat{\Phi}\hat{\Phi}\hat{\mathcal{O}}}\sum\limits_{n=0}^{\infty}c_n 
	\left( \frac{z z_1}{(\tau - \tau_1)^2 + (z - z_1)^2} \right)^{\Delta_{\hat{\Phi}} - \tfrac{1}{2} \Delta_{\hat{\mathcal{O}}}-n} 
	\hat{\mathcal{O}}(\tau_1, z_1)\,.
\end{equation}
This expansion is understood asymptotically in the distance variable. The coefficients $C^\text{UV}$ and scaling dimensions $\Delta_{\hat{\mathcal{O}}}$ are those of the UV bulk theory. The coefficients $c_n$ are normalized by $c_0=1$ and the terms $n\geqslant 1$ describe curvature corrections to the usual OPE in flat space. The sum runs over a complete operator basis, including both primaries and descendants. Convergence of the integral in \eqref{deformation:bulkbdry} requires
\begin{equation}
	\Delta_{\hat{\Phi}} \leqslant 1+ \frac{1}{2}\min\limits_{\substack{\hat{\mathcal{O}}\in\hat{\Phi}\times\hat{\Phi} \\b^{\hat{\mathcal{O}}}_i\neq0}}\Delta_{\hat{\mathcal{O}}}\,.
\end{equation}
If that is not the case, we have to introduce an extra cutoff near the bulk point $(\tau_1,z_1)$ and carefully renormalize the bulk operator to cancel this type of divergence.
In this section, we restrict ourselves to the case where this condition is met, thus no renormalization of bulk operators is required. In this case, the bare bulk operator coincides with the renormalized bulk operator. We will comment on the general case in section~\ref{sec: discussion}.

The bulk-boundary divergence is absorbed by the field-strength renormalization of the boundary operators $\mathcal{O}_i$, as discussed in the previous subsection. We will now verify that this renormalization suffices to remove all bulk-boundary UV divergences arising in \eqref{deformation:bulkbdry}. To do this, we repeat a similar analysis as in the previous subsection. We start by defining the truncated integral 
\begin{equation}\label{def:Ji}
	J^{\hat{\Phi}}_i(\tau_1, z_1; \tau_2; \epsilon)
	:=\int_{(\tau - \tau_2)^2 + z^2 \geqslant \epsilon^2} \frac{d\tau\, dz}{z^2} \braket{\hat{\Phi}(\tau, z) \hat{\Phi}(\tau_1, z_1) \mathcal{O}_i(\tau_2)}_\lambda.
\end{equation}
Assuming the absence of bulk UV divergences, this is a finite integral when $\epsilon>0$. 

By $SO^+(1,2)$ symmetry, the three-point function takes the form discussed in section \ref{subsec:bkbkbdCB}:
\begin{equation}
	    \braket{\hat{\Phi}(\tau, z) \hat{\Phi}(\tau_1, z_1) \mathcal{O}_i(\tau_2)}_\lambda
	= \left( \frac{z_1}{(\tau_1 - \tau_2)^2 + z_1^2} \right)^{\Delta_i} \mathcal{G}^{\hat{\Phi}\hat{\Phi}}_i(\upsilon,\zeta)\, ,
    \label{bkbkbd:PhiPhiO}
\end{equation}
where $\upsilon$ and $\zeta$ are defined in \eqref{def:upsilon and zeta} and reminded here for convenience
\begin{equation} \label{eq:crossratiosbkbkbd}
    \upsilon=\frac{\tau_{12}(\tau-\tau_1)(\tau-\tau_2)+z^2\tau_{12}+z_1^2(\tau_2-\tau)}{z_1(z^2+(\tau-\tau_2)^2)}\,, \qquad 
		\zeta=\frac{z(z_1^2+\tau_{12}^2)}{z_1(z^2+(\tau-\tau_2)^2)}\,.
\end{equation}

\begin{figure}
	\centering
	\begin{tikzpicture}[baseline={(0,-0.2)}]

\node[left] at (-4,1) {$\hat{\Phi}(\tau,z)$};
\draw (-4,1) -- (-2.9,1);

\node at (-2.4,1) {$\mathcal{O}_l$};

\draw (-1.95,1) -- (-1,0);

\node[left] at (-4,-1) {$\hat{\Phi}(\tau_1, z_1)$};
\draw (-4,-1) -- (-2.9,-1);

\node at (-2.4,-1) {$\mathcal{O}_j$};

\draw (-1.95,-1) -- (-1,0);

\filldraw (-1,0) circle (1.5pt);

\draw (-1,0) -- (0.4,0);
\node at (0.8,0) {$\mathcal{O}_i$};

\end{tikzpicture}
	\caption{Conformal block decomposition of $\braket{\hat{\Phi}(\tau, z) \hat{\Phi}(\tau_1, z_1) \mathcal{O}_i(\tau_2)}$, where $\hat{\Phi}(\tau,z)$ is expanded into $\mathcal{O}_l$ and $\hat{\Phi}(\tau_1,z_1)$ into $\mathcal{O}_j$.}
	\label{fig:bkbkbdryope}
\end{figure}
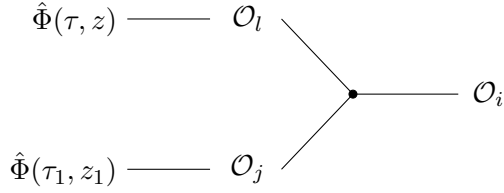

\begin{figure}
	\centering
	\includegraphics[width=0.4\textwidth]{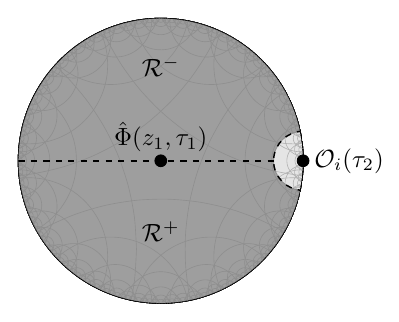}
   \includegraphics[width=0.48\textwidth]{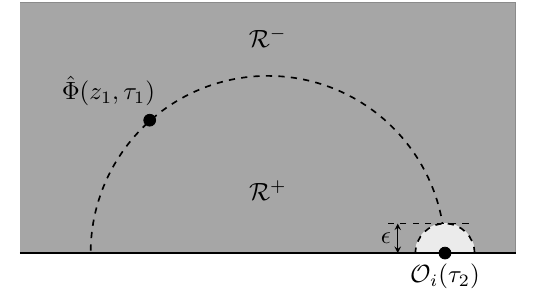}
	\caption{In dark gray, the integration region of (\ref{def:Ji}). We split it into $\mathcal{R}^\pm$ as described in (\ref{def:IntR}).}
	\label{fig:bkbkbdry}
\end{figure}

The three-point function $ \mathcal{G}^{\hat{\Phi}\hat{\Phi}}_i$ has the conformal block expansion (see figure~\ref{fig:bkbkbdryope}):
\begin{equation}
\label{bkbkbd:expPhiPhiO}	\mathcal{G}^{\hat{\Phi}\hat{\Phi}}_i(\upsilon,\zeta) = 
	\begin{cases}
		\sum\limits_{l,j} b^{\hat{\Phi}}_l b^{\hat{\Phi}}_j\, C_{ilj}\, R_{\Delta_l\, \Delta_j}^{\Delta_i}(\upsilon,\zeta)\, , & \quad\text{for }\upsilon<0, \\[1em]
		\sum\limits_{l,j} b^{\hat{\Phi}}_l b^{\hat{\Phi}}_j\, C_{ijl}\, R_{\Delta_l\, \Delta_j}^{\Delta_i}(\upsilon,\zeta)\, , &\quad  \text{for }\upsilon>0\, .
	\end{cases}
\end{equation}
The regimes where $\upsilon<0$ and $\upsilon>0$ are shown in figure~\ref{fig:bkbkbdry}. Note that as in the previous case, the blocks $R_{\Delta_l\, \Delta_j}^{\Delta_i}$ are universal.

\emph{Assuming} that the integral in \eqref{def:Ji} commutes with the block expansion, we write:
\begin{equation}
	\begin{split}
	    J^{\hat{\Phi}}_i(\tau_1, z_1; \tau_2; \epsilon) &= \left( \frac{z_1}{(\tau_1 - \tau_2)^2 + z_1^2} \right)^{\Delta_i} \\
        &\quad \quad\times\sum_{l,j} b^{\hat{\Phi}}_l\, b^{\hat{\Phi}}_j \, \left[ C_{ilj} \mathcal{R}^{\Delta_i;-}_{\Delta_l,\Delta_j}(\tau_1,z_1,\tau_2,\epsilon) + C_{ijl} \mathcal{R}_{\Delta_l,\Delta_j}^{\Delta_i;+}(\tau_1,z_1,\tau_2,\epsilon) \right]\,,
	\end{split}
\end{equation}
where $\mathcal{R}^{+}$ and $\mathcal{R}^{-}$ are the integrated blocks, defined by
\begin{equation}\label{def:IntR}
	\begin{split}
		\mathcal{R}_{\Delta_l,\Delta_j}^{\Delta_i;\pm}(\tau_1,z_1,\tau_2,\epsilon)&:=\int_{\substack{ \text{sgn}(\upsilon)=\pm\\(\tau-\tau_2)^2+z^2\geqslant\epsilon^2 }}
        \frac{d\tau dz}{z^2}R_{\Delta_l,\Delta_j}^{\Delta_i}(\upsilon,\zeta)\,.
	\end{split}
\end{equation}

\paragraph{Regular contributions.}
When $\Delta_j > \Delta_i$, the integral in \eqref{def:IntR} converges without a cutoff.\footnote{In this section, we always assume $\Delta_l>1$. Consequently, for generic $\tau$, there is no divergence around $z=0$.} We leave the details to appendix \ref{subsec:integrated bkbkbd block}.

Both the integrands and the integration regions are $SO^+(1,2)$ invariant, so the final results must be pure numbers (i.e., independent of spacetime coordinates). Moreover, by reflection symmetry, the two integrals yield the same value:
\begin{equation}
	\mathcal{R}_{\Delta_l,\Delta_j}^{\Delta_i;+}(\tau_1,z_1,\tau_2,\epsilon=0)
	= \mathcal{R}_{\Delta_l,\Delta_j}^{\Delta_i;-}(\tau_1,z_1,\tau_2,\epsilon=0)
	= -\frac{1}{2} \, \JJ_{\Delta_i}(\Delta_l, \Delta_j)\, ,\quad (\Delta_j>\Delta_i).
\end{equation}
Here we have introduced the shorthand
\begin{equation}\label{def:CalJ}
	\JJ_{\Delta_i}(\Delta_l, \Delta_j)
	:= -\int_{\text{AdS}} \frac{d\tau\, dz}{z^2} R_{\Delta_l,\Delta_j}^{\Delta_i}(\upsilon,\zeta)\, ,\qquad\quad(\Delta_j>\Delta_i)\, ,
\end{equation}
which denotes the integrated conformal block over the entire AdS space. The minus sign is just a convention which will make the final result more concise. Recall that $\Delta_l$ corresponds to the integrated bulk operator, while $\Delta_j$ corresponds to the fixed bulk operator. The explicit expression of $\JJ_{\Delta_i}(\Delta_l, \Delta_j)$ can be found in \eqref{eq:J}.

\paragraph{Singular contributions.}
When $\Delta_j \leqslant \Delta_i$,  the integrals diverge and we need to impose a cutoff. We define subtraction terms:
\begin{equation}\label{def:Jsubtr}
	\begin{split}
		\JJ^{\pm,\text{subtr}}_{\Delta_i}(\Delta_l, \Delta_j;\tau_1,z_1;\tau_2;\epsilon)
		&:= \int_{\substack{\text{sgn}(\upsilon)=\pm\\(\tau-\tau_2)^2+z^2\leqslant\epsilon^2 }} \frac{d\tau\, dz}{z^2} R_{\Delta_l\, \Delta_j}^{\Delta_i}(\upsilon,\zeta). \\
	\end{split}
\end{equation}
For $\Delta_j>\Delta_i$, the subtraction term is well-defined, and we have
\begin{equation}\label{J:identity}
	\begin{split}
		\mathcal{R}_{\Delta_l\, \Delta_j}^{\Delta_i;\pm}(\tau_1,z_1,\tau_2,\epsilon)=-\frac{1}{2}\JJ_{\Delta_i}(\Delta_l, \Delta_j)-\JJ^{\pm,\text{subtr}}_{\Delta_i}(\Delta_l, \Delta_j;\tau_1,z_1;\tau_2;\epsilon)\,.
	\end{split}
\end{equation}
Note that for fixed $\epsilon>0$, the quantity
$\mathcal{R}_{\Delta_l\, \Delta_j}^{\Delta_i;\pm}(\tau_1,z_1,\tau_2,\epsilon)$
is finite and analytic in $\Delta_j$ when $\mathrm{Re}(\Delta_j)>\mathrm{Re}(\Delta_i)$. One might naively expect the same to hold for other values of $\Delta_j$, since the divergent contribution near $(\tau,z)=(\tau_2,0)$ has been subtracted. However, assuming the absence of bulk UV divergences, this statement is true for the integrated full correlator but not for the integrated $B\partial\partial$ conformal block. 

The reason is that the conformal block
$R_{\Delta_l,\Delta_j}^{\Delta_i}(\upsilon,\zeta)$
behaves as $\zeta^{\Delta_{ij}}$ as $\zeta\to\infty$. This behavior leads to additional divergences in the integral that cannot be absorbed by field renormalization, since the divergent piece is not localized around the boundary point. Nevertheless, for the moment we will proceed as if this issue were absent. Later, we will replace the conformal block with the local block, for which this problem does not arise.

Therefore, we formally assume that the left-hand side of \eqref{J:identity} continues to make sense for $\Delta_j\leqslant\Delta_i$, while the quantities $\JJ_{\Delta_i}(\Delta_l,\Delta_j)$ and $\JJ^{\pm,\text{subtr}}_{\Delta_i}$ on the right-hand side are understood as their analytic continuations in $\Delta_j$. The case $\Delta_j=\Delta_i$ requires special care, since both $\JJ_{\Delta_i}(\Delta_l,\Delta_j)$ and $\JJ^{\pm,\text{subtr}}_{\Delta_i}$ develop simple poles as functions of $\Delta_j$.




Now let us carefully match the bulk-boundary two-point function at coupling $\lambda + \delta\lambda$. The renormalized operators satisfy
\be
\label{eq:bkbd two-pt ren}
\<\hat\F(\t_1,z_1)\OO_i^{\text{(ren)}}(\t_2)\>_{\l+\d\l}=(b^{\hat\F}_{i}+\d b^{\hat\F}_{i})\left(\frac{z_1}{\t_{12}^2+z_1^2}\right)^{\D_i+\d\D_i}\,.
\ee
Recalling the field-strength renormalization \eqref{eq:oprenom} for the boundary operator $\mathcal{O}_i$, 
then the matching condition for the bulk-boundary two-point function is, similarly to the previous case, obtained by identifying two ways to write the same two-point function $\<\hat\F\OO_i^{\text{(bare)}}\>$, namely the RHS of \eqref{deformation:bulkbdry} and using \eqref{eq:oprenom} in the LHS of \eqref{deformation:bulkbdry}. We have
\begin{align}\label{matching:bulkbdry}
&\sum_{j}\sum_{p=0}^{\infty}Z_{ij;p}\partial^p_{\tau_2}\braket{\hat{\Phi}(\tau_1, z_1)\mathcal{O}^{(\text{ren})}_j(\tau_2)}_{\lambda + \delta\lambda} \\
		&\quad\quad=\braket{\hat{\Phi}(\tau_1, z_1)\mathcal{O}_i(\tau_2)}_{\lambda}+ \delta\lambda\sum_{l,j}b^{\hat{\Phi}}_l\, b^{\hat{\Phi}}_j \left( \frac{z_1}{(\tau_1 - \tau_2)^2 + z_1^2} \right)^{\Delta_i}\times\Bigg{[}\frac{ C_{ilj}+C_{ijl}}{2} \JJ_{\Delta_i}(\Delta_l, \Delta_j)\nonumber \\
		&\qquad\quad \qquad \qquad +C_{ilj}\JJ^{+,\text{subtr}}_{\Delta_i}(\Delta_l, \Delta_j;\tau_1,z_1;\tau_2;\epsilon) + C_{ijl} \JJ^{-,\text{subtr}}_{\Delta_i}(\Delta_l, \Delta_j;\tau_1,z_1;\tau_2;\epsilon)\Bigg{]}\,. \nonumber
	\end{align}
We claim that in \eqref{matching:bulkbdry}:
\begin{itemize}
	\item For $j \neq i$, the analytically continued subtraction terms match the $Z_{ij;p}$ contributions:
    \begin{equation}
        \begin{split}
            &\sum_{p=0}^\infty Z_{ij;p}\partial^p_{\tau_2}\braket{\hat{\Phi}(\tau_1, z_1)\mathcal{O}^{(\text{ren})}_j(\tau_2)}_{\lambda + \delta\lambda} =\delta\lambda\sum_{l}b^{\hat{\Phi}}_l\,b^{\hat{\Phi}}_j\,\left( \frac{z_1}{(\tau_1 - \tau_2)^2 + z_1^2} \right)^{\Delta_i} \\
            &\times\Bigg{[} C_{ilj}  \JJ^{+,\text{subtr}}_{\Delta_i}(\Delta_l, \Delta_j;\tau_1,z_1;\tau_2;\epsilon) + C_{ijl} \JJ^{-,\text{subtr}}_{\Delta_i}(\Delta_l, \Delta_j;\tau_1,z_1;\tau_2;\epsilon)\Bigg{]}\,. \\
        \end{split}
    \end{equation}
	\item For $j = i$, we can take the limit
	\begin{equation}
		\begin{split}
			\lim\limits_{\Delta_j\rightarrow\Delta_i}\left(\frac{1}{2}\JJ_{\Delta_i}(\Delta_l, \Delta_j) + \JJ^{\pm,\text{subtr}}_{\Delta_i}(\Delta_l, \Delta_j;\tau_1,z_1;\tau_2;\epsilon)\right).
		\end{split}
	\end{equation}
	The poles cancel out. The $O(1)$ part of the subtraction terms, which are logarithmic, cancel against $Z_{ii;0}$ and $\delta\Delta_i$. So only the finite part of $\JJ_{\Delta_i}$ is left.
\end{itemize}
We leave the justification of the above claims to appendix~\ref{app:HardvsAnal}. 

After all cancellations, we obtain:
\begin{equation}\label{matching:bkbdfinal}
	\begin{split}
		&\delta b^{\hat{\Phi}}_i \left( \frac{z_1}{(\tau_1 - \tau_2)^2 + z_1^2} \right)^{\Delta_i} =\delta\lambda\sum\limits_{l,j} b^{\hat{\Phi}}_l\, b^{\hat{\Phi}}_j \left( \frac{z_1}{(\tau_1 - \tau_2)^2 + z_1^2} \right)^{\Delta_i}\frac{C_{ilj}+C_{ijl}}{2}\left[\JJ_{\Delta_i}(\Delta_l, \Delta_j)\right]_\text{reg}\,,\\
	\end{split}
\end{equation}
where $\left[ \JJ_{\Delta_i}(\Delta_l, \Delta_j) \right]_\text{reg}$ denotes the analytically continued value of $\JJ_{\Delta_i}(\Delta_l, \Delta_j)$ (as a function of $\Delta_j$), with the pole subtracted when $\Delta_j=\Delta_i$:
\begin{equation}\label{def:Jreg}
    [\mathcal{J}_{\Delta_i}(\Delta_l,\Delta_j) ]_\text{reg}:=
    \oint_{\Delta_j} \frac{d\Delta}{2\pi i }\frac{\mathcal{J}_{\Delta_i}(\Delta_l,\Delta)}{\D-\D_j}\,.
\end{equation}
Thus, the flow equation for the BOE coefficients reads:
\begin{equation}\label{eq:flow_BOE_normal}
	\frac{db^{\hat{\Phi}}_i}{d\lambda}
	= \sum_{l,j} b^{\hat{\Phi}}_l\, b^{\hat{\Phi}}_j  \frac{C_{ilj} + C_{ijl}}{2} \left[ \JJ_{\Delta_i}(\Delta_l, \Delta_j) \right]_\text{reg}\,,
\end{equation}
where the explicit expression of $\JJ_{\D_i}(\D_l,\D_j)$ is given in \eqref{eq:J}. We can see from the expression (\ref{eq:J}) that $\JJ_{\D_i}(\D_l,\D_j)$ has poles at $\Delta_{ji}=0,-2,-4\ldots$ The pole at $\Delta_j=\Delta_i$ is expected and is canceled by the pole arising from the subtraction term. The remaining poles are not physical, and we will later see that they do not appear in the integrated local block.

When we choose the boundary operator to be the identity $\mathcal{O}_i = \mathbb{1}$, the flow equation simplifies to
\begin{equation}
	\frac{db^{\hat{\Phi}}_\mathbb{1}}{d\lambda}
	= \sum_{j} \left(b^{\hat{\Phi}}_j\right)^2 \left[ \JJ_{0}(\Delta_j, \Delta_j) \right]_\text{reg}.
\end{equation}
In general, the right-hand side is nonzero, which means that $\hat{\Phi}$ acquires a vacuum expectation value (VEV) under an infinitesimal deformation of the coupling. If the scaling dimension of $\hat{\Phi}$ in the UV bulk theory is $\Delta_{\hat{\Phi}}\geqslant1$, 
its vacuum expectation value (VEV) $\braket{\hat{\Phi}}$ becomes divergent due to a UV divergence in the bulk. Even if $\Delta_{\hat{\Phi}}<1$, i.e. this VEV is finite, the second-order variation of the QFT data—corresponding to the next step in the deformation—will also diverge, because integrating the identity operator produces the (divergent) volume of AdS (IR divergence).\footnote{We thank Edoardo Lauria and Balt van Rees for pointing out this issue.}

However, we are always free to subtract such a contribution from $\hat{\Phi}$ without violating bulk locality. That is, we can shift $\hat{\Phi} \to \hat{\Phi} - \braket{\hat{\Phi}}\mathbb{1}$ without affecting its status as a local bulk operator. As a result, we will impose the condition
\begin{equation}
	b^{\hat{\Phi}}_\mathbb{1} \equiv 0\, .
\end{equation}
for all values of the coupling $\lambda$ (except in the trivial case where $\hat{\Phi}$ itself is the identity operator).

\subsubsection{Equation III: boundary OPE coefficients}\label{subsec:eq3}

The last flow equation to derive is that for the boundary OPE coefficients $C_{ijk}$, which requires analyzing the boundary three-point function:
\begin{equation}\label{bdrythreepoint}
	\braket{\mathcal{O}_i(\tau_1)\mathcal{O}_j(\tau_2)\mathcal{O}_k(\tau_3)}_{\lambda}
	= \begin{cases}
		\frac{C_{ijk}(\lambda)}{\abs{\tau_{12}}^{\Delta_{ijk}(\lambda)}\abs{\tau_{23}}^{\Delta_{jki}(\lambda)}\abs{\tau_{13}}^{\Delta_{ikj}(\lambda)}}\, , &\qquad  [ijk], \\
		\frac{C_{ikj}(\lambda)}{\abs{\tau_{12}}^{\Delta_{ijk}(\lambda)}\abs{\tau_{23}}^{\Delta_{jki}(\lambda)}\abs{\tau_{13}}^{\Delta_{ikj}(\lambda)}}\, , & \qquad [ikj], \\
	\end{cases}
\end{equation}
where we use the shorthands $\tau_{ij} \equiv \tau_i - \tau_j$ and $\Delta_{ijk} \equiv \Delta_i + \Delta_j - \Delta_k$. The symbols $[ijk]$ and $[ikj]$ denote the cyclic orderings of the boundary operators—for example, $[ijk]$ includes $\tau_1 < \tau_2 < \tau_3$, $\tau_2 < \tau_3 < \tau_1$, or $\tau_3 < \tau_1 < \tau_2$.

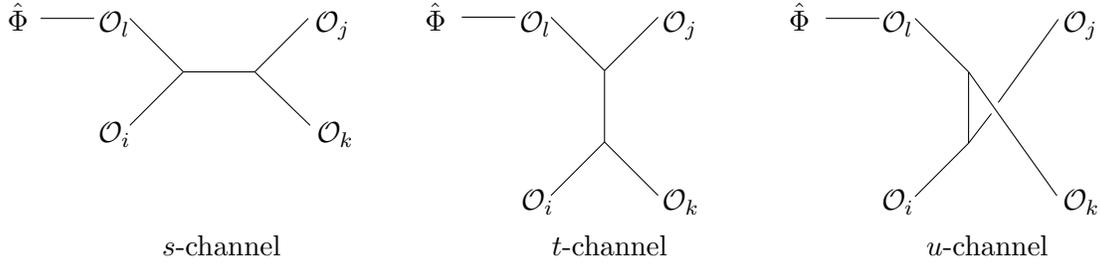
\begin{figure}
\centering
   \begin{tikzpicture}[x=0.75pt,y=0.75pt,yscale=-0.9,xscale=0.9]

\draw    (30,70) -- (60,70) ;
\draw    (266,69.33) -- (296,69.33) ;
\draw    (80,70) -- (110,100) ;
\draw    (150,100) -- (181,130) ;
\draw    (150,100) -- (180,70) ;
\draw    (80,130) -- (110,100) ;
\draw    (316,69.33) -- (346,99.33) ;
\draw    (346,139.33) -- (376,169.33) ;
\draw    (346,99.33) -- (376,69.33) ;
\draw    (316,169.33) -- (346,139.33) ;
\draw    (470,70) -- (500,70) ;
\draw    (520,70) -- (550,100) ;
\draw    (550,100) -- (599.5,169.5) ;
\draw    (550,140) -- (562.8,123.4) ;
\draw    (520,170) -- (550,140) ;
\draw    (566.6,117.2) -- (600.3,70.5) ;
\draw    (110,100) -- (150,100) ;
\draw    (346,99.33) -- (346,139.33) ;
\draw    (550,100) -- (550,140) ;

\draw (10,60) node [anchor=north west][inner sep=0.75pt]    {$\hat{\Phi }$};
\draw (244,60) node [anchor=north west][inner sep=0.75pt]    {$\hat{\Phi }$};
\draw (61,65) node [anchor=north west][inner sep=0.75pt]    {$\mathcal{O}_l$};
\draw (297,65) node [anchor=north west][inner sep=0.75pt]    {$\mathcal{O}_l$};
\draw (448,60) node [anchor=north west][inner sep=0.75pt]    {$\hat{\Phi }$};
\draw (500,65) node [anchor=north west][inner sep=0.75pt]    {$\mathcal{O}_l$};
\draw (601.5,65) node [anchor=north west][inner sep=0.75pt]    {$\mathcal{O}_j$};
\draw (601.5,165) node [anchor=north west][inner sep=0.75pt]    {$\mathcal{O}_k$};
\draw (500,165) node [anchor=north west][inner sep=0.75pt]    {$\mathcal{O}_i$};
\draw (377,65) node [anchor=north west][inner sep=0.75pt]    {$\mathcal{O}_j$};
\draw (377,165) node [anchor=north west][inner sep=0.75pt]    {$\mathcal{O}_k$};
\draw (298,165) node [anchor=north west][inner sep=0.75pt]    {$\mathcal{O}_i$};
\draw (182.4,65) node [anchor=north west][inner sep=0.75pt]    {$\mathcal{O}_{j}$};
\draw (183.4,127) node [anchor=north west][inner sep=0.75pt]    {$\mathcal{O}_{k}$};
\draw (61,127) node [anchor=north west][inner sep=0.75pt]    {$\mathcal{O}_{i}$};
\draw (96.67,190) node [anchor=north west][inner sep=0.75pt]   [align=center] {$s$-channel};
\draw (315,190) node [anchor=north west][inner sep=0.75pt]   [align=center] {$t$-channel};
\draw (525,190) node [anchor=north west][inner sep=0.75pt]   [align=center] {$u$-channel};

\end{tikzpicture}
\caption{Conformal block decomposition of $\braket{\hat{\Phi}(\tau,z)\mathcal{O}_i(\tau_1)\mathcal{O}_j(\tau_2)\mathcal{O}_k(\tau_3)}$, where $\hat{\Phi}(\tau,z)$ is expanded into $\mathcal{O}_l$ and the boundary four-point function admits three possible OPE channels.}
	\label{fig:bkbdrybdrybdryope}
\end{figure}

In what follows, we focus on the case $\tau_1 < \tau_2 < \tau_3$ as the analysis for other orderings proceeds similarly. Under an infinitesimal change of the coupling $\lambda \to \lambda + \delta\lambda$, the boundary three-point function gets deformed as
\begin{equation}\label{deformation:3pt}
	\begin{split}
		\braket{\mathcal{O}^{\text{(bare)}}_i(\tau_1)\mathcal{O}^{\text{(bare)}}_j(\tau_2)\mathcal{O}^{\text{(bare)}}_k(\tau_3)}_{\lambda+\delta\lambda}
		&= \braket{\mathcal{O}_i(\tau_1)\mathcal{O}_j(\tau_2)\mathcal{O}_k(\tau_3)}_{\lambda} \\
		&\quad - \delta\lambda \int \frac{d\tau\, dz}{z^2} \braket{\hat{\Phi}(\tau,z)\mathcal{O}_i(\tau_1)\mathcal{O}_j(\tau_2)\mathcal{O}_k(\tau_3)}_{\lambda}\,.
	\end{split}
\end{equation}
We are thus led to study the integrated four-point function
\begin{equation}
	\begin{split}
		Q_{ijk}^{\hat{\Phi}}(\tau_1, \tau_2, \tau_3; \epsilon)
		:= \int_{(\tau - \tau_i)^2 + z^2 \geqslant \epsilon^2} \frac{d\tau\, dz}{z^2} \braket{\hat{\Phi}(\tau,z)\mathcal{O}_i(\tau_1)\mathcal{O}_j(\tau_2)\mathcal{O}_k(\tau_3)}_{\lambda}\,.
	\end{split}
\end{equation}
This case differs from the previous ones ($B\partial\partial$ and $BB\partial$), where only a single OPE channel was available. Here, the four-point function admits three boundary OPE channels—namely, the $s$-, $t$-, and $u$-channels—as shown in figure~\ref{fig:bkbdrybdrybdryope}.


\begin{figure}
	\centering
	\includegraphics[scale=1]{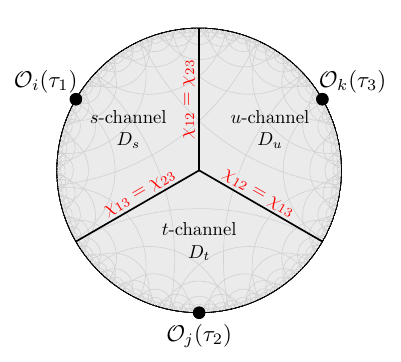}
	\caption{Regions of AdS where the $s$-, $t$-, and $u$-channel OPEs are chosen.}
	\label{fig:stu}
\end{figure}

The three OPE channels each have their own domain of convergence. None of them covers the entire AdS space, so we must select the appropriate channel depending on the region over which the integration is performed. We adopt the following natural prescription. The four-point function admits three natural cross-ratios
$\chi_{12}\, , \chi_{13}\,$and $\chi_{23}$
where the $\chi$-variables are defined in \eqref{def:chi}, and the subscripts indicate the pairs of boundary points involved. Our choice of OPE channels is then determined by the smallest cross-ratio (see also figure~\ref{fig:stu}):
\begin{equation}\label{def:DsDtDu}
	\begin{split}
		s\text{-channel:}&\quad D_s:=\left\{(\tau,z)\ \big{|}\ \chi_{23} \leqslant \chi_{12},\, \chi_{13}\right\}\,, \\
		t\text{-channel:}&\quad D_t:=\left\{(\tau,z)\ \big{|}\ \chi_{13} \leqslant \chi_{12},\, \chi_{23}\right\}\,, \\
		u\text{-channel:}&\quad D_u:=\left\{(\tau,z)\ \big{|}\ \chi_{12} \leqslant \chi_{13},\, \chi_{23}\right\}\,. \\
	\end{split}
\end{equation}

Before proceeding, let us recall the important subtlety of convergence. If we use the standard conformal block expansion — namely, the one described in section~\ref{subsec:bkbdbdbdCB} — there exists a region in the bulk where none of the three OPE channels converge, see figure~\ref{fig:no_stu}. Later, we will resolve this problem by introducing a different type of conformal blocks, referred to as \emph{local blocks}, which ensure convergence in the entire AdS. For the time being, we will proceed without addressing this issue in detail, and our arguments below will not rely on the explicit form of the conformal blocks.

Let us focus on the integration over the s-channel domain $D_s$:
\begin{equation}\label{def:Qs}
	\begin{split}
		\left[Q_{ijk}^{\hat{\Phi}}(\tau_1, \tau_2, \tau_3; \epsilon)\right]_s
		:= \int_{\substack{(\tau,z)\in D_s \\ (\tau-\tau_1)^2+z^2\geqslant\epsilon^2} } \frac{d\tau\, dz}{z^2} \braket{\hat{\Phi}(\tau,z)\mathcal{O}_i(\tau_1)\mathcal{O}_j(\tau_2)\mathcal{O}_k(\tau_3)}_{\lambda}\,,
	\end{split}
\end{equation}
where $\epsilon$ denotes the hard cutoff. 

As described in section~\ref{subsec:bkbdbdbdCB}, the integrand has convergent $s$-channel expansion for $(\tau,z)$ in part of $D_s$:
\begin{equation}
    \begin{split}
        \braket{\hat{\Phi}(\tau,z)\mathcal{O}_i(\tau_1)\mathcal{O}_j(\tau_2)\mathcal{O}_k(\tau_3)}&=\frac{1}{|\tau_{12}|^{\Delta_{ijk}}|\tau_{23}|^{\Delta_{jki}}|\tau_{31}|^{\Delta_{kij}}} \\
        &\times\begin{cases}
			\displaystyle\sum\limits_{m,l}
			b^{\hat{\Phi}}_{l}\,C_{jkm}\,C_{ilm}\,
			K^{\Delta_i\Delta_j\Delta_k}_{\Delta_l\Delta_m}(\mu,\omega)\, ,
			& \mu^2+\omega^2<\mu, \\[8pt]
			\displaystyle\sum\limits_{m,l}
			b^{\hat{\Phi}}_{l}\,C_{jkm}\,C_{lim}\,
			K^{\Delta_i\Delta_j\Delta_k}_{\Delta_l\Delta_m}(\mu,\omega)\, ,
			& \mu<0,
		\end{cases} \\
    \end{split}
\end{equation}
where $\mu$ and $\omega$ are the $SO^+(1,2)$ invariants defined in \eqref{def:mu and omega}, and the conformal block $K^{\Delta_i\Delta_j\Delta_k}_{\Delta_l\Delta_m}(\mu,\omega)$ is given in \eqref{def:bkbdbdbd_sblock_normal}. In terms of $\mu$ and $\omega$, $D_s$ is given by the conditions
\begin{equation}
{D_s}:\quad\mu^2+\omega^2\leqslant1 \quad\text{and}\quad\mu\leqslant\frac{1}{2}\,,
\end{equation}
which is not fully covered by the domain of convergence (see figure~\ref{fig:Ds_mu_omega}). However, we would like to write down the $s$-channel expansion in a heuristic way, which formally covers the whole $D_s$. In the next subsection, we will expand the correlator in the same way, but the building blocks will be replaced by the local blocks.

Recall that the $B\partial\partial\partial$ can be expanded into a sum of integrated $B\partial\partial$ blocks, eq.~\eqref{bkbdbdbd:reduction_bkbdbd}. Plugging this expansion into \eqref{def:Qs}, we get
\begin{equation}
    \begin{split}
        \left[Q_{ijk}^{\hat{\Phi}}(\tau_1, \tau_2, \tau_3; \epsilon)\right]_s
		= \sum_{m}C_{jkm}\int_{\substack{(\tau,z)\in D_s \\ (\tau-\tau_1)^2+z^2\geqslant\epsilon^2} } \frac{d\tau\, dz}{z^2} \int_{\tau_2}^{\tau_3}\!&d\tau_4\,
		\mathcal{B}_{\Delta_j\Delta_k\Delta_m}(\tau_2,\tau_3,\tau_4) \\
        &\times\braket{\hat{\Phi}(\tau,z)\mathcal{O}_i(\tau_1)\mathcal{O}_m(\tau_4)}\,.
    \end{split}
\end{equation}
Now if we first fix $\tau_4$ and integrate over $\tau$ and $z$, for almost all $(\tau,z)\in D_s$ we know how to expand $\braket{\hat{\Phi}(\tau,z)\mathcal{O}_i(\tau_1)\mathcal{O}_m(\tau_4)}$ into conformal blocks: we just need to compute the sign of $\rho_{14}$ and decide whether we should use the first or the second line of \eqref{bkbdbd:exp}. Using this trick, we get
\begin{equation}\label{QsInt:exp_normal}
	\begin{split}
		\left[Q_{ijk}^{\hat{\Phi}}(\tau_1, \tau_2, \tau_3; \epsilon)\right]_s
		&= \sum_{l,m} b^{\hat{\Phi}}_{l}\, C_{jkm}\left(C_{ilm}\, \mathcal{Q}^{\Delta_l,\Delta_m;+}_{\Delta_i,\Delta_j,\Delta_k}(\tau_1, \tau_2, \tau_3; \epsilon)+C_{iml}\, 
        \mathcal{Q}^{\Delta_l,\Delta_m;-}_{\Delta_i,\Delta_j,\Delta_k}(\tau_1, \tau_2, \tau_3; \epsilon)
        \right),
	\end{split}
\end{equation}
where the two terms originate from integrating over the two regions of $D_s$ which are separated by the geodesic joining $\tau_1$ and $\tau_4$, as shown in figure \ref{fig:Ds_mu_omega}. Explicitly,
\begin{equation}\label{def:IntQ_normal}
	\begin{split}
		\mathcal{Q}^{\Delta_l,\Delta_m;\pm}_{\Delta_i,\Delta_j,\Delta_k}&(\tau_1, \tau_2, \tau_3; \epsilon)
		:= \int_{\tau_2}^{\tau_3} d\tau_4\,\mathcal{B}_{\Delta_j\Delta_k\Delta_m}(\tau_2,\tau_3,\tau_4) \\
		&\times \int_{\substack{(\tau,z)\in D_s^\pm \\ (\tau-\tau_1)^2+z^2\geqslant\epsilon^2} } \frac{d\tau\, dz}{z^2} \frac{1}{\abs{\tau_{14}}^{\Delta_i + \Delta_m}} \left( \frac{(\tau - \tau_1)^2 + z^2}{(\tau - \tau_4)^2 + z^2} \right)^{\frac{\Delta_{mi}}{2}} G_{\Delta_l}^{\Delta_i,\Delta_m}(\chi_{14})\,,
	\end{split}
\end{equation}
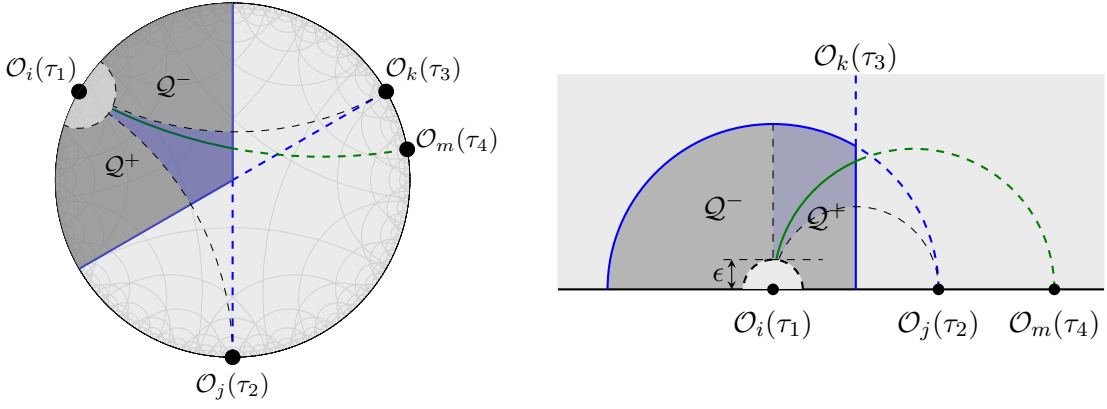
\begin{figure}[t]
    \centering
    \begin{minipage}{0.45\textwidth}
        \centering
        \includegraphics[width=\linewidth]{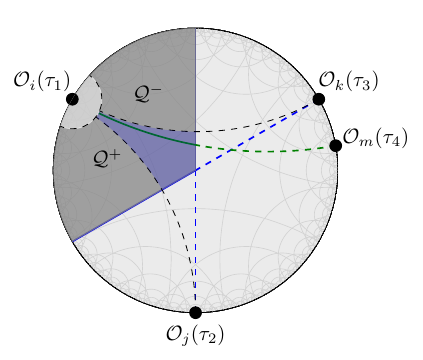}
    \end{minipage}
    \begin{minipage}{0.45\textwidth}
        \centering
            \begin{tikzpicture}[scale=2.2]
\definecolor{medium-gray}{gray}{0.7};
\definecolor{light-gray}{gray}{0.92};
\filldraw[light-gray](0.2,0)--(3.5,0)--(3.5,1.3)--(0.2,1.3)--(0.2,0);
\filldraw[medium-gray,opacity=0.95] (0.5,0) arc (180:60:1)--(2,0)--(0.5,0);
\draw[thick] (0.2,0) -- (3.5,0);
\begin{scope}
\node at (2,1.4) {$\mathcal{O}_k(\tau_3)$};
\draw[thick,blue] (2,0) -- (2,0.862);
\draw[thick,blue,dashed](2,0.862)--(2,1.3);
\draw[thick,blue] (1/2,0) arc (180:60:1);
\clip (1.5,0) -- (1.5,1) arc (90:60:1) -- (2,0) -- cycle;
\fill[blue,opacity=.1, even odd rule] 
    (1.5,0) -- (1.5,1) arc (90:60:1) -- (2,0) -- cycle
    (1.5,0) arc (180:0:0.5);
\end{scope}
\draw[thick,blue,dashed](2.5,0) arc (0:60:1);
\draw[black!50!green,thick] (1.5,0) arc (180:112:0.85);
\draw[dashed,black!50!green,thick] (3.2,0) arc (0:112:0.85);
\fill (1.5,0) circle (1pt) node[below=4pt] {$\mathcal{O}_i(\tau_1)$};
\fill (2.5,0) circle (1pt) node[below=4pt] {$\mathcal{O}_j(\tau_2)$};
\fill (3.2,0) circle (1pt) node[below=4pt] {$\mathcal{O}_m(\tau_4)$};
\draw[dashed] (1.5,0) arc (180:0:0.5);
\draw[dashed] (1.5,0) -- (1.5,1);
\clip(0,0.007)--(3.5,0.007)--(3.5,1.3)--(0,1.3);
\filldraw[light-gray] (1.5,0) circle (0.18);
\draw[dashed,thick] (1.68,0) arc (0:180:0.18);
\fill (1.5,0) circle (1pt);
\node[black] at (1.82,0.44) {$\mathcal{Q}^+$};
\node[black] at (1.2,0.5) {$\mathcal{Q}^-$};
\draw[dashed] (1.2,0.18)--(1.8,0.18);
\draw[stealth-stealth] (1.25,0)--(1.25,0.18);
\node at (1.17,0.09) {$\epsilon$};
\end{tikzpicture}
    \end{minipage}
    \caption{In dark gray, the regions of integration for $\mathcal{Q}^+$ and $\mathcal{Q}^-$ as defined in (\ref{def:IntQ_normal}), separated by the green geodesic which joins $\tau_1$ and $\tau_4$. In the Poincar\'e half-plane, we show the frame in which $\tau_3=\infty$. The light blue region is part of the blue region in figure \ref{fig:no_stu}, where none of the channels converge. The dark gray region is bounded by the blue geodesics defined by $\chi_{12}=\chi_{23}$ and $\chi_{13}=\chi_{23}$.}
    \label{fig:Ds_mu_omega}
\end{figure}
\newline
Note that $D_s^\pm$ depends on $\tau_4$, so the above is a nested double integral.
We now study the properties of $\mathcal{Q}^\pm$. When $\Delta_m > \Delta_i$, the integrals converge even without the cutoff $\epsilon$, similarly to the $BB\partial$ case discussed in section~\ref{subsec:eq2}. In this case, conformal symmetry implies that
\begin{equation}\label{def:K}
	\mathcal{Q}^{\Delta_l,\Delta_m;+}_{\Delta_i,\Delta_j,\Delta_k}(\tau_1, \tau_2, \tau_3; 0)
		=\mathcal{Q}^{\Delta_l,\Delta_m;-}_{\Delta_i,\Delta_k,\Delta_j}(\tau_1, \tau_3, \tau_2; 0):= -\frac{\mathcal{K}_{\Delta_i,\Delta_j,\Delta_k}(\Delta_l, \Delta_m)}{\abs{\tau_{12}}^{\Delta_{ijk}} \abs{\tau_{13}}^{\Delta_{ikj}} \abs{\tau_{23}}^{\Delta_{jki}}}\,.
\end{equation}
where $\mathcal{K}_{\Delta_i,\Delta_j,\Delta_k}(\Delta_l, \Delta_m)$ is a numerical coefficient. Notice the difference in the subscripts of the two $\QQ$'s. As a function of $\Delta_m$, $\mathcal{K}$ is meromorphic, with a single simple pole located at $\Delta_m = \Delta_i$, corresponding to the point at which the integrated block begins to diverge.

With a cutoff $\epsilon$, the integrated blocks $\mathcal{Q}^{\pm}$ can be written as their value at $\epsilon = 0$ minus a subtraction term:
\begin{equation}\label{def:Qsubtracted}
	\begin{split}
		\mathcal{Q}^{\Delta_l,\Delta_m;\pm}_{\Delta_i,\Delta_j,\Delta_k}(\tau_1, \tau_2, \tau_3; \epsilon)
		= \mathcal{Q}^{\Delta_l,\Delta_m;\pm}_{\Delta_i,\Delta_j,\Delta_k}(\tau_1, \tau_2, \tau_3; 0)
		- \mathcal{Q}^{\Delta_l,\Delta_m;\pm\,\text{subtr}}_{\Delta_i,\Delta_j,\Delta_k}(\tau_1, \tau_2, \tau_3; \epsilon)\,,
	\end{split}
\end{equation}
where the subtraction term is defined as
\begin{equation}\label{def:Qsubtraction}
	\begin{split}
		\mathcal{Q}^{\Delta_l,\Delta_m;\pm,\text{subtr}}_{\Delta_i,\Delta_j,\Delta_k}&(\tau_1, \tau_2, \tau_3; \epsilon) =\int_{\tau_2}^{\tau_3} d\tau_4\,\mathcal{B}_{\Delta_j\Delta_k\Delta_m}(\tau_2,\tau_3,\tau_4) \\
		&\times \int_{\substack{(\tau,z)\in D_s^\pm \\ (\tau-\tau_1)^2+z^2\leqslant\epsilon^2} } \frac{d\tau\, dz}{z^2} \frac{1}{\abs{\tau_{14}}^{\Delta_i + \Delta_m}} \left( \frac{(\tau - \tau_1)^2 + z^2}{(\tau - \tau_4)^2 + z^2} \right)^{\frac{\Delta_{mi}}{2}} G_{\Delta_l}^{\Delta_i,\Delta_m}(\chi_{14})\,. \\
	\end{split}
\end{equation}
For fixed $\e$, since the left-hand side of \eqref{def:Qsubtracted} is analytic in $\Delta_m$, it can be obtained by analytic continuation of the right-hand side, which consists of two meromorphic functions in $\Delta_m$. When $\Delta_m = \Delta_i$, the finiteness of the left-hand side implies that the poles in the two terms on the right-hand side must cancel.

Now we match the integrated $B\partial\partial\partial$ correlator with the renormalized three-point function. The renormalized three-point function satisfies (assuming $\t_1<\t_2<\t_3$)
\be
\<\mathcal{O}^{(\text{ren})}_i(\tau_1)\mathcal{O}^{(\text{ren})}_j(\tau_2)\mathcal{O}^{(\text{ren})}_k(\tau_3)\>_{\l+\d\l}
=
\frac{C_{ijk}+\d C_{ijk}}{|\t_{12}|^{\D_{ijk}+\d\D_{ijk}}
|\t_{23}|^{\D_{jki}+\d\D_{jki}}
|\t_{13}|^{\D_{kij}+\d\D_{kij}}\,,
}
\ee
with $\d\D_{ijk}\equiv\d\D_i+\d\D_j-\d\D_k$.
Recalling the definition of field-strength renormalization from \eqref{eq:oprenom}, the matching condition reads:
\begin{equation}\label{matching:bdry3pt}
	\begin{split}
		&\sum_{m} \sum_{p=0}^{\infty} Z_{im;p}\, \partial^p_{\tau_1} \braket{\mathcal{O}^{(\text{ren})}_m(\tau_1)\mathcal{O}^{(\text{ren})}_j(\tau_2)\mathcal{O}^{(\text{ren})}_k(\tau_3)}_{\lambda+\delta\lambda} + (i, \tau_1 \leftrightarrow j, \tau_2) + (i, \tau_1 \leftrightarrow k, \tau_3) \\
		&= \braket{\mathcal{O}_i(\tau_1)\mathcal{O}_j(\tau_2)\mathcal{O}_k(\tau_3)}_{\lambda} \\
		&\quad - \delta\lambda \sum_{l,m}b^{\hat{\Phi}}_{l} C_{jkm} C_{ilm} \mathcal{Q}^{\Delta_l,\Delta_m;+}_{\Delta_i,\Delta_j,\Delta_k}(\tau_1, \tau_2, \tau_3; \epsilon) 
		- \delta\lambda \sum_{l,m}b^{\hat{\Phi}}_{l} C_{jkm} C_{iml} \mathcal{Q}^{\Delta_l,\Delta_m;-}_{\Delta_i,\Delta_j,\Delta_k}(\tau_1, \tau_2, \tau_3; \epsilon) \\
		&\quad + (ijk123 \rightarrow jki231) + (ijk123 \rightarrow kij312)\,,
	\end{split}
\end{equation}
where the shorthands denote the cyclic permutation of coordinates and indices. For example,
\begin{equation}
    \begin{split}
        (i, \tau_1 \leftrightarrow j, \tau_2)&\equiv\sum_{m} \sum_{p=0}^{\infty} Z_{jm;p}\, \partial^p_{\tau_2} \braket{\mathcal{O}^{(\text{ren})}_i(\tau_1)\mathcal{O}^{(\text{ren})}_m(\tau_2)\mathcal{O}^{(\text{ren})}_k(\tau_3)}_{\lambda+\delta\lambda}\,, \\
        (ijk123 \rightarrow jki231)&\equiv- \delta\lambda \sum_{l,m}b^{\hat{\Phi}}_{l} C_{kim} C_{jlm} \mathcal{Q}^{\Delta_l,\Delta_m;+}_{\Delta_j,\Delta_k,\Delta_i}(\tau_2, \tau_3, \tau_1; \epsilon)  \\
		&\quad - \delta\lambda \sum_{l,m}b^{\hat{\Phi}}_{l} C_{kim} C_{jml} \mathcal{Q}^{\Delta_l,\Delta_m;-}_{\Delta_j,\Delta_k,\Delta_i}(\tau_2, \tau_3, \tau_1; \epsilon)\,. \\
    \end{split}
\end{equation}
Similarly to the derivation of the second flow equation (for the BOE coefficients), we claim that:
\begin{itemize}
	\item When $\Delta_m \neq \Delta_i$, the subtraction terms in $\mathcal{Q}^{+}$ and $\mathcal{Q}^{-}$ match the field-strength renormalization terms $Z_{im}$.
	
	\item When $\Delta_m = \Delta_i$, the pole in the subtraction term cancels the pole in the analytic continuation of the unsubtracted term (i.e., the $\epsilon = 0$ piece). The finite (O(1)) part of the subtraction term matches the renormalization contributions from $\delta Z_{ii}$ and the anomalous dimension $\delta \Delta_i$.
	
	\item Similar cancellations occur in the $t$- and $u$-channel contributions.
\end{itemize}
We leave the justification of these claims to appendix~\ref{app:HardvsAnal}.
After all cancellations, the remaining finite contributions yield
\begin{equation}\label{matching:bdry3ptfinal}
	\begin{split}
		\delta C_{ijk}
		&= \delta\lambda\sum_{l,m} b^{\hat{\Phi}}_{l}C_{jkm} \Big(C_{ilm} [\mathcal{K}_{\Delta_i,\Delta_j,\Delta_k}(\Delta_l,\Delta_m)]_\text{reg}+ C_{iml} [\mathcal{K}_{\Delta_i,\Delta_k,\Delta_j}(\Delta_l,\Delta_m)]_\text{reg}\Big) \\
		&\quad + (ijk \rightarrow jki) + (ijk \rightarrow kij),
	\end{split}
\end{equation}
where the dependence on $\tau_i$'s is dropped, and $[\mathcal{K}_{\Delta_i,\Delta_j,\Delta_k}(\Delta_l,\Delta_m)]_\text{reg}$ is defined in a similar way as $[\mathcal{J}]_\text{reg}$ in \eqref{def:Jreg}: 
\begin{equation}
    [\mathcal{K}_{\Delta_i,\Delta_j,\Delta_k}(\Delta_l,\Delta_m) ]_\text{reg}:=
		  \oint_{\Delta_m} \frac{d\Delta}{2\pi i }\frac{\mathcal{K}_{\Delta_i,\Delta_j,\Delta_k}(\Delta_l,\Delta)}{\D-\D_m}\,.
\end{equation}
We thus arrive at the flow equation for the boundary OPE coefficients:
\begin{equation}\label{eq:flow_OPE_normal}
	\begin{split}
		\frac{dC_{ijk}}{d\lambda}
		&= \sum_{l,m} b^{\hat{\Phi}}_{l}\, C_{jkm}\left(C_{ilm} [\mathcal{K}_{\Delta_i,\Delta_j,\Delta_k}(\Delta_l, \Delta_m)]_\text{reg}+C_{iml} [\mathcal{K}_{\Delta_i,\Delta_k,\Delta_j}(\Delta_l, \Delta_m)]_\text{reg}\right) \\
		&\quad + (ijk \rightarrow jki) + (ijk \rightarrow kij)\,,
	\end{split}
\end{equation}
where the explicit form of $\KK_{\D_i\D_j\D_k}(\D_l,\D_m)$ is given in \eqref{eq:K}.

\subsubsection{Recap}
\label{subsec:recap}
In this subsection, we briefly recap the above derivations, with emphasis on simplicity rather than rigor. 

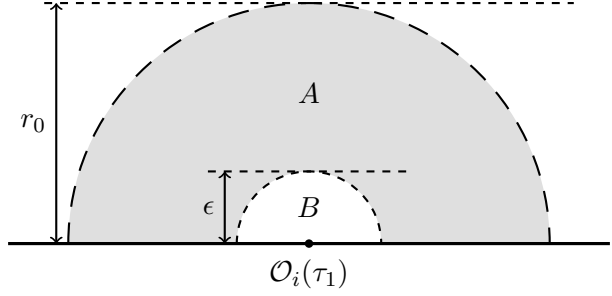
\begin{figure}[t]
  \centering
\begin{tikzpicture}[scale=0.8]
  \def\Rbig{4}     
  \def\Rsmall{1.2} 

  \fill[
    gray!25,
    even odd rule
  ]
  (0,0) ++(-\Rbig,0)
  arc[start angle=180, end angle=0, radius=\Rbig]
  --
  (0,0) ++(\Rbig,0)
  --
  (0,0) ++(\Rsmall,0)
  arc[start angle=0, end angle=180, radius=\Rsmall]
  --
  cycle;

  \draw[
    thick,
    dashed,
    dash pattern=on 8pt off 6pt
  ]
  (0,0) ++(-\Rbig,0)
  arc[start angle=180, end angle=0, radius=\Rbig];

  \node at (0,2.5) {$A$};

  \draw[thick,dashed]
  (0,0) ++(-\Rsmall,0)
  arc[start angle=180, end angle=0, radius=\Rsmall];

  \node at (0,\Rsmall/2) {$B$};

  \draw[very thick] (-5,0) -- (5,0);
  \draw[thick,dashed] (-1.4*
  \Rsmall,\Rsmall) -- (1.4*\Rsmall,\Rsmall);
\draw[thick,dashed] (-1.1*
  \Rbig,\Rbig) -- (1.1*\Rbig,\Rbig);
\draw[<->, thick]
  (-\Rsmall-0.2,0) -- (-\Rsmall-0.2,\Rsmall);
\node[left] at (-\Rsmall-0.2,\Rsmall/2) {$\epsilon$};
\draw[<->, thick]
  (-\Rbig-0.2,0) -- (-\Rbig-0.2,\Rbig);
\node[left] at (-\Rbig-0.2,\Rbig/2) {$r_0$};
  \fill (0,0) circle (2pt);
  \node[below] at (0,-0.1) {$\mathcal O_i(\tau_1)$};

\end{tikzpicture}
  \caption{Schematic regions of the conformal block integrals. The split into ``+'' and ``-'' regions concerning boundary operator ordering is omitted here.}
  \label{fig:schematic region}
\end{figure}

The QFT data consists of the scaling dimensions $\D_i$ and the OPE coefficients $C_{ijk}$ of boundary operators, and the BOE coefficients $b^{\hat\OO}_i$ of bulk operators. To determine how they change infinitesimally as the coupling of the bulk deformation operator $\l$ varies, we need to consider the integral over AdS of three types of correlators, namely the $B\del\del$, $BB\del$ and  $B\del\del\del$ correlators. Identifying these integrals with the correlators of bare operators, e.g., $\<\hat\F\OO_i^{\text{(bare)}}\>_{\l+\d\l}$ for the BOE coefficient flow, and removing divergences through operator renormalization \eqref{eq:oprenom}, we obtain the flow equations. A key intermediate step is to separate kinematics from dynamics in the integrated correlators. Using block decompositions, we are led to consider three types of hard-cutoff regulated integrals of conformal blocks, collected below:
\al{
\mathcal{Y}^{\Delta_l;\pm}_{\Delta_i,\Delta_j}(\tau_1,\tau_2,\epsilon)
&:=\int_{\substack{\text{sgn}(\rho)=\pm \\ (\tau-\tau_i)^2+z^2\geqslant\epsilon^2}}\frac{d\tau dz}{z^2} \frac{1}{\abs{\tau_{12}}^{\Delta_i+\Delta_j}}\left(\frac{(\tau-\tau_1)^2+z^2}{(\tau-\tau_2)^2+z^2}\right)^{\frac{\Delta_j-\Delta_i}{2}} G_{\Delta_l}^{\Delta_i,\Delta_j}(\chi)\,,\nonumber
\\
\mathcal{R}_{\Delta_l,\Delta_j}^{\Delta_i;\pm}(\tau_1,z_1,\tau_2,\epsilon)&:=\int_{\substack{ \text{sgn}(\upsilon)=\pm\\(\tau-\tau_2)^2+z^2\geqslant\epsilon^2 }}
        \frac{d\tau dz}{z^2}R_{\Delta_l,\Delta_j}^{\Delta_i}(\upsilon,\zeta)\,
\\
\mathcal{Q}^{\Delta_l,\Delta_m;\pm}_{\Delta_i,\Delta_j,\Delta_k}(\tau_1, \tau_2, \tau_3; \epsilon)
		&:= \int_{\tau_2}^{\tau_3} d\tau_4\,\mathcal{B}_{\Delta_j\Delta_k\Delta_m}(\tau_2,\tau_3,\tau_4) \nonumber\\
		&\times \int_{\substack{(\tau,z)\in D_s^\pm \\ (\tau-\tau_1)^2+z^2\geqslant\epsilon^2} } \frac{d\tau\, dz}{z^2} \frac{1}{\abs{\tau_{14}}^{\Delta_i + \Delta_m}} \left( \frac{(\tau - \tau_1)^2 + z^2}{(\tau - \tau_4)^2 + z^2} \right)^{\frac{\Delta_{mi}}{2}} G_{\Delta_l}^{\Delta_i,\Delta_m}(\chi_{14})\,.\nonumber
}
Since we assume there is no bulk UV divergences, the only divergence comes from when the integrated bulk operator approaches a boundary operator. In all three cases, we focus, at each time, on a part of AdS$_2$ that surrounds only one boundary operator (region A in figure \ref{fig:schematic region}):
\be
(\t,z) = (\t_1+r\cos\q,r\sin\q)\,, \qquad r\in[\epsilon,r_0]\,,\quad \theta\in[0,\pi]\,,
\ee
where $(\t,z)$ is the position of the integrated bulk operator $\hat\F$ and $\t_1$ is the position of the boundary operator. Because we can rewrite the $BB\del$ block and the $B\del\del\del$ block in terms of $B\del\del$ blocks as in \eqref{eq:Bkbkbd blk from bkbdbd blk} and \eqref{def:bkbdbdbd_sblock_normal}, the analysis of divergences essentially boils down to the integrated $B\del\del$ block. 

Generically, the divergences are of power-law type and they originate from the radial integral. Let us write the $B\del\del$ block as a power series
\be
G_{\D_l}^{\D_i,\D_j}(\chi) = \sum_{n\geqslant0} a_n \chi^{\D_l/2+n}
\ee
and focus on the region close to the boundary operator, corresponding to small $r$. Then for each monomial of $\chi=\sin^2\q+O(r)$, its contribution to $\mathcal{Y}_{\D_i,\D_j}^{\D_l;\pm}$ becomes
\al{\spl{\int_{\substack{\text{sgn}(\rho)=\pm \\ (\tau-\tau_i)^2+z^2 \geqslant \epsilon^2}} \frac{d\tau dz}{z^2}\,
\frac{1}{\abs{\tau_{12}}^{\Delta_i+\Delta_j}}\left(\frac{(\tau-\tau_1)^2+z^2}{(\tau-\tau_2)^2+z^2}\right)^{\frac{\Delta_j-\Delta_i}{2}} \chi^{\D_l/2+n}
\\
\supset\int_0^\pi \frac{d\q}{\sin^2\q} \int_{\e}^{r_0} \frac{dr}{r} 
\left(\sum_{p\geqslant0}d_p(\q,\D_i,\D_j,\D_k)\left(\frac{r}{\t_{21}}\right)^{\D_{ji}+p}
\right)
}}
where $r_0$ is some finite radial distance and the explicit expression of $d_p$ is unimportant here. The $\supset$ symbol above appears because the boundary of the region we actually consider is not exactly a semi-circle with radius $r_0$, but the difference from the integral over the actual region is always finite. The indefinite radial integral for fixed $p$ gives
\be
\int \frac{dr}{r}\left(\frac{r}{\t_{21}}\right)^{\D_{ji}+p}=-\frac{1}{\D_{ij}-p}\left(\frac{r}{\t_{21}}\right)^{p-\D_{ij}}\,.
\ee
When one of the integration limits contain $r=\e$, we obtain the power-law terms $\left(\frac{\e}{\t_{21}}\right)^{p-\D_{ij}}$, which diverge in the limit $\e\to0$ only if $\D_j<\D_i-p$. In the range $\D_j>\D_i$ the integrals are always finite. In other words, as $\hat\F$ approaches $\OO_i$, the integral of the correlator $\<\hat\F\OO_i\OO_j\>$ in the considered domain is finite without cutoff if $\OO_j$ is heavier than $\OO_i$. Similarly, by using BOE in $\<\hat\F\hat\F\OO_i\>$ and using OPE in $\<\hat\F\OO_i\OO_j\OO_k\>$ we see that, term by term in the operator expansion, the integrals are finite when the intermediate exchanged operators are heavier than $\OO_i$. When $\D_j=\D_i$, there is $\log(\e)$ divergence. When $\D_j=\D_i-p$, $\log(\e)$ could have also appeared, but as explained in appendix \ref{app:subsec_IntegralY}, this is not the case because of the properties of $d_p$.

Since the divergences in the $\e\to0$ limit only appear in a certain range of $\D_j$, we can start from $\D_j>\D_i$ with fixed $\e$ where all the integrals in consideration are finite. Then, we write them as a sum of two terms which,  when $\Delta_j\leqslant\Delta_i$, are finite or diverge respectively:
\al{\spl{
\mathcal{Y}^{\Delta_l;\pm,\text{I/II}}_{\Delta_i,\Delta_j}(\tau_1,\tau_2,\tau_3,\epsilon)&=\mathcal{Y}^{\Delta_l;\pm,\text{I/II}}_{\Delta_i,\Delta_j}(\tau_1,\tau_2,\tau_3,0)-\mathcal{Y}^{\Delta_l;\pm,\,\text{I/II, subtr}}_{\Delta_i,\Delta_j}(\tau_1,\tau_2,\epsilon)\,,
\\
\mathcal{R}_{\Delta_l\, \Delta_j}^{\Delta_i;\pm}(\tau_1,z_1,\tau_2,\epsilon)&=-\frac{1}{2}\JJ_{\Delta_i}(\Delta_l, \Delta_j)-\JJ^{\pm,\,\text{subtr}}_{\Delta_i}(\Delta_l, \Delta_j;\tau_1,z_1;\tau_2;\epsilon)\,,
\\
\mathcal{Q}^{\Delta_l,\Delta_m;\pm}_{\Delta_i,\Delta_j,\Delta_k}(\tau_1, \tau_2, \tau_3; \epsilon)
		&= \mathcal{Q}^{\Delta_l,\Delta_m;\pm}_{\Delta_i,\Delta_j,\Delta_k}(\tau_1, \tau_2, \tau_3; 0)
		- \mathcal{Q}^{\Delta_l,\Delta_m;\pm,\,\text{subtr}}_{\Delta_i,\Delta_j,\Delta_k}(\tau_1, \tau_2, \tau_3; \epsilon)\,,
}}
where recall that for $\mathcal{Y}$ we needed to make a further split such that in region I and II there is only one boundary operator, respectively (see appendix \ref{app:subsec_IntegralY}). The last term in each equation above corresponds to an integral in region $B$ in figure \ref{fig:schematic region}. Next we analytically continue to $\D_j\leqslant\D_i$ and finally take the limit $\e\to0$. This leads to a clean separation between finite, physical quantities and divergent pieces. In the scaling dimension flow, in the end there is no finite piece in the integrated block, and the coefficient of the $\log(\e)$ term is physical. In the BOE coefficient flow and OPE coefficient flow, there are finite parts, $\JJ_{\D_i}(\D_l,\D_j)$ and $\mathcal{Q}^{\Delta_l,\Delta_m;\pm}_{\Delta_i,\Delta_j,\Delta_k}(\tau_1, \tau_2, \tau_3; 0)$, and they appear in the final flow equations. The divergences in $\JJ^{\pm,\,\text{subtr}}_{\Delta_i}(\Delta_l, \Delta_j;\tau_1,z_1;\tau_2;\epsilon)$ and $\mathcal{Q}^{\Delta_l,\Delta_m;\pm,\,\text{subtr}}_{\Delta_i,\Delta_j,\Delta_k}(\tau_1, \tau_2, \tau_3; \epsilon)$ include both power-law and log types. When we equate the integrated correlators with the correlators of bare operators, these divergences are exactly canceled by those in the operator renormalization coefficients in \eqref{eq:oprenom} (also see \eqref{eq:zij} and \eqref{zii:result}), as explained in appendix \ref{app:HardvsAnal}.

\subsection{Swappability from locality}\label{subsec:local block}

In the previous subsection, we derived the flow equations for the scaling dimensions $\Delta_i$, the BOE coefficients $b^{\hat{\Phi}}_i$, and the OPE coefficients $C_{ijk}$.  
The derivation relied on a crucial assumption: that AdS bulk integrals are \emph{swappable} with the conformal block expansions of the correlators, schematically
\begin{equation*}
	\int\frac{d\tau\,dz}{z^2}\sum(\ldots)=\sum\int\frac{d\tau\,dz}{z^2}(\ldots)\,.
\end{equation*}
However, this equality does not always hold when using the standard conformal blocks discussed in section~\ref{subsec:Confblock}.  
For instance, in the flow equation for the scaling dimensions, eq.~\eqref{eq:flow_scaling_dim_normal}, a naive large-$\Delta_l$ estimate shows that the sum diverges in general.

To see this explicitly, consider the large-$\Delta_l$ limit of \eqref{eq:flow_scaling_dim_normal}.  
The asymptotics of the relevant coefficients are as follows.  
The behavior of $b^{\hat{\Phi}}_l$ is governed by the UV dimension $\Delta_{\hat{\Phi}}$ of $\hat{\Phi}$ (assuming the theory is UV completed by a CFT), while that of $C_{iil}$ is controlled by $\Delta_i$.  
Finally, the leading term of $a_0(\Delta_l)$ follows directly from its explicit expression.  
We find
\begin{equation}\label{eq:flow_dim_asymp_normal}
	b^{\hat{\Phi}}_l\sim \Delta_l^{\Delta_{\hat{\Phi}}-3/4},\quad 
	C_{iil}\sim 2^{-\Delta_l}\Delta_l^{2\Delta_i-3/4},\quad 
	a_0(0,\Delta_l)\sim \frac{2^{\Delta_l}}{\Delta_l}\, ,
	\qquad(\Delta_l\rightarrow\infty).
\end{equation}
We emphasize that the first two relations hold only in an averaged sense (see appendix~\ref{sec:convergence of flow eqn}), hence our use of the term “naive estimate.”  
Combining these asymptotics, the large-$\Delta_l$ behavior of the sum in \eqref{eq:flow_scaling_dim_normal} scales as
\begin{equation}
2\sum\limits_{\Delta_l\leqslant \Delta_{max}}^{}\Big|b^{\hat{\Phi}}_{l}\,C_{iil}\,a_0(0,\Delta_l)\Big|\sim \int^{\Delta_{max}} d\D_l\, \Delta_l^{2\Delta_i+\Delta_{\hat{\Phi}}-5/2}\,.
\end{equation}
Absolute convergence would then require $\Delta_i<\tfrac{3}{4}-\tfrac{1}{2}\Delta_{\hat{\Phi}}$, a condition that is not generally satisfied.  
A simple counterexample is the free scalar theory deformed by a mass term, where $\hat{\Phi}=\hat{\phi}^2$.  
In that case $\Delta_{\hat{\Phi}}=0$, and the condition is violated once the mass becomes sufficiently large in AdS units.

\vspace{0.3em}
Before proposing a way to tame the divergence, let us first understand its origin.  

Consider the $B\partial\partial$ correlator $\braket{\hat{\Phi}\mathcal{O}_i\mathcal{O}_j}$.  
If we expand it in terms of the normal conformal blocks, the series converges almost everywhere in Euclidean AdS, except when the bulk point lies on the geodesic connecting the two boundary points.  
This special configuration corresponds to $\chi=1$ in \eqref{def:chi}.  
Away from $\chi=1$, one can always draw a semicircle separating the bulk point from the boundary points. The center of the circle is required to be somewhere on the boundary. 
Using the radial quantization picture, one can then argue that the OPE/BOE expansion converges.  
At $\chi=1$, however, no such semicircle exists, and the radial quantization argument fails.  
Indeed, in explicit examples such as free theories, the conformal block expansion is found to diverge at $\chi=1$.

Nevertheless, for the full correlator, the point $\chi=1$ is not special—the correlator remains finite there.  
This does not contradict the divergence of the conformal block expansion because the coefficients $b^{\hat{\Phi}}_l C_{ijl}$ are not positive definite.  
As the bulk point approaches $\chi=1$, the convergence of the conformal block expansion becomes progressively worse, and this is precisely the origin of the failure to swap the sum and the bulk integral.

From this analysis, we learn that the problem lies not in the correlator itself but in the choice of basis used for its expansion.  
The divergence can therefore be resolved by choosing a better basis—one that leads to a convergent expansion throughout AdS (except at genuine physical singularities).

Fortunately, such a basis is known in higher dimensions.  
As explained in \cite{Levine:2023ywq,Meineri:2023mps}, one can expand correlators in terms of \emph{local blocks}.  
However, the higher-dimensional local blocks do not apply to AdS$_2$ via a naive analytic continuation in dimension, due to an additional subtlety that does not arise in higher dimensions: the $SO^+(1,2)$-invariant part of a $B\partial\partial$ correlator is not necessarily $O^+(1,2)$ invariant (i.e.\ parity symmetric). We treated this issue carefully in a separate paper \cite{Loparco:2025aag}. Handling this subtlety requires introducing two types of local blocks—namely, the even and odd local blocks.

We first introduce the local blocks for $B\partial\partial$ correlators in section~\ref{subsec:localblock_bkbdbd} and demonstrate how they resolve the divergence issue in the first flow equation discussed above.  
Subsequently, in sections~\ref{subsec:localblock_bkbkbd} and~\ref{subsec:localblock_bkbdbdbd}, we construct the local blocks for the $BB\partial$ and $B\partial\partial\partial$ correlators from the $B\partial\partial$ local blocks, and use them to formulate refined versions of the second and third flow equations.

\subsubsection{\texorpdfstring{Local blocks for the $B\partial\partial$ correlator}{Local blocks for the Bbb correlator}}\label{subsec:localblock_bkbdbd}

We begin with the $B\partial\partial$ correlator $\mathcal{G}^{\hat{\Phi}}_{ij}(\rho)$ in \eqref{Gbkbdbd:generalform}, omitting the overall scaling prefactor, which is irrelevant for the present discussion. For further technical details, see \cite{Loparco:2025aag}.

We define the even and odd parts of the correlator as
\begin{equation}\label{def:G01}
	\begin{split}
		[\mathcal{G}^{\hat{\Phi}}_{ij}]_e(\chi)
		&:=\frac{1}{2}\left[\mathcal{G}^{\hat{\Phi}}_{ij}(\rho)+\mathcal{G}^{\hat{\Phi}}_{ij}(-\rho)\right], \\
		[\mathcal{G}^{\hat{\Phi}}_{ij}]_o(\chi)
		&:=\frac{1}{2\rho}\left[\mathcal{G}^{\hat{\Phi}}_{ij}(\rho)-\mathcal{G}^{\hat{\Phi}}_{ij}(-\rho)\right].
	\end{split}
\end{equation}
The full correlator can then be reconstructed from these two components:
\begin{equation}\label{Gbkbdbd:evenodd}
	\begin{split}
		\mathcal{G}^{\hat{\Phi}}_{ij}(\rho)
		= [\mathcal{G}^{\hat{\Phi}}_{ij}]_e(\chi)
		+ \rho\, [\mathcal{G}^{\hat{\Phi}}_{ij}]_o(\chi)\,.
	\end{split}
\end{equation}
Note that $\mathcal{G}^{\hat{\Phi}}_{ij}(\rho)$ is not parity symmetric in general:
\(\mathcal{G}^{\hat{\Phi}}_{ij}(\rho)\neq \mathcal{G}^{\hat{\Phi}}_{ij}(-\rho)\).
However, by construction, both
\([\mathcal{G}^{\hat{\Phi}}_{ij}]_e\)
and
\([\mathcal{G}^{\hat{\Phi}}_{ij}]_o\)
are parity symmetric and therefore depend on $\rho$ only through the parity-symmetric variable $\chi$.

The functions $[\mathcal{G}^{\hat{\Phi}}_{ij}]_e(\chi)$ and $[\mathcal{G}^{\hat{\Phi}}_{ij}]_o(\chi)$ are analytic in $\chi$, with a branch point only at $\chi=0$. Under a mild assumption that the growth of $\mathcal{G}^{\hat{\Phi}}_{ij}(\rho)$ at large $\rho$ is bounded by $O(e^{\abs{\rho}^{M}})$, one can show that they admit the following expansions:
\begin{equation}\label{Gbkbdbd:newexpansion}
	\begin{split}
		[\mathcal{G}^{\hat{\Phi}}_{ij}]_e(\chi)
		&=\sum_{l}b^{\hat{\Phi}}_{l}\,
		\frac{C_{ijl}+C_{jil}}{2}\,
		G_{\Delta_l}^{\Delta_i,\Delta_j,(\alpha)}(\chi), \\
		[\mathcal{G}^{\hat{\Phi}}_{ij}]_o(\chi)
		&=\sum_{l}b^{\hat{\Phi}}_{l}\,
		\frac{C_{jil}-C_{ijl}}{2}\,
		\tilde{G}_{\Delta_l}^{\Delta_i,\Delta_j,(\beta)}(\chi),
	\end{split}
\end{equation}
where the functions
\(G_{\Delta_l}^{\Delta_i,\Delta_j,(\alpha)}(\chi)\)
and
\(\tilde{G}_{\Delta_l}^{\Delta_i,\Delta_j,(\beta)}(\chi)\)
are referred to as the \emph{even} and \emph{odd local blocks}, respectively, since they contribute to the even and odd parts of the full correlator $\mathcal{G}^{\hat{\Phi}}_{ij}(\rho)$. 

The explicit form of the even local block is obtained by analytically continuing the higher-dimensional local block in the dimension parameter, as known from \cite{Levine:2023ywq,Meineri:2023mps}.

\begin{equation}\label{localblock:bkbdbd_even}
	\begin{split}
		G^{\Delta_i\Delta_j,(\alpha)}_{\Delta_l}(\chi)
		&=G^{\Delta_i\Delta_j}_{\Delta_l}(\chi)
		-\frac{
			\Gamma\!\left(\Delta_l+\tfrac{1}{2}\right)
			\Gamma\!\left(\tfrac{\Delta_{ij}}{2}+\alpha\right)
			\Gamma\!\left(\tfrac{\Delta_{ji}}{2}+\alpha\right)
		}{
			\Gamma\!\left(\tfrac{\Delta_{lij}}{2}\right)
			\Gamma\!\left(\tfrac{\Delta_{lji}}{2}\right)
			\Gamma\!\left(\tfrac{\Delta_l+1}{2}+\alpha\right)
			\Gamma\!\left(-\tfrac{\Delta_l}{2}+\alpha+1\right)
		} \\
		&\hspace{5cm}\times \chi^\alpha\,
		\pFq{3}{2}{1,\tfrac{\Delta_{ij}}{2}+\alpha,\tfrac{\Delta_{ji}}{2}+\alpha}
		{\tfrac{\Delta_l+1}{2}+\alpha,-\tfrac{\Delta_l}{2}+\alpha+1}{\chi}\,.
	\end{split}
\end{equation}
Here, the first term is the standard conformal block defined in \eqref{def:bkbdbd block}.  
The parameter $\alpha$ is a free parameter appearing in the expansion \eqref{Gbkbdbd:newexpansion}.  
The expansion is uniformly convergent in any compact complex domain of $\chi$ provided that
\begin{equation}\label{evenblock:convergence}
	\begin{split}
		\alpha>\frac{1}{2}\left(\Delta_i+\Delta_j+\Delta_{\hat{\Phi}}\right)\,.
	\end{split}
\end{equation}

The odd local blocks were computed in \cite{Loparco:2025aag}:
\begin{equation}\label{localblock:bkbdbd_odd}
	\begin{split}
		\tilde{G}^{\Delta_i\Delta_j,(\beta)}_{\Delta_l}(\chi)
		&=\frac{\chi^{\Delta_l/2}}{\sqrt{1-\chi}}\,
		{}_2F_1\!\left(\tfrac{\Delta_{lij}}{2},\tfrac{\Delta_{lji}}{2};
		\Delta_l+\tfrac{1}{2};\chi\right) \\
		&\quad -\frac{
			\Gamma(\Delta_l+\tfrac{1}{2})
			\Gamma(\tfrac{\Delta_{ij}+1}{2}+\beta)
			\Gamma(\tfrac{\Delta_{ji}+1}{2}+\beta)
		}{
			\Gamma(\tfrac{\Delta_{lij}+1}{2})
			\Gamma(\tfrac{\Delta_{lji}+1}{2})
			\Gamma(\tfrac{\Delta_l+1}{2}+\beta)
			\Gamma(1-\tfrac{\Delta_l}{2}+\beta)
		} \\
		&\hspace{2.5cm}\times \chi^\beta\,
		\pFq{3}{2}{1,\tfrac{\Delta_{ij}+1}{2}+\beta,\tfrac{\Delta_{ji}+1}{2}+\beta}
		{\tfrac{\Delta_l+1}{2}+\beta,-\tfrac{\Delta_l}{2}+\beta+1}{\chi}\,.
	\end{split}
\end{equation}
The corresponding expansion in \eqref{Gbkbdbd:newexpansion} is convergent when
\begin{equation}\label{oddblock:convergence}
	\begin{split}
		\beta>\frac{1}{2}\left(\Delta_i+\Delta_j+\Delta_{\hat{\Phi}}-1\right)\,.
	\end{split}
\end{equation}
Comparing \eqref{oddblock:convergence} with \eqref{evenblock:convergence}, the additional $-\tfrac{1}{2}$ arises from the extra factor of $\tfrac{1}{\rho}$ in \eqref{Gbkbdbd:evenodd}, which scales as $\chi^{-1/2}$ when $\chi$ is large.  
In higher dimensions, the full correlator is analytic in $\chi$ except for a branch point at $\chi=0$, so no square-root singularity $\sqrt{1-\chi}$ appears, and consequently the odd local block is absent.

When $\abs{\chi}\leqslant1$, the standard conformal block expansion \eqref{bkbdbd:exp} is recovered by taking the limit $\alpha,\beta\rightarrow\infty$ in the local block expansion \eqref{Gbkbdbd:newexpansion}.\footnote{In this limit, one can easily verify that the second terms of the local blocks vanish, leaving only the standard conformal blocks. This is not true for $\abs{\chi}>1$, since in that regime the second term grows exponentially fast as $\alpha,\beta\rightarrow\infty$.} We would like to emphasize that the purpose of keeping $\alpha$ and $\beta$ finite is to ensure uniform convergence for all $\chi$ in the Euclidean region, including $\chi=1$ (where the standard conformal block expansion does not converge).

Now let us see how the introduction of local blocks resolves the divergence issue.  
Using the modified expansion given by \eqref{Gbkbdbd:evenodd} and \eqref{Gbkbdbd:newexpansion}, and repeating the analysis of section~\ref{subsec:eq1}, we find that the coefficients entering the flow equation for the scaling dimensions \eqref{eq:flow_scaling_dim_normal} are modified as
\begin{equation}
	a_0(0,\Delta_l)\quad\longrightarrow\quad 
	a_0^{(\alpha)}(0,\Delta_l)
	=\frac{2^{2-2\alpha}\pi\Gamma\!\left(\Delta_l+\tfrac{1}{2}\right)\Gamma(2\alpha-1)}
	{(\Delta_l-1)
		\Gamma\!\left(\tfrac{\Delta_l}{2}\right)
		\Gamma\!\left(\tfrac{\Delta_l}{2}+1\right)
		\Gamma\!\left(\tfrac{\Delta_l-1}{2}+\alpha\right)
		\Gamma\!\left(\alpha-\tfrac{\Delta_l}{2}\right)}\,.
\end{equation}
Since $C_{iil}=C_{ili}=C_{lii}$, the correlator in this case is even under $\rho\!\to\!-\rho$, and therefore the relevant contribution to $a_0^{(\alpha)}(0,\Delta_l)$ arises solely from the even local block $G^{\Delta_i\Delta_j,(\alpha)}_{\Delta_l}(\chi)$.

The large-$\Delta$ asymptotics of $a_0^{(\alpha)}(0,\Delta_l)$ is
\begin{equation}
	a_0^{(\alpha)}(0,\Delta)
	\sim \frac{2^\Delta}{\Delta}\,
	\Delta^{\tfrac{3}{2}-2\alpha}\,
	\sin\!\left(\alpha-\tfrac{1}{2}\Delta\right)\,.
\end{equation}
Comparing with \eqref{eq:flow_dim_asymp_normal}, we see that the local blocks introduce an additional suppression factor $\Delta^{\tfrac{3}{2}-2\alpha}$.  
Consequently, the convergence condition for the flow equation \eqref{eq:flow_scaling_dim_normal} is improved to
\begin{equation}\label{eq:flow_dim_local_condition}
	\alpha>\Delta_i+\tfrac{1}{2}\Delta_{\hat{\Phi}}\,.
\end{equation}
which is exactly the convergence condition for the local block expansion, eq.~\eqref{evenblock:convergence}. Since $\alpha$ is an auxiliary parameter, one may always choose it large enough—depending on $\Delta_i$ and $\Delta_{\hat{\Phi}}$—to guarantee absolute convergence of the expansion for every operator $\mathcal{O}_i$.

Therefore, the corrected first flow equation is 
\begin{mdframed}[backgroundcolor=shadecolor,linewidth=0pt]
\begin{equation}\label{eq:flow_dim_local}
				\frac{d\Delta_i}{d\lambda}
				= \sum_{l} b^{\hat{\Phi}}_{l}\, C_{iil}\, \mathcal{I}^{(\alpha)}(\Delta_l)
\end{equation}
\end{mdframed}
where the coefficient $\mathcal{I}^{(\alpha)}(\Delta_l)$ is given by
\begin{equation}\label{def:I_local}
	\begin{split}
		\mathcal{I}^{(\alpha)}(\Delta_l)
		&=
		\frac{\pi\,
			\Gamma(\Delta_l+\tfrac{1}{2})\,
			\Gamma(2\alpha-1)}
		{
			2^{\,2\alpha-3}(\Delta_l-1)\,
			\Gamma(\tfrac{\Delta_l}{2})\,
			\Gamma(\tfrac{\Delta_l}{2}+1)\,
			\Gamma(\tfrac{\Delta_l-1}{2}+\alpha)\,
			\Gamma(\alpha-\tfrac{\Delta_l}{2})
		}\,.
	\end{split}
\end{equation}
The free parameter $\alpha$ is required to satisfy \eqref{eq:flow_dim_local_condition}.

\subsubsection{\texorpdfstring{Local blocks for the $BB\partial$ correlator}{Local blocks for the BBb correlator}}\label{subsec:localblock_bkbkbd}

For the $BB\partial$ correlator 
$\braket{\hat{\Phi}(\tau,z)\hat{\mathcal{O}}(\tau_1,z_1)\mathcal{O}_i(\tau_2)}$,  
we apply the BOE to $\hat{\mathcal{O}}(\tau_1,z_1)$, thereby reducing the correlator to a sum of $B\partial\partial$ correlators:
\begin{equation}\label{Bkbkbd:reduction_bkbdbd}
	\begin{split}
		\braket{\hat{\Phi}(\tau,z)\hat{\mathcal{O}}(\tau_1,z_1)\mathcal{O}_i(\tau_2)}
		=\sum_{j}b^{\hat{\mathcal{O}}}_{j}\,
		\mathcal{D}\!\left(\Delta_j;\tau_1,z_1,\tau_1'\right)
		\braket{\hat{\Phi}(\tau,z)\mathcal{O}_j(\tau_1')\mathcal{O}_i(\tau_2)}\,,
	\end{split}
\end{equation}
where the explicit form of the operator $\DD$ is given in \eqref{eq:Doperator}.
We adopt a canonical choice of $\tau_1'$ such that the three points $(\tau_1,z_1)$, $(\tau_1',0)$, and $(\tau_2,0)$ lie on the same geodesic:
\begin{equation}
	\begin{split}
		\left(\tau_1-\tfrac{\tau_1'+\tau_2}{2}\right)^2+z_1^2
		=\left(\tfrac{\tau_1'-\tau_2}{2}\right)^2 
		\quad\Rightarrow\quad
		\tau_1'=\frac{\tau_1^2+z_1^2-\tau_1\tau_2}{\tau_1-\tau_2}\,.
	\end{split}
\end{equation}
The local-block expansion of $\braket{\hat{\Phi}(\tau,z)\mathcal{O}_j(\tau_1')\mathcal{O}_i(\tau_2)}$ was derived in the previous subsection, and is given by
\begin{equation}\label{Bkbdbd_review}
	\begin{split}
		\braket{\hat{\Phi}(\tau,z)\mathcal{O}_j(\tau_1')\mathcal{O}_i(\tau_2)}
		&=\frac{1}{\abs{\tau_1'-\tau_2}^{\Delta_i+\Delta_j}}
		\left(\frac{(\tau-\tau_1')^2+z^2}{(\tau-\tau_2)^2+z^2}\right)^{\frac{\Delta_{ij}}{2}}
		\mathcal{G}^{\hat{\Phi}}_{ji}(\rho_{1'2})\,, \\
		\mathcal{G}^{\hat{\Phi}}_{ji}(\rho)
		&=\sum_{l}b^{\hat{\Phi}}_{l}
		\frac{C_{ijl}+C_{jil}}{2}\,
		G_{\Delta_l}^{\Delta_j,\Delta_i,(\alpha_{ij})}(\chi_{1'2}) \\
		&\quad+\rho_{1'2}\sum_{l}b^{\hat{\Phi}}_{l}
		\frac{C_{ijl}-C_{jil}}{2}\,
		\tilde{G}_{\Delta_l}^{\Delta_j,\Delta_i,(\alpha_{ij})}(\chi_{1'2})\,,
	\end{split}
\end{equation}
where $\rho_{1'2}$ and $\chi_{1'2}$ are the $SO^+(1,2)$ invariants constructed from the points $(\tau,z)$, $(\tau_1',0)$, and $(\tau_2,0)$. The free parameter $\alpha_{ij}$ is the same as before: it needs to satisfy
\begin{equation}\label{eq:flow_BOE_local_condition}
	\begin{split}
		\alpha_{ij}>\frac{1}{2}\left(\Delta_i+\Delta_j+\Delta_{\hat{\Phi}}\right)
	\end{split}
\end{equation}
for the convergence at $\rho_{1'2}=0$.

Combining \eqref{Bkbkbd:reduction_bkbdbd} with \eqref{Bkbdbd_review}, we get the local-block expansion of 
$\braket{\hat{\Phi}(\tau,z)\hat{\mathcal{O}}(\tau_1,z_1)\mathcal{O}_i(\tau_2)}$:
\begin{equation}
	\begin{split}
		\braket{\hat{\Phi}(\tau,z)\hat{\mathcal{O}}(\tau_1,z_1)\mathcal{O}_i(\tau_2)}
		&=\left(\frac{z_1}{(\tau_1-\tau_2)^2+z_1^2}\right)^{\Delta_i}
		\mathcal{G}^{\hat{\Phi}\hat{\mathcal{O}}}_i(\upsilon,\zeta)\,, \\
		\mathcal{G}^{\hat{\Phi}\hat{\mathcal{O}}}_i(\upsilon,\zeta)
		&=\sum_{l,j}b^{\hat{\Phi}}_{l}b^{\hat{\mathcal{O}}}_{j}
		\frac{C_{ijl}+C_{jil}}{2}\,
		R_{\Delta_l\Delta_j}^{\Delta_i,(\alpha_{ij})}(\upsilon,\zeta) \\
		&\quad+\upsilon\sum_{l,j}b^{\hat{\Phi}}_{l}b^{\hat{\mathcal{O}}}_{j}
		\frac{C_{ijl}-C_{jil}}{2}\,
		\tilde{R}_{\Delta_l\Delta_j}^{\Delta_i,(\alpha_{ij})}(\upsilon,\zeta)\,,
	\end{split}
\end{equation}
where $R_{\Delta_l\Delta_j}^{\Delta_i,(\alpha)}(\upsilon,\zeta)$ and 
$\tilde{R}_{\Delta_l\Delta_j}^{\Delta_i,(\alpha)}(\upsilon,\zeta)$ denote the even and odd local blocks of the $BB\partial$ correlator, respectively. They are defined by
\begin{equation}\label{def:bkbkbd_local}
	\begin{split}
		R_{\Delta_l\Delta_j}^{\Delta_i,(\alpha)}(\upsilon,\zeta)
		&\assign\sum_{n=0}^{\infty}
		\frac{(-1)^n}{n!\,4^n\,(\Delta_j+\tfrac{1}{2})_{n}}\partial_t^{2n}
		\left[
		\frac{G_{\Delta_l}^{\Delta_j,\Delta_i,(\alpha)}\!\left(\tfrac{\zeta^2}{(\upsilon-t)^2+\zeta^2}\right)}
		{\left[(\upsilon-t)^2+\zeta^2\right]^{\Delta_{ji}/2}}
		\right]_{t=0}\!, \\
		\tilde{R}_{\Delta_l\Delta_j}^{\Delta_i,(\alpha)}(\upsilon,\zeta)
		&\assign\frac{1}{\upsilon}
		\sum_{n=0}^{\infty}
		\frac{(-1)^n}{n!\,4^n\,(\Delta_j+\tfrac{1}{2})_{n}}\partial_t^{2n}
		\left[
		\frac{(\upsilon-t)
			\tilde{G}_{\Delta_l}^{\Delta_j,\Delta_i,(\alpha)}\!\left(\tfrac{\zeta^2}{(\upsilon-t)^2+\zeta^2}\right)}
		{\left[(\upsilon-t)^2+\zeta^2\right]^{(\Delta_{ji}+1)/2}}
		\right]_{t=0}\!.
	\end{split}
\end{equation}
Both $R_{\Delta_l\Delta_j}^{\Delta_i,(\alpha)}(\upsilon,\zeta)$ and 
$\tilde{R}_{\Delta_l\Delta_j}^{\Delta_i,(\alpha)}(\upsilon,\zeta)$ are invariant under the $O^+(1,2)$ isometry:
\begin{equation}
	\begin{split}
		R_{\Delta_l\Delta_j}^{\Delta_i,(\alpha)}(\upsilon,\zeta)
		=R_{\Delta_l\Delta_j}^{\Delta_i,(\alpha)}(-\upsilon,\zeta)\,,\qquad
		\tilde{R}_{\Delta_l\Delta_j}^{\Delta_i,(\alpha)}(\upsilon,\zeta)
		=\tilde{R}_{\Delta_l\Delta_j}^{\Delta_i,(\alpha)}(-\upsilon,\zeta)\,.
	\end{split}
\end{equation}

Now let us revisit the second flow equation, which considers the integrated $BB\partial$ correlator. For the same reason as in the $B\partial\partial$ case, the standard conformal-block expansion does not converge at $\upsilon=0$, corresponding to configurations where the three points lie on a common geodesic.  
Hence, we expect the second flow equation expressed in terms of integrated \emph{normal} blocks to fail in general. We will see later that it is indeed the case in some free-theory examples. The resolution is the same: work with local blocks.

The renormalization argument of section~\ref{sec:FlowEqNormal} continues to hold upon replacing normal blocks by local blocks. For block-by-block cancellation of field-strength renormalization constants and anomalous dimensions, one must choose the same parameter $\alpha_{ij}$ both in the computation of $\delta Z_{ij}$ for the $B\partial\partial$ case and for the $\mathcal{O}_j$ contribution in \eqref{Bkbkbd:reduction_bkbdbd}.

The remaining task is to compute the analytic continuation of the integrated block.  
We start from $\Delta_j>\Delta_i$, where the integral converges without a hard cutoff.  
Since the integration is over the full AdS, only the even part contributes; consequently, the coefficient $\mathcal{J}$ in \eqref{eq:flow_BOE_normal} (the integrated normal block) is replaced by the integrated \emph{even} local block:
\begin{equation}
	\begin{split}
		\mathcal{J}_{\Delta_i}^{(\alpha)}(\Delta_l,\Delta_j)
		:=-\int_{\text{AdS}}\frac{dz\, d\tau}{z^2}\,
		R_{\Delta_l\Delta_j}^{\Delta_i,(\alpha)}(\upsilon,\zeta)\quad(\Delta_j>\Delta_i),
	\end{split}
\end{equation}
for which the final result is given by (see \eqref{eq:J alpha in app})
\begin{align}\label{def:J_local}
		\mathcal{J}_{\Delta_i}^{(\alpha)}(\Delta_l,\Delta_j)&=-\frac{2 \sqrt{\pi } \Gamma \left(\alpha -\tfrac{1}{2}\right) \Gamma \left(\Delta_j+\tfrac{1}{2}\right) \Gamma \left(\tfrac{\Delta_l+1}{2}\right) \Gamma \left(\alpha +\tfrac{\Delta_{ij}}{2}\right)}
		{(\Delta_l-1) (\Delta_{j}-\Delta_{i}) \Gamma \left(\alpha -\tfrac{\Delta_l}{2}\right) \Gamma \left(\tfrac{\Delta_i+\Delta_j+1}{2}\right) \Gamma \left(\tfrac{\Delta_{lji}}{2}\right) \Gamma \left(\alpha +\tfrac{\Delta_{lij}-1}{2}\right)}
		\nonumber\\
		&\times
		\,\pFq{3}{2}{ \tfrac{\Delta_{lij}}{2}, \tfrac{\Delta_l-1}{2}, \alpha+\tfrac{\Delta_{l}-1}{2}}{\Delta_l+\tfrac{1}{2}, \alpha+\tfrac{\Delta_{lij}-1}{2}}{1}\, .
\end{align}
The second flow equation is then modified to
\begin{mdframed}[backgroundcolor=shadecolor,linewidth=0pt]
\begin{equation}\label{eq:flow_BOE_local}
				\frac{db^{\hat{\Phi}}_i}{d\lambda}
				= \sum_{l,j} b^{\hat{\Phi}}_l\, b^{\hat{\Phi}}_j \,
				\frac{C_{ilj}+C_{ijl}}{2}\,
				\left[\mathcal{J}_{\Delta_i}^{(\alpha_{ij})}(\Delta_l,\Delta_j)\right]_{\mathrm{reg}}
\end{equation}
\end{mdframed}
where $[\mathcal{J}_{\Delta_i}^{(\alpha_{ij})}(\Delta_l,\Delta_j) ]_\text{reg}$ is defined by

\begin{equation}\label{def:Jreg_alpha}
[\mathcal{J}^{(\alpha)}_{\Delta_i}(\Delta_l,\Delta_j) ]_\text{reg}:=
    \oint_{\Delta_j} \frac{d\Delta}{2\pi i }\frac{\mathcal{J}^{(\alpha)}_{\Delta_i}(\Delta_l,\Delta)}{\D-\D_j}\,.
\end{equation}
Note that we allow an infinite family of free parameters $\alpha_{ij}$ obeying the condition \eqref{eq:flow_BOE_local_condition}. Under these constraints, the unphysical poles of $\mathcal{J}_{\Delta_i}(\Delta_l,\Delta_j)$ at $\Delta_{ji}=-2,-4,\ldots$ are now shifted to $\Delta_{ji}=-2n-2\alpha_{ij}$ with $n=1,2,3,\ldots$.

\subsubsection{\texorpdfstring{Local blocks for $B\partial\partial\partial$ correlators}{Local blocks for Bbbb correlators}}\label{subsec:localblock_bkbdbdbd}

Consider the $B\partial\partial\partial$ correlator 
$\braket{\hat{\Phi}(\tau,z)\mathcal{O}_i(\tau_1)\mathcal{O}_j(\tau_2)\mathcal{O}_k(\tau_3)}$
with fixed $\tau_1<\tau_2<\tau_3$.  
We focus on the $s$-channel domain
\begin{equation}
	\begin{split}
		D_s=\left\{(\tau,z)\in\mathbb{R}\times\mathbb{R}_+\ \big|\ 
		\chi_{23}\leqslant\chi_{12},\,\chi_{13}\right\}\,,
	\end{split}
\end{equation}
where $\chi_{ij}$ denotes the cross-ratio defined in \eqref{def:chi}.

In the $s$-channel expansion, we first apply the OPE to $\mathcal{O}_j\mathcal{O}_k$,  
which reduces the correlator to a sum of integrated $B\partial\partial$ correlators:
\begin{equation}\label{Bkbdbdbd:reduction_bkbdbd}
	\begin{split}
		&\braket{\hat{\Phi}(\tau,z)\mathcal{O}_i(\tau_1)\mathcal{O}_j(\tau_2)\mathcal{O}_k(\tau_3)} \\
		&=\sum_{m}C_{jkm}
		\int_{\tau_2}^{\tau_3}d\tau_4\,\mathcal{B}_{\Delta_j\Delta_k\Delta_m}(\tau_2,\tau_3,\tau_4)\,
		\braket{\hat{\Phi}(\tau,z)\mathcal{O}_i(\tau_1)\mathcal{O}_m(\tau_4)}\,.
	\end{split}
\end{equation}

Up to this point, the construction is standard.  
From the perspective of radial quantization, the convergence of the OPE requires the existence of a geodesic semicircle that separates $(\tau_2,0)$ and $(\tau_3,0)$ from all other operators.  
When the bulk point $(\tau,z)$ lies within the $s$-channel domain $D_s$, this condition is satisfied, and such a semicircle indeed exists.

A subtlety arises, however, when we further expand the bulk operator $\hat{\Phi}$ into boundary operators.  
For each $m$ and fixed $\tau_4$ in \eqref{Bkbdbdbd:reduction_bkbdbd}, the standard conformal-block expansion of the $B\partial\partial$ correlator $\braket{\hat{\Phi}(\tau,z)\mathcal{O}_i(\tau_1)\mathcal{O}_m(\tau_4)}$ fails to converge when $\chi_{14}=1$.  
Since $\tau_4$ is integrated over the interval $[\tau_2,\tau_3]$, convergence requires
\begin{equation}
	\begin{split}
		\chi_{14}<1\,,\qquad \forall\,\tau_4\in[\tau_2,\tau_3]\,.
	\end{split}
\end{equation}
Geometrically, this means that as $\tau_4$ varies from $\tau_2$ to $\tau_3$,  
the geodesic connecting $(\tau_1,0)$ and $(\tau_4,0)$ must never intersect the bulk point $(\tau,z)$.  
Unfortunately, this condition does not hold for all bulk points in $D_s$; it fails within the triangular region shown in figure~\ref{fig:no_stu}.

The resolution is analogous to the $BB\partial$ case.  
We replace the standard conformal-block expansion of 
$\braket{\hat{\Phi}(\tau,z)\mathcal{O}_i(\tau_1)\mathcal{O}_m(\tau_4)}$ 
by its local-block expansion.  
With this replacement, the series over local blocks converges even at $\chi_{14}=1$,  
ensuring that the integrated representation remains valid throughout the entire $s$-channel domain.

Using the local-block expansion, we obtain
\begin{equation}\label{bkbdbdbd:exp_local}
	\begin{split}
		&\braket{\hat{\Phi}(\tau,z)\mathcal{O}_i(\tau_1)\mathcal{O}_j(\tau_2)\mathcal{O}_k(\tau_3)} \\
		&=\sum_{l,m}b^{\hat{\Phi}}_{l}C_{jkm}
		\int_{\tau_2}^{\tau_3}d\tau_4\,\mathcal{B}_{\Delta_j\Delta_k\Delta_m}(\tau_2,\tau_3,\tau_4)\,
		\frac{1}{\abs{\tau_{14}}^{\Delta_i+\Delta_m}}
		\left(\frac{(\tau-\tau_1)^2+z^2}{(\tau-\tau_4)^2+z^2}\right)^{\frac{\Delta_{mi}}{2}} \\
        &\quad\quad \times\left[\frac{C_{iml}+C_{mil}}{2}\,
		G_{\Delta_l}^{\Delta_i,\Delta_m,(\alpha_{im})}(\chi_{14})+
		\frac{C_{mil}-C_{iml}}{2}\,\rho_{14}\,
		\tilde{G}_{\Delta_l}^{\Delta_i,\Delta_m,(\alpha_{im})}(\chi_{14})\right]\,. \\
	\end{split}
\end{equation}

When we integrate over the bulk coordinates $(\tau,z)$,  
the sign of $\rho_{14}$ in \eqref{bkbdbdbd:exp_local} naturally divides the integration domain into the two regions $D_s^{+}$ and $D_s^{-}$ defined in section~\ref{subsec:eq3}.  
The building blocks of the integral, which were denoted by $\mathcal{Q}^{\Delta_l,\Delta_m;\pm}_{\Delta_i,\Delta_j,\Delta_k}$ in \eqref{QsInt:exp_normal}, are now replaced by
\begin{equation}\label{def:IntQ_local}
    \begin{split}
        \mathcal{Q}&^{\Delta_l,\Delta_m;\pm(\alpha)}_{\Delta_i,\Delta_j,\Delta_k}(\tau_1, \tau_2, \tau_3; \epsilon):=\int_{\tau_2}^{\tau_3} d\tau_4\,\mathcal{B}_{\Delta_j,\Delta_k,\Delta_m}(\tau_2,\tau_3,\tau_4) \\
		&\quad \quad \quad \times \int_{\substack{(\tau,z)\in D_s^\pm \\ (\tau-\tau_1)^2+z^2\geqslant\epsilon^2} } \frac{d\tau\, dz}{z^2} \frac{1}{\abs{\tau_{14}}^{\Delta_i + \Delta_m}} \left( \frac{(\tau - \tau_1)^2 + z^2}{(\tau - \tau_4)^2 + z^2} \right)^{\frac{\Delta_{mi}}{2}} G_{\Delta_l}^{\Delta_i,\Delta_m,(\alpha)}(\chi_{14})\,, \\
    \end{split}
\end{equation}
and 
\begin{equation}\label{def:IntQ_local2}
    \begin{split}
         \tilde{\mathcal{Q}}^{\Delta_l,\Delta_m;\pm(\alpha)}_{\Delta_i,\Delta_j,\Delta_k}(\tau_1, \tau_2, \tau_3; \epsilon):=   \mathcal{Q}&^{\Delta_l,\Delta_m;\pm(\alpha)}_{\Delta_i,\Delta_j,\Delta_k}(\tau_1,\tau_2,\tau_3;\epsilon)\Big|_{G_{\Delta_l}^{\Delta_i,\Delta_m,(\alpha)}(\chi_{14})\ \rightarrow\ \abs{\rho_{14}}\tilde{G}_{\Delta_l}^{\Delta_i,\Delta_m,(\alpha)}(\chi_{14})}\,. \\
    \end{split}
\end{equation}
Compared with the definition of $\mathcal{Q}^{\Delta_l,\Delta_m;\pm}_{\Delta_i,\Delta_j,\Delta_k}$ in \eqref{def:IntQ_normal},  
the only modification here is that the normal block in the integrand is replaced by the local block.  
Substituting \eqref{def:IntQ_local} into \eqref{bkbdbdbd:exp_local}, we find
\begin{equation}
	\begin{split}
		\left[Q_{ijk}^{\hat{\Phi}}\right]_s
		&= \sum_{l,m} b^{\hat{\Phi}}_{l}\, C_{jkm}\, C_{ilm}\, \frac{\mathcal{Q}^{\Delta_l,\Delta_m;+(\alpha_{im})}_{\Delta_i,\Delta_j,\Delta_k}+\mathcal{Q}^{\Delta_l,\Delta_m;-(\alpha_{im})}_{\Delta_i,\Delta_j,\Delta_k}+\tilde{\mathcal{Q}}^{\Delta_l,\Delta_m;+(\alpha_{im})}_{\Delta_i,\Delta_j,\Delta_k}-\tilde{\mathcal{Q}}^{\Delta_l,\Delta_m;-(\alpha_{im})}_{\Delta_i,\Delta_j,\Delta_k}}{2} \\
		&\quad + \sum_{l,m} b^{\hat{\Phi}}_{l}\, C_{jkm}\, C_{iml}\, \frac{\mathcal{Q}^{\Delta_l,\Delta_m;+(\alpha_{im})}_{\Delta_i,\Delta_j,\Delta_k}+\mathcal{Q}^{\Delta_l,\Delta_m;-(\alpha_{im})}_{\Delta_i,\Delta_j,\Delta_k}-\tilde{\mathcal{Q}}^{\Delta_l,\Delta_m;+(\alpha_{im})}_{\Delta_i,\Delta_j,\Delta_k}+\tilde{\mathcal{Q}}^{\Delta_l,\Delta_m;-(\alpha_{im})}_{\Delta_i,\Delta_j,\Delta_k}}{2},
	\end{split}
\end{equation}
where $\alpha_{im}$ is a free auxiliary parameter for each fixed pair of indices $i$ and $m$, and the minus signs originate from the sign of $\rho_{14}$ in the integral.

The remaining procedure follows the same steps as in section~\ref{subsec:eq3}.  
All subtraction terms cancel against the field-strength renormalization and anomalous-dimension contributions.  
In the end, the third flow equation involves the regularized version of the integrated local blocks.  
We leave the technical details to appendices \ref{subsec:intbkbdbdbd} and \ref{subsec:bkbdbdbd integral transform} and present only the final result here.  

We define even and odd versions of the coefficient $\mathcal{K}$, which arise from using $G$ or $\tilde G$ in (\ref{def:IntQ_local2})\footnote{The relation between $\mathcal{K}$ and $\mathcal{Q}$ is precisely as in (\ref{def:K}).}:
\begin{equation}
    \begin{aligned}
\mathcal{K}^{(\alpha)e}_{\Delta_i\Delta_j\Delta_k}(\Delta_l,\Delta_m)&:=\int\frac{d\Delta}{2\pi i}K^{(\alpha)}_{\Delta_i,\Delta_m}(\Delta,\Delta_l)\left(\mathcal{K}_{\Delta_i\Delta_j\Delta_k}(\Delta,\Delta_m)+\mathcal{K}_{\Delta_i\Delta_k\Delta_j}(\Delta,\Delta_m)\right)\,,\\
\mathcal{K}^{(\alpha)o}_{\Delta_i\Delta_j\Delta_k}(\Delta_l,\Delta_m)&:=\int\frac{d\Delta}{2\pi i}\tilde {K}^{(\alpha)}_{\Delta_i,\Delta_m}(\Delta,\Delta_l)\left(\mathcal{K}_{\Delta_i\Delta_j\Delta_k}(\Delta,\Delta_m)-\mathcal{K}_{\Delta_i\Delta_k\Delta_j}(\Delta,\Delta_m)\right)\,,
\label{eq:evenoddK}
\end{aligned}
\end{equation}
where the integral kernels $K$ and $\tilde K$ were introduced in \cite{Loparco:2025aag} and are reported in appendix \ref{app:local block kernel}, the contour of integration over $\Delta$ wraps the poles at $\Delta=\Delta_l$ and at $\Delta=2\alpha+2\mathbb{N}_0$ and the explicit form of $\mathcal{K}$ is in (\ref{eq:K}). In practice, the integral can be parametrized through a straight vertical contour, as we discuss in more detail in \ref{subsec:bkbdbdbd integral transform}.

The full coefficient is the sum of the even and odd parts
\begin{equation}\label{K:local_decomp_even_odd}
    \mathcal{K}^{(\alpha)}_{\Delta_i\Delta_j\Delta_k}(\Delta_l,\Delta_m):=\frac{1}{2}\left[\mathcal{K}^{(\alpha)e}_{\Delta_i\Delta_j\Delta_k}(\Delta_l,\Delta_m)+\mathcal{K}^{(\alpha)o}_{\Delta_i\Delta_j\Delta_k}(\Delta_l,\Delta_m)\right]\,.
\end{equation}
With these definitions, the third flow equation reads
\begin{mdframed}[backgroundcolor=shadecolor,linewidth=0pt]
\begin{align}
\label{eq:flow_OPE_local}
				\frac{dC_{ijk}}{d\lambda}
				&= \sum_{l,m} b^{\hat{\Phi}}_{l}\, C_{jkm}
				\Big(
				C_{ilm}\,[\mathcal{K}^{(\alpha_{im})}_{\Delta_i,\Delta_j,\Delta_k}(\Delta_l,\Delta_m)]_{\mathrm{reg}}
				+ C_{iml}\,[\mathcal{K}^{(\alpha_{im})}_{\Delta_i,\Delta_k,\Delta_j}(\Delta_l,\Delta_m)]_{\mathrm{reg}}
				\Big) \nonumber\\
				&\qquad + (ijk \rightarrow jki) + (ijk \rightarrow kij)
\end{align}
\end{mdframed}
where once again we need to subtract the pole at $\Delta_m=\Delta_i$:
\begin{equation}
    [\mathcal{K}^{(\alpha)}_{\Delta_i,\Delta_j,\Delta_k}(\Delta_l,\Delta_m) ]_\text{reg}:=
		  \oint_{\Delta_m} \frac{d\Delta}{2\pi i }\frac{\mathcal{K}^{(\alpha)}_{\Delta_i,\Delta_j,\Delta_k}(\Delta_l,\Delta)}{\D-\D_m}\,.
\end{equation}
The parameter $\alpha_{pm}$ is bounded analogously to the previous cases,
\begin{equation}
\alpha_{pm}>\frac{\Delta_{\hat{\Phi}}+\Delta_p+\Delta_m}{2}\,,
\end{equation}
where $p$ can be $i$, $j$, or $k$, depending on the channel. This condition ensures that, for each fixed $m$, the sum over $l$ converges absolutely. Once the sum over $l$ has been performed, the remaining sum over $m$ is expected to converge exponentially fast.\footnote{For the sum over $m$, the exponential decay arises because the domain of the bulk integral lies at a finite distance away from the geodesic connecting the external boundary operators. Propagation of the state across this distance leads to an exponential suppression of the form $e^{-\Delta \times \text{distance}}$.
}

In a parity-preserving theory, the third flow equation simplifies. Consider the following equivalent rewriting of (\ref{eq:flow_OPE_local}):
\begin{equation}
 \begin{aligned}
 				\frac{dC_{ijk}}{d\lambda}
 				&= \sum_{l,m} b^{\hat{\Phi}}_{l}\, C_{jkm}
 				\Big(
 				\frac{C_{ilm}+C_{mli}}{2}\,[\mathcal{K}^{(\alpha_{im})e}_{\Delta_i,\Delta_j,\Delta_k}(\Delta_l,\Delta_m)]_{\mathrm{reg}}
 				+ \frac{C_{ilm}-C_{mli}}{2}\,[\mathcal{K}^{(\alpha_{im})o}_{\Delta_i,\Delta_j,\Delta_k}(\Delta_l,\Delta_m)]_{\mathrm{reg}}
 				\Big) \\
 				&\qquad + (ijk \rightarrow jki) + (ijk \rightarrow kij)
                \label{eq:parityCijkeq}
     \end{aligned}
     \end{equation}
When operators have definite parity, either $C_{ilm}=C_{mli}$ or $C_{ilm}=-C_{mli}$. Therefore, one of the two terms in the first line of (\ref{eq:parityCijkeq}) vanishes. 

\subsection{Poles of the universal coefficients}\label{subsec:poles}
We have derived three flow equations, \eqref{eq:flow_dim_local}, \eqref{eq:flow_BOE_local}, and \eqref{eq:flow_OPE_local}, which form a closed system under the evolution. Here, we discuss the pole structures of the coefficients that appear in these equations:
\[
\mathcal{I}^{(\alpha)}, \qquad
\mathcal{J}^{(\alpha)}, \qquad
\text{and} \qquad
\mathcal{K}^{(\alpha)}\,.
\]

\subsubsection{\texorpdfstring{$\mathcal{I}$}{I}}
For the first equation, the coefficient $\mathcal{I}^{(\alpha)}(\Delta_l)$ is given by \eqref{def:I_local}, which exhibits three types of poles:
\begin{equation}
	\begin{split}
		(a)\ \Delta_l = -\frac{1}{2}-n,\qquad (b)\ \Delta_l = 1,\qquad (c)\ \alpha = \frac{1}{2}-n\,,
	\end{split}
\end{equation}
with $n$ being non-negative integers. 

The poles of type ($a$) reflect the fact that the conformal block itself has poles at these values of $\Delta_l$.  The dimension-$\Delta_l$ representations of the 1D conformal group becomes reducible there, and the divergence comes from the contribution of the null states. In a unitary theory, these poles are never reached because $\Delta_l\geqslant0$.

The poles of type ($b$) and ($c$) arise from the structure of the integrated $B\partial\partial$ local block. When we integrate out the bulk point $(\tau,z)$ (in {Poincaré coordinates}), near the boundary point $(\tau_1,0)$ we can use the polar coordinates
\begin{equation}
	\begin{split}
		\tau=\tau_1+r\cos\theta,\qquad z=r\sin\theta,\qquad0\leqslant\theta\leqslant\pi\,,
	\end{split}
\end{equation}
and the AdS$_2$ measure becomes $\frac{drd\theta}{r\sin^2\theta}$. In the case when the two boundary operators are identical, the $r$-integral diverges logarithmically. The $\theta$-integral, corresponding to the coefficient of the logarithmic divergence, contains terms of the following form: 
\begin{equation*}
	\begin{split}
		\int_{0}^{\pi} d\theta\, (\sin\theta)^{\Delta_l-2}\,{}_2F_1(\ldots;\sin^2\theta),
		\qquad
		\int_{0}^{\pi} d\theta\, (\sin\theta)^{2\alpha-2}\,{}_3F_2(\ldots;\sin^2\theta),
	\end{split}
\end{equation*}
which originate from the first and second terms of the local block, respectively.  
Thus the total contribution begins to diverge at $\Delta_l=1$ and $\alpha=\tfrac{1}{2}$.\footnote{Typically, when integrating a power function against a smooth function, e.g. $\int_{0}^{1} dx\, x^{\mu}\,\varphi(x)$, the analytic continuation in the exponent $\mu$ develops poles at negative integers \cite{GelfandShilov1964}. This explains the poles in $\alpha$. The appearance of a pole at $\Delta_l=1$ but without further poles afterwards is special to this setup, and can be verified explicitly from the calculation.}

The pole of type ($b$) corresponds to the case that a boundary operator hits marginality. Its residue is given by
\begin{equation}
    \text{Res}_{\Delta_l=1}\mathcal{I}^{(\alpha)}(\Delta_l)=2\,.
\end{equation}
When it happens, the contribution from this operator dominates the flow equation of the scaling dimension. Consequently, the scaling dimensions will have square-root behavior with respect to the coupling: $\Delta_i(\lambda)\propto\sqrt{\lambda-\lambda_c}$, where $\lambda_c$ is the critical coupling at which one of the scaling dimensions is equal to 1. We will discuss this point in detail in section \ref{sec:relevant and marginal boundary op}.

Another interesting feature of $\mathcal{I}^{(\alpha)}(\Delta_l)$ is the presence of the factor $\Gamma(\alpha-\Delta_l/2)^{-1}$. Suppose we choose, for example, $\alpha=\Delta_i+\lfloor\tfrac{1}{2}\Delta_{\hat{\Phi}}\rfloor+1$, where $\Delta_i$ is the scaling dimension of the external boundary operator. Then almost all operators in the “double-trace family”, with dimensions
\[
\Delta_l = 2\Delta_i + 2\lfloor\tfrac{1}{2}\Delta_{\hat{\Phi}}\rfloor + 2n\, , \qquad (n=1,2,\ldots),
\]
do not contribute. In free theories, this specific choice leads to a truncation of the double-trace channel; in weakly coupled theories (where the CFT data are close to those of the free theory), it leads instead to a suppression of the double-trace contributions. See section \ref{sec:Large N factorization} for more details.

\subsubsection{\texorpdfstring{$\mathcal{J}$}{J}}

For the second equation, the coefficient $\mathcal{J}_{\Delta_i}^{(\alpha)}(\Delta_l,\Delta_j)$ is given by \eqref{def:J_local}. In the expression for $\mathcal{J}$, the following poles originate from the poles of the conformal/local blocks:
\begin{equation}
	\begin{split}
		(a)\ \Delta_l=-\frac{1}{2}-n,\qquad
		(b)\ \Delta_j=-\frac{1}{2}-n,\qquad
		(c)\ \alpha=\frac{1}{2}\Delta_{ji}-n\,,
	\end{split}
\end{equation}
while the following poles arise from divergences of the integral:
\begin{equation}
	\begin{split}
		(e)\ \Delta_l=1,\qquad
		(f)\ \Delta_l=-1-2n,\qquad
		(g)\ \alpha=\frac{1}{2}-n,\qquad
		(h)\ \Delta_j=\Delta_i\,.
	\end{split}
\end{equation}

In a unitary theory, the poles of types ($a$), ($b$), and ($f$) are never reached. The poles of type ($c$) are also avoided because we have imposed the condition
\begin{equation}
	\begin{split}
		\alpha>\frac{1}{2}(\Delta_i+\Delta_j+\Delta_{\hat{\Phi}})\geqslant\frac{1}{2}\Delta_{ji}\,,
	\end{split}
\end{equation}
where we have used the unitarity condition $\Delta_i,\Delta_{\hat{\Phi}}\geqslant 0$.

The poles of type ($g$) are not physical, since they can always be avoided by choosing $\alpha$ sufficiently large.

Let us now examine the pole at $\Delta_l=1$ (type ($e$)). The residue of $\mathcal{J}_{\Delta_i}^{(\alpha)}(\Delta_l,\Delta_j)$ at this pole is given by
\begin{equation}
	\begin{split}
		\text{Res}_{\Delta_l=1}\mathcal{J}_{\Delta_i}^{(\alpha)}(\Delta_l,\Delta_j)
		=-\frac{2\sqrt{\pi}\Gamma\!\left(\Delta_j+\tfrac{1}{2}\right)}
		{(\Delta_j-\Delta_i)\Gamma\!\left(\tfrac{1+\Delta_j+\Delta_i}{2}\right)
			\Gamma\!\left(\tfrac{1+\Delta_j-\Delta_i}{2}\right)}\,.
	\end{split}
\end{equation}
We see that the singular term is independent of $\alpha$. This is because, when integrating the local block with $\Delta_l=1$, only the part of the conformal block contributes to the logarithmic divergence. This singularity leads to the square-root behavior of the BOE coefficients when a boundary operator becomes marginal (see section~\ref{sec:relevant and marginal boundary op}). 

Next, we consider the pole at $\Delta_j=\Delta_i$ (type ($h$)). Its residue is
\begin{equation}\label{eq:J_residue}
	\begin{split}
		\text{Res}_{\Delta_j=\Delta_i}\mathcal{J}_{\Delta_i}^{(\alpha)}(\Delta_l,\Delta_j)
		&= -\mathcal{I}^{(\alpha)}(\Delta_l)\,,
	\end{split}
\end{equation}
where $\mathcal{I}^{(\alpha)}(\Delta_l)$ is the coefficient appearing in the first flow equation, given in \eqref{def:I_local}. This relation is not surprising: near $\Delta_i=\Delta_j$, the divergent part of the integral arises from the leading term of the BOE,
\[
\hat{\Phi}(\tau_1,z_1)
= b^{\hat{\Phi}}_j\, z_1^{\Delta_j}\,\mathcal{O}_j(\tau_1) + \ldots\,,
\]
where $\hat{\Phi}(\tau_1,z_1)$ is the unintegrated bulk operator. The computation then reduces to integrating the $B\partial\partial$ correlator $\langle \hat{\Phi}\,\mathcal{O}_i\,\mathcal{O}_j\rangle$.  
In this case, the residue of the subtraction term at $\Delta_j=\Delta_i$ coincides with the coefficient of the logarithmic divergence (block by block), which explains \eqref{eq:J_residue}. As we will see later, this pole plays an important role in the mechanism of level repulsion (see section~\ref{subsec:level repulsion}).

\subsubsection{\texorpdfstring{$\mathcal{K}$}{K}}

For the third equation, the coefficient
$\mathcal{K}^{(\alpha)}_{\Delta_i,\Delta_k,\Delta_j}(\Delta_l,\Delta_m)$
has not been obtained in closed form. Nevertheless, based on its construction, we can still infer its expected pole structure. Since $\mathcal{K}$ is decomposed into even and odd parts, cf.\ \eqref{K:local_decomp_even_odd}, which originate from the integrated even and odd local blocks, respectively, we discuss the pole structure of these two contributions separately.

For the even part $\mathcal{K}^{(\alpha)e}_{\Delta_i\Delta_j\Delta_k}(\Delta_l,\Delta_m)$, we expect the following poles arising from the even local block itself:
\begin{equation}
	\begin{split}
		(a)\ \Delta_l=-\frac{1}{2}-n,\quad
		(b)\ \Delta_m=-\frac{1}{2}-n,\quad
		(c)\ \alpha=\frac{1}{2}\Delta_{mi}-n,\quad
		(d)\ \alpha=\frac{1}{2}\Delta_{im}-n\,.
	\end{split}
\end{equation}
The poles originating from divergent integrals are similar to those in the case of $\mathcal{J}$:
\begin{equation}
	\begin{split}
		(e)\ \Delta_l=1,\qquad
		(f)\ \Delta_l=1-2n,\qquad
		(g)\ \alpha=\frac{1}{2}-n,\qquad
		(h)\ \Delta_m=\Delta_i\,.
	\end{split}
\end{equation}
The actual result may be more regular, since the residues at some of these poles could vanish. Here we simply list all poles that are expected on general grounds.

For the odd part $\mathcal{K}^{(\beta)o}_{\Delta_i\Delta_j\Delta_k}(\Delta_l,\Delta_m)$, the poles coming from the odd local block itself are expected to be
\begin{equation}
\begin{aligned}
&(a')\ &\Delta_l&= -\frac{1}{2}-n &\qquad &(b')&\Delta_m &=-\frac{1}{2}-n\\
&(c')\ & \beta&= \frac{1}{2}\Delta_{mi}-\frac{1}{2}-n&\qquad &(d')&\beta &= \frac{1}{2}\Delta_{im}-\frac{1}{2}-n\, .
\end{aligned}
\end{equation}
The poles arising from divergent integrals are similar to the even case:
\begin{equation}
	\begin{split}
		(e')\ \Delta_l=1,\qquad
		(f')\ \Delta_l=1-2n,\qquad
		(g')\ \beta=\frac{1}{2}-n\,.
	\end{split}
\end{equation}
The pole at $\Delta_m=\Delta_i$ does not appear for the odd part because it gets canceled.

In a unitary theory, the poles of types ($a$)–($d$), ($f$), and ($g$) are all avoidable, for the same reasons discussed in the analysis of $\mathcal{J}$. The same conclusion applies to the odd part.

We now focus on the pole at $\Delta_l=1$. Since this pole arises from the divergent integral of the standard conformal block, its residue does not depend on the free parameters $\alpha$ and $\beta$:
\begin{equation}\label{eq:K_residue_Dl_evenodd}
	\begin{split}
		\text{Res}_{\Delta_l=1}\mathcal{K}^{(\alpha)e}_{\Delta_i\Delta_j\Delta_k}(\Delta_l,\Delta_m)
		&= \frac{2^{\Delta_{im}}}{\Delta_{im}}
		\left[
		\hyperF{\Delta_{mi}}{\Delta_{mjk}}{2\Delta_m}{\tfrac{1}{2}}
		+\hyperF{\Delta_{mi}}{\Delta_{mkj}}{2\Delta_m}{\tfrac{1}{2}}
		\right], \\
		\text{Res}_{\Delta_l=1}\mathcal{K}^{(\beta)o}_{\Delta_i\Delta_j\Delta_k}(\Delta_l,\Delta_m)
		&= \frac{2^{\Delta_{im}}}{\Delta_{im}}
		\left[
		\hyperF{\Delta_{mi}}{\Delta_{mjk}}{2\Delta_m}{\tfrac{1}{2}}
		-\hyperF{\Delta_{mi}}{\Delta_{mkj}}{2\Delta_m}{\tfrac{1}{2}}
		\right].
	\end{split}
\end{equation}
In terms of the full coefficient
$\mathcal{K}^{(\alpha)}_{\Delta_i\Delta_j\Delta_k}(\Delta_l,\Delta_m)$
defined in \eqref{K:local_decomp_even_odd}, we find
\begin{equation}\label{eq:K_residue_Dl_full}
	\begin{split}
		\text{Res}_{\Delta_l=1}
		\mathcal{K}^{(\alpha)}_{\Delta_i\Delta_j\Delta_k}(\Delta_l,\Delta_m)
		= \frac{2^{\Delta_{im}}}{\Delta_{im}}\,
		\hyperF{\Delta_{mi}}{\Delta_{mjk}}{2\Delta_m}{\tfrac{1}{2}}\,.
	\end{split}
\end{equation}
The right-hand side coincides with the coefficient appearing in the ordinary differential equation describing the marginal deformation in one-dimensional CFTs \cite{Behan:2017mwi} (up to a minus sign due to different conventions). Technically, this agreement arises because we are integrating a one-dimensional CFT four-point function
$\langle \mathcal{O}_l \mathcal{O}_i \mathcal{O}_j \mathcal{O}_k \rangle$,
where $\mathcal{O}_l$ is the integrated operator with $\Delta_l=1$. This integral precisely corresponds to the first-order correction to the three-point function under a marginal deformation. As in the previous case, this pole leads to a square-root behavior of the OPE coefficients when a boundary operator becomes marginal (see section~\ref{sec:relevant and marginal boundary op}).

Next, we consider the pole at $\Delta_m=\Delta_i$. The residue of the even   part of $\mathcal{K}$ is given by  
\begin{equation}\label{eq:K_residue}
	\begin{split}
		\text{Res}_{\Delta_m=\Delta_i}\mathcal{K}^{(\alpha)e}_{\Delta_i\Delta_j\Delta_k}(\Delta_l,\Delta_m)
		&= -\mathcal{I}^{(\alpha)}(\Delta_l)\,.
	\end{split}
\end{equation}
 Once again, $\mathcal{I}^{(\alpha)}$ denotes the coefficient appearing in the first flow equation. The origin of this behavior is similar to that in the case of $\mathcal{J}$. Here we spell out the argument in more detail, since we do not have an explicit formula for $\mathcal{K}$. We use the OPE to reduce the $B\partial\partial\partial$ block to an integral of the $B\partial\partial$ block, cf.\ \eqref{Bkbdbdbd:reduction_bkbdbd}. The residue of the even part of $\mathcal{K}$ at $\Delta_m=\Delta_i$ then follows from the fact that the subtracted integral does not possess such a pole:
\begin{equation}
	\begin{split}
		&-\frac{\mathcal{K}^{(\alpha)e}_{\Delta_i\Delta_j\Delta_k}(\Delta_l,\Delta_m)}
		{\abs{\tau_{12}}^{\Delta_{ijk}}\abs{\tau_{13}}^{\Delta_{ikj}}\abs{\tau_{23}}^{\Delta_{jki}}}
		+\frac{2a_0^{(\alpha)}(\Delta_{im},\Delta_l)}{\Delta_{im}}
		\frac{\epsilon^{\Delta_{mi}}}
		{\abs{\tau_{12}}^{\Delta_{mjk}}\abs{\tau_{13}}^{\Delta_{mkj}}\abs{\tau_{23}}^{\Delta_{jkm}}} 
	\end{split}
\end{equation}
is regular at $\Delta_m=\Delta_i$.
Here $a_0^{(\alpha)}$ denotes the coefficient of the leading subtraction term in the integrated $B\partial\partial$ local block, which is related to $\mathcal{I}^{(\alpha)}$ via
$\mathcal{I}^{(\alpha)}(\Delta_l)=2a_0^{(\alpha)}(0,\Delta_l)$. By canceling the residue at $\Delta_m=\Delta_i$, we obtain \eqref{eq:K_residue}. 

For the odd part of $\mathcal{K}$, the situation is different: technically, the factor $2a_0$ is replaced by $a_0-a_0=0$, and therefore no pole arises at $\Delta_m=\Delta_i$.

In terms of the full coefficient $\mathcal{K}^{(\alpha)}_{\Delta_i\Delta_j\Delta_k}(\Delta_l,\Delta_m)$, we have
\begin{equation}\label{eq:K_residue_full}
	\begin{split}
		\text{Res}_{\Delta_m=\Delta_i}
		\mathcal{K}^{(\alpha)}_{\Delta_i\Delta_j\Delta_k}(\Delta_l,\Delta_m)
		= -\frac{1}{2}\mathcal{I}^{(\alpha)}(\Delta_l)\,.
	\end{split}
\end{equation}
As in the previous case, this pole will play an important role in the mechanism of level repulsion (see section~\ref{subsec:level repulsion}).

\subsection{\texorpdfstring{Summary of integrated blocks in AdS$_2$}{Summary of integrated blocks in AdS2}}
\label{sec:integrals of blocks}

Here we collect the final expressions for both the integrated normal and local blocks. The technical details are delegated to appendix \ref{app:integrated blocks}.

Using normal blocks, the integral for the $B\del\del$ block is
\begin{equation}
\II(\D_l)
\assign
\int_{AdS} G_{\D_l}^{\D_1 \D_1}(\chi)\bigg|_{\text{log coef}}
=\frac{2 \sqrt{\pi } \Gamma \left(\Delta_l +\frac{1}{2}\right)}{(\Delta_l -1) \Gamma \left(\frac{\Delta_l}{2}\right) \Gamma \left(\frac{\Delta_l}{2}+1\right)}\,.
\label{eq:I}
\end{equation}
The integral for the $BB\del$ block is
\begin{equation}
    \JJ_{\D_i}(\D_l,\D_j)
\assign -\int_{AdS} R_{\D_l,\D_j}^{\D_i}(\upsilon,\z)   
=
-
    \frac{
    2^{\Delta_{lji}-2}
\Gamma\!\left( \frac{\Delta_l - 1}{2} \right)
\Gamma\!\left( \Delta_l + \frac{1}{2} \right)
\Gamma\!\left( \Delta_j + \frac{1}{2} \right)
\Gamma\!\left( \frac{\Delta_{ji}}{2} \right)
}{
\Gamma\!\left( \frac{\Delta_l}{2} + 1 \right)
\Gamma\!\left( \frac{\Delta_j + \Delta_i + 1}{2} \right)
\Gamma(\Delta_{lji})
}\,.
    \label{eq:J}
\end{equation}
The poles at $\Delta_i=\Delta_j+2n$ are unphysical, except for the one at $n=0$ which is canceled by field-strength renormalization. We are not aware of a natural procedure to eliminate these poles, but they do not appear when we use local blocks.

The integral for the $B\del\del\del$ block is 
\begin{align}
&\KK_{\D_i \D_j \D_k}(\D_l,\D_m)
\assign - \int_{R} K_{\D_l\D_m}^{\D_i\D_j\D_k}(\m,\w)\label{eq:K}\\
=&C\Bigg[\int_{s,c}\frac{\Gamma(\frac{1-c}{2}-s)\Gamma(s-\frac{1}{2})\Gamma(c+2s)\Gamma(\Delta_{ijk}-c)\Gamma(\frac{\Delta_l}{2}-s)\Gamma(\Delta_l-c)\Gamma(c+\Delta_{mi})}{2^{2(s+1)}\Gamma(1-\frac{c}{2})\Gamma(\frac{1+\Delta_l}{2}+s)\Gamma(c+\Delta_l)\Gamma(\Delta_i+\Delta_m-c)}\nonumber\\
&+\int_u\frac{\pi\csc(\pi(2u+\Delta_{mi}))\Gamma(\frac{\Delta_l}{2}-u)}{2^{2u}(1-2u)\Gamma(u+\frac{\Delta_l+1}{2})}\int_0^1d\tau\ \frac{\tau^{\Delta_{ikj}-2u}\ _2\tilde F_1\left(\begin{matrix}1-\Delta_{jmk} & \Delta_{kmj}\\ 1-2u+\Delta_{ikj}\end{matrix};\tau\right)}{\max(\sqrt{1-\tau^2},\sqrt{(2-\tau)\tau})^{1-2u}}\Bigg]\nonumber
\end{align}
where the region $R$ is schematically one-sixth of AdS$_2$, see figure \ref{fig:pizza slices}, $\tilde{F}$ is a regularized hypergeometric function, $\int_{s}\equiv\int\frac{ds}{2\pi i}$ and the integration contours over $s$, $c$ and $u$ are vertical contours satisfying the constraints listed in (\ref{eq:Kcontours}). Visualizations of the contours and the poles are available in the ancillary Mathematica notebook \texttt{testflows.nb}. Finally, here
\begin{equation}
    C\equiv-\frac{2^{\Delta_l}\Gamma(2\Delta_m)\Gamma(\Delta_l+\frac{1}{2})}{\Gamma(\Delta_{jmk})\Gamma(\Delta_{lmi})}
\end{equation}
and when $\Delta_m=\Delta_i$ one has to throw away the pole and keep the finite part.

With local blocks, the integrals are functions of an auxiliary parameter $\alpha$, which can be chosen in a way which optimizes the convergence of the sums in the flow equations. Explicitly, we have
\begin{equation}
    \II^{(\a)}(\Delta_l)=\frac{2^{3-2\alpha}\pi\Gamma(\Delta_l+\frac{1}{2})\Gamma(2\alpha-1)}{(\Delta_l-1)\Gamma(\frac{\Delta_l}{2})\Gamma(\frac{\Delta_l}{2}+1)\Gamma(\frac{\Delta_l-1}{2}+\alpha)\Gamma(\alpha-\frac{\Delta_l}{2})}\,,
\label{eq:I alpha}
\end{equation}
which is derived in \eqref{eq:Local Bbb app}
and
\begin{equation}
\begin{aligned}
    \JJ_{\D_i}^{(\a)}(\D_l,\D_j)
    =
    &-\frac{2 \sqrt{\pi } \Gamma \left(\alpha -\frac{1}{2}\right) \Gamma \left(\D_j+\frac{1}{2}\right) \Gamma \left(\frac{\D_l+1}{2}\right) \Gamma \left(\alpha +\frac{\D_{ij}}{2}\right)}
    {(\D_l-1) (\D_{j}-\D_{i}) \Gamma \left(\alpha -\frac{\D_l}{2}\right) \Gamma \left(\frac{\D_i+\D_j+1}{2}\right) \Gamma \left(\frac{\D_{lji}}{2}\right) \Gamma \left(\alpha +\frac{\D_{lij}-1}{2}\right)}\times
    \\
    &\times
    \,\pFq{3}{2}{ \frac{\Delta_{lij}}{2}, \frac{\Delta_l-1}{2}, \alpha+\frac{\Delta_{l}-1}{2}}{\Delta_l+\frac{1}{2}, \alpha+\frac{\Delta_{lij}-1}{2}}{1}\, .
\end{aligned}
\label{eq:J alpha}
\end{equation}
When $\Delta_i=\Delta_j$ one must throw away the simple pole and keep the finite part.

Finally, 
\begin{equation}
    \mathcal{K}^{(\alpha)}_{\Delta_i\Delta_j\Delta_k}(\Delta_l,\Delta_m):=\frac{\mathcal{K}^{(\alpha)e}_{\Delta_i\Delta_j\Delta_k}(\Delta_l,\Delta_m)+\mathcal{K}^{(\alpha)o}_{\Delta_i\Delta_j\Delta_k}(\Delta_l,\Delta_m)}{2}
    \label{eq:K alpha}
\end{equation}
with 
\begin{equation}
    \begin{aligned}
\mathcal{K}^{(\alpha)e}_{\Delta_i\Delta_j\Delta_k}(\Delta_l,\Delta_m)&:=\int\frac{d\Delta}{2\pi i}K^{(\alpha)}_{\Delta_i,\Delta_m}(\Delta,\Delta_l)\left(\mathcal{K}_{\Delta_i\Delta_j\Delta_k}(\Delta,\Delta_m)+\mathcal{K}_{\Delta_i\Delta_k\Delta_j}(\Delta,\Delta_m)\right)\,,\\
\mathcal{K}^{(\alpha)o}_{\Delta_i\Delta_j\Delta_k}(\Delta_l,\Delta_m)&:=\int\frac{d\Delta}{2\pi i}\tilde {K}^{(\alpha)}_{\Delta_i,\Delta_m}(\Delta,\Delta_l)\left(\mathcal{K}_{\Delta_i\Delta_j\Delta_k}(\Delta,\Delta_m)-\mathcal{K}_{\Delta_i\Delta_k\Delta_j}(\Delta,\Delta_m)\right)\,,
\end{aligned}
\end{equation}
where $K$ and $\tilde K$ are the kernels that turn normal blocks into even and odd local blocks (see appendix \ref{app:local block kernel}), and the constraints on the contours of integration over $\Delta$ are discussed in appendix \ref{subsec:bkbdbdbd integral transform}, eq. (\ref{eq:CppKR1R2}).


\section{Tests in free theory}
\label{sec:test of flow eqn}
In this section we test the flow equations (\ref{eq:dimension flow} - \ref{eq:OPE flow}) in the theory of a free massive scalar in AdS$_2$. We treat the mass as a tuneable coupling and study the evolution of some scaling dimensions, BOE and OPE coefficients.
Despite this being a free theory, the flow equations are not obviously satisfied, so the checks we perform are non-trivial. All checks presented in this section are contained in the attached Mathematica notebook \texttt{testflows.nb}.
\subsection{Setup}
We consider a free massive scalar in AdS$_2$
\begin{equation}\label{def:freescalar}
		S=\frac{1}{2}\int d^2x \sqrt{g}\left[g^{\mu\nu}\partial_\mu\hat\phi\partial_\nu\hat\phi+m^2\phih^2\right]\,,
\end{equation}
with Dirichlet boundary conditions 
\begin{equation}\label{Dbc}
	\begin{split}
		\phih(\tau,z=0)=\phih(\tau,z=+\infty)=0\, .
	\end{split}
\end{equation}
The bulk two-point function of $\hat{\phi}(x)$ is the Green's function of the AdS$_2$ Laplacian:
\begin{equation}\label{eq:bulktwopt:free}
\braket{\hat{\phi}(x)\hat{\phi}(y)} = \frac{\Gamma(\Delta_\f)}{2\sqrt{\pi}\Gamma\left(\Delta_\f + \frac{1}{2}\right)}\eta^{\Delta_\f}\, _{2}F_1\left(\Delta_\f,\frac{1}{2};\Delta_\f + \frac{1}{2};\eta^2\right)\, ,\qquad \qquad (0\leqslant \eta <1)\, ,
\end{equation}
where $\eta$ is the two-point invariant defined in \eqref{def:eta} for $x= (\tau_1,z_1)$ and $y= (\tau_2,x_2)$. We have the standard relation between $\Delta_\phi$ and the mass 
\begin{equation}\label{eq:DefDelta}
    \Delta_\f(\Delta_\f-1)= m^2\, ,\qquad \qquad (\Delta_\f\geqslant 1)\, .
\end{equation}
Imposing Dirichlet boundary conditions means picking a specific branch of the solutions to \eqref{eq:DefDelta} and imposing Neumann boundary conditions leads to the other branch
\al{\spl{
{\rm D:} \quad \D_\f = \frac12 + \frac12 \sqrt{1+4m^2}\,,
\\
{\rm N:} \quad \D_\f = \frac12 - \frac12 \sqrt{1+4m^2}\,.
}
\label{eq:Delta_phi D vs N}}
The BOE of the bulk operator $\hat{\phi}(\tau,z)$ contains only one primary boundary operator which we denote by $\phi$, whose two-point function is unit-normalised
\begin{equation}
   \braket{\phi(\tau_1)\phi(\tau_2)} = \frac{1}{|\tau_1-\tau_2|^{2\Delta_\f}}\, .
\end{equation}
Moreover, in this work, we will consider the bulk operator $\hat{\phi}^2(\tau,z)$, whose boundary operator expansion contains more primary operators. In particular, by Wick theorem, all double trace primaries $[\phi^2]_n$ of the form \cite{Mikhailov:2002bp,Penedones:2010ue}
\begin{equation}
    [\f^2]_n\propto \sum_{k=0}^n\frac{(-1)^k(2n)!(2\Delta_\f)_{2n}}{k!(2n-k)!(2\Delta_\f)_k(2\Delta_\f)_{2n-k}}:(\partial^k\f(\tau))(\partial^{2n-k}\f(\tau)):\, ,
\end{equation}
contribute, together with their descendants. The scaling dimension of such an operator is $\Delta_{[\f^2]_n} = 2\Delta_\f+2n$.

\subsection{Flow equations}
From the flow equation point of view, the coupling $\lambda$ is $\frac{1}{2}m^2$ and the deformation operator $\hat{\Phi}$ is $\hat\f^2$. Since all the QFT data is known (see appendix \ref{app:free theory data}), 
we can check the flow equations at arbitrary values of $m^2$.

\subsubsection{Scaling dimension flow}
We will start by testing the flow of the scaling dimensions of the primaries $\phi$ and $[\phi^2]_p$ as we tune the mass. The scaling dimension of $\phi$ is fixed by the relation $\Delta_\phi(\Delta_\phi-1)=m^2$ and $[\phi^2]_p$ instead has scaling dimension $\Delta_{[\phi^2]_p}=2\Delta_\phi+2p$. The derivatives can be computed trivially
\begin{equation}
\D_\f(\D_\f-1) = m^2
\ \Rightarrow\ 
\frac{d \D_\f}{dm^2} = \frac{1}{2\D_\f-1}\,, \qquad \qquad \frac{d \D_{[\f^2]_p}}{d m^2} = \frac{2}{2\D_\f-1}\,.
\label{eq:analyticsDeltaO}
\end{equation}
Applying (\ref{eq:dimension flow}) to the specific cases at hand, the flow equations to test are\footnote{Notice the necessary factor of $\frac{1}{2}$ coming from the definition of the coupling in the action $\lambda=\frac{m^2}{2}$.}
\begin{equation}
\frac{d \Delta_\f}{d m^2} =
\frac12\sum_{n=0}^{\infty} b^{\hat\phi^2}_{[\f^2]_n} C_{\f\f[\f^2]_n} 
\II^{(\a)}(2\D_\f+2n)\,,
\label{eq:flowofDeltaphi}
\end{equation}
and
\begin{equation}
\frac{d \Delta_{[\f^2]_p}}{d m^2}
=
\frac12\sum_{n=0}^{\infty} b^{\hat\phi^2}_{[\f^2]_n} C_{[\f^2]_p[\f^2]_p[\f^2]_n} \II^{(\a)}(2\D_\f+2n)
\label{eq:flowofDeltaOn}\,,
\end{equation}
where $\mathcal{I}^{(\alpha)}(\Delta)$ is the integrated block given in (\ref{eq:I alpha}).
\begin{figure}[t!]
  \centering
  \begin{subfigure}[t]{0.49\textwidth}
    \includegraphics[width=\textwidth]{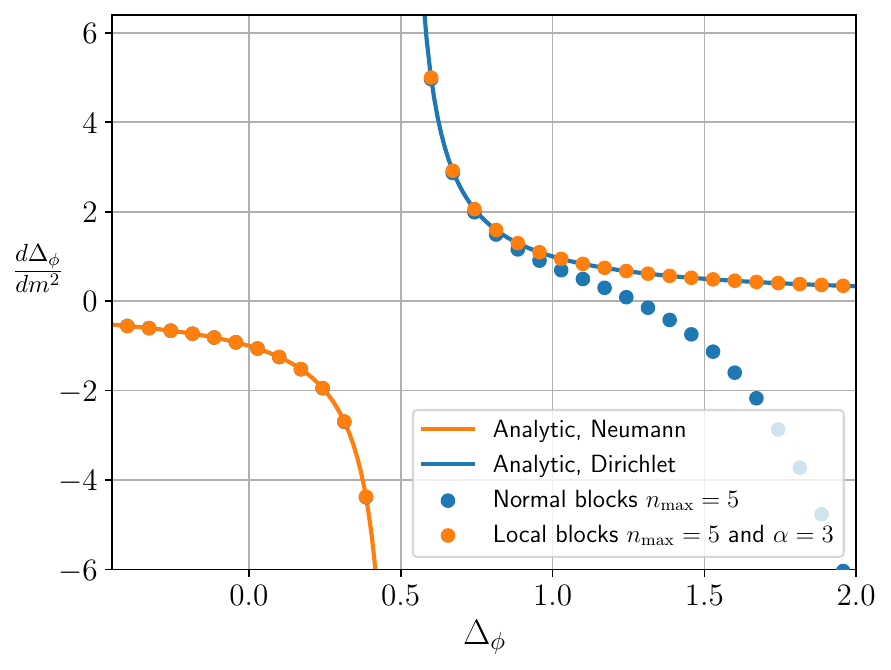}
    \caption{}
  \end{subfigure}
  \hfill
  \begin{subfigure}[t]{0.49\textwidth}
    \includegraphics[width=\textwidth]{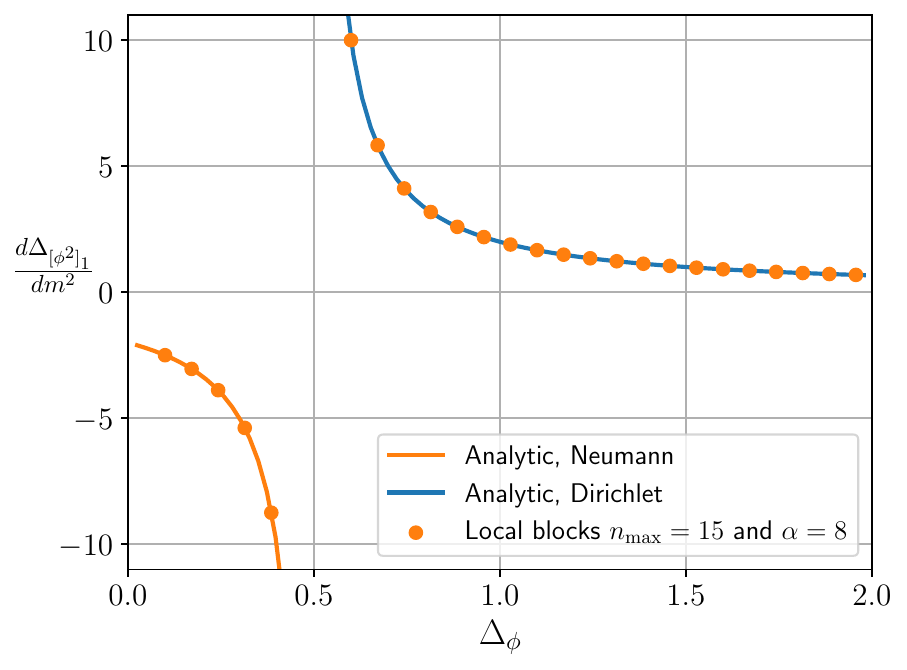}
    \caption{}
  \end{subfigure}
  \caption{Comparison between the truncated flow equations (\ref{eq:flowofDeltaphi}),  (\ref{eq:flowofDeltaOn}) and  the analytic expressions (\ref{eq:analyticsDeltaO}). In figure (a), the sum over integrated normal blocks diverges if $\Delta_\phi>\frac{3}{4}$, we discuss the detailed reasons in appendix \ref{subsec:asymptoticsdeltaflow}. For the same reasons, in (b) the sum over normal blocks never converges.}
  \label{fig:flowofDeltaO}
\end{figure}

\begin{figure}[!t]
    \centering \includegraphics[width=0.5\linewidth]{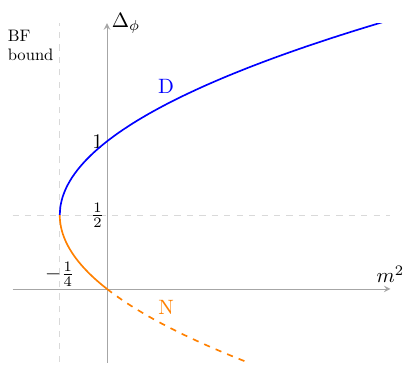}
    \caption{The Neumann (orange) and Dirichlet (blue) boundary conditions for the free scalar field. In the interval $-\frac{1}{4}<m^2<0$, both boundary conditions correspond to unitary theories, while for $m^2>0$ only Dirichlet is unitary.}
    \label{fig:deltam2}
\end{figure}

\begin{figure}[t!]
  \centering
  \begin{subfigure}[t]{0.49\textwidth}
    \includegraphics[width=\textwidth]{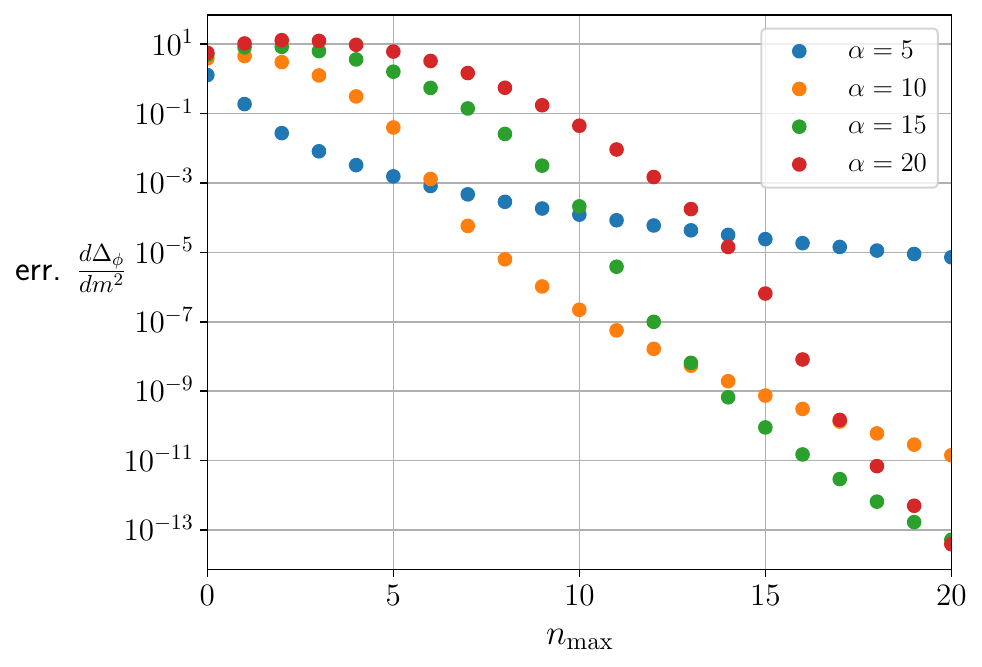}
    \caption{}
  \end{subfigure}
  \hfill
  \begin{subfigure}[t]{0.49\textwidth}
    \includegraphics[width=\textwidth]{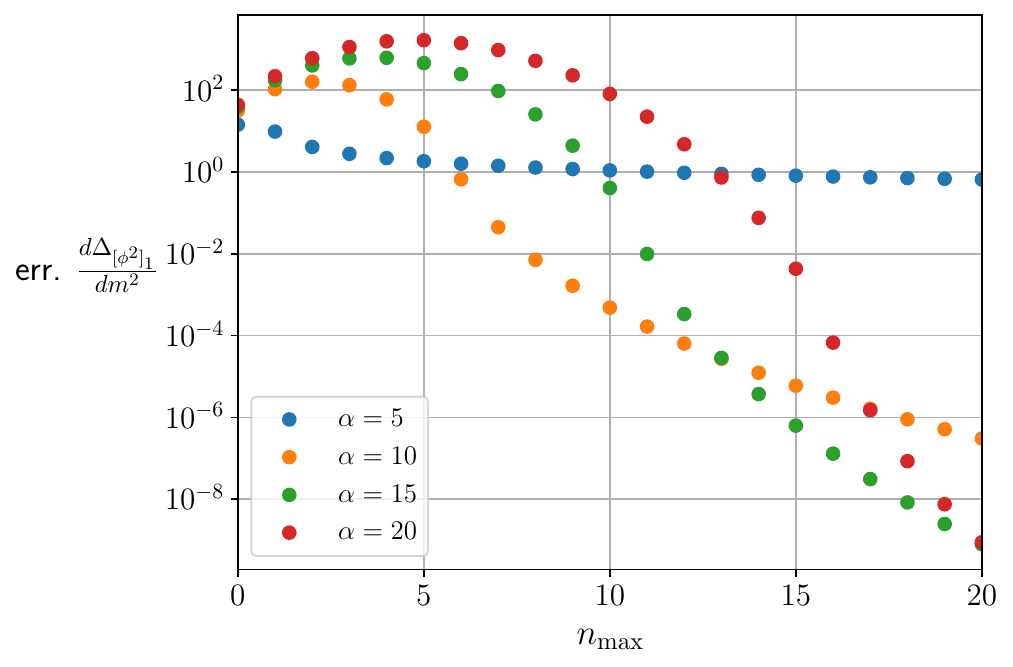}
    \caption{}
  \end{subfigure}
  \caption{Relative errors between the truncated flow equations (\ref{eq:flowofDeltaphi}),  (\ref{eq:flowofDeltaOn}) and  the analytic expressions (\ref{eq:analyticsDeltaO}), defined as $\frac{|\text{numerics}-\text{analytics}|}{|\text{analytics}|}$ for increasing truncation $n_{\max}$. Higher values of $\alpha$ lead to better convergence for higher $n_{\max}$. Vice versa for lower values of $\alpha$. To generate these plots we chose $\Delta_\phi=\frac{3\sqrt{3}}{2}$.}
  \label{fig:relerrorsDeltaflow}
\end{figure}
In practice, we truncate the sums at some $n=n_{\max}$ and we choose a value of $\alpha$. We use the QFT data of free scalar theory, which we report in appendix \ref{app:free theory data}.

In figure \ref{fig:flowofDeltaO} we compare the analytic expressions in (\ref{eq:analyticsDeltaO}) to the values obtained by truncating the flow equations (\ref{eq:flowofDeltaphi}) and (\ref{eq:flowofDeltaOn}). We also compare to the result obtained using integrated normal blocks instead of integrated local blocks. In addition, by comparing with the derivative of \eqref{eq:Delta_phi D vs N}, here we see that the dimension flow of the free scalar theory with both Dirichlet ($\D_\f>\frac{1}{2}$) and Neumann ($\D_\f<\frac{1}{2}$) boundary conditions are reproduced.  The divergences in figure \ref{fig:flowofDeltaO} happen at $\D_\f=\frac{1}{2}$, which is exactly the point when the leading boundary operator $\f^2$ in the BOE of the deformation operator $\hat\f^2$ becomes marginal. This is the turning point of the scaling dimensions in figure \ref{fig:deltam2}, so the divergences are ``coordinate singularities''. The main message is that while we set up our flow equations with Dirichlet boundary conditions, they automatically capture the flow involving \emph{relevant} boundary operators as well. We will discuss more about marginal and relevant boundary operators in section \ref{sec:relevant and marginal boundary op}.

In figure \ref{fig:relerrorsDeltaflow} we show the relative error between the result obtained by truncating the flow equations and the analytic expression as a function of the truncation and for various choices of the parameter $\alpha$.

\subsubsection{BOE coefficient flow}
To test the flow equation of the BOE coefficients, we consider $b^{\hat\f}_{\f}$ and $b^{\hat\f^2}_{[\f^2]_l}$. Their explicit expressions are collected in appendix \ref{app:free theory data}.
The flow equations for these coefficients are (using (\ref{eq:BOE flow}) and the integrated block (\ref{eq:J alpha})) 
\al{
\frac{d b^{\hat\f }_{\f}}{d m^2} 
&=
\frac{1}{2} \sum_{l=0}^\infty b^{\hat\f^2}_{ [\f^2]_l} b^{\hat\f }_{\f} C_{\f\f[\f^2]_l} \JJ^{(\a)}_{\D_\f}(2\D_\f+2l,\D_\f)
\label{eq:flow of b_phi_O}\\
\frac{d b^{\hat\f^2}_{\f^2}}{d m^2} 
&=
\frac{1}{2} \sum_{j=0}^\infty\sum_{l=0}^\infty b^{\hat\f^2}_{ [\f^2]_l} b^{\hat\f^2}_{[\f^2]_j} C_{[\f^2]_l[\f^2]_j \f^2} 
\JJ^{(\a_j)}_{2\D_\f}(2\D_\f+2l,2\D_\f+2j)\,.
\label{eq:flow of b_phi^2_O^2}
}
In practice, we again truncate the sums and explore the relation between the chosen truncation and the accuracy with which we reproduce the derivatives of the BOE coefficients as we know them analytically. For the term with $j=0$ in (\ref{eq:flow of b_phi^2_O^2}), which has the property $\Delta_j=\Delta_i$, we need to use the finite part of the integrated local block (\ref{def:Jreg}).

Taking a derivative of the expressions in appendix \ref{app:free theory data} we have 
\begin{equation}
\begin{aligned}
    \frac{d b^{\hat\f }_{\f}}{d m^2}&=
    \sqrt{\frac{\Gamma(\Delta_\phi)}{2\sqrt{\pi}\Gamma(\Delta_\phi+\frac{1}{2})}}
    \frac{\psi ^{(0)}\left(\Delta _{\phi }\right)-\psi ^{(0)}\left(\Delta _{\phi }+\frac{1}{2}\right)}{2\left(2 \Delta _{\phi }-1\right) }\,,
    \\
    \frac{d b^{\hat\f^2}_{\f^2}}{d m^2}&=
    \frac{ 
    \Gamma \left(\Delta _{\phi }\right) \left(\psi ^{(0)}\left(\Delta _{\phi }\right)-\psi ^{(0)}\left(\Delta _{\phi }+\frac{1}{2}\right)\right)}{\sqrt{2\pi}\left(2 \Delta _{\phi }-1\right) \Gamma \left(\Delta _{\phi }+\frac{1}{2}\right)}\,,
    \end{aligned}
    \label{eq:analyticdbdm}
\end{equation}
where $\psi^{(n)}(x)\equiv\frac{d^n}{dx^n}\frac{\Gamma'(x)}{\Gamma(x)}$ is the polygamma function and $\bb{\hat{\phi}^2}{\phi^2 }=\bb{\hat{\phi}^2}{[\phi^2]_{n=0} }$.

\begin{figure}[t!]
  \centering
  \begin{subfigure}[t]{0.49\textwidth}
    \includegraphics[width=\textwidth]{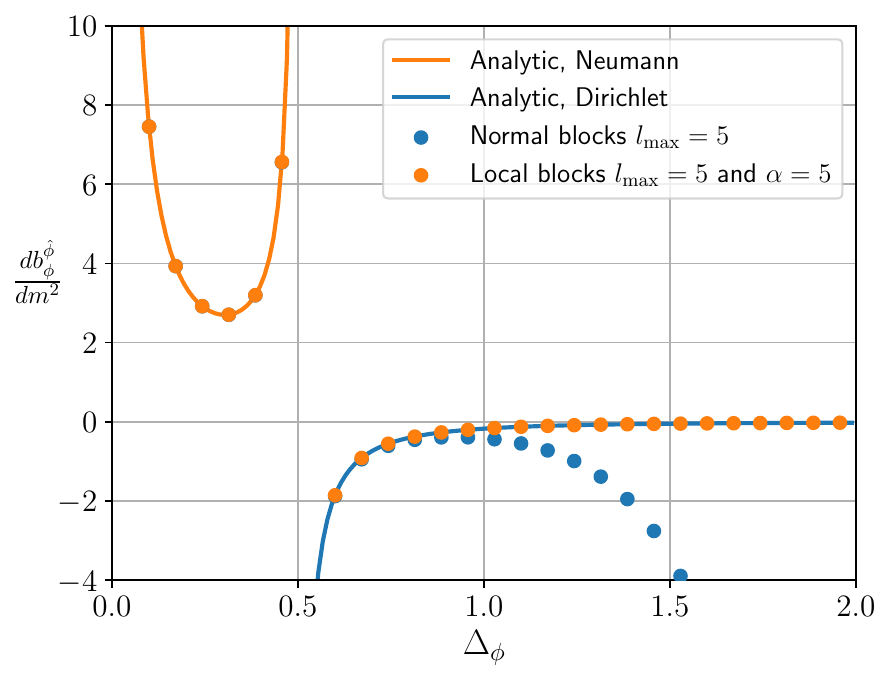}
    \caption{}
  \end{subfigure}
  \hfill
  \begin{subfigure}[t]{0.49\textwidth}
    \includegraphics[width=\textwidth]{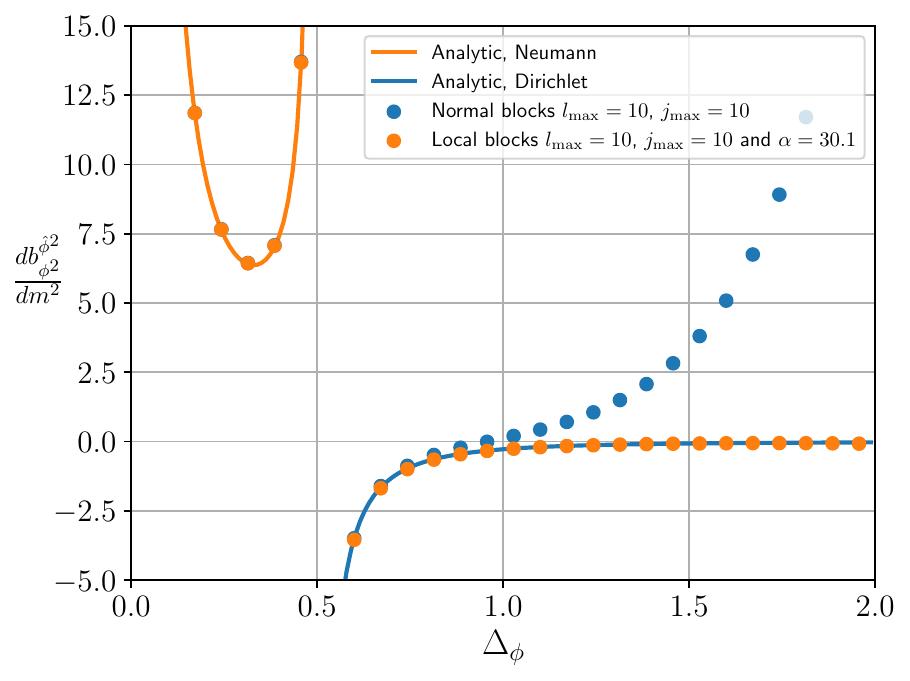}
    \caption{}
  \end{subfigure}
  \caption{Comparison between the truncated flow equations (\ref{eq:flow of b_phi_O}), (\ref{eq:flow of b_phi^2_O^2}) and the analytic expressions (\ref{eq:analyticdbdm}).}
  \label{fig:flow of b_phi_O}
\end{figure}

\begin{figure}[t!]
  \centering
  \begin{subfigure}[t]{0.49\textwidth}
    \includegraphics[width=\textwidth]{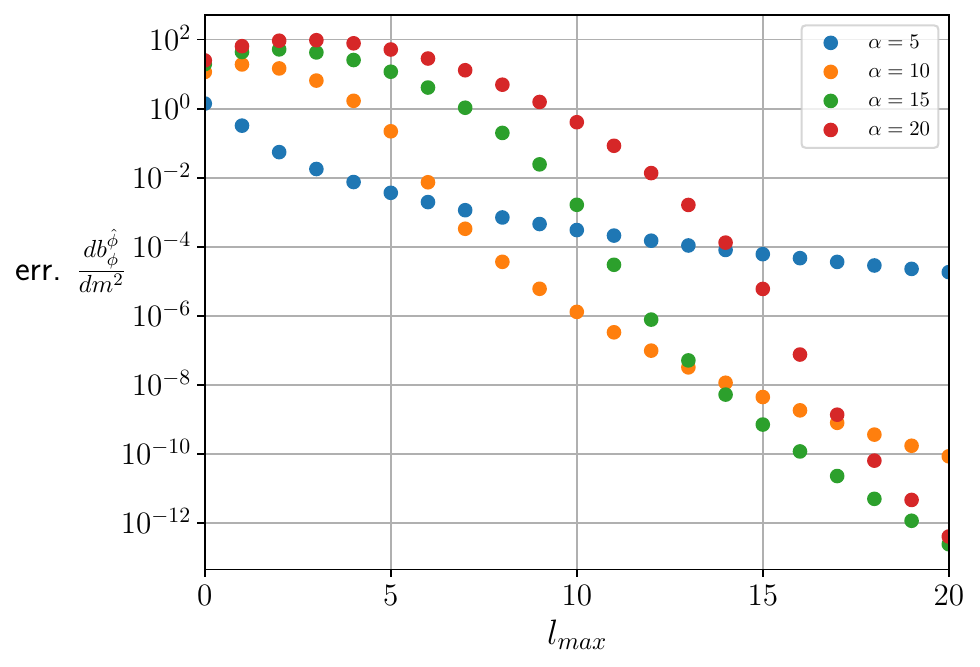}
    \caption{}
  \end{subfigure}
  \hfill
  \begin{subfigure}[t]{0.49\textwidth}
    \includegraphics[width=\textwidth]{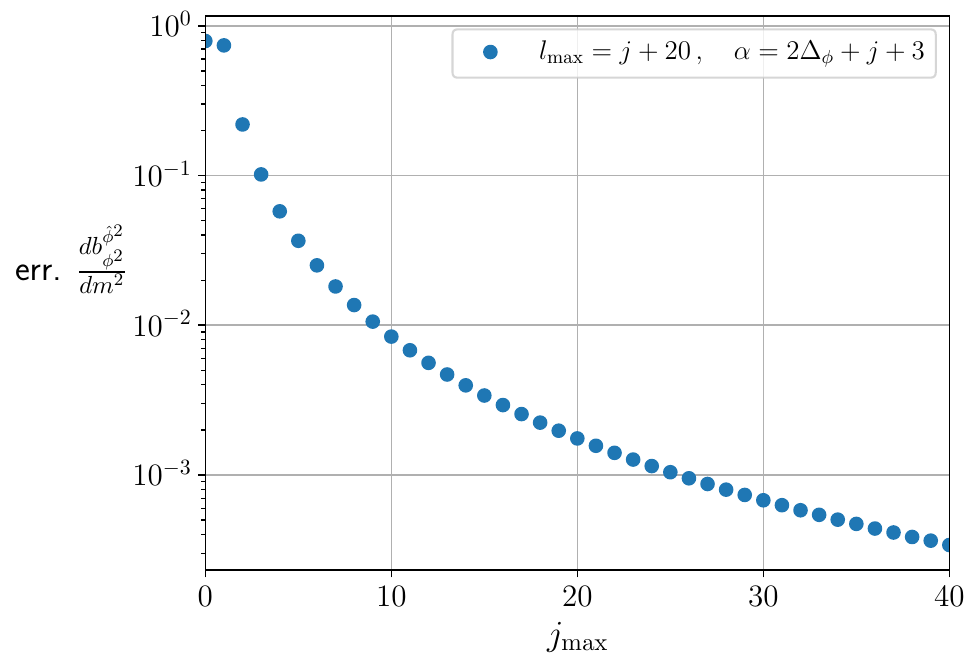}
    \caption{}
    \label{fig:plot2bphi2}
  \end{subfigure}
  \caption{Relative errors between the truncated flow equations (\ref{eq:flow of b_phi_O}),  (\ref{eq:flow of b_phi^2_O^2}) and  the analytic expressions (\ref{eq:analyticdbdm}), defined as $\frac{|\text{numerics}-\text{analytics}|}{|\text{analytics}|}$ for increasing truncation $l_{\max}$ and $j_{\max}$. In (a), we chose $\Delta_\phi=\frac{3\sqrt{3}}{2}$ and higher values of $\alpha$ lead to better convergence for higher $l_{\max}$. In (b), we chose $\Delta_\phi=2.1$ and we let $\alpha_j$ and $l_{\text{max}}$ increase with $j$ as indicated in the legend.}
  \label{fig:flow error of b_phi_O}
\end{figure}

The comparisons with the truncated flow equations are shown in figure \ref{fig:flow of b_phi_O} and \ref{fig:flow error of b_phi_O}. For the flow equation of $b^{\hat\phi}_\phi$ (\ref{eq:flow of b_phi_O}), which involves only a single sum, we observe a comparable rate of convergence to the scaling dimension flow. The flow equation of $b^{\hat\phi^2}_{\phi^2}$ instead involves two sums. As such, choosing the parameters $\alpha_j$, as well as the truncations over the two sums to achieve a good rate of convergence is far less trivial.

To do that, we recall that the role of $\alpha$ is to cure the convergence of the sum over $\{\Delta_l\}$, the set of operators appearing in the BOE of the deforming operator $\hat\Phi$ (see the discussion in \ref{subsec:localblock_bkbkbd}).
As such, we can choose a different $\alpha_j$ and a different truncation $l_{\max}$ for each $\Delta_j$. By studying the large $\Delta_l$ behavior of the integrated blocks $\mathcal{J}$ in appendix \ref{sec:convergence of flow eqn},  we found the lower bound 
\begin{equation}
    \alpha_j>\frac{\Delta_i+\Delta_j+\Delta_{\hat\Phi}}{2}\,,
\end{equation}
where $\Delta_{\hat\Phi}$ is the UV dimension of $\hat\Phi$. This lower bound immediately forces us to choose an $\alpha_j$ that increases with $\Delta_j$, which in turn requires an increasing truncation $\Delta_{l\max}$. After some heuristic attempts, we find that a good choice to achieve convergence for various choices of $\Delta_\phi$ in the case (\ref{eq:flow of b_phi^2_O^2}) is
\begin{equation}
\begin{aligned}
    \alpha_j&=\frac{\Delta_i+\Delta_j+\Delta_{\hat\Phi}}{2}+3=2\Delta_\phi+j+3\,,\\
    \Delta_{l\max}&=\Delta_j+40\,.
\end{aligned}
\end{equation}
This is the choice that leads to figure \ref{fig:plot2bphi2}.

One of the aspects of the scheme we use to implement our flow equations is to always ignore the contributions from the identity. In other words, we always set $b^{\hat\Phi}_{\mathbb{1}}=0$. This corresponds to consistently subtracting the vev at every order in perturbation theory. Nevertheless, we can check that our flow equations reproduce the value of the vev of $\hat\phi^2$ in the free scalar theory. We have 
\begin{equation}
    \langle\hat\phi^2(x)\rangle=b^{\hat\phi^2}_{\mathbb{1}}=-\frac{\psi^{(0)}(\Delta_\phi)}{2\pi}+c\,, \qquad \frac{db^{\hat\f^2 }_{\mathbb{1}}}{dm^2}=-\frac{1}{2\pi}\frac{\psi^{(1)}(\Delta_\phi)}{(2\Delta_\phi-1)}\,,
\end{equation}
where $c$ is a constant. We obtained this result by taking the limit $x\rightarrow y$ (which corresponds to $\eta\rightarrow 1$) in the bulk two-point function \eqref{eq:bulktwopt:free} and subtracting the divergence. As this is a logarithmic divergence, the constant depends on the details of the subtraction, but as the derivative matters here, this is irrelevant.  
The associated flow equation is
\begin{equation}
\frac{db^{\hat\f^2 }_{\mathbb{1}}}{dm^2}=
\frac{1}{2} \sum_{l=0}^\infty b^{\hat\f^2}_{[\f^2]_l} b^{\hat\f^2}_{[\f^2]_l} C_{[\f^2]_l[\f^2]_l \mathbb{1}} 
\JJ^{(\a)}_{0}(2\D_\f+2l,2\D_\f+2l)\,.
\end{equation}
\begin{figure}[t!]
  \centering
  \begin{subfigure}[t]{0.49\textwidth}
    \includegraphics[width=\textwidth]{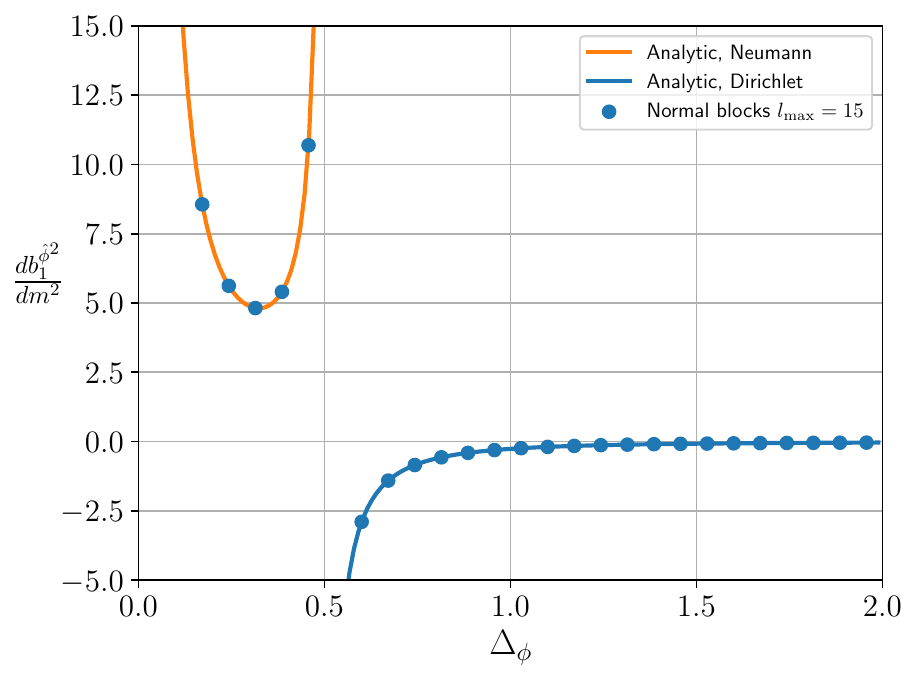}
    \caption{}
  \end{subfigure}
  \hfill
  \begin{subfigure}[t]{0.49\textwidth}
    \includegraphics[width=\textwidth]{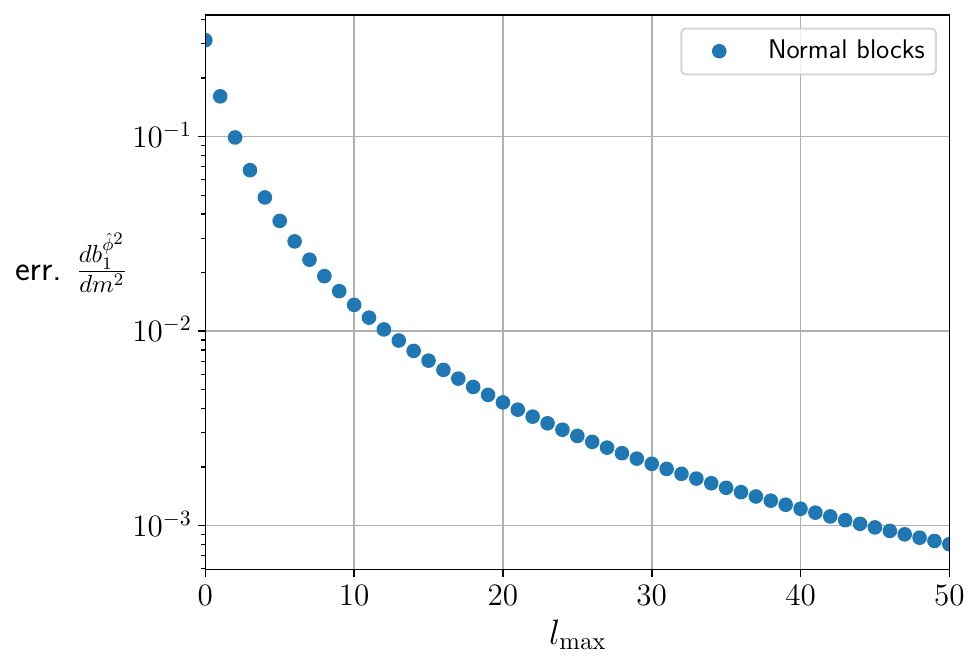}
    \caption{}
  \end{subfigure}
  \caption{Checks of the flow of $b^{\hat\phi^2}_{\mathbb{1}}$, the vev of $\hat\phi^2$. In subfigure b, we  chose $\Delta_\phi=\frac{3\sqrt{3}}{2}$.}
  \label{fig:flow of identity}
\end{figure}
In figure \ref{fig:flow of identity}, we show that the flow equation  works  to evolve this vev (using $C_{[\f^2]_l[\f^2]_l \mathbb{1}}=1$). Let us emphasize that, in general, the consistent scheme is to always set $b^{\hat\Phi}_{\mathbb{1}}=0$.
\subsubsection{OPE coefficient flow}
\label{sec:OPEflowFB}
Testing the flow equations of the OPE coefficients is significantly harder, because we know the relevant integrated blocks only through some Mellin Barnes-type integral representations (see equations (\ref{eq:K}), (\ref{eq:K alpha}),  (\ref{eq:evenoddK})). Straight contours for these integrals are only available for special values of the scaling dimensions of the operators involved. For general cases, we analytically continued each integral with the help of the Mathematica package \texttt{MB.m} \cite{Czakon:2005rk}, and then performed the numerical integrations with the CUBA library \cite{Hahn:2004fe}, specifically with the Cuhre algorithm. Here we present some example cases and the results we obtained, reminding the reader that they can be reproduced with the ancillary Mathematica notebook \texttt{testflows.nb}.

The first example we check is the trivial statement that $\frac{dC_{\phi\phi\mathbb{1}}}{dm^2}=0$. This is obvious because the boundary operators are defined to have unit-normalized two-point function throughout the flow. The associated flow equation, though, is far from obvious, and is obtained by studying the various OPE channels in the six regions in which we are splitting the integral over AdS. In the flow of $C_{ijk}$, we generally call the $s$, $t$ and $u$ channels those in which we fuse the operators $\mathcal{O}_l$ in the BOE of the bulk deformation $\hat\Phi$ with $\mathcal{O}_i$, $\mathcal{O}_j$ and $\mathcal{O}_k$ respectively. For example, for the $s-$channel, we can indicate the integrated block $\mathcal{K}_{\Delta_i\Delta_j\Delta_k}(\Delta_l,\Delta_m)$ as follows\\

\begin{figure}
    \centering
    \includegraphics[scale=0.6]{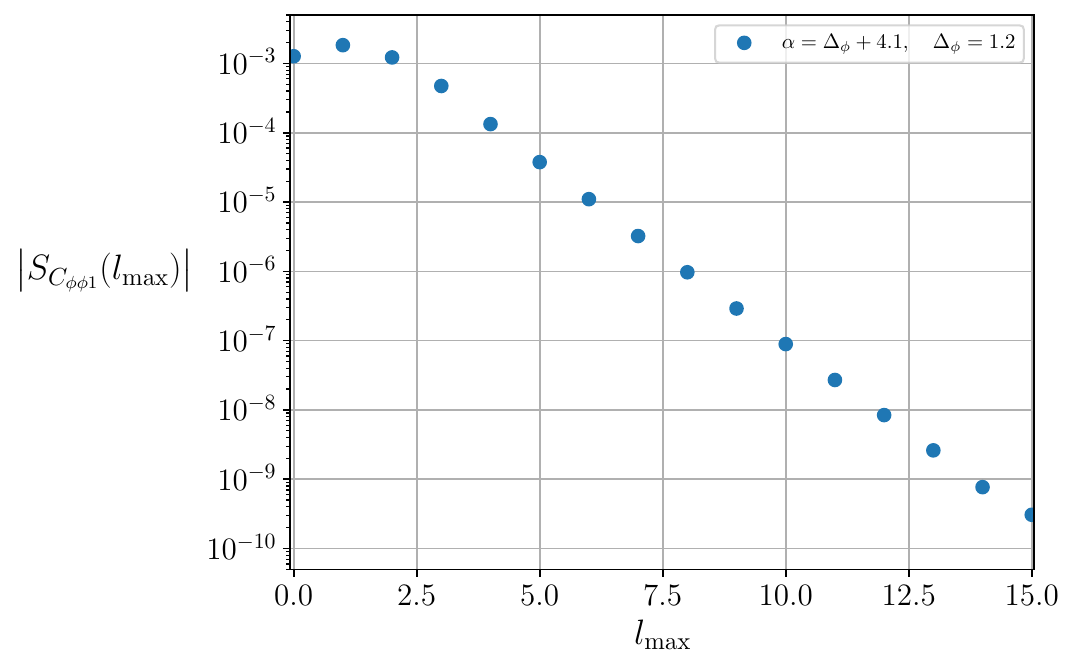}
    \caption{Plot of the partial sums obtained by truncating the series in eq. (\ref{eq:dCoo1}).}
    \label{fig:COOidentity}
\end{figure}

\begin{center}\includegraphics[scale=2]{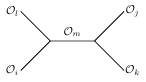}\\

\end{center}
In the case we are studying of $C_{\phi\phi\mathbb{1}}$, the $s$ and $t$ channel coincide, and are different than the $u$ channel, as shown in the following:
\begin{center}\includegraphics[scale=2]{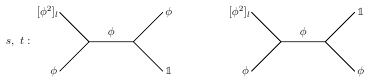}\\
\includegraphics[scale=2]{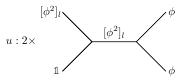}
\end{center}
In equations, we have
\begin{equation}
\begin{aligned}
    \frac{dC_{\phi\phi\mathbb{1}}}{dm^2}=2\sum_{l=0}^\infty b^{\hat\phi^2}_{[\phi^2]_l}C_{\phi[\phi^2]_l\phi}\Big[&\mathcal{K}^{(\alpha)e}_{\Delta_\phi,\Delta_\phi,0}(\Delta_{[\phi^2]_l},\Delta_\phi)+\mathcal{K}^{(\alpha)e}_{\Delta_\phi,0,\Delta_\phi}(\Delta_{[\phi^2]_l},\Delta_\phi)\\
    &+\mathcal{K}_{0,\Delta_\phi,\Delta_\phi}(\Delta_{[\phi^2]_l},\Delta_{[\phi^2]_l})\Big]
    \label{eq:dCoo1}
\end{aligned}
\end{equation}
where we used that $C_{ii\mathbb{1}}=C_{i\mathbb{1}i}=1$ and the fact that all operators involved are parity-even. Notice that in the second line, corresponding to the $u$ channel contribution, for each $\mathcal{O}_m$ there is only one $\mathcal{O}_l$ to be summed. This allows us to use for this term normal integrated blocks in place of local integrated blocks, which are less expensive to evaluate numerically.  In figure \ref{fig:COOidentity} we plot the partial sums obtained by truncating (\ref{eq:dCoo1}) up to some $l_{\max}$, which we indicate as $S_{C_{\phi\phi\mathbb{1}}}(l_{\max})$, in which one can see exponential convergence\footnote{We expect the sums over boundary operators to converge like power-laws rather than exponentially. If we split the sums over the two lines in (\ref{eq:dCoo1}), they in fact individually converge like power-laws.}. 

A less trivial check can be carried out by studying the OPE coefficient $C_{\phi\phi\phi^2}=\sqrt{2}$. In this case, the derivative is again zero. The associated OPE flow equation involves summing over the channels represented in the following figures:
\begin{center}\includegraphics[scale=2]{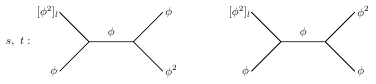}\\
\qquad\quad\includegraphics[scale=2]{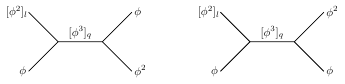}\\
\includegraphics[scale=2]{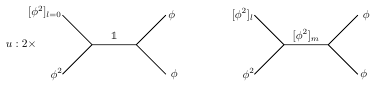}
\end{center}
Notice that the channels involving triple trace operators are controlled by the OPE coefficients $C_{\phi[\phi^2]_l[\phi^3]_q}C_{\phi\phi^2[\phi^3]_q}$, which we reported in the ancillary Mathematica notebook \texttt{testflows.nb} and which importantly are zero when $l>\lfloor\frac{q}{2}\rfloor$. That means the sums over $l$ truncate for those terms, allowing us to use normal blocks in place of local blocks. Moreover, $[\phi^3]_q$ is parity-odd when $q$ is odd, meaning $C_{\phi[\phi^2]_l[\phi^3]_q}=(-1)^qC_{\phi[\phi^3]_q[\phi^2]_l}$. Overall, the flow equation for $C_{\phi\phi\phi^2}$ reads
\begin{equation}
\begin{aligned}
    &\frac{dC_{\phi\phi\phi^2}}{dm^2}=2\sum_{l=0}^\infty b^{\hat\phi^2}_{[\phi^2]_l}C_{\phi[\phi^2]_l\phi}C_{\phi\phi^2\phi}\left(\mathcal{K}^{(\alpha)e}_{\Delta_\phi\Delta_\phi\Delta_{\phi^2}}(\Delta_{[\phi^2]_l},\Delta_\phi)+\mathcal{K}^{(\alpha)e}_{\Delta_\phi\Delta_{\phi^2}\Delta_\phi}(\Delta_{[\phi^2]_l},\Delta_\phi)\right)\\
    &+2\sum_{q=0}^\infty\sum_{l=0}^{\lfloor\frac{q}{2}\rfloor}b^{\hat\phi^2}_{[\phi^2]_l}C_{\phi[\phi^2]_l[\phi^3]_q}C_{\phi\phi^2[\phi^3]_q}\left(\mathcal{K}_{\Delta_\phi\Delta_\phi\Delta_{\phi^2}}(\Delta_{[\phi^2]_l},\Delta_{[\phi^3]_q})+(-1)^q\mathcal{K}_{\Delta_\phi\Delta_{\phi^2}\Delta_\phi}(\Delta_{[\phi^2]_l},\Delta_{[\phi^3]_q})\right)\\
    &+2b^{\hat\phi^2}_{\phi^2}\mathcal{K}_{\Delta_{\phi^2}\Delta_\phi\Delta_\phi}(\Delta_{\phi^2},0)+2\sum_{n=0}^\infty\sum_{m=0}^\infty b^{\hat\phi^2}_{[\phi^2]_n}C_{\phi\phi[\phi^2]_m}C_{\phi^2[\phi^2]_n[\phi^2]_m}\mathcal{K}^{(\alpha_m)e}_{\Delta_{\phi^2}\Delta_\phi\Delta_\phi}(\Delta_{[\phi^2]_n},\Delta_{[\phi^2]_m})
    \label{eq:dCphiphiphi2}
\end{aligned}
\end{equation}
In this case the rate of convergence is different for the various sums. Moreover, the last sum requires us to choose a value of $\alpha_m$ which grows with $\Delta_m$. In practice, choosing $l_{\max}=10$, $q_{\max}=20$, $m_{\max}=13$, $n_{\max}=20$, $\alpha=\Delta_\phi+3.1$ and $\alpha_m=2\Delta_\phi+m+3.1$,
we obtain the following values, where each number stands for the result of the partial sum of each line of (\ref{eq:dCphiphiphi2}):
\begin{equation}
\begin{aligned}
    \frac{dC_{\phi\phi\phi^2}}{dm^2}\Big|_{\text{numerical}}&=-0.52230(7)+1.73624(8)-1.21394(3)\\
                                                            &=(-2)\times 10^{-6}\,,
\end{aligned}
\end{equation}
where the digits in parenthesis are uncertain because the size of the last summand that we considered in each partial sum was of order $\sim10^{-7}$. Uncertainties due to numerical integration were instead kept below  $\sim10^{-9}$. Details are in our ancillary Mathematica notebook \texttt{testflows.nb}.

We remind the reader the full result should be $\frac{dC_{\phi\phi\phi^2}}{dm^2}=0$.
\section{Discussion}
\label{sec: discussion}

We shall now discuss several aspects of the flow equations. Some of the ideas discussed here are preliminary and should be thought of as an invitation to future study.

\subsection{Relevant and marginal boundary operators}
\label{sec:relevant and marginal boundary op}

So far we have assumed that all boundary operators are irrelevant, 
\emph{i.e.} $\Delta_i>1$.
Here, we discuss the consequences of the presence of marginal and relevant boundary operators.

Let us first consider the situation where the dimension $\Delta_1(\lambda)$ of the lightest boundary operator approaches 1 from above as $\lambda \to \lambda_c$ (also from above).
Notice that the kinematical function $\II$   in \eqref{eq:dimension flow}, whose form is explicitly given in \eqref{eq:I alpha}, diverges as $\II^{(\alpha)}(\Delta)\approx 2/(\Delta-1)$ in this limit.
Therefore,\footnote{We assume $c>0$ so that the theory exists for $\lambda>\lambda_c$. When $c<0$, it means that marginality is approached when $\lambda$ approaches $\lambda_c$ from below. } 
\begin{equation}
    \frac{d\Delta_1}{d\lambda}\approx\frac{c}{\Delta_1-1}
    \qquad \Rightarrow \qquad 
    \Delta_1(\lambda) \approx 1 \pm \sqrt{2c(\lambda-\lambda_c)}
    \,,\qquad c= 2 \bb{\hat\Phi}{1}(\lambda_c) C_{111}(\lambda_c) >0 \,.
    \label{eq:Delta1marginal}
    \end{equation}
This is the characteristic signature of fixed point annihilation \cite{Kaplan:2009kr, Gorbenko:2018ncu, Copetti:2023sya}. There are two consistent sets of QFT data, corresponding to the $\pm$ branches of the square root, that merge at $\lambda =\lambda_c$. For $\lambda<\lambda_c$, the QFT data becomes complex.
Notice that (generically) all scaling dimensions show the same square-root behavior,
 \begin{equation}
    \frac{d\Delta_i}{d\lambda}\approx \frac{2 \bb{\hat\Phi}{1} C_{ii1} }{\Delta_1-1} \approx \pm \frac{
    2 \bb{\hat\Phi}{1} C_{ii1} }{\sqrt{2c(\lambda-\lambda_c)}}
     \qquad \Rightarrow \qquad \Delta_i(\lambda)\approx \Delta_i(\lambda_c) \pm \frac{C_{ii1}(\lambda_c) }{C_{111}(\lambda_c)} \sqrt{2c(\lambda-\lambda_c)}\,.
    \end{equation}
The same is true for BOE and OPE coefficients. As $\lambda\rightarrow\lambda_c$, we have
\begin{equation}\label{eq:flow_BOE_OPE_marginal}
    \begin{split}
        \frac{db^{\hat{\Phi}}_i}{d\lambda}&\approx-\frac{b^{\hat{\Phi}}_1}{\Delta_1-1}\sum_{j}\frac{C_{ij1}+C_{ji1}}{2} b^{\hat{\Phi}}_j \,F_{\Delta_i}(\Delta_j), \\
        \frac{dC_{ijk}}{d\lambda}&\approx-\frac{b^{\hat{\Phi}}_1}{\Delta_1-1}\sum_{m}C_{jkm}\left(C_{i1m} H_{\Delta_i,\Delta_j,\Delta_k}(\Delta_m)+C_{im1} H_{\Delta_i,\Delta_k,\Delta_j}(\Delta_m)\right) \\
        &\qquad + (ijk \rightarrow jki) + (ijk \rightarrow kij)\,. \\
    \end{split}
\end{equation}
Here, $F_{\Delta_i}(\Delta_j)$ and $H_{\Delta_i,\Delta_j,\Delta_k}(\Delta_m)$ are the regularized residues of the flow-equation coefficients at $\Delta_l=1$ derived in section \ref{subsec:poles}:
\begin{equation}
    \begin{split}
        F_{\Delta_i}(\Delta_j)&=\oint_{\Delta_j}\frac{d\Delta}{2\pi i}\frac{1}{\Delta-\Delta_j}\frac{2\sqrt{\pi}\Gamma\!\left(\Delta+\tfrac{1}{2}\right)}
		{(\Delta-\Delta_i)\Gamma\!\left(\tfrac{1+\Delta+\Delta_i}{2}\right)
			\Gamma\!\left(\tfrac{1+\Delta-\Delta_i}{2}\right)}\,, \\
        H_{\Delta_i,\Delta_j,\Delta_k}(\Delta_m)&=\oint_{\Delta_m}\frac{d\Delta}{2\pi i}\frac{1}{\Delta-\Delta_m}\frac{2^{\Delta_{i}-\Delta}}{\Delta-\Delta_i}\,
		\hyperF{\Delta-\Delta_i}{\Delta+\Delta_j-\Delta_k}{2\Delta}{\tfrac{1}{2}}\,. \\
    \end{split}
\end{equation}
Plugging \eqref{eq:Delta1marginal} into the approximate flow equations \eqref{eq:flow_BOE_OPE_marginal}, we get the square-root behaviors,
\begin{equation}
    \begin{split}
     b^{\hat{\Phi}}_i(\lambda)&=b^{\hat{\Phi}}_i(\lambda_c)\mp\sqrt{\frac{c}{2}(\lambda-\lambda_c)}\left[\sum_{j}\frac{C_{ij1}+C_{ji1}}{2C_{111}}b^{\hat{\Phi}}_jF_{\Delta_i}(\Delta_j)\right]_{\lambda=\lambda_c}, \\
     C_{ijk}(\lambda)&=C_{ijk}(\lambda_c) \\
     &\mp\sqrt{\frac{c}{2}(\lambda-\lambda_c)}\left[\sum_{m}\frac{C_{jkm}}{C_{111}}\left(C_{i1m} H_{\Delta_i,\Delta_j,\Delta_k}(\Delta_m)+C_{im1} H_{\Delta_i,\Delta_k,\Delta_j}(\Delta_m)\right)\right]_{\lambda=\lambda_c} \\
     &\qquad + (ijk \rightarrow jki) + (ijk \rightarrow kij)\,. \\
    \end{split}
\end{equation}
The massive free scalar discussed in the previous section provides an explicit example of this phenomenon. 
Using $\lambda =\frac{1}{2}m^2 =\frac{1}{2} \Delta_\phi(\Delta_\phi-1)$, we can write the dimension of the boundary operator $\phi^2$ as follows,
\begin{equation}
    \Delta_{\phi^2} = 2\Delta_\phi = 1\pm \sqrt{8\lambda+1}\,,\label{eq:5.3}
\end{equation}
which shows that $\lambda>-\frac{1}{8}\equiv \l_c^{\f^2}$.

As is clear from \eqref{eq:Delta1marginal}, in the merging event, one of the QFTs must have a relevant boundary operator. 
Are the flow equations still valid for such theories?
In the example of the free massive scalar, the flow equations hold for $\Delta_\phi<1/2$ if one uses analytically continued kinematical functions $\mathcal{I,J,K}$.
How can this be? Recall that the AdS integrals of the deforming operator $\hat{\Phi}$ diverge if $\Delta_l<1$ and $\bb{\hat{\Phi}}{l}\neq 0$.
More precisely, if we introduce a cutoff near the AdS boundary, we find that
\begin{equation}
    \int_{z>a}\frac{dzd\t}{z^2}\langle \hat{\Phi}(z,\tau)\dots \rangle
    = \sum_{\Delta_l<1}
   \frac{ a^{\Delta_l-1} }{1-\Delta_l}\bb{\hat{\Phi}}{l}
    \int d\t \langle \mathcal{O}_l(\tau)\dots \rangle
    +({\rm finite\ when\ } a\to 0) \,.
\end{equation}
These power-law divergences 
can be removed by adding to the action boundary counter terms involving the boundary relevant operators. 
This produces a fine-tuned change in the QFT data that preserves conformal symmetry in the presence of relevant boundary deformations. As explained in \cite{Lauria:2023uca}, if one imposes bulk covariance, there is no freedom to turn on relevant boundary deformations independently of the bulk deformation.
This is what the analytic continuation prescription automatically implements.  Therefore, our flow equations are also valid in the presence of relevant boundary operators. 

\subsubsection{Weakly relevant boundary operators}
\label{subsec:weakly relevant boundary op}

Finally, let us comment on the special case, where $\Delta_m(\lambda_c)=1$ and $\bb{\hat{\Phi}}{m}(\lambda_c)=0$.
In this case, the QFT data is real for both $\lambda<\lambda_c$ and $\lambda>\lambda_c$.
When $\Delta_m=1-\delta$ with $\delta\ll1$, the operator $\mathcal{O}_m$ is weakly relevant and we can use it to seed a short boundary RG flow that ends in a nearby QFT preserving the AdS isometries. Adding $g\int d\t \mathcal{O}_m(\tau)$ to the action, 
we find the beta function \cite{Zamolodchikov:1987ti,Cardy_1996} 
\begin{equation}
   \beta_g = -(1-\Delta_m)g + \pi C_{mmm} g^2+ O(g^3)\,,
\end{equation}
which has a perturbative fixed point at $g_* = \delta/(\pi C_{mmm})+O(\delta^2)$.
This means that $\lambda=\lambda_c$ is a bifurcation point of the flow equations. For the same bulk theory, there are two nearby sets of QFT data: the original one with $\Delta_m=1-\delta$ and a new one with $\Delta_m=1+\delta+O(\delta^2)$. 
Therefore, the QFT data at $\lambda=\lambda_c$ can be deformed in two different ways for $\lambda\neq \lambda_c$. 
Let us denote by $\{ \Delta_i,  \bb{\hat{\Phi}}{i},  C_{ijk} \}$ and $\{ \tilde{\Delta}_i,  \tilde{b}^{\hat{\Phi}}_{i},  \tilde{C}_{ijk} \}$ the two sets of QFT data, which by construction are equal at $\lambda=\lambda_c$.
It is convenient to write
\begin{align}
    \Delta_m&=1+\gamma(\lambda-\lambda_c)+O(\lambda-\lambda_c)^2\,,\qquad
    \bb{\hat{\Phi}}{m} = \beta(\lambda-\lambda_c)+O(\lambda-\lambda_c)^2\,,\\
 \tilde{\Delta}_m&=1+\tilde{\gamma}(\lambda-\lambda_c)+O(\lambda-\lambda_c)^2\,,\qquad
    \tilde{b}^{\hat{\Phi}}_{m} = \tilde{\beta}(\lambda-\lambda_c)+O(\lambda-\lambda_c)^2\,.
    \end{align}   
Then, the flow equations at $\lambda=\lambda_c$ imply
\begin{align}
    \gamma-\tilde{\gamma}&=\left(\frac{\beta}{\gamma}-\frac{\tilde{\beta}}{\tilde{\gamma}}\right)c_1\,,\qquad c_1=C_{mmm} {\rm Res}_{\Delta=1} \II^{(\alpha)}(\Delta)\,,\\
    \beta-\tilde{\beta}&=\left(\frac{\beta}{\gamma}-\frac{\tilde{\beta}}{\tilde{\gamma}}\right)c_2\,,\qquad c_2=\sum_j \bb{\hat{\Phi}}{j}C_{jmm}{\rm Res}_{\Delta=1}\mathcal{J}^{(\alpha)}_1(\Delta,\Delta_j)\,.
\end{align}
Given the untilded set of QFT data, we can solve these equations for $\tilde{\gamma}$ and $\tilde{\beta}$ and then find the full tilded set of QFT data perturbatively in an expansion in $\lambda-\lambda_c$.
For example, the derivatives of the scaling dimensions at $\lambda-\lambda_c$ are simply given by
\begin{align}
    \frac{d\Delta_i}{d\lambda}- \frac{d\tilde{\Delta}_i}{d\lambda}
    = \frac{C_{iim}}{C_{mmm}}( \gamma-\tilde{\gamma})\,.
    \label{eq:deriv of new scaling dim}
\end{align}

As explained above, we expect $\tilde{\g}=-\g$ for a short boundary RG flow. This, together with the equations above, leads to
\begin{equation}
    \beta=\frac{\gamma(\gamma+c_2)}{c_1}\,,\qquad
    \tilde{\beta}=\frac{\gamma(\gamma-c_2)}{c_1}\,.
    \label{eq:betabetatilde}
\end{equation}
Notice that the first equation is a constraint on the untilded set of QFT data.

The scenario $\Delta_m(\lambda_c)=1$ and $\bb{\hat{\Phi}}{m}(\lambda_c)=0$ is non-generic but it happens in 
the free boson theory. Let us focus on that example. We have $\lambda=\frac{m^2}{2}$, and when
 $\Delta_\phi=1/4$, the boundary operator $\phi^4$ hits marginality $\Delta_{\phi^4}=1$, meanwhile
$\bb{\hat{\Phi}}{\phi^4}=0$.
In fact, in this example, $\bb{\hat{\Phi}}{\phi^4}=0$ for any $m^2$ and 
\begin{equation}
   1\overset{!}{=} \Delta_{\phi^4}= 2-2\sqrt{1+4m^2}\, ,\qquad \Rightarrow\qquad m^2_{c,\phi^4} = -\frac{3}{16}\, ,
\end{equation}
where we used \eqref{eq:5.3}. Expanding $\Delta_{\phi^4}$ close to $m^2_{c,\phi^4}$, we get 
\begin{equation}
    \Delta_{\phi^4}= 1-8(m^2-m^2_{c,\phi^4})+O(m^2-m^2_{c,\phi^4})^2\, .
\end{equation}
Therefore, using $\lambda=\frac{m^2}{2}$, we have $\gamma=-16$ and $\beta=0$. One may also compute
\begin{equation}
    c_1=2C_{\phi^4\phi^4\phi^4}=12\sqrt{6}\,,\qquad
    c_2=\sum_{n=0}^\infty \bb{\hat{\Phi}}{[\phi^2]_n}C_{[\phi^2]_n\phi^4\phi^4}{\rm Res}_{\Delta=1}\mathcal{J}^{(\alpha)}_1\left(\Delta,\frac{1}{2}+2n\right)=16\,,
\end{equation}
where the result of $C_{[\phi^2]_n\phi^4\phi^4}$ is given in \eqref{eq:C_phi2phi4phi4}.
Reassuringly, this is consistent with the first equation in \eqref{eq:betabetatilde}. The second equation predicts $\tilde{\beta}=\frac{64}{3}\sqrt{\frac{2}{3}}$. This leads to $\tilde{\g}=16=-\g$ as predicted above.

\begin{figure}[!t]
    \centering  \includegraphics[width=0.7\linewidth]{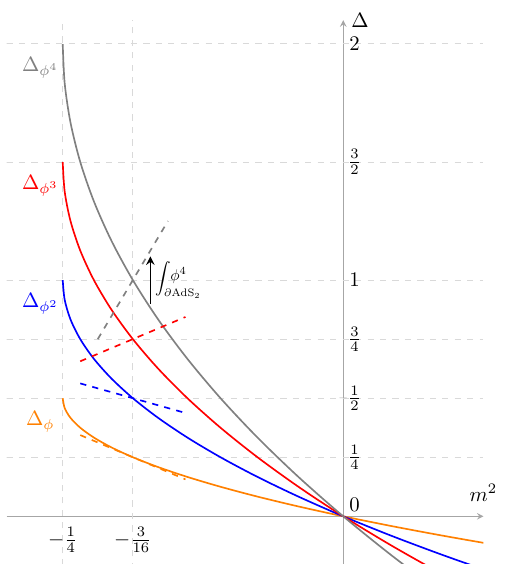}
    \caption{The scaling dimensions of $\f$, $\f^2$, $\f^3$, $\f^4$ in the free scalar theory with Neumann boundary conditions (solid) and of $\tilde\f$, $\tilde\f^2$, $\tilde\f^3$, $\tilde\f^4$ in LRI (dashed, linear approximation). When the boundary operator $\f^4$ is marginal, the QFT data of the two theories coincides. When it is weakly relevant, it can induce a short boundary RG flow connecting the two theories, as indicated by the arrow. LRI is the IR fixed point of this flow.}
    \label{fig:weakly relevant op}
\end{figure}

In principle, one can study the expansion in $m^2-m^2_{c,\phi^4}$ of the flow equations to determine the 
new set of QFT data $\{ \tilde{\Delta}_i,  \tilde{b}^{\hat{\Phi}}_{i},  \tilde{C}_{ijk} \}$ that merges with the free boson theory at  $m^2=m^2_{c,\phi^4}$. Notice that this QFT data describes an interacting theory (although with interactions restricted to the boundary). In fact, it is known that this theory is equivalent to the  1D long-range Ising  (LRI) fixed point \cite{Behan:2017emf, Benedetti:2024wgx}. Let us use \eqref{eq:deriv of new scaling dim} to compute the leading corrections to the scaling dimensions in the interacting theory. For example,
\begin{align}
    \tilde{\Delta}_{\phi}-\frac{1}{4}&=
    -2(m^2-m^2_{c,\phi^4}) +O(m^2-m^2_{c,\phi^4})^2\,,\\
 \tilde{\Delta}_{\phi^2}-\frac{1}{2}&=
    -\frac{4}{3}(m^2-m^2_{c,\phi^4}) +O(m^2-m^2_{c,\phi^4})^2\,,\\
    \tilde{\Delta}_{\phi^3}-\frac{3}{4}&=
    2(m^2-m^2_{c,\phi^4}) +O(m^2-m^2_{c,\phi^4})^2\,.
\end{align}
To this order, we confirm the expectation $\tilde{\Delta}_{\phi}+\tilde{\Delta}_{\phi^3}=1$ \cite{Paulos:2015jfa}. This is illustrated in figure \ref{fig:weakly relevant op}. We leave for the future a more detailed study of this theory using our flow equations. 

\subsection{Level repulsion}
\label{subsec:level repulsion}

The scaling dimensions $\Delta_i$ of boundary operators are the eigenvalues of the QFT Hamiltonian conjugate to global time in AdS \cite{Paulos:2016fap}.
It is well known that the eigenvalues of a Hamiltonian generically do not cross when the Hamiltonian varies with a continuous parameter. 
Assuming unitarity, here we show that 
the flow equations imply level repulsion generically. The argument is similar to that of \cite[sec.3.2]{Behan:2017mwi}.

Consider two boundary operators $\OO_1$ and $\OO_2$. Suppose that at some point along the flow $\l=\l_0$, we have $0<\D_{12}\ll |\D_{ij}|$ for all other pairs $ij$.
We further assume unitarity which implies
\al{
C_{ijk}^*=C_{jik}\,.
\label{eq:c = c* in lv rps}
}
In this setup we want to argue that
\begin{equation}
\frac{d^2 \D_{12}(\l)}{d\l^2} = \frac{\g}{\D_{12}(\l)^3}+O(\D_{12}^{-2})\,,
\label{eq:level repulsion eqn}
\end{equation}
with a positive constant $\g$. One can then view $\l$ as time and $\D_{12}(\l)$ as the position of a particle moving in one dimension and experiencing a repulsive potential that diverges at the origin $\D_{12}=0$. Thus $\D_{12}$ has to stay positive and the level crossing is avoided, no matter the value of $\frac{d\D_{12}}{d\l}$ at $\l=\l_0$. To this end, consider the second-order derivative of the scaling dimension which can be obtained using the scaling dimension flow equation \eqref{eq:dimension flow}
\begin{equation}
\frac{d^2 \D_1}{d \l^2}
= \sum_{{l}} 
\left(
\bb{\hat{\Phi}}{l}(\l) \frac{dC_{11 l}(\l)}{d\l} \II^{(\a)}(\D_l)
+
\frac{d \bb{\hat{\Phi}}{l}(\l)}{d\l} C_{11 l}(\l) \II^{(\a)}(\D_l)
+ \bb{\hat{\Phi}}{l}(\l) C_{11 l}(\l) \frac{d \II^{(\a)}(\D_l)}{d\l} 
\right)\,.
\label{eq:sec order deriv of Delta}
\end{equation}
We analyze these three terms in the limit $\D_2\to\D_1$.


For the first term, using the OPE coefficient flow equation \eqref{eq:OPE flow} we have
\al{\spl{
\frac{dC_{11 l}(\l)}{d\l} &= 
\sum_{m}\sum_{ p} 
  \bb{\hat{\Phi}}{p}
    \Bigg[\left(
   C_{p1m}C_{m1l} \KK^{(\a_m)}_{\D_1\D_1\D_l}(\D_p,\D_m)
     + C_{1pm}C_{m1l}\KK^{(\a_m)}_{\D_1\D_l\D_1}(\D_p,\D_m)
     \right)
    \\
    &\hspace{2.3cm}+
    \left(
    C_{p1m}C_{ml1}\KK^{(\a_m)}_{\D_1\D_l\D_1}(\D_p,\D_m)
     + C_{1pm}C_{ml1}\KK^{(\a_m)}_{\D_1\D_1\D_l}(\D_p,\D_m)
     \right)\\ 
     &\hspace{2.3cm}+ 
    \left(
    C_{plm}C_{m11}\KK^{(\a_m)}_{\D_l\D_1\D_1}(\D_p,\D_m)
     + C_{lpm}C_{m11}\KK^{(\a_m)}_{\D_l\D_1\D_1}(\D_p,\D_m)
     \right)\Bigg]\, .
     \label{eq:dC11ldetailed}
}}
Recall from \eqref{eq:K_residue_full} that we have the following kinematical result
\begin{equation}
\KK^{(\a)}_{\D_1\D_2\D_3}(\D_p,\D_m) = \frac{-1}{\D_m-\D_1}\frac{\II^{(\a)}(\D_p)}{2}+O(1) = \KK^{(\a)}_{\D_1\D_3\D_2}(\D_p,\D_m)\,.
\end{equation}
Without special selection rules, we expect that the three-point functions $\<\OO_1 \OO_1 \OO_l\>$ and $\<\OO_1 \OO_2 \OO_l\>$ are non-zero, thus the sum over $m$ contains both $\D_m=\D_1$ and $\D_m=\D_2$. In our analytic continuation regularization scheme, the pole at $\D_m=\D_1$ is removed. On the other hand, the (almost) divergence at $\D_m = \D_2 \to \D_1$ is physical and dominates $\KK$ as well as $d C_{11l}/d\l$. In the $s$ and $t$ channel (first two lines of the equation above) the pole exists for any $l$, but the $u$-channel pole (third line) $\frac{1}{\D_m - \D_l}$ only contributes when $m=1,l=2$ or $m=2,l=1$. Therefore, near $\D_2=\D_1$ the first term of \eqref{eq:sec order deriv of Delta} becomes
\begin{equation}\spl{
\sum_l \bb{\hat{\Phi}}{l}(\l) \frac{dC_{11 l}(\l)}{d\l} \II^{(\a)}(\D_l)
\approx
\frac{\W_{12}^2 + \W_{12}(\bb{\hat{\Phi}}{1} C_{211} \II^{(\a)}(\D_1)-\bb{\hat{\Phi}}{2} C_{111} \II^{(\a)}(\D_2))}{2\D_{12}}  \,,
\label{eq:sec order deriv of Delta aprox}}
\end{equation}
where
\be
\W_{12} \equiv \sum_{p}\bb{\hat{\Phi}}{p} (C_{12 p} + C_{21 p}) \II^{(\a)}(\D_{p})\,.
\label{eq:def of Omega fcn}
\ee
Note that $\Omega_{12} = \Omega_{21}$ and $\W_{12}\in\mathbb{R}$ because of \eqref{eq:c = c* in lv rps}. For simplicity, we chose $\a_1$, $\a_2$ in \eqref{eq:dC11ldetailed} to be equal to $\a$ in \eqref{eq:sec order deriv of Delta}.

For the second term in \eqref{eq:sec order deriv of Delta}, using BOE flow equation \eqref{eq:BOE flow} we have 
\begin{equation}
\sum_{{l}}  
\frac{d \bb{\hat{\Phi}}{l}}{d\l} C_{11 l} \II^{(\a)}(\D_l)
=\frac12
\sum_{l,k,j} \bb{\hat{\Phi}}{k}\bb{\hat{\Phi}}{j}(C_{kjl}+C_{klj}) C_{11l}\II^{(\a)}(\D_l)
\JJ^{(\a_k)}_{\D_l}(\D_k,\D_j)\, .
\end{equation}
Recall that $\JJ$ has the following pole structure near $\D_j=\D_l$ (see (\ref{eq:J_residue}))
\begin{equation} 
\JJ^{(\a)}_{\D_l}(\D_k,\D_j) = 
-\frac{1}{\D_j-\D_l}\II^{(\a)}(\Delta_k)+O(1)\,,
\end{equation}
so the sums over $j$ and $l$ are dominated by $\D_j=\D_1$, $\D_l=\D_2$ as well as $\D_j=\D_2$, $\D_l=\D_1$. It follows that (choosing $\a_k=\a$ for $k=1,2$)
\begin{equation}
\sum_{{l}} 
\frac{d\bb{\hat{\Phi}}{l}(\l)}{d\l} C_{11 l}(\l) \II^{(\a)}(\D_l)\approx
\frac{\W_{12} (
\bb{\hat{\Phi}}{2} C_{111} \II^{(\a)}(\D_1)
-\bb{\hat{\Phi}}{1} C_{112} \II^{(\a)}(\D_2)
 )}{2\D_{12}}\,.
\end{equation}

The third term in \eqref{eq:sec order deriv of Delta} does not have the pole structure $\sum_\D\frac{1}{\D - \D_1}$ so it is in fact subleading. We thus obtain 
\begin{equation}
\frac{d^2\Delta_1}{d\lambda^2} \approx 
\frac{\W_{12}^2 + \W_{12}(\bb{\hat{\Phi}}{1} C_{211} +\bb{\hat{\Phi}}{2} C_{111})(\II^{(\a)}(\D_1)- \II^{(\a)}(\D_2))}{2\D_{12}} 
\approx
\frac{\Omega_{12}^2}{2\Delta_{12}}\,,
\end{equation}
using the fact that $\II^{(\a)}(\D)$ is a smooth function of $\Delta$ for $\Delta>1$.
$d^2\D_2/d\l^2$ can be computed similarly, and altogether we have
\begin{equation}
\frac{d^2\D_{12}}{d\l^2}
\approx
\frac{\W_{12}^2}{\D_{12}}     \,,
\label{eq:sec order deriv of Delta aprox 2}
\end{equation}
for small $\D_{12}$.

Next we want to express the RHS of \eqref{eq:sec order deriv of Delta aprox 2} in terms of $\D_{12}$, which requires determining the scaling of $\W_{12}$ with $\D_{12}$ as $\D_{12}\to0$. Let us consider the derivative $\frac{d\W_{12}}{d\l}$ (again, the derivative of $\II^{(\a)}$ is subleading as before)
\begin{equation}
\frac{d\W_{12}}{d\l} \approx
\sum_{l } \bb{\hat{\Phi}}{l} \frac{d(C_{12l} +C_{21l}) }{d\l} \II^{(\a)}(\D_{l})
+
\sum_{l } \frac{d \bb{\hat{\Phi}}{l}}{d\l} (C_{12l} +C_{21l})\II^{(\a)}(\D_{l})\,.
\end{equation}
Using the flow equations  \eqref{eq:OPE flow} and  \eqref{eq:BOE flow}, and keeping only the singular terms in the limit $\D_{12}\to 0$, we find
\begin{align}
\spl{
\frac{d\W_{12}}{d\l} &\approx
\frac{\W_{12}\sum_l(\bb{\hat{\Phi}}{l} (C_{22l}-C_{11l})\II^{(\a)}(\D_{l})  + (\bb{\hat{\Phi}}{1}C_{122}+\bb{\hat{\Phi}}{2}C_{211} )(\II^{(\a)}(\D_{1})-\II^{(\a)}(\D_{2}))}{\D_{12}} \\
&\approx
\frac{\W_{12} \sum_l\bb{\hat{\Phi}}{l} (C_{22l}-C_{11l})\II^{(\a)}(\D_{l})  }{\D_{12}} = - 
\frac{\W_{12}   }{\D_{12}} \frac{d \D_{12}}{d\l}
\ \ \ \Rightarrow\ \ \ 
\W_{12}\sim \frac{c}{\D_{12}}\,.
}
\label{eq:self consistency ansatz}
\end{align}
Here $c$ is an integration constant that is real, because $\D_{12}$ and $\W_{12}$ are both real.

Plugging \eqref{eq:self consistency ansatz} into \eqref{eq:sec order deriv of Delta aprox 2}, we get
\begin{equation}
\frac{d^2 \D_{12}(\l)}{d\l^2} = \frac{c^2}{\D_{12}(\l)^3}\, .
\end{equation}
The solution is
\begin{equation}
\D_{12}(\l) = \pm \sqrt{c_1(\l-c_2)^2+\frac{c^2}{c_1}}\,,
\end{equation}
where $c_1(>0)$ and $c_2$ are integration constants. The minimum of $\D_{12}$ is attained at $\lambda=c_2$. In order for $\D_{12}$ to be able to approach 0, we  need $c_1\gg c^2$. 
The slopes of $\D_1(\l)$ and $\D_2(\l)$ get exchanged before and after the turning point. In conclusion, the repulsive potential in \eqref{eq:level repulsion eqn} leads to avoided level crossing.

The above argument implicitly assumes that $\Omega_{12}$, defined in \eqref{eq:def of Omega fcn}, is nonzero.  
	However, in certain special situations, $\Omega_{12}$ may vanish for all values of $\lambda$, in which case level crossing becomes possible.  
	A natural mechanism for this to occur is the presence of an additional global symmetry that is preserved by the deformation operator $\hat{\Phi}$.  
	If $\mathcal{O}_1$ and $\mathcal{O}_2$ correspond to states belonging to different symmetry sectors, the selection rule implies that
	\begin{equation}
		\begin{split}
			C_{12p}=C_{21p}=0
		\end{split}
	\end{equation}
	for all $\mathcal{O}_p$ appearing in the BOE of $\hat{\Phi}$, and consequently $\Omega_{12}=0$.  
	Indeed, in systems with global symmetries, there is no obstruction to level crossing between states in distinct symmetry sectors.  
	The level repulsion, in this case, occurs independently within each symmetry sector.

\subsection{Renormalization of bulk operators}
\label{subsec:renormbulk}

A local bulk operator is encoded in the coefficients $\bb{\hat{\Psi}}{i}$, where $\hat{\Psi}$ denotes a generic local (scalar) bulk operator.
So far, we wrote the flow equation \eqref{eq:BOE flow} that controls $\bb{\hat{\Phi}}{i}$. Here, we discuss flow equations for a generic local bulk operator.

Let us return to the origin of the flow equations and write
\begin{equation}
    \frac{d}{d\lambda}\langle \hat{\Psi}(\tau_1, z_1) \mathcal{O}_i(\t_2) \rangle_\l =-
    \int\frac{d\t dz}{z^2}\langle \hat{\Phi}(\tau,z)
    \hat{\Psi}(\tau_1, z_1) \mathcal{O}_i(\t_2)  \rangle_\l\,.
    \label{eq:flow general BOE}
\end{equation}
We have already discussed the convergence properties of this integral near the conformal boundary of AdS ($z\to 0$) and how this is related to the presence of relevant or marginal boundary operators. Now, we focus on possible divergences from the coincident limit $(\t,z)\to (\t_1,z_1)$. These can be understood using the bulk OPE 
\begin{equation}
	\hat{\Phi}(\tau, z) \hat{\Psi} (\tau_1, z_1) 
	= \sum_{\hat{\OO}} C^\text{UV}_{\hat{\Phi}\hat{\Psi}\hat{\OO} }\sum_{n=0}^\infty c_n
	\left( \frac{z z_1}{(\tau - \tau_1)^2 + (z - z_1)^2} \right)^{\frac{1}{2}(\Delta_{\hat{\Phi}}+\Delta_{\hat{\Psi}}-\Delta_{\hat{\OO}})-n} 
	\hat{\mathcal{O}}(\tau_1, z_1)\,,
\end{equation}
where the sum runs over all local operators (primary and descendants) of the bulk UV CFT.
The coefficients $c_n$ are normalized by $c_0=1$ and the terms $n\ge 1$ describe curvature corrections to the usual OPE in flat space.
In general, there are also  operators with spin in the bulk OPE of two scalars. However, if we choose a regulator rotationally symmetric around the point $(\t_1,z_1)$ their contribution integrates to zero.
The most singular term corresponds to 
$\hat{\mathcal{O}} = \mathbb{1}$ but this does not contribute to \eqref{eq:flow general BOE} because $\langle \mathcal{O}_i(\tau_2) \rangle=0$. If $\hat{\Psi} = \hat{\Phi}$ is the only bulk relevant operator (\emph{i.e.} $\Delta_{\hat{\Phi}}<2$) then there are no divergences from the coincident limit. More generally, we will have a divergence from each bulk operator with $\Delta_{\hat{\OO}}< \Delta_{\hat{\Phi}}+\Delta_{\hat{\Psi}}-2$. Let us regulate the integral in 
\eqref{eq:flow general BOE} by imposing $(\t-\t_1)^2+(z-z_1)^2>\epsilon^2$. Then, we find
\begin{equation}
    \frac{d}{d\lambda}\langle \hat{\Psi}(\tau_1, z_1) \mathcal{O}_i(\t_2) \rangle_\l =
    \sum_{\hat{\OO}}
    C^\text{UV}_{\hat{\Phi}\hat{\Psi}\hat{\OO}} \sum_{n=0}^\infty \tilde{c}_n
    \epsilon^{\Delta_{\hat{\OO}}-\Delta_{\hat{\Phi}}-\Delta_{\hat{\Psi}}+2+2n}
    \langle \hat{\mathcal{O}}(\tau_1, z_1) \mathcal{O}_i(\t_2) \rangle_\l
    +{\rm finite}\,.
\end{equation}
This means that under an infinitesimal change in the bulk coupling $\lambda$, the bulk operator $\hat{\Psi}$ can mix with other bulk operators $\hat{\mathcal{O}}$ with a smaller scaling dimension obeying $\Delta_{\hat{\OO}}<
\Delta_{\hat{\Psi}}+\Delta_{\hat{\Phi}}-2$. 
This reasoning suggests 
flow equations involving a UV cutoff,
\begin{equation}
     \frac{d \bb{\hat{\Psi}}{i}}{d\lambda}
    =\lim_{\epsilon\to 0} \left[\sum_{j,l} b^{\hat\Phi}_l b^{\hat\Psi}_j \frac{C_{l ji}+C_{lij}}{2} \JJ_{\D_i}^{(\a,\epsilon)}(\D_l,\D_j) -
    \sum_{\hat{\OO}}
    \bb{\hat\OO}{i}
    C^\text{UV}_{\hat{\Phi}\hat{\Psi}\hat{\OO}} 
    \sum_{n=0}^\infty 
    \frac{  \tilde{c}_n}
    {\epsilon^{\Delta_{\hat{\Phi}}+\Delta_{\hat{\Psi}}-\Delta_{\hat{\OO}}-2-2n}
   }  
     \right]\,,
     \label{eq: div BOE flow}
\end{equation}
where $\JJ_{\D_i}^{(\a,\epsilon)}(\D_l,\D_j)$ is a regulated version of the kinematical integral $\JJ_{\D_i}^{(\a)}(\D_l,\D_j)$ using $(\t-\t_1)^2+(z-z_1)^2>\epsilon^2$.

We can avoid these divergences if we choose a convenient basis for the space of local bulk operators.
Let us order the bulk operators $\hat{\OO}_u$ by increasing bulk scaling dimension: $\D_{\hat{\OO}_{u+1}}\ge \D_{\hat{\OO}_{u}}$.
Each bulk operator is encoded into a vector $\bb{\hat{\OO}_u}{i}$.
Then introduce an inner product
$b^{\hat{\OO}}\cdot  b^{\hat{\Psi}}\equiv\sum_{i,j} \bb{\hat{\OO}}{i}\bb{\hat{\Psi}}{j} G_{ij}$ and impose that 
$b^{\hat{\OO}_u}\cdot  b^{\hat{\OO}_v}=0$ if $u\neq v$ (for example, using the Gram–Schmidt process).
In general $b^{\hat{\OO}_u}$ is no longer orthogonal to all the $b^{\hat{\OO}_v}$'s once the flow starts, but now we can take the flow equation for $\bb{\hat{\OO}_u}{i}$ and project it onto the space orthogonal to the vectors $\bb{\hat{\OO}_v}{i}$ with $v<u$,
\begin{equation}
     \frac{d \bb{\hat{\OO}_u}{k}}{d\lambda}
     \to
     \sum_i [P^u]_{ki} \frac{d \bb{\hat{\OO}_u}{i}}{d\lambda}
    = \sum_{j,l} b^{\hat\Phi}_l b^{\hat{\OO}_u}_j 
    \sum_i [P^u]_{ki}
    \frac{C_{l ji}+C_{lij}}{2} \JJ_{\D_i}^{(\a)}(\D_l,\D_j) \,,
\end{equation}
%
where $[P^u]$ is the  projector
\begin{equation}
    [P^u]_{ki} =\delta_{ki} - 
\sum_{v<u}\frac{ \bb{\hat{\OO}_v}{k}}{b^{\hat{\OO}_v}\cdot b^{\hat{\OO}_v}}
\sum_j G_{ij}
\bb{\hat{\OO}_v}{j}\,,
\end{equation}
so that $\sum_i [P^u]_{ki} \bb{\hat{\OO}_v}{i}=0$ for all $v<u$.
This gives us a cutoff independent flow equation.

Let us give a more explicit construction. Order the boundary primary operators $\mathcal{O}_i$
by increasing scaling dimension: $\Delta_{i+1}\ge \D_i$.
Then, we can choose a basis of bulk operators such that $b_i^{\hat\OO_u}=0$ if $u>i$.
In addition, we can normalize the bulk operators such that $b_i^{\hat\OO_i}=1$.
Equation \eqref{eq: div BOE flow} can then be written as
\begin{equation}
  \d \bb{\hat\OO_u}{i}
    =\d \l  \left[\sum_{j,l} b^{\hat\Phi}_l b^{\hat\OO_u}_j \frac{C_{l ji}+C_{lij}}{2} \JJ_{\D_i}^{(\a,\epsilon)}(\D_l,\D_j) -
    \sum_{v< u}
    \bb{\hat\OO_v}{i}
    C_{vu}(\epsilon) \right]\,.
\end{equation}
The equations with $i\in \{0,1,\dots,u\}$ (here $\hat\OO_0=\OO_0=\mathbb{1}$) are special  because $b^{\hat\OO_u}_i$ with $i\le u$ are fixed by our renormalization conditions. We can impose these renormalization conditions recursively.
We start from $b^{\hat\OO_0}_i=\delta_{i,0}$, which is automatically satisfied.
Then, we consider $\hat\OO_1$ and 
redefine the variations of the BOEs as follows
\begin{equation}
  \d \bb{\hat\OO_u}{i} \to \d \bb{\hat\OO_u}{i}-\d \bb{\hat\OO_u}{1}   \bb{\hat\OO_1}{i} \,,\qquad u\ge 1\,.
\end{equation}
This ensures that $\d \bb{\hat\OO_u}{1}=0$ 
for all $u\ge 1$ because we chose $\bb{\hat\OO_1}{1}=1$. The next step is 
\begin{equation}
  \d \bb{\hat\OO_u}{i} \to \d \bb{\hat\OO_u}{i}-\d \bb{\hat\OO_u}{2}   \bb{\hat\OO_2}{i} \,,\qquad u\ge 2\,.
\end{equation}
Now we have $\d \bb{\hat\OO_u}{2}=0$ 
for all $u\ge 2$. Proceeding this way, we can impose all renormalization conditions and, at the same time, remove all divergences. For example, we obtain
\begin{equation}
  \frac{d\bb{\hat\OO_1}{i}}{d\l}
    = \sum_{j,l} b^{\hat\Phi}_l b^{\hat\OO_1}_j \left[\frac{C_{l ji}+C_{lij}}{2} \JJ_{\D_i}^{(\a)}(\D_l,\D_j) -\bb{\hat\OO_1}{i}\frac{C_{l j1}+C_{l1j}}{2} \JJ_{\D_1}^{(\a)}(\D_l,\D_j)\right]\,,
    \label{eq:BOErenO1}
\end{equation}
for the properly renormalized operator $\hat\OO_1$. Similarly, we find
\begin{align}
  \frac{d\bb{\hat\OO_2}{i}}{d\l}
    = \sum_{j,l} b^{\hat\Phi}_l b^{\hat\OO_2}_j &\left[ \frac{C_{l ji}+C_{lij}}{2} \JJ_{\D_i}^{(\a)}(\D_l,\D_j) -\bb{\hat\OO_1}{i}\frac{C_{l j1}+C_{l1j}}{2} \JJ_{\D_1}^{(\a)}(\D_l,\D_j)\right.\\
   &\left. 
   -\bb{\hat\OO_2}{i}
   \frac{C_{l j2}+C_{l2j}}{2} \JJ_{\D_2}^{(\a)}(\D_l,\D_j) +\bb{\hat\OO_2}{i}
   \bb{\hat\OO_1}{2}\frac{C_{l j1}+C_{l1j}}{2} \JJ_{\D_1}^{(\a)}(\D_l,\D_j)
    \right]\,,\nonumber
\end{align}
for the properly renormalized operator $\hat\OO_2$.
By iterating this procedure, we can make the flow equations for all BOE coefficients cut-off free. 
 In fact, the subtraction of the vev $\langle \hat{\Psi} \rangle$ from local bulk operators can be seen as a special case of this method that makes all non-trivial bulk operators orthogonal to the bulk identity operator.

In the future, it will be important to study if this orthogonalization scheme is compatible with numerical truncation algorithms for the flow equations.

There are also bulk operators with spin. We think a conceptually similar treatment can be applied to those, but we leave a detailed study for the future.

\subsubsection{Marginally relevant bulk deformation}

In asymptotically free theories, the deforming operator $\hat{\Phi}$ is marginally relevant, \emph{i.e.}  it has $\Delta_{\hat{\Phi}}=2$. 
This leads to logarithmic divergences in \eqref{eq: div BOE flow}. More precisely, this gives
$\frac{ d \bb{\hat{\Phi}}{i}}{d\lambda} \sim \bb{\hat{\Phi}}{i} \log \epsilon +\dots  $. Such divergences can be removed 
with the  strategy discussed above.

For simplicity, consider the case where $\hat{\Phi}$ is the most  relevant bulk operator (above the identity). Then, we can normalize $\hat{\Phi}$ by the condition $\bb{\hat{\Phi}}{1}=1$, where $\OO_1$ is the boundary operator with lowest scaling dimension above the identity. Recall that $\bb{\hat{\Phi}}{0}=0$ where $\OO_0 =\mathbb{1}$ is the boundary identity operator.
The flow equations can then be written as in \eqref{eq:BOErenO1},
\begin{equation}
     \frac{d \bb{\hat{\Phi}}{i}}{d\lambda}
    = \sum_{j,l} b^{\hat\Phi}_l b^{\hat\Phi}_j
    \left[ 
    \frac{C_{l ji}+C_{lij}}{2} \JJ_{\D_i}^{(\a)}(\D_l,\D_j)-b^{\hat\Phi}_i
    \frac{C_{l j1}+C_{l1j}}{2} \JJ_{\D_1}^{(\a)}(\D_l,\D_j)
    \right]\,,\qquad i\ge 2\,.
    \label{eq: BOE flow marginal}
\end{equation}
Together with the renormalization conditions $\bb{\hat{\Phi}}{0}=0$ and $\bb{\hat{\Phi}}{1}=1$, this equation defines the evolution of $\bb{\hat{\Phi}}{i}$ directly in the continuum limit.

The $O(N)$  model on AdS$_2$ (see
\cite{Carmi:2018qzm, Copetti:2023sya}) is the obvious example to apply equation \eqref{eq: BOE flow marginal}.

\subsubsection{Irrelevant bulk deformation}
Deforming a CFT by an irrelevant deformation jeopardizes the UV completion of the QFT. However, this is  standard  in Effective Field Theory (EFT) because these come with a UV cutoff. What happens if we try to deform the QFT by an irrelevant bulk operator from the perspective of the flow equations?\footnote{We thank Francesco Riva for asking this question.}

Let us analyze the bulk UV divergences. If the bulk OPE $\hat{\Phi}\times \hat{\Phi}$ contains operators $\hat\OO$ such that  $\Delta_{\hat\F}<\Delta_{\hat{\OO}} < 2\Delta_{\hat{\F}}-2$, then this will spoil our recursive renormalization scheme described above. Generically, this will require more and more renormalization choices at each order in perturbation theory. It may be possible to introduce a UV cutoff and treat the theory as an EFT but this is beyond the scope of this paper.

Suppose that the bulk OPE $\hat{\Phi}\times \hat{\Phi} = \mathbb{1} + \hat{\Phi} + \sum  \hat{\OO}$ with $\Delta_{\hat{\OO}} > 2\Delta_{\hat{\F}}-2$. Then, the only UV divergences are caused by the identity and the deforming operator itself.  In this case, it seems that we can use the same flow equation as for a marginally relevant deformation and there is no need for a UV cutoff. This is reminiscent of the $T\bar{T}$ deformation of 2D QFTs \cite{Smirnov:2016lqw}.
It would be interesting to explore this connection in more detail in the future.

\subsection{Bulk locality}

In section \ref{subsec:local block}, we used analyticity of the correlators  to obtain expansions in blocks with improved convergence. As shown in \cite{Loparco:2025aag}, this follows from locality and unitarity. Here, we discuss the implications of the existence of a bulk stress tensor which is a stronger form of locality.\footnote{Notice that our flow equation would still be valid in a setup where  AdS$_2$  is (rigidly) embedded into a higher dimensional AdS$_d$. 
In this case, there is no conserved stress tensor in AdS$_2$.  }

The correlation functions of the stress tensor obey Ward identities. In \cite{Meineri:2023mps}, it was shown that these can be used to derive the following sum rules:
\begin{align}
    c_{\rm UV} &= \sum_l  (\bb{\hat{\Theta}}{l} )^2 f(\D_l)\, ,\\
    \D_i &= \sum_l \bb{\hat{\Theta}}{l} C_{iil } \,\kappa^{(\a)}(\D_l)\, ,
\end{align}
where $\hat{\Theta}=\hat{T}^\mu_\mu$ is the trace of the stress tensor, $c_{\rm UV}$ is the central charge of the CFT that describes the short distance limit of the bulk QFT, and $f(\D)$ and $\kappa^{(\a)}(\D)$ are explicitly known.
It would be interesting to explore the interplay between our flow equations and these sum rules. For instance, can we show that they are preserved under the evolution in $\lambda$ produced by the flow equations? 
The violation of these sum rules can also be used as a measure of the error of an approximate numerical solution of the flow equations.

Notice that for a 2D QFT with a single relevant deformation, the trace of the stress tensor is proportional to the relevant operator: $\hat{\Theta}\propto \hat{\Phi}$ \cite{Zamolodchikov:1986gt}. 
Therefore, we can choose to normalize $\hat{\Phi}$ so that $\hat{\Theta}=  \hat{\Phi}$. In this scheme, the sum rules above constrain the central object $\bb{\hat{\Phi}}{l}$.

\subsection{Flat space limit} 
The flat-space limit (FSL) of a gapped QFT in AdS  
corresponds to the regime where the energy gap above the vacuum is much larger than the inverse of the AdS radius. 
This implies that all scaling dimensions of boundary operators tend to infinity \cite{Susskind:1998vk, Polchinski:1999ry, Penedones:2010ue, Paulos:2016fap}.
Recent studies have provided strong evidence that the flat-space limit of conformal correlators on the boundary of AdS morph into flat-space scattering amplitudes in the FSL \cite{Paulos:2016fap,Hijano:2019qmi,Komatsu:2020sag,Li:2021snj,Cordova:2022pbl,vanRees:2022zmr,vanRees:2023fcf}. The relation between $B\del\del$ correlators and flat-space form factors has also been explored in \cite{Meineri:2023mps,Levine:2023ywq}.

If we are interested in center-of-mass energy $E$ of the same order as the mass gap $m$ in flat space, then we should take the limit
\be
\D_1,\D_i \to\infty\,,
\qquad \frac{\D_i}{\D_1}=\frac{E}{m}  \text{ fixed}.
\label{eq:flat-space limit}
\ee
The mass spectrum, scattering amplitudes and form factors of the QFT in flat space constrain the QFT data $\{\D_i,C_{ijk},\bb{\hat{\OO}}{i}\}$ in this limit. It would be interesting to show explicitly that the flow equations admit such limiting solutions.

The kinematical factors $\II,\JJ,\KK$ should simplify significantly in the FSL. 
For $\II$ it is straightforward to show that 
\be
\II^{(\a)}(\D) = 2^{\D+1} \frac{\sin \left(\pi  \left(\alpha -\frac{\D}{2}\right)\right)\Gamma (2 \alpha -1)  }{\sqrt{\pi }\D^{2 \alpha-\frac{1}{2} }} \left(1+O(\D^{-1})\right)\,.
\ee
The large $\D$ expansions of $\JJ$ and $\KK$ are more non-trivial.  
Nevertheless, 
the flow equations may simplify using the right variables in this asymptotic limit.

Of course, the flow equations need to be supplemented by an appropriate initial condition for the first order ODEs. In general, the bulk RG flow can experience phase transitions  at finite  AdS radius \cite{Aharony:2012jf,Copetti:2023sya,Ciccone:2024guw}. 
However, after the supposed last phase transition the evolution is expected to be continuous and this is where the asymptotic flow equations may be applied. 

One  open question concerning the FSL is the origin of anomalous thresholds in the boundary conformal correlators. The anomalous thresholds correspond to poles or branch points in the flat-space scattering amplitude which do not correspond to any intermediate propagating on-shell particles. On the other hand, the counterpart of the imaginary part of the scattering amplitude in conformal correlators is the spectral density, which only has support on physical operators. Numerical study of the triangle Witten diagram in general dimensions suggests the onset of the flat-space anomalous thresholds always lies within the support of the conformal spectral density (which becomes dense in the FSL) \cite{Komatsu:2020sag}. In AdS$_2$/CT$_1$, the magnitude of the spectral density of the box Witten diagram is unbounded from above in a region of scaling dimensions covering the locus of the corresponding anomalous threshold \cite{Cordova:2022pbl}. Therefore, to uncover the emergence of anomalous thresholds in the FSL, it is crucial to be able to extract the subleading terms of the spectral density in this unbounded region. The flow equations in the FSL may provide insight into this problem, because they have access to the spectral density through the dimension and the OPE coefficient flows.

Given that higher-point functions can be constructed from the QFT data, the flow equations can provide a window for studying the multi-particle S-matrix elements in flat space. 

\begin{figure}
\begin{subfigure}{.48\textwidth}
    \centering
    \includegraphics[height=6.7cm]{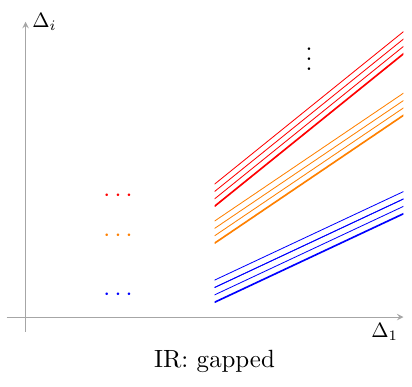}
\end{subfigure}%
\begin{subfigure}{.48\textwidth}
    \centering
    \raisebox{0.04cm}{\includegraphics[height=6.7cm]{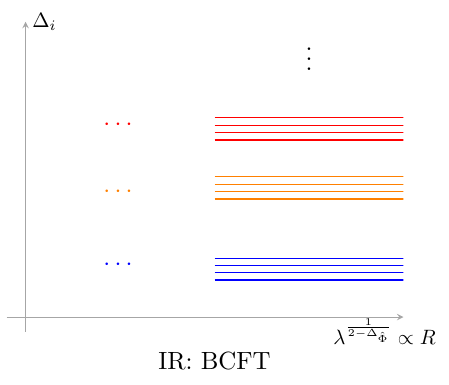}}
\end{subfigure}
 \caption{The expected asymptotic behavior of $\Delta_i(\lambda)$ for large $\lambda$. The IR theory can be gapped, in which case $\Delta_i\sim m_iR$ grow linearly as in the left figure. The slopes of the colored lines are associated to the masses of the stable particles in the IR theory in the unit of the mass gap. Thinner lines represent descendant states (states in which the particle is not at rest). In the left figure, we use $\Delta_1$ instead of $R$ for the horizontal axis in order to remove scheme dependence as otherwise, the asymptotic behavior of the curves might not be linear. If the IR theory is instead not gapped, we reach a BCFT phase, where the $\Delta_i$'s asymptote to constants. }
 	\label{fig:fsl}
\end{figure}

Another possible endpoint of the bulk RG flow is a BCFT in the IR. This would give rise to a solution of the flow equations where $\D_i$ tend to constants as $\l \to \infty$. In particular, the IR boundary spectrum should contain the displacement operator with $\Delta=2$. Furthermore, the boundary operators should organize into representations of the Virasoro algebra.
In this case, generically, we cannot probe the flat space limit with an S-matrix but we can probe it with the bulk OPE (see for instance \cite[sec. 5.1]{Carmi:2018qzm}). In figure \ref{fig:fsl}, we contrast these two possible IR behaviors of the boundary scaling dimensions.



\subsection{Large $N$ factorization}

\label{sec:Large N factorization}

Some QFT's enjoy the property of large $N$ factorization. In this case, the Hilbert space of the QFT has the structure of a Fock space built from a set of single-particle states (sometimes called single-trace or single-twist). Quantum chromodynamics with $SU(N)$ gauge group is a famous example: when $N\to \infty$ mesons and glueballs are free ``single-particle" states.

For such theories, it should be possible to reduce the QFT data to $\{\D_i,C_{ijk},\bb{\hat{\OO}}{i}\}$ with $i$/$\hat{\OO}$ running solely over single-particle boundary/bulk operators.
There have been past attempts at finding such consistent truncations of the conformal bootstrap equations - see for instance \cite{Caron-Huot:2020adz, Carmi:2020ekr}. 
It would be fantastic if our flow equations could be written in a closed form using only single-particle QFT data.\footnote{We thank Antonio Antunes and Shota Komatsu for this suggestion.} We could call them \emph{dispersive} flow equations.

We leave a careful exploration of this idea for the future, but let us make a few observations.
Firstly, notice that it is easy to write a dispersive  version of the first flow equation \eqref{eq:dimension flow}, which we repeat here for convenience:
\begin{equation}
    \frac{d\Delta_i}{d\lambda}=\sum_{l}b^{\hat\Phi}_l C_{iil} \II^{(\alpha)}(\Delta_l)\,.
\label{eq:dimension flow large N}
\end{equation}
The key observation is that $\II^{(\alpha)}(\Delta)$ contains a factor $1/\Gamma(\a-\D/2)$.  Therefore, choosing $\a=\D_i+w$ with $w\in\mathbb{N},w>\frac{1}{2}\D_{\hat\F}$ eliminates the two-particle operators $\OO_i \partial^{2n}\OO_i$ with $n\ge w$ from the sum over $l$. The other multi-particle operators $\OO_l$ do not contribute because $C_{iil}$ is suppressed in the large $N$ limit.
Secondly, notice that $\JJ_{\D_i}^{(\a)}(\D_l,\D_j)$ and $\KK^{(\a)}_{\D_i\D_j\D_k}(\D_l,\D_m)$ in equations \eqref{eq:BOE flow} and \eqref{eq:OPE flow} also contain the factor $1/\Gamma(\a-\D_l/2)$, as can be seen from \eqref{eq:local block kernel}. Therefore, we can make the sums over $l$ dispersive by choosing $2\a_{ij}=\D_i+\D_j+2w$ in \eqref{eq:BOE flow} and $2\a_{pm}=\D_p+\D_m+2w$ with $p$ equal to $i,j$ or $k$ appropriately chosen in each term of equation \eqref{eq:OPE flow} ($w\in\mathbb{N},w>\frac{1}{2}\D_{\hat\F}$).

The remaining challenge is to eliminate multi-particle operators from the sums over $j$ in
\eqref{eq:BOE flow} and $m$ in \eqref{eq:OPE flow}.
For \eqref{eq:BOE flow}, this may be possible using doubly local blocks\footnote{We may try to define doubly local blocks for the $BB\partial$ correlator by integrating normal blocks twice against the kernels $K$ and $\tilde K$ introduced in \cite{Loparco:2025aag}. } to expand the $BB\partial$ correlator.
For \eqref{eq:OPE flow}, one can try to build local blocks for the $B\partial\partial\partial$ correlator starting from  Polyakov blocks for the boundary four-point function \cite{Polyakov:1974gs, Caron-Huot:2020adz, Paulos:2020zxx}.

As the zeroth-order example, let us consider again the free massive scalar theory. The generalized free field theory on the boundary is just the leading-order in the large $N$ limit. In the dimension flow of the boundary operator $\f$ with dimension $\D_\f$, if we set $\a=\D_\f+w$ and $\D_l=2\D_\f+2l$, then the flow equation \eqref{eq:dimension flow large N} truncates to a finite sum with $l\in[0,w-1]$. Namely an infinite tower of double-twist operators are removed from the flow equation through the appropriate choice of $\a$. Furthermore, for any choice of allowed $w$, the finite sums all give the same final result.

A full fledged example is the $O(N)$ linear sigma model in AdS$_2$
\be
S_{O(N)} =\int_{\text{AdS}_2} \left(\frac{1}{2}\del_\m\hat\f^i \del^\m\hat\f^i + \frac{1}{2}m^2 \hat\f^i\hat\f^i + \frac{\l}{2N} (\hat\f^i \hat\f^i)^2\right)\,, \quad i=1,\ldots,N\,.
\ee
In the large $N$ limit, the $O(N)$ singlet operator $\hat\OO\equiv\hat\f^i\hat\f^i$ creates meson-like single-particle states. An infinite tower of operators, denoted as $\s_n$, $n=0,1,2,\ldots$, appear in the BOE of $\hat\OO$. The spectrum of $\s_n$ and the BOE coefficients $b^{\hat\OO}_{\s_n}$ can be extracted from the poles and the residues of the spectral density of the bulk two-point function $\<\hat\OO \hat\OO\>$, which was computed in \cite{Carmi:2018qzm} by resumming the bubble diagrams. Each $\s_n$ receives $O(1)$ anomalous dimension as $\l$ becomes large enough (see \cite[Fig.5]{Carmi:2018qzm}), but the anomalous dimensions of multi-particle states like $[\s_n^2]_k$ are suppressed by $1/N$. It would be interesting to apply the dispersive flow equations to this model. For example, consider the dimension flow of $\D_{\s_n}$, which requires the data $b^{\hat\OO^2}_{[\s_n^2]_k}$ and $C_{\s_n \s_n [\s_n]^2_k}$. By setting $\a=\D_{\s_n}+1$, the dispersive flow equation takes a simpler form than the general one
\be
\frac{d\D_{\s_n}}{d\l}=\sum_{n,k}
b^{\hat\OO^2}_{[\s_n^2]_k}
C_{\s_n \s_n [\s_n]^2_k}
\II^{(\a)}(2\D_{\s_n}+2k)
\ \to\ 
\frac{d\D_{\s_n}}{d\l}=\sum_{n}
b^{\hat\OO^2}_{\s_n^2}
C_{\s_n \s_n \s_n^2}
\II^{(\D_{\s_n}+1)}(2\D_{\s_n})\,.
\label{eq:dispersive flow of O(N)}
\ee
One can use the flow equations starting from the free massive scalar and compare the dimension flow with the results in \cite{Carmi:2018qzm}. Alternatively, the data needed in \eqref{eq:dispersive flow of O(N)} can be extracted from the spectral density of the bulk two-point function $\<\hat\OO^2 \hat\OO^2\>$ and the boundary four-point function $\<\s_n \s_n \s_n \s_n\>$.


\subsection{Numerical algorithms}

\label{discussion numerics}

In order to use the flow equations to study interacting QFTs, we will need to devise numerical algorithms. The simplest idea is to truncate the sums over the primary boundary operators to $\D_i<\D_{\rm max}$ and study
the variation of the QFT data as the cutoff $\D_{\rm max}$ is increased. This is similar to the spirit of Hamiltonian truncation but without the need for renormalization. We leave the exploration of this simple cutoff idea for the future.

A more refined version of the previous idea is to think of $\D_{\rm max}$ as a parameter that separates the discrete low-energy part of the spectrum from the continuum high-energy part.
In reality, the entire spectrum is discrete but, for large $\Delta$, we can appeal to the Eigenstate-Thermalization-Hypothesis (ETH) to approximately describe the QFT data by smooth functions of $\Delta$.

Consider, for example, the first flow equation \eqref{eq:dimension flow}. It can be written as 
\begin{equation}
    \frac{d\Delta_i}{d\lambda}=\sum_{l:\,\D_l\le \D_{\rm max}}b^{\hat\Phi}_l C_{iil} \II^{(\alpha)}(\Delta_l) +\int_{\D_{\rm max}}^\infty d\D f_{ii}(\Delta) \II^{(\alpha)}(\Delta)\,,
\end{equation}
with
\begin{equation}
    f_{ij}(\Delta) = \sum_{l:\,\D_l> \D_{\rm max}}b^{\hat\Phi}_l C_{lij} \,\delta(\Delta-\D_l)\,.
\end{equation}
The approximation step is to describe $f_{ij}(\Delta)$ by a smooth function of $\Delta$. For example, we could use the average of the free theory result. It should also be possible to derive flow equations to evolve these functions with $\lambda$. The exploration of these ideas is beyond the scope of this paper.

\subsection{Some  future applications}

There have been many interesting papers studying different aspects of QFT in AdS, from confinement \cite{Callan:1989em, Aharony:2012jf, Ciccone:2024guw,Ciccone:2025dqx,DiPietro:2025ozw}, fermions \cite{Giombi:2021cnr},
scattering amplitudes from the flat space limit
\cite{Penedones:2010ue,Paulos:2016fap, Mazac:2018mdx,Hijano:2019qmi, Komatsu:2020sag,Li:2021snj, Cordova:2022pbl, vanRees:2022zmr,vanRees:2023fcf}, RG flows \cite{Hogervorst:2021spa, Antunes:2021abs, Lauria:2023uca,Antunes:2024hrt} and large $N$ vector models
\cite{Carmi:2018qzm, Ankur:2023lum,Copetti:2023sya}.
It will be interesting to revisit these works from the perspective of the flow equations.

There are many open avenues for exploration. Here is a possible roadmap:
\begin{itemize}
\item Use the flow equations to derive perturbative expansions of QFT data. 
This is important to understand in detail the initial condition for a (truncated) numerical approach. Notice that solvable theories usually have a degenerate spectrum of boundary scaling dimensions. We expect these degeneracies to be lifted by generic interactions but the flow equations  are written assuming no degeneracy. 
    \item The next step is to use the flow equations non-perturbatively to find the QFT data of an interacting theory. The simplest target is probably the Long-Range Ising theory discussed in section \ref    {subsec:weakly relevant boundary op}.
    \item Integrable RG flows from BCFT minimal models to gapped phases are a good next target.
    Most likely, integrability is broken by the AdS background geometry and it is only recovered in the flat space limit \cite{Antunes:2025iaw}. In practice, we can compare the known results from integrability with the predictions for the mass spectrum and the scattering amplitudes from our flow equations.
    It is natural to start with the scaling Lee-Yang model because it only has one  relevant bulk operator. Moreover, we can compare the QFT data with Hamiltonian truncation \cite{Hogervorst:2021spa} and with integrability in the flat space limit \cite{Cardy:1989fw}.
 Ising Field Theory (IFT) with both thermal and magnetic deformations,  is not integrable, which makes our new approach more interesting.
    \item  RG flows between two nontrivial BCFTs described by minimal models (for example, tricritical to Ising).

    \item Asymptotically free theories like the $O(N)$ model in AdS$_2$ \cite{Carmi:2018qzm}.
    Here, it is interesting to study the large $N$ limit using the \emph{dispersive} flow equations discussed in \ref{sec:Large N factorization}. 
    
     \item 2D gauge theories
like adjoint QCD$_2$ (see for instance \cite{Dempsey:2022uie})  for small number of colors,  and in the planar limit again using the dispersive flow equations.

\item Derive flow equations for 1D conformal defects in CFTs with marginal deformations. This is very close to the spirit of \cite{Behan:2017mwi} but the marginal operator is in the higher dimensional bulk. A famous example in this class is the Maldacena-Wilson line in supersymmetric Yang-Mills (SYM) theory. If we restrict to defect operators and the bulk lagrangian density, then there should be flow equations with the same form as ours but with different kinematical functions $\mathcal{I},\mathcal{J},\mathcal{K}$. This is an interesting model to study because much is known about this theory using perturbation theory, integrability  and bootstrap \cite{Giombi:2017cqn,Giombi:2018qox,Cavaglia:2021bnz}.

    \item Derive flow equations for  QFTs on $AdS_{d+1}$ with $d>1$. The challenge here is purely technical. In order to close the flow equations, we need to evolve all boundary QFT data, including scaling dimensions and OPE coefficients, of \emph{spinning} operators. For a $d$-dimensional boundary, operators are classified by irreducible representations of $SO(d)$. Thus, the technical difficulty increases with $d$. We are optimistic that the most interesting cases $d=2,3$ can be treated.
    
    \item The dream application for our approach is to follow Yang-Mills with Neumann 
    boundary conditions\footnote{It would also be interesting to study Dirichlet boundary conditions and test the scenario of fixed point merger and annihilation suggested by  perturbative computations \cite{Ciccone:2024guw}. } from the perturbative regime of small AdS radius to the strongly coupled regime of large AdS radius (see \cite{Ciccone:2024guw, DiPietro:2025ozw} for more detail). This would provide a new approach to confinement.
    It would also be very interesting to study the 
    flux tube theory in AdS$_3$ and AdS$_4$ \cite{Gabai:2025hwf}. From our point of view, the QFT data is the same as for a QFT on AdS$_2$ if we restrict ourselves to operators on the 1D line defect on the boundary of AdS. However, the kinematical functions $\mathcal{I},\mathcal{J},\mathcal{K}$ in the flow equations will be different because the bulk deformation needs to be integrated over AdS$_3$ or AdS$_4$.


\end{itemize}

\section*{Acknowledgments} We would like to thank Ant\'{o}nio Antunes, Fabiana De Cesare, Kelian H\"{a}ring, Nat Levine, Marco Meineri, Dalimil Maz\'{a}\v{c}, Slava Rychkov, Miguel Paulos, Petr Kravchuk, Edoardo Lauria, Kamran Salehi Vaziri, Marco Serone, Bernardo Zan for useful discussions. We also thank Sebastian Harris for providing us with the explicit result of the GFF OPE coefficients involving triple-twist operators $C_{\f [\f^2]_m [\f^3]_q} C_{[\f^3]_q [\f^2]_n \f}$.

We also thank the participants of the workshop QFT in AdS 2025 in Trieste, as well as the participants of the Bootstrap 2025 conference in São Paulo, for stimulating discussions. GM and JQ thank Riken iTHEMS and the Yukawa Institute for Theoretical Physics at Kyoto University. Discussions during “Progress of Theoretical Bootstrap” were useful in completing this work.

This work was performed in part at Aspen Center for Physics, which is supported by National Science Foundation grant PHY-2210452 and by a grant from the Simons Foundation (1161654, Troyer).

ML, GM and JP are supported by the Simons Foundation grant 488649 (Simons Collaboration on the Nonperturbative Bootstrap) and the Swiss National Science Foundation through the project
200020\_197160 and through the National Centre of Competence in Research SwissMAP.
The research of ML was also supported by the Italian Ministry of University and Research (MUR) under the FIS grant BootBeyond (CUP: D53C24005470001) and by the
INFN “Iniziativa Specifica” ST\&FI. The work of JQ was supported by World Premier International Research Center Initiative (WPI), MEXT, Japan. JQ also acknowledges support by Simons Foundation grant 994310
(Simons Collaboration on Confinement and QCD Strings), under which a portion of this work was performed. The work of XZ is supported by an ANR grant from the Tremplin - ERC Starting Grant funding scheme.

\appendix

\newpage

\section{Details on the series representation of conformal blocks}
We give more details regarding the computation of the various conformal blocks presented in section \ref{subsec:Confblock} of the main text. 

\subsection{\texorpdfstring{Bulk-bulk ($BB$) conformal block}{Bulk-bulk (BB) conformal block}}
\label{app:BKBK block}
The conformal block for the bulk-bulk two-point function can be derived using the BOE \eqref{eq:BOE}. We choose the configuration
\begin{equation}
	\begin{split}
		x_1=(0,\eta)\, ,\qquad x_2=(0,1)\, ,\qquad0<\eta<1\, .
	\end{split}
\end{equation}
Using the BOE \eqref{eq:BOE} once and the bulk-to-boundary propagator \eqref{def:bkbdtwopt} we have
\begin{equation}
	\begin{split}                  
 \braket{\hat{\mathcal{O}}_1(0,\eta)\hat{\mathcal{O}}_2(0,1)}&=
 \sum\limits_{i}\bb{\hat{\mathcal{O}}_1}{i}\sum\limits_{n=0}^{\infty}\frac{(-1)^n(\Delta_i)_n}{n! (2\Delta_i)_{2n}}\eta^{\Delta_i+2n}\partial_\tau^{2n}\braket{\mathcal{O}_i(\tau)\hat{\mathcal{O}}_2(0,1)}\Big{|}_{\tau=0} \\
&=\sum\limits_{i}\bb{\hat{\mathcal{O}}_1}{i}\bb{\hat{\mathcal{O}}_2}{i}\sum\limits_{n=0}^{\infty}\frac{(-1)^n(\Delta_i)_n}{n! (2\Delta_i)_{2n}}\eta^{\Delta_i+2n}\partial_\tau^{2n}\left[\frac{1}{\tau^2+1}\right]^{\Delta_i}\Big{|}_{\tau=0} \\
		&=\sum\limits_{i}\bb{\hat{\mathcal{O}}_1}{i}\bb{\hat{\mathcal{O}}_2}{i}\sum\limits_{n=0}^{\infty}\frac{(-1)^n(\Delta_i)_n}{n! (2\Delta_i)_{2n}}\eta^{\Delta_i+2n}\ \frac{(-1)^n\ (\Delta_i)_n\ (2n)!}{n!} \\
		&=\sum\limits_{i}\bb{\hat{\mathcal{O}}_1}{i}\bb{\hat{\mathcal{O}}_2}{i} \eta^{\Delta_i}\hyperF{\Delta_i}{\frac{1}{2}}{\Delta_i+\frac{1}{2}}{\eta^2}\, . \\
	\end{split}
\end{equation}
This is the result quoted in equation \eqref{def:bkbk block} in the main text. 

\subsection{\texorpdfstring{Bulk-boundary-boundary ($B\partial\partial$) conformal block}{Bulk-boundary-boundary (Bbb) conformal block}}
\label{app:BKBDBD block}

We consider the bulk-boundary-boundary three-point function $	\braket{\hat{\mathcal{O}}(\tau,z)\mathcal{O}_i(\tau_1)\mathcal{O}_j(\tau_2)} $ at the special configuration 
\begin{equation}\label{eq:ConfigBppap}
		(\tau,z)=(0,\xi),\qquad \tau_1=1,\qquad \tau_2=\infty\, .
\end{equation}
As usual, an operator $\mathcal{O}_i$ of weight $\Delta_i$ is defined at infinity as 
\be 
\mathcal{O}_i(\infty)\equiv\lim\limits_{\tau\rightarrow+\infty}[\tau^{2\Delta_i}\mathcal{O}_i(\tau)]\, .\label{eq:OpeInfinity}
\ee
Using the boundary expansion \eqref{eq:BOE} to expand the bulk operator in terms of boundary operators, we have
\begin{align}
		\braket{\hat{\mathcal{O}}(0,\xi)\mathcal{O}_i(1)\mathcal{O}_j(\infty)}  &=\sum\limits_{l}\bb{\hat{\mathcal{O}}}{l}\sum\limits_{n=0}^{\infty}\frac{(-1)^n(\Delta_l)_n}{n! (2\Delta_l)_{2n}}\xi^{\Delta_l+2n}\partial_\tau^{2n}\braket{\mathcal{O}_l(\tau)\mathcal{O}_i(1)\mathcal{O}_j(\infty)}\Big{|}_{\tau=0} \nonumber\\
		&=\sum\limits_{l}\bb{\hat{\mathcal{O}}}{l}C_{ijl}\sum\limits_{n=0}^{\infty}\frac{(-1)^n(\Delta_l)_n}{n! (2\Delta_l)_{2n}}\xi^{\Delta_l+2n}\ \partial_\tau^{2n}\frac{1}{|1-\tau|^{\Delta_l+\Delta_{i}-\D_j}}\Big{|}_{\tau=0} \nonumber\\
		&=\sum\limits_{l}\bb{\hat{\mathcal{O}}}{l}C_{ijl}\sum\limits_{n=0}^{\infty}\frac{(-1)^n(\Delta_l)_n}{n! (2\Delta_l)_{2n}}\xi^{\Delta_l+2n}(\Delta_l+\Delta_i-\Delta_j)_{2n} \nonumber\\
		&=\sum\limits_{l}\bb{\hat{\mathcal{O}}}{l}C_{ijl}\xi^{\Delta_l}\hyperF{\frac{\Delta_{lij}}{2}}{\frac{\Delta_{lij}+1}{2}}{\Delta_l+\frac{1}{2}}{-\xi^2}, 
    \label{eq:bkbdbd using BOE}
\end{align}
where we used the boundary three-point function \eqref{def:bdbdbdthreept} from the first to the second line\footnote{Note that we also took $\mathcal{O}_2$ to infinity before acting with derivatives, using 
\begin{equation}
    \braket{\mathcal{O}_l(\tau)\mathcal{O}_i(1)\mathcal{O}_j(\infty)} = \frac{1}{|1-\tau|^{\Delta_l+\Delta_i-\Delta_j}}\, .
\end{equation}}. 
We thus get that a bulk-boundary-boundary three-point function can be decomposed as 
\begin{equation}\label{cb:threeptspecial}
		\braket{\hat{\mathcal{O}}(0,\xi)\mathcal{O}_i(1)\mathcal{O}_j(\infty)}
=\sum\limits_{l}\bb{\hat{\mathcal{O}}}{l}C_{ijl}g_{\Delta_l}^{\Delta_i\Delta_j}(\xi), 
\end{equation}
with 
\begin{equation}
g_{\Delta_l}^{\Delta_i\Delta_j}(\xi)=\xi^{\Delta_l}\hyperF{\frac{\Delta_{lij}}{2}}{\frac{\Delta_{lij}+1}{2}}{\Delta_l+\frac{1}{2}}{-\xi^2}.
\end{equation}
To obtain the block for a generic three-point configuration, we use the fact that the three-point function $\braket{\hat{\OO}(\tau,z)\OO_i(\tau_1)\OO_j(\tau_2)}$ is fixed by $SL(2;\mathbb{R})$ invariance up to an arbitrary function of the cross ratio $\xi$:
\begin{equation}\label{bkbdbd:generalformap}
	\begin{split}
		\braket{\hat{\mathcal{O}}(\tau,z)\mathcal{O}_i(\tau_1)\mathcal{O}_j(\tau_2)}=\frac{1}{\abs{\tau_1-\tau_2}^{\Delta_i+\Delta_j}}\left(\frac{(\tau-\tau_1)^2+z^2}{(\tau-\tau_2)^2+z^2}\right)^{\frac{\Delta_j-\Delta_i}{2}}\mathcal{G}_{ij}^{\hat{\mathcal{O}}}(\xi)\, ,
	\end{split}
\end{equation}
where $\xi$ is the cross-ratio defined by mapping the generic configuration to \eqref{eq:ConfigBppap} via an $SL(2;\mathbb{R})$ transformation. The explicit form of $\xi$ is given by
\begin{equation}\label{def:xi}
	\begin{split}
		\xi=\abs{\frac{(\tau_1-\tau_2)z}{(\tau-\tau_1)(\tau-\tau_2)+z^2}}\, .
	\end{split}
\end{equation}
The function $\mathcal{G}_{ij}^{\hat{\mathcal{O}}}(\xi)$ can be obtained by comparing \eqref{eq:bkbdbd using BOE} and \eqref{bkbdbd:generalformap}, 
which yields
\begin{equation}\label{cb:threeptgeneral}
\mathcal{G}_{ij}^{\hat{\mathcal{O}}}(\xi)=\sum\limits_{l}\bb{\hat{\mathcal{O}}}{l}C_{ijl}(1+\xi^2)^{\frac{\Delta_i-\Delta_j}{2}}g^{\D_i\D_j}_{\D_l}(\xi),
\end{equation}
with $g^{\D_i\D_j}_{\D_l}(\xi)$ defined in \eqref{cb:threeptspecial}. Using the identity 
\begin{equation}
\hyperF{a}{b}{c}{z}= (1-z)^{-a}\hyperF{a}{c-b}{c}{\frac{z}{z-1}}\, ,\label{eq:2F1identity1}
\end{equation}
we can define
\begin{equation}
    \chi \assign \frac{\xi^2}{1+\xi^2}\, .
\end{equation}
and rewrite \eqref{cb:threeptgeneral} as 
\begin{equation}\label{cb:threeptgeneral:chi}
		\mathcal{G}_{ij}^{\hat{\mathcal{O}}}(\xi)=\sum\limits_{l}\bb{\hat{\mathcal{O}}}{l}C_{ijl}
  G^{\D_i\D_j}_{\D_l}(\chi)\, ,
\end{equation}
where $ G^{\D_i\D_j}_{\D_l}(\chi)$ is the conformal block presented in equation \eqref{def:bkbdbd block}. The cross ratio $\chi$ can be equivalently defined as in \eqref{def:chi}.

\subsection{\texorpdfstring{Bulk-Bulk-boundary ($BB\partial$) conformal block}{Bulk-Bulk-boundary (BBb) conformal block}}
\label{app:BKBKBD block}

We consider the following bulk-bulk-boundary configuration
\begin{equation}
	x_1=(\tau_1,z_1),\qquad x_2=(\tau_2,z_2),\qquad x_3=(\tau_3,0)\, .
\end{equation}
By $SL(2;\mathbb{R})$ invariance and using the BOE, the three-point function has the form presented in \eqref{bkbkbd:generalform} and reproduced here for convenience\footnote{Note that in the way we are setting up the computation here, the decomposition in \eqref{eq:BBoApTC1} is valid for $\upsilon<0$ as in \eqref{bkbkbd:exp}. This does not change the conclusion and the derivation of the block: the only difference is the ordering in the OPE coefficient.}
\begin{equation}
\label{eq:BBoApTC1}
\braket{\hat{\OO}_1(\tau_1,z_1)\hat{\OO}_2(\tau_2,z_2)\OO_i(\tau_3)} = \left(\frac{z_2}{(\tau_2-\tau_3)^2 + z_2^2}\right)^{\Delta_i}
\sum_{l,j}\bb{\hat{\mathcal{O}}_1}{l}\bb{\hat{\mathcal{O}}_2}{j}C_{lji}R^{\D_i}_{\D_l\D_j}(\upsilon,\zeta)\,.
\end{equation}
Here $R^{\D_i}_{\D_l\D_j}(\upsilon,\zeta)$ is the conformal block, and $(\upsilon, \zeta)$ is a pair of cross ratios to be determined. To determine the conformal block, we expand $\braket{\hat{\mathcal{O}}_{ 1}(\tau_1,z_1)\hat{\mathcal{O}}_{2}(\tau_2,z_2)\mathcal{O}_i(\tau_3)}$ using the BOE \eqref{eq:BOE} twice, such that 
\begin{equation}
	\begin{split}
\braket{\hat{\mathcal{O}}_{ 1}(\tau_1,z_1)\hat{\mathcal{O}}_{2}(\tau_2,z_2)\mathcal{O}_i(\tau_3)}&=\sum\limits_{l,j}
\bb{\hat{\mathcal{O}}_1}{l}\bb{\hat{\mathcal{O}}_2}{j}
\sum\limits_{n=0}^{\infty}\sum\limits_{m=0}^{\infty}\frac{(-1)^{n+m}(\Delta_l)_n(\Delta_j)_m}{n!\,m!\, (2\Delta_l)_{2n}(2\Delta_j)_{2m}}z_1^{\Delta_l+2n}z_2^{\Delta_j+2m} \\
		&\quad\times\partial_{\tau_1}^{2n}\partial_{\tau_2}^{2m}\frac{C_{lji}}{\abs{\tau_{12}}^{\Delta_l+\Delta_j-\Delta_i}\abs{\tau_{13}}^{\Delta_l+\Delta_i-\Delta_j}\abs{\tau_{23}}^{\Delta_j+\Delta_i-\Delta_l}}\, , \label{eq:BulkBulkBoundary3pt339}\\
	\end{split}
\end{equation}
with $\tau_{ij} \equiv \tau_i-\tau_j$. Comparing \eqref{eq:BBoApTC1} and \eqref{eq:BulkBulkBoundary3pt339}, we obtain the conformal block $R^{\D_i}_{\D_l\D_j}(v,\zeta)$ written as double sum
\begin{equation}
	\begin{split}
	R^{\D_i}_{\D_l\D_j}&=
  \left(\frac{z_2}{(\tau_2-\tau_3)^2+z_2^2}\right)^{-\Delta_i}
\sum\limits_{n=0}^{\infty}\sum\limits_{m=0}^{\infty}\frac{(-1)^{n+m}(\Delta_l)_n(\Delta_j)_m}{n!\,m!\, (2\Delta_l)_{2n}(2\Delta_j)_{2m}}z_1^{\Delta_l+2n}z_2^{\Delta_j+2m} \\
		&\quad \times\partial_{\tau_1}^{2n}\partial_{\tau_2}^{2m}\frac{1}{\abs{\tau_{12}}^{\Delta_l+\Delta_j-\Delta_i}\abs{\tau_{13}}^{\Delta_l+\Delta_i-\Delta_j}\abs{\tau_{23}}^{\Delta_j+\Delta_i-\Delta_l}}\, .
	\end{split}
\end{equation}
The above expression is complicated, but by $SL(2;\mathbb{R})$ invariance, it can be computed in the following simpler configuration
\begin{equation}
	(\tau_1,z_1)=(\upsilon,\zeta),\qquad (\tau_2,z_2)=(0,1),\qquad \tau_3=\infty\, ,\label{eq:A21}
 \end{equation}
 and the cross ratios (they are the same as in \eqref{def:upsilon and zeta})
 \begin{equation}
 \upsilon = 
\frac{(z_1^2+\t_{12}\t_{13})\t_{23}+z_2^2\t_{31}}{z_2 (z_1^2 + \t_{13}^2)}\,
,
\qquad\quad 
\z = \frac{z_1(z_2^2+\tau_{23}^2)}{z_2(z_1^2+\tau_{13}^2)}\,  .
\end{equation}
Note that to take the limit $\tau_3\rightarrow\infty$, we do not need to dress the operator with $\tau_3^{2\Delta_i}$, as this is taken care of by the prefactor in \eqref{eq:BBoApTC1}.
At this configuration we have
\begin{equation}
\label{eq:A22app}
\begin{split}
R^{\D_i}_{\D_l\D_j}(\upsilon,\z)
  =
  &\zeta^{\Delta_l}(\upsilon^2)^{-\frac{\Delta_{lji}}{2}}
\sum\limits_{n=0}^{\infty}\sum\limits_{m=0}^{\infty}\frac{(-1)^{n+m}(\Delta_{lji})_{2m+2n}}{n!\,m!\,\left(\Delta_l+\frac{1}{2}\right)_n\,\left(\Delta_j+\frac{1}{2}\right)_m\,4^{n+m}}\frac{\z^{2n}}{\upsilon^{2n+2m}}\,.
\end{split}
\end{equation}
In fact, using
\be
(2x)_{2n}=(x)_n \left(x+\frac12\right)_n 4^n\,,
\ee
and the series representation of the Appell $F_4$ function 
\begin{equation}
    F_4(a,b;c,d;x,y) = \sum_{m,n=0}^\infty \frac{(a)_{m+n}\,(b)_{m+n}}{m!\,n!\,(c)_m\,(d)_n }\,x^my^n\qquad (\sqrt{|x|} + \sqrt{|y|}<1)\, ,
\end{equation}
we can write the block as 
\begin{equation}
\label{eq:A22app2}
\begin{split}
R^{\D_i}_{\D_l\D_j}(\upsilon,\z)
  =
  &
  \left(\upsilon^2\right)^{-\frac{\Delta_{ji}}{2}}
  \left(\frac{\z^2}{\upsilon^2}\right)^{\frac{\D_l}{2}}
  F_4\left(\frac{\Delta_{lji}}{2},\frac{\Delta_{lji}+1}{2};\Delta_l+\frac{1}{2},\Delta_j+\frac{1}{2};-\frac{\zeta^2}{\upsilon^2},-\frac{1}{\upsilon^2}\right)
  \,,
\end{split}
\end{equation}
but this is not the most useful form for this work because the convergence domain $\abs{\zeta}+1<\abs{\upsilon}$ does not cover the entire Euclidean AdS$_2$.\footnote{
There is a change of variables for the Appell function that is convergent in another area \cite{Ananthanarayan:2020xut}
\begin{equation}
\begin{aligned}
F_4(a,b;c,d;x,y)&=\frac{\Gamma(d)\Gamma(b-a)}{\Gamma(d-a)\Gamma(b)}(-y)^{-a}F_4\left(a,a-d+1;c,a-b+1;\frac{x}{y},\frac{1}{y}\right)\\
    &+\frac{\Gamma(d)\Gamma(a-b)}{\Gamma(d-b)\Gamma(a)}(-y)^{-b}F_4\left(b,b-d+1;c,b-a+1;\frac{x}{y},\frac{1}{y}\right)\,.
\end{aligned}
\end{equation}
In our case the domain of convergence in this form is
$|\zeta|+|\upsilon|<1$, which still does not cover the entire EAdS$_2$.
}


To make progress, we need to resum \eqref{eq:A22app} to get a convergent series\footnote{Note that we have relabelled some indices just to make this computation more transparent.}:
\begin{equation}\label{eq:A25app}
	\begin{split}	&\sum\limits_{n=0}^{\infty}\sum\limits_{m=0}^{\infty}\frac{(-1)^{n+m}(\Delta_{lji})_{2m+2n}}{n!\,m!\,\left(\Delta_l+\frac{1}{2}\right)_n\,\left(\Delta_j+\frac{1}{2}\right)_m\,4^{n+m}}\frac{\zeta^{2n}}{\upsilon^{2n+2m}} \\
		&=\sum\limits_{m=0}^{\infty}\frac{(-1)^{m}(\Delta_{lji})_{2m}}{m!\,\left(\Delta_j+\frac{1}{2}\right)_m}\frac{1}{(2\upsilon)^{2m}}\sum\limits_{n=0}^{\infty}\frac{(-1)^n\,(\Delta_{lji}+2m)_{2n}}{n!\,\left(\Delta_l+\frac{1}{2}\right)_n}\left(\frac{\zeta}{2\upsilon}\right)^{2n} \\
		&=\sum\limits_{m=0}^{\infty}\frac{(-1)^{m}(\Delta_{lji})_{2m}}{m!\,\left(\Delta_j+\frac{1}{2}\right)_m}\frac{1}{(2\upsilon)^{2m}}\,\pFq{2}{1}{\frac{\Delta_{lji}}{2}+m,\frac{\Delta_{lji}+1}{2}+m}{\Delta_l+\frac{1}{2}}{-\frac{\zeta^2}{\upsilon^2}} \\
		&=\sum\limits_{m=0}^{\infty}\frac{(-1)^{m}(\Delta_{lji})_{2m}}{m!\,\left(\Delta_j+\frac{1}{2}\right)_m}\frac{1}{(2\upsilon)^{2m}}\left(\frac{\upsilon^2}{\zeta^2+\upsilon^2}\right)^{\frac{\Delta_{lji}}{2}+m}
        \,\pFq{2}{1}{\frac{\Delta_{lji}}{2}+m,\frac{\Delta_{lij}}{2}-m}{\Delta_l+\frac{1}{2}}{\frac{\zeta^2}{\zeta^2+\upsilon^2}}\, , \\
	\end{split}
\end{equation}
where we used \eqref{eq:2F1identity1} to obtain the last line. We see that the convergence domain is extended to $\upsilon>1$ and arbitrary $\zeta>0$. Then we expand the ${}_2F_1$ function appearing in \eqref{eq:A25app} and resum over $m$ to obtain
\begin{equation}
	\begin{split}
&\sum\limits_{n=0}^{\infty}\sum\limits_{m=0}^{\infty}\frac{(-1)^{n+m}(\Delta_{lji})_{2m+2n}}{n!\,m!\,\left(\Delta_l+\frac{1}{2}\right)_n\,\left(\Delta_j+\frac{1}{2}\right)_m\,4^{n+m}}\frac{\zeta^{2n}}{\upsilon^{2n+2m}}= \\
 & \left(\frac{\upsilon^2}{\zeta^2+\upsilon^2}\right)^{\frac{\Delta_{lji}}{2}}\sum_{n=0}^{\infty}\frac{\left(\frac{\Delta_{lji}}{2}\right)_n\left(\frac{\Delta_{lij}}{2}\right)_n}{n!\,\left(\Delta_l+\frac{1}{2}\right)_n}\left(\frac{\zeta^2}{\zeta^2+\upsilon^2}\right)^{n}
 \pFq{3}{2}{\frac{\Delta_{lji}+1}{2},\frac{\Delta_{lji}}{2}+n,1-\frac{\Delta_{lij}}{2}}{\Delta_j+\frac{1}{2},1-\frac{\Delta_{lij}}{2}-n}{-\frac{1}{\zeta^2+\upsilon^2}} \, .\label{eq:DoubleblockBBdexpansion}\\
	\end{split}
\end{equation}
Using \eqref{eq:A22app} together with \eqref{eq:DoubleblockBBdexpansion} and slightly massaging the expression, we finally obtain 
\begin{align}
		R^{\Delta_i}_{\Delta_l\Delta_j}(\upsilon, \zeta)&=\frac{\zeta^{\Delta_l}}{(\zeta^2+\upsilon^2)^{\frac{\Delta_{lji}}{2}}}
		\sum_{n=0}^{\infty}\frac{\left(\frac{\Delta_{lji}}{2}\right)_n\Gamma(1-\frac{\Delta_{ilj}}{2})\Gamma(\Delta_j+\frac{1}{2})}{n!\,\left(\Delta_l+\frac{1}{2}\right)_n}\left(\frac{-\zeta^2}{\zeta^2+\upsilon^2}\right)^{n}\label{eq:A30ap}\\
		&\qquad\quad \times
		\regpFq{3}{2}{\frac{\Delta_{lji}+1}{2},\frac{\Delta_{lji}}{2}+n,1-\frac{\Delta_{lij}}{2}}{\Delta_j+\frac{1}{2},1-\frac{\Delta_{lij}}{2}-n}{-\frac{1}{\zeta^2+\upsilon^2}}.\nonumber
\end{align}
which is convergent for $\upsilon^2,\zeta^2\in(0,\infty)$, covering the full AdS$_2$.

 \section{Integral representations of BOE and OPE}

\label{app:OPE and BOE blocks}

In this appendix we review the basics of the OPE blocks introduced in \cite{Czech:2016xec} (see also \cite{deBoer:2016pqk}). We also discuss the generalization to the BOE and introduce the BOE blocks. 

The idea is to repackage the sum over descendants that appear both in the OPE and the BOE expansion into an integral representation. In particular, the generic form of the OPE of two operators in CFT is 
\begin{equation}
    \OO_i(x)\OO_j(y) = \sum_kC_{ijk}\frac{\OO_k(y)}{(x-y)^{\D_i+\D_j-\D_k}}+ \text{ conformal descendants}\, .
\end{equation}
In principle, the sum is over all primaries $\OO_k$. There is also a second (infinite) sum over all conformal descendants, and this is the one that can be repackaged in terms of an integral of $\OO_k$ against a suitable kernel, which is the essence of an OPE block. In detail, it amounts to finding the suitable kernel  $F_{ijk}$ and integration domain $\mathcal{R}$ such that we can write 
\begin{equation}
    \OO_i(x)\OO_j(y) = \sum_kC_{ijk}\int _{\mathcal{R}}d^dzF_{ijk}(x,y,z)\OO_k(z)\, .
\end{equation}
The kernel $F_{ijk}(x,y,z)$ is completely fixed by conformal invariance. 
\subsection{OPE blocks}
Examining how the kernel $F_{ijk}$ must transform under global conformal transformation, we reach the conclusion that the OPE in CFT has the following integral representation \cite[sec.2, App.B]{Czech:2016xec}
\begin{align}
\mathcal{O}_i(x) \mathcal{O}_j(y)= 
& \sum_k C_{i j k} n_{ijk}\int_\RR d^d z 
\<\!\<\OO_i(x) \OO_j(y) \tilde\OO_k(z)\>\!\>\mathcal{O}_k(z)\,,
\label{eq:OPE int rep general d}
\end{align}
where the bracket $\<\!\<\ldots\>\!\>$ denotes non-physical correlators that have no OPE coefficient. Moreover, the operators with tilde are shadow transforms of local operators in the CFT. In particular, the operators $\tilde{\mathcal{O}}$ have ``shadow" dimension $\Delta_{\tilde{\mathcal{O}}}=\tilde{\D}_\OO=1-\Delta_{\mathcal{O}}$ in one dimension. The object\footnote{Our convention differs from \cite{Czech:2016xec} by not including the factor $|x-y|^{\D_i+\D_j}$ on the RHS.}
\begin{equation}
\BB_k^{ij}(x,y;d) \equiv 
n_{ijk}\int_\RR d^d z 
\<\!\<\OO_i(x) \OO_j(y) \tilde\OO_k(z)\>\!\>\mathcal{O}_k(z)
\end{equation}
is called the OPE block. Its form is determined by conformal invariance up to an overall coefficient $n_{ijk}$, which can be fixed by taking the OPE limit. If  $x$ and $y$ are timelike separated,  the integration region $\RR$ is the Lorentzian causal diamond with vertices at $x$ and $y$.
If $x$ and $y$ are spacelike separated, it is not known how to define the region $\RR$ in general dimension,
as discussed in \cite[sec.6]{Czech:2016xec}.

For our setup, we need to construct the one-dimensional analogue of \eqref{eq:OPE int rep general d}, which reads
\begin{equation}
\mathcal{O}_i(\tau_1)\mathcal{O}_j(\tau_2)= 
 \sum_k C_{i j k}\BB_k^{ij}(\t_1,\t_2)  \,,
\label{eq: OPE int rep}
\end{equation}
with the OPE block defined as 
\begin{equation}
\BB_k^{ij}(\t_1,\t_2) \equiv 
n_{ijk}\int_{\tau_1}^{\tau_2} d \tau 
\<\!\<\OO_i(\t_1) \OO_j(\t_2) \tilde\OO_k(\t)\>\!\>\mathcal{O}_k(\tau)\,.
\label{eq:OPE block B def}
\end{equation}
The normalization factor, with the given integration range, can be fixed by imposing
\begin{equation}
C_{ijk}=\<\OO_i(0)\OO_j(1)\OO_k(\infty)\>\overset{!}{=}
C_{ijk}n_{ijk} \int_{0}^{1} d \tau 
\<\!\<\OO_i(0) \OO_j(1) \tilde\OO_k(\t)\>\!\> \<\mathcal{O}_k(\tau) \OO_k(\infty)\>\,,
\end{equation}
which gives
\begin{equation}
n_{ijk}=\frac{\Gamma\left(2 \Delta_k\right)}{\Gamma\left(\Delta_{kij}\right) \Gamma\left(\Delta_{kji}\right)} \,.
\label{eq:OPE block normalisation}
\end{equation}
The only non-trivial ingredient in \eqref{eq:OPE block B def} is the integration range, which in 1D is the same for both Lorentzian and Euclidean signature. This choice guarantees that when inserting \eqref{eq: OPE int rep} in correlators we only have blocks and no shadow blocks. To see this, let us compute the ``two-point function'' of the OPE block \eqref{eq:OPE block B def} with the four points placed at $0,z,1,\infty$ respectively ($0<z<1$):
\al{\spl{
\<\BB_s^{ij}(0,z)\BB_s^{kl}(1,\infty)\>
&=
n_{ijs}n_{kls}
\int d\t  d\t'
\<\!\<\OO_i(0) \OO_j(z) \tilde\OO_s(\t) \>\!\>
\<\!\<\OO_k(1) \OO_l(\infty) \tilde\OO_s(\t') \>\!\>
\<\OO_s(\t) \OO_s(\t')\>
\\
&=n_{ijs}n_{kls}
\int_0^z d\t \int_1^\infty d\t'
\frac{1}{z^{\D_{ij\tilde{s}}}\t^{\D_{\tilde{s}ij}}(z-\t)^{\D_{\tilde{s}ji}}}
\frac{1}{(\t'-1)^{\D_{\tilde{s}kl}}}
\frac{1}{(\t'-\t)^{2\D_s}}\,.
}
\label{eq:OPE block 2-pt fcn}}
A priori this can be a linear combination of the one-dimensional four-point conformal block and its shadow. However, in the OPE limit $z\to 0$, $\t$ also goes to zero because of the chosen integration range, and we have
\begin{equation}
\<\BB_s^{ij}(0,z)\BB_s^{kl}(1,\infty)\> \overset{z\to0}{\sim} z^{\D_s-\D_i-\D_j}\,,
\end{equation}
which is precisely the same scaling as a one-dimensional conformal block.\footnote{
In our convention the kinematical prefactor $\left|\frac{x_{14}}{x_{24}}\right|^{\Delta_{21}}\left|\frac{x_{14}}{x_{13}}\right|^{\Delta_{34}} \frac{1}{\left|x_{12}\right|^{\Delta_1+\Delta_2}\left|x_{34}\right|^{\Delta_3+\Delta_4}}$ is included, leading to $z^{-\D_i-\D_j}$.
}

In contrast, if the limits of integration in \eqref{eq:OPE block B def} are finite distance away from $[\t_1,\t_2]$, the analog of \eqref{eq:OPE block 2-pt fcn} (using tilde to denote the object defined as $\BB_{s}^{ij}$ but with this different integration range) 
will scale as
\[
\langle \tilde{\BB}_s^{ij}(0,z)\tilde{\BB}_s^{kl}(1,\infty)\rangle 
\overset{z\to0}{\sim} 
\left\{
\begin{array}{rl}
  z^{\Delta_s-\Delta_i-\Delta_j}\, ,       & \quad \text{for }|\tau|= O(z), \\[6pt]
  z^{\tilde\Delta_s-\Delta_i-\Delta_j}\, , & \quad \text{for }|\tau| = O(1).
\end{array}
\right.
\]
which now contains both the block and the shadow block. In particular, when the integration range is $(-\infty,\infty)$, the ``two-point function''  is proportional to the conformal partial wave since, e.g., the $\t'$ integral is just a shadow transform and then the $\t$ integral gives the integral representation of the conformal partial wave.

\subsection{BOE blocks}
Similar to \eqref{eq: OPE int rep}, the BOE has the following integral representation
\begin{equation}
\hat\OO(\t,z) =\sum_k \bb{\hat{\mathcal{O}}}{k} n_k
\int_{\t-iz}^{\t+iz} \frac{d\t'}{i} \<\!\<\hat\OO(\t,z)\tilde\OO_k(\t')\>\!\> \OO_k(\t')
\label{eq:BOE int rep}\,,
\end{equation}
with
\begin{equation}
n_k = \frac{\G(\D_k+\frac12)}{\sqrt{\pi}\G(\D_k)}\,.
\end{equation}
One can verify this agrees with the usual BOE \eqref{eq:BOE} by Taylor expanding $\OO(\t')$ in $\t'-\t\equiv i t z$ near 0 and integrating in $t\in[-1,1]$ term by term.\footnote{The factors of $i$ are important to reproduce $(-1)^n$ in \eqref{eq:BOE}.} The BOE block is then defined as
\begin{equation}
\CC_{k}(\t,z)\equiv n_k
\int_{\t-iz}^{\t+iz} \frac{d\t'}{i} \<\!\<\hat\OO(\t,z)\tilde\OO_k(\t')\>\!\> \OO_k(\t')\,.
\label{eq:BOE block app}
\end{equation}
One can consider the ``three-point function'' $\<\CC_k(0,z)\OO_i(-1)\OO_j(1)\>$, repeat the analysis similar to the OPE block case (send $z\to0$) and conclude that indeed \eqref{eq:BOE int rep} only produces blocks and not shadow blocks inside correlators.

\subsection{Consistency check}
As a consistency check, we compute the $B\del\del$ block using both the OPE block and the BOE block separately.

First, we can perform the OPE of the boundary operators $\OO(\tau_1)\OO(\tau_2)$ using \eqref{eq: OPE int rep} and 
\eqref{eq:OPE block normalisation}. We obtain
\begin{align}
\braket{\hat{\mathcal{O}}(\tau,z)\mathcal{O}_i\left(\tau_1\right) \mathcal{O}_j\left(\tau_2\right)}&= 
 \sum_k  \frac{C_{i j k}\Gamma\left(2 \Delta_k\right)}{\Gamma\left(\Delta_{kij}\right) \Gamma\left(\Delta_{kji}\right)}\int_{\tau_1}^{\tau_2} d \tau' 
\<\!\<\OO_i(\t_1) \OO_j(\t_2) \tilde\OO_k(\t')\>\!\>\braket{\hat{\mathcal{O}}(\tau,z)\mathcal{O}_k(\tau')}\,,\nonumber\\ 
&= \sum_kC_{i j k} \frac{\Gamma\left(2 \Delta_k\right)}{\Gamma\left(\Delta_{kij}\right) \Gamma\left(\Delta_{kji}\right)}\nonumber\\&\quad\times\int_{\tau_1}^{\tau_2}d\tau'\frac{1}{|\tau_{12}|^{\Delta_{ij\tilde{k}}}|\tau'-\tau_1|^{\Delta_{\tilde{k}ij}}|\tau'-\tau_2|^{\Delta_{\tilde{k}ji}}} \bb{\hat\OO}{k} \left(\frac{z}{z^2+(\tau-\tau')^2}\right)^{\Delta_k}\, ,
\end{align}
where we used the following notation for the shadow transform of $\OO_k$: $\Delta_{ij\tilde{k}} = \Delta_i+\Delta_j-\tilde{\Delta}_k =  \Delta_i+\Delta_j-(1-\Delta_k)$. To simplify the computation, and hence the comparison, we choose the following configuration
\be 
(\tau=0\,,  z=\xi)\, ,\qquad \text{ and }\qquad \tau_1= 1\,,\qquad \tau_2=\infty\, ,\label{eq:ConfigA3}
\ee 
such that 
\begin{align}
\braket{\hat{\mathcal{O}}(0,\xi)\mathcal{O}_i\left(1\right) \mathcal{O}_j\left(\infty\right)}
&= \sum_k\bb{\hat{\mathcal{O}}}{k}C_{i j k} \frac{\Gamma\left(2 \Delta_k\right)}{\Gamma\left(\Delta_{kij}\right) \Gamma\left(\Delta_{kji}\right)}\int_{1}^{\infty}d\tau'\frac{1}{|\tau'-1|^{\Delta_{\tilde{k}ij}}} \left(\frac{\xi}{\xi^2+(\tau')^2}\right)^{\Delta_k}\, .
\end{align}
Note that $\hat\OO$ never touches the integration region of $\t'$ for any $\xi>0$, so the OPE block representation is valid in the entire EAdS$_2$. Performing the integral, we obtain
\begin{align}
\braket{\hat{\mathcal{O}}(0,\xi)\mathcal{O}_i\left(1\right) \mathcal{O}_j\left(\infty\right)}= 
& \sum_k \bb{\hat{\mathcal{O}}}{k}C_{i j k}\, \xi^{\Delta_k}{}_2F_1\left(\frac{\Delta_{ikj}}{2},\frac{1+\Delta_{ikj}}{2};\Delta_k+\frac{1}{2};-\xi^2\right)\,,\label{eq:BbbA5answer}
\end{align}
which agrees exactly with \eqref{cb:threeptspecial}.

Equivalently, we can use \eqref{eq:BOE int rep} to write 
\begin{equation}
    \braket{\hat{\mathcal{O}}(\tau,z)\mathcal{O}_i\left(\tau_1\right) \mathcal{O}_j\left(\tau_2\right)} = \sum_k \bb{\hat{\mathcal{O}}}{k}\frac{\G(\D_k+\frac12)}{\sqrt{\pi}\G(\D_k)}
\int_{\t-iz}^{\t+iz} \frac{d\t'}{i} \<\!\<\hat\OO(\t,z)\tilde\OO_k(\t')\>\!\> \braket{\OO_k(\t')\OO_i(\tau_1)\OO_j(\tau_2)}\, .
\end{equation}
In the configuration \eqref{eq:ConfigA3}, this gives 
\begin{equation}
    \braket{\hat{\mathcal{O}}(0,\xi)\mathcal{O}_i\left(1\right) \mathcal{O}_j\left(\infty\right)} = \sum_k \bb{\hat{\mathcal{O}}}{k}\frac{\G(\D_k+\frac12)}{\sqrt{\pi}\G(\D_k)}
\int_{-i\xi}^{+i\xi} \frac{d\t'}{i} \left(\frac{\xi}{\xi^2+(\tau')^2}\right)^{1-\Delta_k}\frac{C_{ijk}}{|1-\tau'|^{\Delta_{ikj}}}\, .
\end{equation}
Performing the same change of variables and integrating terms by term before resumming, we obtain again \eqref{cb:threeptspecial} as needed.

\section{From hard cutoff to analytic continuation\label{app:HardvsAnal}}
We now explain in detail the claims made in sections~\ref{subsec:eq2} and \ref{subsec:eq3}: in eqs.~\eqref{matching:bulkbdry} and \eqref{matching:bdry3pt}, the field-strength renormalization terms, subtraction terms, poles, and anomalous-dimension terms cancel, leaving only the corrections to the OPE/BOE coefficients and the (finite part of the) analytically continued integrals.

\vspace{1em}
\noindent\textbf{Bulk-boundary correlator}
Let us first consider eq.~\eqref{matching:bulkbdry}. The left-hand side can be written more concretely as, using \eqref{eq:Zansatz} and \eqref{eq:bkbd two-pt ren},
\begin{equation}
	\begin{split}
		\text{LHS of } \eqref{matching:bulkbdry}&=Z_{ii;0}\braket{\hat{\Phi}(\tau_1,z_1)\mathcal{O}^{(\text{ren})}_i(\tau_2)}_{\lambda+\delta\lambda}+\sum\limits_{j\neq i}\sum_{p=0}^\infty Z_{ij;p}\partial^p_{\tau_2}\braket{\hat{\Phi}(\tau_1,z_1)\mathcal{O}^{(\text{ren})}_j(\tau_2)}_{\lambda+\delta\lambda} \\
		&=\left(\frac{z_1}{\tau_{12}^2+z_1^2}\right)^{\Delta_i(\lambda)}\left[b^{\hat{\Phi}}_i+b^{\hat{\Phi}}_i \delta Z_{ii;0}+b^{\hat{\Phi}}_i\delta\Delta_i \log\left(\frac{z_1}{\tau_{12}^2+z_1^2}\right)+\delta b^{\hat{\Phi}}_i\right] \\
		&\quad+\sum\limits_{j\neq i} \sum_{p=0}^\infty b^{\hat{\Phi}}_i \delta Z_{ij;p}\partial^p_{\tau_2}\left(\frac{z_1}{\tau_{12}^2+z_1^2}\right)^{\Delta_j}\,. \\
	\end{split}
\end{equation}

We now match these terms with the right-hand side of \eqref{matching:bulkbdry}: 
\begin{align}
		\text{RHS of } \eqref{matching:bulkbdry}&=b^{\hat{\Phi}}_i\left( \frac{z_1}{(\tau_1 - \tau_2)^2 + z_1^2} \right)^{\Delta_i(\lambda)} +\delta\lambda\sum_{l,j}b^{\hat{\Phi}}_l\, b^{\hat{\Phi}}_j \left( \frac{z_1}{(\tau_1 - \tau_2)^2 + z_1^2} \right)^{\Delta_i(\lambda)}\nonumber \\
		&\qquad\qquad\times\Bigg{[} C_{ilj} \left(\frac{1}{2}\JJ_{\Delta_i}(\Delta_l, \Delta_j)+\JJ^{+,\text{subtr}}_{\Delta_i}(\Delta_l, \Delta_j;\tau_1,z_1;\tau_2;\epsilon)\right)\nonumber \\
		&\qquad\qquad+C_{ijl} \left(\frac{1}{2}\JJ_{\Delta_i}(\Delta_l, \Delta_j)+\JJ^{-,\text{subtr}}_{\Delta_i}(\Delta_l, \Delta_j;\tau_1,z_1;\tau_2;\epsilon)\right)\Bigg{]}\,. 
\end{align}
The terms in the right-hand side of the equation should be understood as the analytically continued values:
\begin{equation}
	\begin{split}
		&\frac{1}{2}\JJ_{\Delta_i}(\Delta_l, \Delta_j)+\JJ^{\pm,\text{subtr}}_{\Delta_i}(\Delta_l, \Delta_j;\tau_1,z_1;\tau_2;\epsilon) \\
		&:=\frac{1}{2}\left[\JJ_{\Delta_i}(\Delta_l, \Delta_j)\right]_\text{reg}+\left[\JJ^{\pm,\text{subtr}}_{\Delta_i}(\Delta_l, \Delta_j;\tau_1,z_1;\tau_2;\epsilon)\right]_\text{reg}\,,
	\end{split}
\end{equation}
where “reg” denotes the analytically continued value and, when $\Delta_j=\Delta_i$, the finite part. As noted in section~\ref{subsec:eq2}, the integrated block with cutoff $\epsilon$ is finite and analytic in $\Delta_j$, so the poles cancel out at $\Delta_j=\Delta_i$.

For $j\neq i$, we can formally rewrite the contribution from the subtraction terms as 
\begin{equation}\label{Jsubtr:rewrite1}
	\begin{split}
		&\delta\lambda\sum_{l}b^{\hat{\Phi}}_l\, b^{\hat{\Phi}}_j \left( \frac{z_1}{(\tau_1 - \tau_2)^2 + z_1^2} \right)^{\Delta_i} \\
		&\qquad\quad\times\Bigg{[} C_{ilj}\JJ^{+,\text{subtr}}_{\Delta_i}(\Delta_l, \Delta_j;\tau_1,z_1;\tau_2;\epsilon)
		+C_{ijl} \JJ^{-,\text{subtr}}_{\Delta_i}(\Delta_l, \Delta_j;\tau_1,z_1;\tau_2;\epsilon)\Bigg{]} \\
		&``="\delta\lambda\, b^{\hat{\Phi}}_j \int_{(\tau-\tau_2)^2+z^2\leqslant\epsilon^2} \frac{d\tau\, dz}{z^2}\mathcal{D}\left(\Delta_j;\tau_1,z_1,\tau_1'\right)\braket{\hat{\Phi}(\tau,z)\mathcal{O}_j(\tau_1')\mathcal{O}_i(\tau_2)} \\
		&``="\delta\lambda\, b^{\hat{\Phi}}_j \mathcal{D}\left(\Delta_j;\tau_1,z_1,\tau_1'\right)\int_{(\tau-\tau_2)^2+z^2\leqslant\epsilon^2} \frac{d\tau\, dz}{z^2}\braket{\hat{\Phi}(\tau,z)\mathcal{O}_j(\tau_1')\mathcal{O}_i(\tau_2)} \\
		&=b^{\hat{\Phi}}_j \mathcal{D}\left(\Delta_j;\tau_1,z_1,\tau_1'\right)\sum\limits_{p=0}^{\infty}Z_{ij};p\braket{\mathcal{O}_j(\tau_1')\partial^p\mathcal{O}_j(\tau_2)} \\
		&=\sum\limits_{p=0}^{\infty}\d Z_{ij};p\braket{\hat{\Phi}(\tau_1,z_1)\partial^p\mathcal{O}_j(\tau_2)}. \\
	\end{split}
\end{equation}
In the first equality we used the definition \eqref{def:Jsubtr} of $\mathcal{J}^\text{substr}$ and the BOE
\begin{equation}\label{BOE:Phi}
	\begin{split}
		\hat{\Phi}(\tau_1,z_1)=\sum_{j}b^{\hat{\Phi}}_j\mathcal{D}\left(\Delta_j;\tau_1,z_1,\tau_1'\right)\mathcal{O}_j(\tau_1')\,.
	\end{split}
\end{equation}
Here $\mathcal{D}$ is a differential operator for the BOE of the bulk operators, which is fixed kinematically by conformal invariance.\footnote{\label{fn:BOE diff op}An explicit form of $\mathcal{D}$ is given by
\begin{equation*}
	\begin{split}
		\mathcal{D}\left(\Delta_j;\tau_1,z_1,\tau_1',\partial_{\tau_1'}\right)=\sum\limits_{n=0}^{\infty}\sum_{m=0}^{\infty}\frac{(-1)^n}{4^n\,n! (\Delta_j+\tfrac{1}{2})_{n}}z_1^{\Delta_j+2n}\frac{(\tau_1-\tau_1')^m}{m!}\partial_{\tau_1'}^{2n+m}\,.
	\end{split}
\end{equation*}
} 
The right-hand side is to be interpreted as an analytic-continuation, block-by-block identity (hence the quotation marks “$=$”). In the second equality we use that $\mathcal{D}$ depends only on $\tau_1$, $z_1$, and $\tau_1'$, so it commutes with the $(\tau,z)$-integral. In the third equality we use the renormalization of the boundary two-point function $\braket{\mathcal{O}_j\mathcal{O}_i}$, eq.~\eqref{matching:bdrybdry} as well as \eqref{Y:prop}. In the last line we use the BOE of $\Phi$ inside $\braket{\Phi(\tau_1,z_1)\mathcal{O}_j(\tau_2)}$.

We see that, after the above manipulations, the subtraction terms with $j \neq i$ in the right-hand side of \eqref{matching:bulkbdry} match exactly the field-strength renormalization terms in the left-hand side, so they cancel. Moreover, from the derivation it is clear that this cancellation occurs block by block.

For the case $j = i$, since $\mathcal{J}^\text{subtr}$ is a purely kinematical function, we may evaluate it via analytic continuation. Starting with $\Delta_j$ slightly larger than $\Delta_i$, we obtain, up to an $O\!\left(\epsilon^{1-\Delta_{ij}}\right)$ error,
\begin{equation}\label{Jsubtr:rewrite2}
	\begin{split}
		&\left( \frac{z_1}{(\tau_1 - \tau_2)^2 + z_1^2} \right)^{\Delta_i}\JJ^{+,\text{subtr}}_{\Delta_i}(\Delta_l, \Delta_j;\tau_1,z_1;\tau_2;\epsilon) \\
		&=\left( \frac{z_1}{(\tau_1 - \tau_2)^2 + z_1^2} \right)^{\Delta_i}\JJ^{-,\text{subtr}}_{\Delta_i}(\Delta_l, \Delta_j;\tau_1,z_1;\tau_2;\epsilon) \\
		&=\frac{1}{2} \mathcal{D}\!\left(\Delta_j;\tau_1,z_1,\tau_1'\right)\int_{(\tau-\tau_2)^2+z^2\leqslant\epsilon^2} \frac{d\tau\, dz}{z^2}\Bigg{[}\frac{1}{\abs{\tau_1'-\tau_2}^{\Delta_j+\Delta_i}} \\
		&\qquad\qquad\qquad\qquad\qquad\qquad\qquad\times\left(\frac{(\tau-\tau_1')^2+z^2}{(\tau-\tau_2)^2+z^2}\right)^{\tfrac{\Delta_i-\Delta_j}{2}} G_{\Delta_l}^{\Delta_j,\Delta_i}(\chi(\tau,z,\tau_1',\tau_2))\Bigg{]} \\
		&=-\mathcal{D}\!\left(\Delta_j;\tau_1,z_1,\tau_1'\right)\frac{1}{\abs{\tau_1'-\tau_2}^{\Delta_j+\Delta_i}}\frac{a_0(\Delta_{ij},\Delta_l)}{\Delta_{ij}}\abs{\tfrac{\tau_1'-\tau_2}{\epsilon}}^{\Delta_{ij}} \\
		&=-\left(\frac{z_1}{(\tau_{1}-\tau_2)^2+z_1^2}\right)^{\Delta_j}\frac{a_0(\Delta_{ij},\Delta_l)}{\Delta_{ij}}\epsilon^{-\Delta_{ij}} \\
		&=\left(\frac{z_1}{(\tau_{1}-\tau_2)^2+z_1^2}\right)^{\Delta_i}a_0(0,\Delta_l)\left[\frac{1}{\Delta_{ji}}+\log\epsilon+\log\!\left(\frac{z_1}{(\tau_{1}-\tau_2)^2+z_1^2}\right)\right]+O(\Delta_{ij})\,. \\
	\end{split}
\end{equation}
Here, the differential operator $\mathcal{D}$ is the same as in \eqref{Jsubtr:rewrite1}, and $G^{\Delta_l}_{\Delta_j,\Delta_i}$ denotes the $B\partial\partial$ conformal block defined in \eqref{def:bkbdbd block}. In the first equality, we use the fact that the leading divergence ($\epsilon^{-\Delta_{ij}}$) is identical for both $\mathcal{J}^\text{+,subtr}$ and $\mathcal{J}^\text{-,subtr}$. In the second equality, we rewrite the $BB\partial$ block as the action of $\mathcal{D}$ on the $B\partial\partial$ block. In the third equality, we keep only the leading divergent term. In the fourth step, we use the identity
\begin{equation}
	\begin{split}
		\mathcal{D}\!\left(\Delta_j;\tau_1,z_1,\tau_1'\right)\frac{1}{\abs{\tau_1'-\tau_2}^{2\Delta_j}}
		=\left(\frac{z_1}{(\tau_{1}-\tau_2)^2+z_1^2}\right)^{\Delta_j}\,.
	\end{split}
\end{equation}
Finally, the last line of \eqref{Jsubtr:rewrite2} follows from the Laurent expansion in $\Delta_{ij}$.

Recalling the renormalization constant $\delta Z_{ii;0}$ and the anomalous dimension $\delta\Delta_i$ from \eqref{zii:result}, we can now match the terms appearing in the last line of \eqref{Jsubtr:rewrite2}. The first term, corresponding to the pole, is canceled by the same pole from $\mathcal{J}_{\Delta_i}$ (the un-subtracted integral). The second term matches the $\delta Z_{ii;0}$ term, while the third term matches the $\delta\Delta_i$ term. 

Therefore, after all cancellations, the only surviving contributions in \eqref{matching:bulkbdry} are the ones in \eqref{matching:bkbdfinal}.

\vspace{1em}
\noindent\textbf{Boundary three-point function}
Let us now consider eq.~\eqref{matching:bdry3pt}. The left-hand side can be expressed as
\begin{equation}
	\begin{split}
		\text{LHS of } \eqref{matching:bdry3pt}
		&=\frac{1}{\abs{\tau_{12}}^{\Delta_{ijk}(\lambda)} \abs{\tau_{23}}^{\Delta_{jki}(\lambda)} \abs{\tau_{13}}^{\Delta_{ikj}(\lambda)}}\Bigg{[}C_{ijk}(\lambda)+\delta C_{ijk}\\
		&\quad+\delta\Delta_i\log\abs{\frac{\tau_{23}}{\tau_{12}\tau_{13}}}+\delta\Delta_j\log\abs{\frac{\tau_{13}}{\tau_{12}\tau_{23}}}+\delta\Delta_k\log\abs{\frac{\tau_{12}}{\tau_{13}\tau_{23}}} \Bigg{]} \\
		&\quad+\sum_{m}\sum_{p=0}^{\infty}\delta Z_{im;p}\partial^p_{\tau_1}\frac{1}{\abs{\tau_{12}}^{\Delta_{mjk}} \abs{\tau_{23}}^{\Delta_{jkm}} \abs{\tau_{13}}^{\Delta_{mkj}}} \\
		&\quad+\sum_{m}\sum_{p=0}^{\infty}\delta Z_{jm;p}\partial^p_{\tau_2}\frac{1}{\abs{\tau_{12}}^{\Delta_{imk}} \abs{\tau_{23}}^{\Delta_{mki}} \abs{\tau_{13}}^{\Delta_{ikm}}} \\
		&\quad+\sum_{m}\sum_{p=0}^{\infty}\delta Z_{km;p}\partial^p_{\tau_3}\frac{1}{\abs{\tau_{12}}^{\Delta_{ijm}} \abs{\tau_{23}}^{\Delta_{jmi}} \abs{\tau_{13}}^{\Delta_{imj}}}\,. \\
	\end{split}
\end{equation}

The matching with the right-hand side of \eqref{matching:bdry3pt} proceeds analogous to the bulk–boundary case:
\begin{align}
		\text{RHS of } \eqref{matching:bdry3pt}
		&=\frac{C_{ijk}(\lambda)}{\abs{\tau_{12}}^{\Delta_{ijk}(\lambda)} \abs{\tau_{23}}^{\Delta_{jki}(\lambda)} \abs{\tau_{13}}^{\Delta_{ikj}(\lambda)}} \\
		&- \delta\lambda \sum_{l,m}b^{\hat{\Phi}}_{l} C_{jkm} \left[C_{ilm} \mathcal{Q}^{\Delta_l,\Delta_m;+}_{\Delta_i,\Delta_j,\Delta_k}(\tau_1, \tau_2, \tau_3; \epsilon) + C_{iml} \mathcal{Q}^{\Delta_l,\Delta_m;-}_{\Delta_i,\Delta_j,\Delta_k}(\tau_1, \tau_2, \tau_3; \epsilon)\right] \nonumber\\
		&- \delta\lambda \sum_{l,m}b^{\hat{\Phi}}_{l} C_{kim} \left[C_{jlm} \mathcal{Q}^{\Delta_l,\Delta_m;+}_{\Delta_j,\Delta_k,\Delta_i}(\tau_2, \tau_3, \tau_1; \epsilon)+C_{jml} \mathcal{Q}^{\Delta_l,\Delta_m;-}_{\Delta_j,\Delta_k,\Delta_i}(\tau_2, \tau_3, \tau_1; \epsilon)\right] \nonumber \\
		&- \delta\lambda \sum_{l,m}b^{\hat{\Phi}}_{l} C_{ijm} \left[C_{klm} \mathcal{Q}^{\Delta_l,\Delta_m;+}_{\Delta_k,\Delta_i,\Delta_j}(\tau_3, \tau_1, \tau_2; \epsilon)+C_{kml} \mathcal{Q}^{\Delta_l,\Delta_m;-}_{\Delta_k,\Delta_i,\Delta_j}(\tau_3, \tau_1, \tau_2; \epsilon)\right]\,. \nonumber
\end{align}
The subtracted integrals $\mathcal{Q}^{\pm}$ are treated in the same way as the $\mathcal{J}$-integrals. We rewrite them as the difference between the analytically continued un-subtracted integral and the analytically continued subtraction term:
\begin{equation}\label{prop:QsubtrReg}
	\begin{split}
		\mathcal{Q}^{\Delta_l,\Delta_m;\pm}_{\Delta_i,\Delta_j,\Delta_k}(\tau_1, \tau_2, \tau_3; \epsilon)
		&= \left[\mathcal{Q}^{\Delta_l,\Delta_m;\pm}_{\Delta_i,\Delta_j,\Delta_k}(\tau_1, \tau_2, \tau_3; 0)\right]_\text{reg} \\
		&\quad -\left[\mathcal{Q}^{\Delta_l,\Delta_m;\pm,\text{subtr}}_{\Delta_i,\Delta_j,\Delta_k}(\tau_1, \tau_2, \tau_3; \epsilon)\right]_\text{reg}.
	\end{split}
\end{equation}
Here, “reg” again denotes the analytic continuation in $\Delta_m$, and its finite part at the pole $\Delta_m = \Delta_i$. As in the analysis of the $\mathcal{J}$-integrals, the poles cancel between the two contributions.

By \eqref{def:K}, the first term in the right-hand side of \eqref{prop:QsubtrReg} is
\begin{equation}
	\begin{split}
		\left[\mathcal{Q}^{\Delta_l,\Delta_m;+}_{\Delta_i,\Delta_j,\Delta_k}(\tau_1, \tau_2, \tau_3; 0)\right]_\text{reg}
		&= \frac{[\mathcal{K}_{\Delta_i,\Delta_j,\Delta_k}(\Delta_l, \Delta_m)]_\text{reg}}{\abs{\tau_{12}}^{\Delta_{ijk}} \abs{\tau_{13}}^{\Delta_{ikj}} \abs{\tau_{23}}^{\Delta_{jki}}}\,, \\
		\left[\mathcal{Q}^{\Delta_l,\Delta_m;-}_{\Delta_i,\Delta_j,\Delta_k}(\tau_1, \tau_2, \tau_3; 0)\right]_\text{reg}
		&= \frac{[\mathcal{K}_{\Delta_i,\Delta_k,\Delta_j}(\Delta_l, \Delta_m)]_\text{reg}}{\abs{\tau_{12}}^{\Delta_{ijk}} \abs{\tau_{13}}^{\Delta_{ikj}} \abs{\tau_{23}}^{\Delta_{jki}}}\,. \\
	\end{split}
\end{equation}
As we will see later, terms of this type survive after all cancellations. Let us now turn to the subtraction terms $\mathcal{Q}^\text{subtr}$.

For $m \neq i$, the subtraction terms near $(\tau_1,0)$ (i.e.\ in the $s$-channel) can be rewritten as
\begin{equation}
	\begin{split}
		&\delta\lambda\sum_{l} b^{\hat\F}_l C_{jkm}\Big[C_{ilm} \mathcal{Q}^{\Delta_l,\Delta_m;+,\text{subtr}}_{\Delta_i,\Delta_j,\Delta_k}(\tau_1, \tau_2, \tau_3; \epsilon) + C_{iml} \mathcal{Q}^{\Delta_l,\Delta_m;-,\text{subtr}}_{\Delta_i,\Delta_j,\Delta_k}(\tau_1, \tau_2, \tau_3; \epsilon)\Big] \\
		&``="\delta\lambda\, C_{jkm} \int_{\tau_2}^{\tau_3} d\tau_4\,\mathcal{B}_{\Delta_j\Delta_k\Delta_m}(\tau_2,\tau_3,\tau_4)\int_{(\tau-\tau_1)^2+z^2\leqslant\epsilon^2}\frac{d\tau\,dz}{z^2}\braket{\hat{\Phi}(\tau,z)\mathcal{O}_i(\tau_1)\mathcal{O}_m(\tau_4)} \\
		&= C_{jkm}\int_{\tau_2}^{\tau_3} d\tau_4\,\mathcal{B}_{\Delta_j\Delta_k\Delta_m}(\tau_2,\tau_3,\tau_4)\sum_{p=0}^{\infty}Z_{im;p}\,\partial^p_{\tau_1}\braket{\mathcal{O}_m(\tau_1)\mathcal{O}_m(\tau_4)} \\
		&=\sum_{p=0}^{\infty}\d Z_{im;p}\,\partial^p_{\tau_1}\braket{\mathcal{O}_m(\tau_1)\mathcal{O}_j(\tau_2)\mathcal{O}_k(\tau_3)}\,. \\
	\end{split}
\end{equation}
Here the integration kernel $\mathcal{B}_{\Delta_j\Delta_k\Delta_m}$ is defined in \eqref{eq:Bkernel}. The first equality should again be understood as a block-by-block analytic continuation. In the second equality we used the renormalization of the boundary two-point function $\braket{\mathcal{O}_i\mathcal{O}_m}$ from \eqref{matching:bdrybdry}. In the last step we applied the OPE of $\mathcal{O}_j\mathcal{O}_k$ inside the correlator $\braket{\mathcal{O}_m\mathcal{O}_j\mathcal{O}_k}$. This shows that  in the matching condition \eqref{matching:bdry3pt}, the $s$-channel subtraction terms with $m\neq i$ cancel precisely against the field-strength renormalization terms $\delta Z_{im;p}$ (which are equal to $ Z_{im;p}$ for $m\neq i$).

For $m=i$, we proceed as before by starting from $\Delta_m$ slightly larger than $\Delta_i$. This yields the $s$-channel subtraction term, accurate up to an $O(\epsilon^{1-\Delta_{im}})$ error:
\begin{equation}\label{Qsubtr:rewrite}
	\begin{split}
		&\mathcal{Q}^{\Delta_l,\Delta_m;+,\text{subtr}}_{\Delta_i,\Delta_j,\Delta_k}(\tau_1, \tau_2, \tau_3; \epsilon) =\mathcal{Q}^{\Delta_l,\Delta_m;-,\text{subtr}}_{\Delta_i,\Delta_j,\Delta_k}(\tau_1, \tau_2, \tau_3; \epsilon) \\
		&=\frac{1}{2}\int_{\tau_2}^{\tau_3}d\tau_4 \Bigg{[} \mathcal{B}_{\Delta_j\Delta_k\Delta_m}(\tau_2,\tau_3,\tau_4) \\
		&\qquad \times\int_{(\tau-\tau_1)^2+z^2\leqslant\epsilon^2}\frac{d\tau\, dz}{z^2}\frac{1}{\abs{\tau_{14}}^{\Delta_i+\Delta_m}}\left(\frac{(\tau-\tau_1)^2+z^2}{(\tau-\tau_4)^2+z^2}\right)^{\tfrac{\Delta_m-\Delta_i}{2}} G_{\Delta_l}^{\Delta_i,\Delta_m}(\chi_{14}) \Bigg{]} \\
		&=-\int_{\tau_2}^{\tau_3}d\tau_4 \,\mathcal{B}_{\Delta_j\Delta_k\Delta_m}(\tau_2,\tau_3,\tau_4)\,\frac{1}{\abs{\tau_{14}}^{\Delta_i+\Delta_m}}\frac{a_0(\Delta_{im},\Delta_l)}{\Delta_{im}}\abs{\tfrac{\tau_{14}}{\epsilon}}^{\Delta_{im}} \\
		&=-\frac{a_0(\Delta_{im},\Delta_l)}{\Delta_{im}}\frac{\epsilon^{\Delta_{mi}}}{\abs{\tau_{12}}^{\Delta_{mjk}}\abs{\tau_{13}}^{\Delta_{mkj}}\abs{\tau_{23}}^{\Delta_{jkm}}} \\
		&=\frac{a_0(0,\Delta_l)}{\abs{\tau_{12}}^{\Delta_{ijk}}\abs{\tau_{13}}^{\Delta_{ikj}}\abs{\tau_{23}}^{\Delta_{jki}}}\left[\frac{1}{\Delta_{mi}}+\log\epsilon+\log\abs{\frac{\tau_{23}}{\tau_{12}\tau_{13}}}\right]+O(\Delta_{im})\,.
	\end{split}
\end{equation}
The logic is essentially the same as in the bulk–boundary case; the only difference is that here we use the OPE of boundary operators instead of the BOE of a bulk operator. This corresponds to a different identity:
\begin{equation}
	\begin{split}
		\int_{\tau_2}^{\tau_3}d\tau_4 \,\mathcal{B}_{\Delta_j\Delta_k\Delta_m}(\tau_2,\tau_3,\tau_4) \frac{1}{\abs{\tau_{14}}^{2\Delta_m}}
		=\frac{1}{\abs{\tau_{12}}^{\Delta_{mjk}}\abs{\tau_{13}}^{\Delta_{mkj}}\abs{\tau_{23}}^{\Delta_{jkm}}}\,.
	\end{split}
\end{equation}
The matching proceeds as follows: in the last line of \eqref{Qsubtr:rewrite}, the first term cancels against the pole from the un-subtracted integral $\mathcal{Q}^{\Delta_l,\Delta_m;\pm}_{\Delta_i,\Delta_j,\Delta_k}(\tau_1,\tau_2,\tau_3;0)$. The second term matches the $\delta Z_{ii;0}$ term, while the third term matches the $\delta\Delta_i$ term. 

The same reasoning applies to the $t$- and $u$-channel integrated blocks. After all cancellations, the only surviving contributions in \eqref{matching:bdry3pt} are precisely those shown in \eqref{matching:bdry3ptfinal}.

\section{Integrated conformal blocks}
\label{app:integrated blocks}
The main technical ingredients needed to derive the flow equations (\ref{eq:dimension flow}-\ref{eq:OPE flow}) is the computation of integrals of conformal blocks over the full volume of EAdS$_2$. In this appendix we report the steps we took to compute said integrals, specifically for the $B\partial\partial$, $BB\partial$ and $B\partial\partial\partial$ blocks. 

\subsection{\texorpdfstring{Bulk-boundary-boundary ($B\partial\partial$)}{Bulk-boundary-boundary (Bbb)}}

\label{app:subsec_IntegralY}


First, we would like to compute the truncated integral $\mathcal{Y}^{\Delta_l;\pm}_{\Delta_i,\Delta_j}(\tau_1,\tau_2,\epsilon)$ defined in \eqref{def:Y} and reminded here for convenience
\begin{equation}\label{def:Yapp}
	\begin{split}
		\mathcal{Y}^{\Delta_l;\pm}_{\Delta_i,\Delta_j}(\tau_1,\tau_2,\epsilon):=\int_{\substack{\text{sgn}(\rho)=\pm \\ (\tau-\tau_i)^2+z^2\geqslant\epsilon^2}}\frac{d\tau dz}{z^2} \frac{1}{\abs{\tau_{12}}^{\Delta_i+\Delta_j}}\left(\frac{(\tau-\tau_1)^2+z^2}{(\tau-\tau_2)^2+z^2}\right)^{\frac{\Delta_j-\Delta_i}{2}} G_{\Delta_l}^{\Delta_i,\Delta_j}(\chi)\,.
	\end{split}
\end{equation}
Formally we can  separate it as
$$
\mathcal{Y}^{\Delta_l;\pm}_{\Delta_i,\Delta_j}(\tau_1,\tau_2,\epsilon)=\mathcal{Y}^{\Delta_l;\pm}_{\Delta_i,\Delta_j}(\tau_1,\tau_2,0)-\mathcal{Y}^{\Delta_l;\pm,\text{subtr}}_{\Delta_i,\Delta_j}(\tau_1,\tau_2,\epsilon)\,,
$$
where subtraction terms are given by
\begin{equation}\label{def:Ysubtr}
	\begin{split}
		&\mathcal{Y}^{\Delta_l;\pm,\text{subtr}}_{\Delta_i,\Delta_j}(\tau_1,\tau_2,\epsilon) :=\int_{\substack{\text{sgn}(\rho)=\pm \\ (\tau-\tau_i)^2+z^2\leqslant\epsilon^2}}\frac{d\tau dz}{z^2} \frac{1}{\abs{\tau_1-\tau_2}^{\Delta_i+\Delta_j}}\left(\frac{(\tau-\tau_1)^2+z^2}{(\tau-\tau_2)^2+z^2}\right)^{\frac{\Delta_j-\Delta_i}{2}} G_{\Delta_l}^{\Delta_i,\Delta_j}(\chi)\,.
	\end{split}
\end{equation}
The integration domain of the subtraction term contains two disjoint regimes: one near $(\tau_1,0)$ and the other near $(\tau_2,0)$. At first glance, the integral must diverge in at least one of them. A careful definition of the subtraction term is as follows. The integral over the regime near $(\tau_1,0)$ is defined by a genuine integral for $\Delta_j>\Delta_i$, and by analytic continuation in $\Delta_j$ for $\Delta_j<\Delta_i$; while the integral over the regime near $(\tau_2,0)$ is defined by a genuine integral for $\Delta_j<\Delta_i$, and by analytic continuation in $\Delta_j$ for $\Delta_j>\Delta_i$.

Having defined $\mathcal{Y}^{\Delta_l;\pm}_{\Delta_i,\Delta_j}(\tau_1,\tau_2,\epsilon)$ and $\mathcal{Y}^{\Delta_l;\pm,\text{subtr}}_{\Delta_i,\Delta_j}(\tau_1,\tau_2,\epsilon)$, we now look at the un-truncated integral $\mathcal{Y}^{\Delta_l;\pm}_{\Delta_i,\Delta_j}(\tau_1,\tau_2,0)$. Suppose $\mathcal{Y}^{\Delta_l;\pm}_{\Delta_i,\Delta_j}(\tau_1,\tau_2,0)$ is a convergent integral, then it must be zero by conformal invariance.\footnote{For a convergent integral, the final result is kinematically a CFT two-point function of operators with different scaling dimensions. This must vanish.} Consequently, we will have
\begin{equation}\label{Y:prop}
    \mathcal{Y}^{\Delta_l;\pm}_{\Delta_i,\Delta_j}(\tau_1,\tau_2,\epsilon)=-\mathcal{Y}^{\Delta_l;\pm,\text{subtr}}_{\Delta_i,\Delta_j}(\tau_1,\tau_2,\epsilon)\,.
\end{equation}
However, in our case, the integral is divergent for all values of $\Delta_l$, 
so \eqref{Y:prop} cannot be justified by analytic continuation. Nevertheless, \eqref{Y:prop} can still be justified since both sides are well-defined. The trick is presented as follows.







\begin{figure}
\centering
    \begin{tikzpicture}[scale=3]

\draw (0,0) -- (4,0);

\def\centerx{1.75}
\def\centery{0}
\def\radius{0.75}



\fill (1,0) circle (0.7pt) node[below=4pt] {$\tau_1$};
\fill (2.5,0) circle (0.7pt) node[below=4pt] {$\tau_2$};
\fill (3.5,0) circle (0.7pt) node[below=4pt] {$\tau_3$};


\draw[olive, thick] (0.8,0) arc (180:0:0.2);
\draw[olive, thick] (2.3,0) arc (180:0:0.2);

\def\centerx{1.75}
\def\centery{0}
\def\radius{0.75}

\draw[dotted, thick,red] (\centerx,\centery) ++(180:\radius) arc (180:164.7:\radius);

\draw[solid, thick, red] (\centerx,\centery) ++(164.7:\radius) arc (164.7:15.3:\radius);

\draw[dotted, thick,red] (\centerx,\centery) ++(15.3:\radius) arc (15.3:0:\radius);
\draw[cyan, thick] (1.5,0) arc (180:0:1);

\node[above, cyan] at (2.5,1) {$\chi_{13} = \chi_{23}$};

\node[above left] at (2,1) {$\text{\small R:} -,\text{I}$};
\node[above right] at (2.5,0.45) {$\text{\small R:} -,\text{II}$};
\node[above right] at (1.7,0.25) {$\text{\small R:} +,\text{II}$};
\node[above right] at (1.15,0.25) {$\text{\small R:} +,\text{I}$};

\begin{scope}[xshift=0cm, yshift=0.7cm, scale=0.8]
  \draw[black] (0,0) rectangle (1.2,0.8);
  
  \draw[olive, thick] (0.2,0) arc (180:0:0.4);
  
  \draw[red, dashed, thick] (2.1,0) ++(165:1.5) arc (165:180:1.5);
  
  \draw[red, solid, thick] (2.1,0) ++(165:1.5) arc (165:150:1.5);
  
  \node[align=center] at (0.45,0.15) {\tiny subtr \\[-0.4em]\textrm{\scriptsize{\(-,\text{I}\)}}};
  \node[align=center] at (0.76,0.15) {\tiny subtr \\ [-0.4em]\textrm{\scriptsize{\(+,\text{I}\)}}};

\end{scope}
\draw[-stealth, thick] (0.6,0.6) -- (0.9,0.25);

\end{tikzpicture}
\caption{The four regimes labeled by $(\pm,\text{I/II})$. The red semicircle is the geodesic connecting two boundary points. The green semicircles are the $\epsilon$-cutoffs. The blue semicircle is the line where $\chi_{13}=\chi_{23}$.}
	\label{fig:bkbdbd_four_regime}
\end{figure}
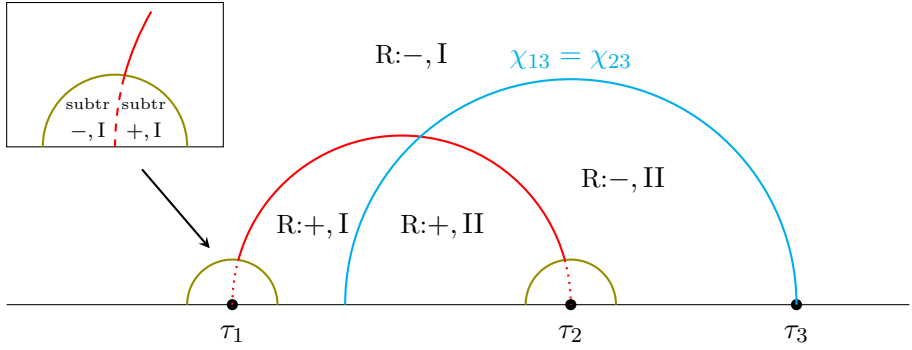

To avoid the difficulty of defining the un-truncated integral, we introduce an auxiliary point $(\tau_3,0)$, which is interpreted as the insertion of the identity operator there. This introduces two extra conformal invariants: $\chi_{13}$ and $\chi_{23}$, defined in the same way as \eqref{def:chi} and the subscripts denote the choice of the boundary points. Then we divide AdS into two parts, separated by the line $\chi_{13}=\chi_{23}$. Also taking into account that AdS is already divided by the geodesic between $(\tau_1,0)$ and $(\tau_2,0)$, the whole AdS is divided into four regimes as shown in figure~\ref{fig:bkbdbd_four_regime}. In each regime, the equation
\begin{equation}
	\begin{split}
		\mathcal{Y}^{\Delta_l;\pm,\text{I/II}}_{\Delta_i,\Delta_j}(\tau_1,\tau_2,\tau_3,\epsilon)=\mathcal{Y}^{\Delta_l;\pm,\text{I/II}}_{\Delta_i,\Delta_j}(\tau_1,\tau_2,\tau_3,0)-\mathcal{Y}^{\Delta_l;\pm,\,\text{I/II subtr}}_{\Delta_i,\Delta_j}(\tau_1,\tau_2,\epsilon)\,
	\end{split}
\end{equation}
makes sense: the left-hand side is always well-defined, and the terms in the right-hand side are understood as analytically continued values in $\Delta_i$ and $\Delta_j$. The original truncated integrals and the subtraction terms are given by
\begin{equation}
	\begin{split}
		\mathcal{Y}^{\Delta_l;\pm}_{\Delta_i,\Delta_j}(\tau_1,\tau_2,\epsilon)&=\mathcal{Y}^{\Delta_l;\pm,\text{I}}_{\Delta_i,\Delta_j}(\tau_1,\tau_2,\tau_3,\epsilon)+\mathcal{Y}^{\Delta_l;\pm,\text{II}}_{\Delta_i,\Delta_j}(\tau_1,\tau_2,\tau_3,\epsilon)\,, \\
		\mathcal{Y}^{\Delta_l;\pm,\text{subtr}}_{\Delta_i,\Delta_j}(\tau_1,\tau_2,\epsilon)&=\mathcal{Y}^{\Delta_l;\pm,\text{I subtr}}_{\Delta_i,\Delta_j}(\tau_1,\tau_2,\epsilon)+\mathcal{Y}^{\Delta_l;\pm,\text{II subtr}}_{\Delta_i,\Delta_j}(\tau_1,\tau_2,\epsilon)\,. \\
	\end{split}
\end{equation}
Now it follows from $SO^+(1,2)$ that the analytically continued, un-truncated integrals in each of the four regimes must have the form
\begin{equation}
	\begin{split}
		\mathcal{Y}^{\Delta_l;\pm,\text{I/II}}_{\Delta_i,\Delta_j}(\tau_1,\tau_2,\tau_3,0)=\frac{A^{\pm,\text{I/II}}(\Delta_i,\Delta_j;\Delta_l)}{\abs{\tau_{12}}^{\Delta_i+\Delta_j}\abs{\tau_{13}}^{\Delta_{ij}}\abs{\tau_{23}}^{\Delta_{ji}}}\,.
	\end{split}
\end{equation}
So we have
\begin{equation}
	\begin{split}
		\mathcal{Y}^{\Delta_l;\pm}_{\Delta_i,\Delta_j}(\tau_1,\tau_2,\epsilon)&=\frac{A^{\pm,\text{I}}+A^{\pm,\text{II}}}{\abs{\tau_{12}}^{\Delta_i+\Delta_j}\abs{\tau_{13}}^{\Delta_{ij}}\abs{\tau_{23}}^{\Delta_{ji}}}-\mathcal{Y}^{\Delta_l;\pm,\text{subtr}}_{\Delta_i,\Delta_j}(\tau_1,\tau_2,\epsilon)\,. \\
	\end{split}
\end{equation}
Because both the left-hand side and the subtraction term do not depend on $\tau_3$, it must be true that for $\Delta_i\neq\Delta_j$
\begin{equation}
	\begin{split}
		A^{\pm,\text{I}}+A^{\pm,\text{II}}=0\,,
	\end{split}
\end{equation}
which implies \eqref{Y:prop} for $\Delta_i\neq\Delta_j$. As the left-hand side of \eqref{Y:prop} is analytic in $\Delta_i$ at $\Delta_i=\Delta_j$, the result must also hold when $\Delta_i=\Delta_j$ by analytic continuation. This implies that the singularities contained in the right-hand side of \eqref{Y:prop} must cancel. This finishes the proof of \eqref{Y:prop}.\footnote{\label{fn:IR div}Note that the conditions $\text{Re}(\Delta_i),\text{Re}(\Delta_j)>1$ are still needed for the truncated integral to be convergent.}

Now the problem is reduced to computing the subtraction term, which can be done order by order in $\epsilon$. We focus on the case $\tau_1<\tau_2$ and write an Ansatz for the expansion in $\epsilon$ as (see section \ref{subsec:recap} for the motivation of this Ansatz)
\begin{equation}\label{Ysubtr:result}
	\begin{split}
		\mathcal{Y}^{\Delta_l;\pm,\text{subtr}}_{\Delta_i,\Delta_j}(\tau_1,\tau_2,\epsilon)&=-\frac{1}{\tau_{21}^{\Delta_i+\Delta_j}}\sum_{p=0}^{\infty}\frac{a_p^{\pm}(\Delta_{ij},\Delta_l)}{\Delta_{ij}-p}\left(\frac{\epsilon}{\tau_{21}}\right)^{p-\Delta_{ij}} \\
        &\quad-\frac{1}{\tau_{21}^{\Delta_i+\Delta_j}}\sum_{p=0}^{\infty}\frac{a_p^{\pm}(\Delta_{ji},\Delta_l)}{\Delta_{ji}-p}\left(\frac{\epsilon}{\tau_{21}}\right)^{p-\Delta_{ji}}\,, \\ 
	\end{split}
\end{equation}
where the minus sign is here for convenience. The first term comes from the integral over the regime near $(\tau_1,0)$, and the second term comes from the one near $(\tau_2,0)$.

The first coefficients $a^{+}_0$ and $a^{-}_0$ are the same and can be computed by matching the expansion \eqref{Ysubtr:result} against its definition \eqref{def:Ysubtr}. As we are interested in computing the contribution from the subtraction, we can parametrize $\tau=\tau_1 + R\cos\theta$ and $z= R\sin\theta$ for small $R$ and with $\theta\in(0,\frac{\pi}{2})$. For small $R$, this implies $\chi= \sin^2\theta$ and performing the integral over $R$, we obtain
\begin{equation}
	\begin{split}
		a^{\pm}_0(\Delta_{ij},\Delta_l)&=\int_0^{\tfrac{\pi}{2}}\frac{d\theta}{\sin^2\theta}\,G_{\Delta_l}^{\Delta_i,\Delta_j}\left(\sin^2\theta\right) \\
		&=\frac{\sqrt{\pi}\,\Gamma(\tfrac{\Delta_l-1}{2})}{2\,\Gamma(\tfrac{\Delta_l}{2})}\,\pFq{3}{2}{\tfrac{1}{2}\Delta_{lij},\tfrac{1}{2}\Delta_{lji},\tfrac{\Delta_l-1}{2}}{\Delta_l+\tfrac{1}{2},\tfrac{1}{2}\Delta_l}{1}\,.
	\end{split}
\end{equation}
Later, we will need the specific value of $a^{\pm}_0$ and its derivative at $\Delta_i=\Delta_j$:
\begin{equation}\label{a0:normal}
	\begin{split}
		a^{\pm}_0(0,\Delta_l)=\frac{\sqrt{\pi}\Gamma(\Delta_l+\tfrac{1}{2})}{(\Delta_l-1)\Gamma(\tfrac{\Delta_l}{2})\Gamma(\tfrac{\Delta_l}{2}+1)},\qquad \frac{d a^{\pm}_0(\Delta_{ij},\Delta_l)}{d\Delta_{ij}}\big{|}_{\Delta_{ij}=0}=0\,.
	\end{split}
\end{equation}
The calculation of $a^{\pm}_p$ for $p\geqslant1$ is more complicated, but we do not really need the precise form. As required by the finiteness of the subtraction term (which follows from \eqref{Y:prop} and the finiteness of the truncated integral), we must have
\begin{equation}\label{ap:prop}
	\begin{split}
		a^{\pm}_p(p,\Delta_l)=0\, ,\qquad (p=1,2,3,\ldots)\,.
	\end{split}
\end{equation}
A direct consequence of \eqref{ap:prop} is that the contributions from $p\geqslant1$ never create a logarithmic term ($\log\epsilon/\tau_{21}$) when $\Delta_{ij}\rightarrow p$. It follows that the primary operators with different scaling dimensions never mix in the computation of anomalous dimensions, even if their scaling dimensions differ by an integer.

The last property we would like to establish is
\begin{equation}\label{apm:toshow}
	a_p^{+}(\Delta_{ij},\Delta_l)=(-1)^{p}\,a_p^{-}(\Delta_{ij},\Delta_l)\,.
\end{equation}
These coefficients arise from part of the subtraction term near $(\tau_1,0)$ (see the zoomed-in region in figure~\ref{fig:bkbdbd_four_regime}):
\begin{equation}
	\begin{split}
		\mathcal{Y}^{\Delta_l;\pm,\text{I subtr}}_{\Delta_i,\Delta_j}(\tau_1,\tau_2,\epsilon)
		=-\frac{1}{\tau_{21}^{\Delta_i+\Delta_j}}
		\sum_{p=0}^{\infty}
		\frac{a_p^{\pm}(\Delta_{ij},\Delta_l)}{\Delta_{ij}-p}
		\left(\frac{\epsilon}{\tau_{21}}\right)^{p-\Delta_{ij}}\,.
	\end{split}
\end{equation}
Recall that in the definition of the subtraction term \eqref{def:Ysubtr}, the function 
$G^{\Delta_i,\Delta_j}_{\Delta_l}(\chi)$ is expressed as a sum over $\chi^{\Delta_l/2+n}$, as in section \ref{subsec:recap}. To prove \eqref{apm:toshow}, it therefore suffices to verify the same property after substituting
\[
G^{\Delta_i,\Delta_j}_{\Delta_l}(\chi)\ \longrightarrow\ \chi^\beta\,.
\]
Furthermore, by translation and scale invariance, we can set $\tau_1=0$ and $\tau_2=1$ without loss of generality.  
We thus aim to establish the property for the following integrals:
\begin{equation}\label{def:subtr_simple}
	\begin{split}
		I_{\pm}(\epsilon)
		&:=\int_{\substack{\text{sgn}(\rho)=\pm \\ \tau^2+z^2\leqslant\epsilon^2}}
		\frac{d\tau\,dz}{z^2}
		\left(\frac{\tau^2+z^2}{(\tau-1)^2+z^2}\right)^{\frac{\Delta_j-\Delta_i}{2}}
		\left(\frac{z^2}{(\tau^2+z^2)((\tau-1)^2+z^2)}\right)^\beta \\
		&=\sum\limits_{p=0}^{\infty} c_p^{\pm}\epsilon^{p+\Delta_{ji}}\,.
	\end{split}
\end{equation}
In this setup,
\begin{equation}
	\begin{split}
		\text{sgn}(\rho)=\text{sgn}\!\left((1-\tau)\tau-z^2\right)\,.
	\end{split}
\end{equation}
Introducing polar coordinates
\begin{equation}
	\begin{split}
		\tau=r\cos\theta,\qquad z=r\sin\theta
		\quad(0\leqslant r\leqslant\epsilon,\ 0\leqslant\theta\leqslant\pi)\,,
        \label{eq:polarcoordsrth}
	\end{split}
\end{equation}
we obtain
\begin{equation}
	\begin{split}
		I_{\pm}(\epsilon)
		&=\int_{\substack{\text{sgn}(\rho)=\pm \\ r\leqslant\epsilon}}
		dr\,d\theta\,
		r^{\Delta_{ji}-1}
		\frac{(\sin^{2}\theta)^{\beta-1}}{(1+r^2-2r\cos\theta)^{\beta+\Delta_{ji}/2}} \\
		&=\sum_{p=0}^{\infty}
		\int_{\substack{\text{sgn}(\rho)=\pm \\ r\leqslant\epsilon}}
		dr\,d\theta\,r^{\Delta_{ji}+p-1}
		(\sin^{2}\theta)^{\beta-1}
		C^{(\beta+\Delta_{ji}/2)}_{p}(\cos\theta)\,.
	\end{split}
\end{equation}
If the integration domains of $I_{\pm}$ were both quarter disks (i.e.~defined by $\text{sgn}(\cos\theta)=\pm$),  
then
\begin{equation}
	\begin{split}
		c_p^{+}=(-1)^{p}\,c_p^{-}
	\end{split}
\end{equation}
would follow immediately from the parity property of Gegenbauer polynomials,
\begin{equation}
	\begin{split}
		C^{(\beta+\Delta_{ji}/2)}_{p}(x)=(-1)^p\,C^{(\beta+\Delta_{ji}/2)}_{p}(-x)\,.
	\end{split}
\end{equation}
However, the actual integration domains are slightly more complicated, and thus an additional step is required.  
We split each integral into two pieces:
\begin{equation}
	I_+(\epsilon)\ =\ 
	\begin{tikzpicture}[scale=2, baseline=(current bounding box.center)]
		\draw[thick] (0,0) -- (0.4,0);
		\draw[thick, domain=0.16:0.40] plot (\x, {sqrt(0.16-\x*\x)});
		\draw[thick, domain=0:0.16] plot (\x, {sqrt(\x*(1-\x))});
	\end{tikzpicture}
	\ =\
	\begin{tikzpicture}[scale=2, baseline=(current bounding box.center)]
		\draw[thick] (0,0) -- (0.4,0);
		\draw[thick, domain=0:0.4] plot (\x, {sqrt(0.16-\x*\x)});
		\draw[thick, domain=0:0.4] plot (0,\x);
	\end{tikzpicture}
	\ -\
	\begin{tikzpicture}[scale=2, baseline=(current bounding box.center)]
		\draw[thick, domain=0:0.4] plot (0,\x);
		\draw[thick, domain=0:0.16] plot (\x, {sqrt(0.16-\x*\x)});
		\draw[thick, domain=0:0.16] plot (\x, {sqrt(\x*(1-\x))});
	\end{tikzpicture}
\end{equation}
and similarly
\begin{equation}
	I_-(\epsilon)\ =\ 
	\begin{tikzpicture}[scale=2, baseline=(current bounding box.center)]
		\draw[thick] (0,0) -- (-0.4,0);
		\draw[thick, domain=-0.4:0.16] plot (\x, {sqrt(0.16-\x*\x)});
		\draw[thick, domain=0:0.16] plot (\x, {sqrt(\x*(1-\x))});
	\end{tikzpicture}
	\ =\
	\begin{tikzpicture}[scale=2, baseline=(current bounding box.center)]
		\draw[thick] (0,0) -- (-0.4,0);
		\draw[thick, domain=-0.4:0] plot (\x, {sqrt(0.16-\x*\x)});
		\draw[thick, domain=0:0.4] plot (0,\x);
	\end{tikzpicture}
	\ +\
	\begin{tikzpicture}[scale=2, baseline=(current bounding box.center)]
		\draw[thick, domain=0:0.4] plot (0,\x);
		\draw[thick, domain=0:0.16] plot (\x, {sqrt(0.16-\x*\x)});
		\draw[thick, domain=0:0.16] plot (\x, {sqrt(\x*(1-\x))});
	\end{tikzpicture}
\end{equation}
The parity argument for the Gegenbauer polynomials applies to the first pieces.  
For the second piece,
\begin{tikzpicture}[scale=0.8]
	\draw[thick, domain=0:0.4] plot (0,\x);
	\draw[thick, domain=0:0.16] plot (\x, {sqrt(0.16-\x*\x)});
	\draw[thick, domain=0:0.16] plot (\x, {sqrt(\x*(1-\x))});
\end{tikzpicture},
it suffices to show that it contributes only to $c_p$ with odd $p$ in \eqref{def:subtr_simple}.  
To see this, we write its explicit integral form:
\begin{equation}
	\delta I
	=\int_{\arccos\epsilon}^{\tfrac{\pi}{2}}d\theta\int_{\cos\theta}^{\epsilon}dr\,
	r^{\Delta_{ji}-1}\frac{(\sin^{2}\theta)^{\beta-1}}{(1+r^2-2r\cos\theta)^{\beta+\Delta_{ji}/2}}\,.
\end{equation}
Evaluating this integral yields
\begin{equation}
	\begin{split}
		\delta I
		&=\sum\limits_{n=0}^{\infty}\int_{\arccos\epsilon}^{\tfrac{\pi}{2}}d\theta
		\int_{\cos\theta}^{\epsilon}dr\,
		r^{\Delta_{ji}+n-1}(\sin^{2}\theta)^{\beta-1}
		C^{(\beta+\Delta_{ji}/2)}_{n}(\cos\theta) \\
		&=\sum\limits_{n=0}^{\infty}\int_{\arccos\epsilon}^{\tfrac{\pi}{2}}d\theta\,
		\frac{\epsilon^{\Delta_{ji}+n}-\cos^{\Delta_{ji}+n}\theta}{\Delta_{ji}+n}
		(\sin^{2}\theta)^{\beta-1}
		C^{(\beta+\Delta_{ji}/2)}_{n}(\cos\theta) \\
		&=\sum\limits_{n=0}^{\infty}\int_{0}^{\epsilon}dx\,
		\frac{\epsilon^{\Delta_{ji}+n}-x^{\Delta_{ji}+n}}{\Delta_{ji}+n}
		(1-x^2)^{\beta-3/2}
		C^{(\beta+\Delta_{ji}/2)}_{n}(x)\,.
	\end{split}
\end{equation}
Expanding the factors in powers of $x$, we find
\begin{equation}
	\begin{split}
		(1-x^2)^{\beta-3/2}&=\sum\limits_{l}\#\,x^{2l}\,,\\
		C^{(\beta+\Delta_{ji}/2)}_{n}(x)&=\sum\limits_{m=0}^{[n/2]}\#\,x^{n-2m}\,.
	\end{split}
\end{equation}
Term by term, the integrand thus contains powers of the form
\[
\epsilon^{\Delta_{ji}+n} x^{2l+n-2m},\qquad x^{\Delta_{ji}+2n+2l-2m},
\]
which are always $\Delta_{ji}$ plus an even integer.  
Integrating $x$ from $0$ to $\epsilon$ increases the power by one, yielding $\Delta_{ji}$ plus an \emph{odd} integer.  
Hence $\delta I$ only contributes to odd $c_p$, completing the argument and proving \eqref{apm:toshow}.

\subsubsection{Local block}
We want to repeat the same analysis for the integral of the local bulk-boundary-boundary block. As argued in \ref{subsec:localblock_bkbdbd}, we only care about the even local block. 

We start from the regulated integral of the bulk-boundary-boundary correlator defined in eq. (\ref{def:Iij}), $I^{\hat\Phi}_{ij}(\tau_1,\tau_2,\epsilon)$. Now we expand the correlator in even blocks and integrate block by block. We can define the integrated blocks as $\mathcal{Y}^{\Delta_l(\alpha);\pm}_{\Delta_i,\Delta_j}$ precisely in the same way as (\ref{def:Y}) but with even local blocks in place of normal blocks. Following the same arguments as the previous section, these objects admit the expansion
\begin{equation}
	\begin{split}
		\mathcal{Y}^{\Delta_l(\alpha);\pm}_{\Delta_i,\Delta_j}(\tau_1,\tau_2,\epsilon)&=\frac{1}{\abs{\tau_{12}}^{\Delta_i+\Delta_j}}\sum_{p=0}^{\infty}\frac{a_p^{(\alpha)\pm}(\Delta_{ij},\Delta_l)}{\Delta_{ij}-p}\abs{\frac{\tau_{21}}{\epsilon}}^{\Delta_{ij}}\left(\frac{\epsilon}{\tau_{21}}\right)^p +(\Delta_i\leftrightarrow \Delta_j)\, .
	\end{split}
\end{equation}
with the following properties
\begin{align}
		a^{(\alpha)\pm}_p(p,\Delta_l)&=0\, ,&&\text{for }p =1,2,3,\ldots\,, \\
        a^{(\alpha)+}_p(\Delta_{ij},\Delta_l)&=(-1)^p\,a^{(\alpha)-}_p(\Delta_{ij},\Delta_l),&&\text{for all }p\,. 
\end{align}
We just need to find the new form of $a^{(\alpha)}_0(0,\Delta):=a^{(\alpha)+}_0(0,\Delta)=a^{(\alpha)-}_0(0,\Delta)$.

To do that, we find it easiest to start from the basic dispersive definition of the local block from \cite{Meineri:2023mps,Levine:2023ywq} with $\D_i=\D_j=\Delta$
\begin{align}
		G^{\D\D,(\alpha)}_{\D_l}(\chi)&:=\chi^\alpha\sin\left[\frac{\pi}{2}(\Delta_l-2\alpha)\right]  \int_{-\infty}^{0}\frac{d\chi'}{\pi}\frac{(-\chi')^{\Delta_l/2-\alpha}}{\chi'-\chi}\hyperF{\frac{\Delta_l}{2}}{\frac{\Delta_l}{2}}{\Delta_l+\frac{1}{2}}{\chi'}\\ 
        &= -\frac{\sin\left[\frac{\pi}{2}(\Delta_l-2\alpha)\right]}{\pi}\label{eq:C9}\\ 
        &\phantom{=}\times \sum_{n=0}^\infty\frac{\left(\frac{\D_l}{2}\right)_n\left(\frac{\D_l+1}{2}\right)_n}{n!\left(\D_l+\frac{1}{2}\right)_n}\frac{\Gamma(\alpha)\Gamma(\frac{\D_l}{2}-\alpha+n+1)}{\Gamma(\frac{\D_l}{2}+n+1)}\chi^{\alpha}\,{}_2F_1\left(\alpha,1;\frac{\D_l}{2}+n+1;1-\chi\right)\nonumber\, .
\end{align}
To obtain \eqref{eq:C9} from the first line, we first changed variables as $\chi'=z/(z-1)$ with $z\in(0,1)$. We then used the hypergeometric identity \eqref{eq:2F1identity1},
expanded the resulting ${}_2F_1$ function into a power series, exchanged the sum and integral, computed the integral over $z$ and used the hypergeometric identity \eqref{eq:2F1identity1} on the result again. 

In the configuration $\tau_1=0$ and $\tau_2=1$, we get that the contribution of the operator $\mathcal{O}_l$ to $I^{\hat\Phi}_{ii}$ is
\begin{align}
    \left[I^{\hat\Phi}_{ii}(0,1,\epsilon)\right]_l &= b^{\hat\Phi}_lC_{iil}\int_{(\epsilon-reg)}\frac{d\tau dz}{z^2}G_{\D_l}^{\D\D,(\alpha)}(\chi) \\ 
    &\sim 2b^{\hat\Phi}_lC_{iil}\int_\epsilon^1\frac{dr}{r}\int_0^{\pi}\frac{d\theta}{\sin^2\theta}G_{\D_l}^{\D\D,(\alpha)}(\sin^2\theta)\, ,
\end{align}
where we use the same polar coordinates as in equation \eqref{eq:polarcoordsrth}, expanded for small $r$, and the $\sim$ symbol stands to indicate that we simply want to match the coefficient of the logarithm.

Let us first compute the integral over $\theta$. We have 
\begin{align}
    \int_0^\pi \frac{d\theta}{\sin^2\theta}(\sin\theta)^{2\alpha}\,{}_2F_1\left(\alpha,1;\frac{\D_l}{2}+n+1;\cos^2\theta\right) = \frac{\sqrt{\pi}(\D_l+2n)\Gamma\left(\alpha-\frac{1}{2}\right)}{(\D_l+2n-1)\Gamma(\alpha)}\, .
\end{align}
We can then write 
\begin{align}
  \left[I^{\hat\Phi}_{ii}(0,1,\epsilon)\right]_l
    &\sim -2\int_\epsilon^1\frac{dr}{r}\frac{\sin\left[\frac{\pi}{2}(\Delta_l-2\alpha)\right]}{\pi}\\ 
        &\phantom{=}\times \sum_{n=0}^\infty\frac{\left(\frac{\D_l}{2}\right)_n\left(\frac{\D_l+1}{2}\right)_n}{n!\left(\D_l+\frac{1}{2}\right)_n}\frac{\Gamma(\alpha)\Gamma(\frac{\D_l}{2}-\alpha+n+1)}{\Gamma(\frac{\D_l}{2}+n+1)}\frac{\sqrt{\pi}(\D_l+2n)\Gamma\left(\alpha-\frac{1}{2}\right)}{(\D_l+2n-1)\Gamma(\alpha)}\\
        &= \frac{\pi\Gamma\left(\D_l+\frac{1}{2}\right)\Gamma\left(2\alpha-1\right)}{2^{2\alpha-4}(\D_l-1)\Gamma\left(\frac{\D_l}{2}\right)\Gamma\left(\frac{\D_l}{2}+1\right)\Gamma\left(\frac{\D_l-1}{2}+\alpha\right)\Gamma\left(\alpha-\frac{\D_l}{2}\right)}
\log\frac{1}{\epsilon}\, .
\end{align}
Matching with (\ref{eq:Iiiepslog}), we get
\begin{equation}
    a_0^{(\alpha)}(0,\Delta)=\frac{2^{2-2\alpha}\pi\Gamma\left(\D_l+\frac{1}{2}\right)\Gamma\left(2\alpha-1\right)}{(\D_l-1)\Gamma\left(\frac{\D_l}{2}\right)\Gamma\left(\frac{\D_l}{2}+1\right)\Gamma\left(\frac{\D_l-1}{2}+\alpha\right)\Gamma\left(\alpha-\frac{\D_l}{2}\right)}\,.
\end{equation}
Note that in the limit $\alpha\rightarrow \infty$, we obtain
\begin{equation}
\lim_{\alpha\rightarrow\infty}a^{(\alpha)}_0(0,\Delta_l)=a_0(0,\Delta_l)\, ,
\end{equation}
where $a_0(0,\Delta_l)$ is given in (\ref{apm:prop1}). Comparing with (\ref{eq:flow_scaling_dim_normal}), we finally get
\begin{equation}
    \mathcal{I}^{(\alpha)}(\Delta_l)=2a^{(\alpha)}_0(0,\Delta_l)\,.
    \label{eq:Local Bbb app}
\end{equation}
\subsection{\texorpdfstring{Bulk-bulk-boundary ($BB\del$)}{Bulk-bulk-boundary (BBb)}}
\label{subsec:integrated bkbkbd block}
In this section, we would like to perform the integral over the $BB\partial$ block in order to derive \eqref{eq:J} in the main text. In particular, our goal is to compute 
\begin{equation}
    \JJ_{\D_i}(\D_l,\D_j)
\assign -\int_{AdS} R_{\D_l,\D_j}^{\D_i}(\upsilon,\z)  \, .
    \label{eq:Japp}
\end{equation}
As justified in appendix \ref{app:HardvsAnal}, the only term that survives in the flow equations is the analytic continuation of the integral of the conformal block 
\eqref{def:bkbkbd block}, which converges for $\Delta_j>\Delta_i$. Let us report the block here:
\begin{align}
		R^{\Delta_i}_{\Delta_l\Delta_j}(\upsilon, \zeta)&=\frac{\zeta^{\Delta_l}}{(\zeta^2+\upsilon^2)^{\frac{\Delta_{lji}}{2}}}
		\sum_{n=0}^{\infty}\frac{\left(\frac{\Delta_{lji}}{2}\right)_n\Gamma(1-\frac{\Delta_{ilj}}{2})\Gamma(\Delta_j+\frac{1}{2})}{n!\,\left(\Delta_l+\frac{1}{2}\right)_n}\left(\frac{-\zeta^2}{\zeta^2+\upsilon^2}\right)^{n}\\
		&\qquad\quad \times
		\regpFq{3}{2}{\frac{\Delta_{lji}+1}{2},\frac{\Delta_{lji}}{2}+n,1-\frac{\Delta_{lij}}{2}}{\Delta_j+\frac{1}{2},1-\frac{\Delta_{lij}}{2}-n}{-\frac{1}{\zeta^2+\upsilon^2}}.\nonumber
\end{align}
This representation of the conformal block is convergent for all $\zeta^2,\upsilon^2\in(0,\infty)$, but it is not the best suited representation to actually perform the integral over AdS. To obtain a representation that is more suited for our goal, we will use\footnote{This identity can be found \href{http://functions.wolfram.com/07.27.03.0117.01}{here} for example. We also used 
\begin{equation}
\ _3\tilde F_2\left(\begin{matrix}
    a_1 & a_2 & a_3\\
    & b_1 & b_2
\end{matrix};z\right)=\frac{1}{\Gamma(b_1)\Gamma(b_2)}\ _3F_2\left(\begin{matrix}
    a_1 & a_2 & a_3\\
    & b_1 & b_2
\end{matrix};z\right)\,.
\end{equation}}
\begin{align}
    \pFq{3}{2}{a_1, a_2, c}{b, c-n}{z}
    =\sum_{m=0}^n\binom{n}{m}\frac{z^m}{(c-n)_m}\frac{(a_1)_m(a_2)_m}{(b)_m}
    \,\pFq{2}{1}{a_1+m,a_2+m}{b+m}{z}
    \,.
\end{align}
We obtain
\begin{align}
\spl{
		R^{\D_i}_{\D_l\D_j}(\upsilon, \zeta)&=\frac{\zeta^{\Delta_l}}{(\zeta^2+\upsilon^2)^{\frac{\Delta_{lji}}{2}}} \sum_{n=0}^{\infty}\sum_{m=0}^n \binom{n}{m}(-1)^m\frac{\zeta^{2n}}{\left(\zeta^2+\upsilon^2\right)^{n+m}}\\
&\times \frac{\left(\frac{\Delta_{lji}}{2}\right)_n\left(\frac{\Delta_{lij}}{2}\right)_n}{n!\,\left(\Delta_l+\frac{1}{2}\right)_n}\frac{\left(\frac{\Delta_{lji}+1}{2}\right)_m\left(\frac{\Delta_{lji}}{2}+n\right)_m}{\left(1-\frac{\Delta_{lij}}{2}-n\right)_m\left(\Delta_j + \frac{1}{2}\right)_m}
  \\
		&\times\,
        \pFq{2}{1}{\frac{\Delta_{lji}+1}{2}+m,  \frac{\Delta_{lji}}{2} + n+m}{\Delta_j+\frac{1}{2}+m}{-\frac{1}{\zeta^2+\upsilon^2}}\, .
}\label{eq:RD44}
\end{align}

The sum over $m$ and $n$ is absolutely convergent for all $\zeta^2,\upsilon^2\in(0,\infty)$, which allows us to perform the integral term by term. For each fixed $(n,m)$ pair, we have
\begin{equation}
	\begin{split}
		&\int_0^{\infty}\frac{d\zeta}{\zeta^2}\int_{\mathbb{R}}d\upsilon\,\frac{\zeta^{\Delta_l+2n}}{(\zeta^2+\upsilon^2)^{\frac{\Delta_{lji}}{2}+n+m}}\,\pFq{2}{1}{\tfrac{\Delta_{lji}+1}{2}+m,  \tfrac{\Delta_{lji}}{2} + n+m}{\Delta_j+\tfrac{1}{2}+m}{-\tfrac{1}{\zeta^2+\upsilon^2}} \\
		=&\int_0^{\infty}\frac{d\zeta}{\zeta^2}\int_{\mathbb{R}}d\upsilon\,\frac{\zeta^{\Delta_l+2n}}{(1+\zeta^2+\upsilon^2)^{\frac{\Delta_{lji}}{2}+n+m}}\,\pFq{2}{1}{\tfrac{\Delta_{ijl}}{2},  \tfrac{\Delta_{lji}}{2} + n+m}{\Delta_j+\tfrac{1}{2}+m}{\tfrac{1}{1+\zeta^2+\upsilon^2}} \\
		=&\sum_{k=0}^{\infty}\frac{(\tfrac{\Delta_{ijl}}{2})_k\,(\tfrac{\Delta_{lji}}{2} + n+m)_k}{k!\,(\Delta_j+\tfrac{1}{2}+m)_k}\, \int_0^{\infty}\frac{d\zeta}{\zeta^2}\int_{\mathbb{R}}d\upsilon\,\frac{\zeta^{\Delta_l+2n}}{(1+\zeta^2+\upsilon^2)^{\frac{\Delta_{lji}}{2}+n+m+k}} \\
		=&\frac{\sqrt{\pi}\Gamma(\tfrac{\Delta_{ji}}{2}+m)\Gamma(\tfrac{\Delta_l-1}{2}+n)}{2\Gamma(\tfrac{\Delta_{lji}}{2} + n+m)\,}\frac{\Gamma(\tfrac{\Delta_l+1}{2})\Gamma(\Delta_j+\tfrac{1}{2}+m)}{\Gamma(\tfrac{\Delta_i+\Delta_j+1}{2})\Gamma(\tfrac{\Delta_{lji}+1}{2}+m)}\,, \\
	\end{split}
\end{equation}
where, in the first step, we used the hypergeometric identity
\begin{equation}
    \hyperF{a}{b}{c}{z}=(1-z)^{-b}\,\hyperF{c-a}{b}{c}{\tfrac{z}{z-1}}\,;
\end{equation}
and in the second step, we expanded the hypergeometric function and performed the integral term by term. The integral is convergent when 
\begin{equation}
    \Delta_l>1,\qquad\Delta_j>\Delta_i\,.
\end{equation}
Then taking into account all the factors in \eqref{eq:RD44} and explicitly summing over $m$ and $n$, we obtain  
\begin{equation}\label{eq:Ibkbkbd}
	\begin{split}
		\JJ_{\D_i}(\D_l,\D_j)&=-\frac{2^{\Delta _{lji}-2} \Gamma \left(\frac{\Delta _l-1}{2}\right) \Gamma \left(\Delta _l+\frac{1}{2}\right) \Gamma \left(\Delta _j+\frac{1}{2}\right) \Gamma \left(\frac{\Delta _j-\Delta _i}{2}\right)}{\Gamma \left(\frac{\Delta _l}{2}+1\right) \Gamma \left(\Delta_{lji}\right) \Gamma \left(\frac{\Delta _j+\Delta _i+1}{2}\right)}\, .
	\end{split}
\end{equation}
This is the result presented in the main text in equation \eqref{eq:J}.

\subsection{\texorpdfstring{Bulk-boundary-boundary-boundary ($B\del\del\del$)}{Bulk-boundary-boundary-boundary (Bbbb)}}
\label{subsec:intbkbdbdbd}
Let us now consider the $B\del\del\del$ correlator $\<\hat\Phi(\t,z)\OO_i(\t_1)\OO_j(\t_2)\OO_k(\t_3)\>$. The novelty of this case compared to the previous two is that now there are three different OPE channels that require different expansions. Therefore, we need to split AdS into three regions as in figure \ref{fig:stu} and integrate the corresponding block in each region. In fact, each region has a subregion where none of the BOE channels converges, essentially the blue region from figure \ref{fig:no_stu_a}, and we need to use the local blocks to cure this issue.  The integral diverges when $\Delta_m>\Delta_i$. As argued in appendix \ref{app:HardvsAnal}, the coefficient that effectively enters the third flow equation is the analytic continuation of the integral from the regime in which it converges to all values of the scaling dimensions.

\begin{figure}[t!]
  \centering
  \begin{subfigure}[t]{0.85 \textwidth}
    \includegraphics[width=\textwidth]{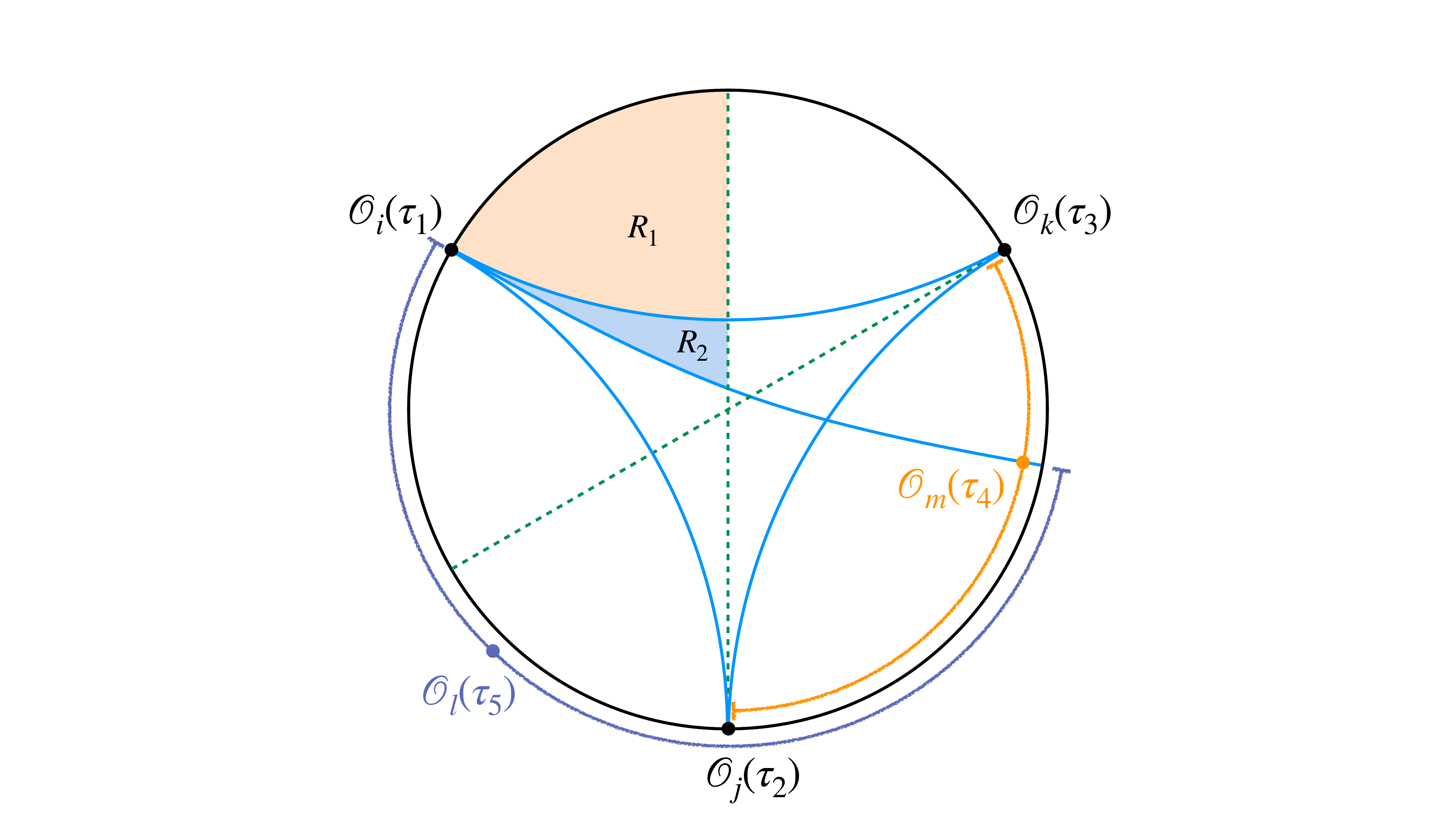}
    \caption{}
  \end{subfigure}
  \hfill
  \begin{subfigure}[t]{0.85\textwidth}
    \includegraphics[width=\textwidth]{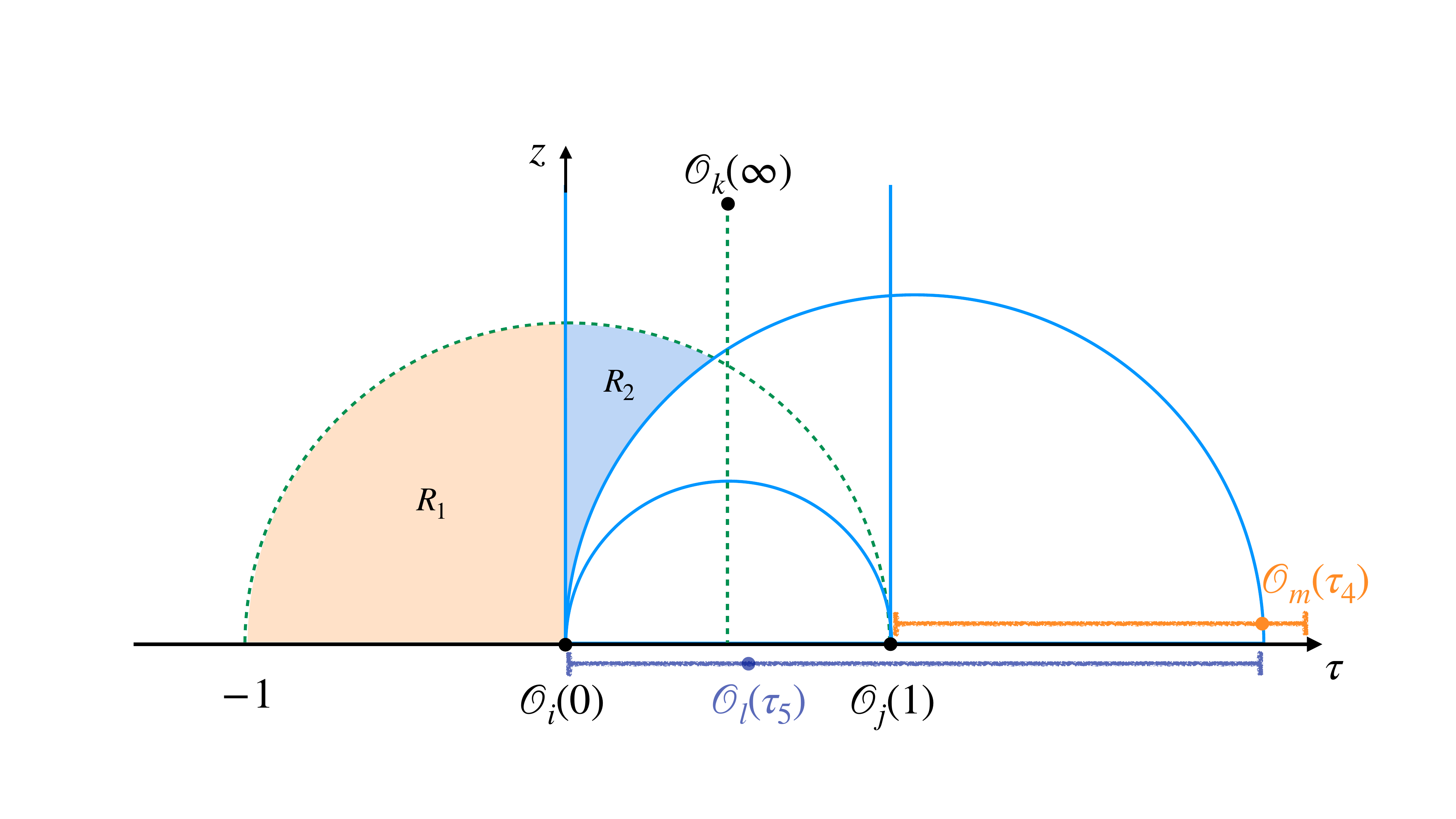}
    \caption{}
  \end{subfigure}
  \caption{The split of AdS$_2$ into six regions in (\textbf{a}) the Poincar\'{e} disk and in (\textbf{b}) the Poincar\'{e} patch in the frame (\ref{eq:frameR1}). 
  The shaded region is denoted as region $R$ and is further split into $R_1$ (orange) and $R_2$ (light blue). The dotted lines are geodesics connecting $\mathcal{O}_j$ and $\mathcal{O}_k$ to the point that is antipodal to their insertion. The dark purple and orange lines are the integration regimes of $\t_4$ and $\t_5$ in \eqref{eq:int rep of bk-bd^3 block}. The blue lines are geodesics connecting the boundary operators. This is a convenient way to split AdS because different operator orderings appear in the OPE coefficients when the geodesics connecting $\tau_1$ and $\tau_4$ is crossed.}
  \label{fig:pizza slices}
\end{figure}

As explained in section \ref{sec:OPE on the boundary}, due to operator ordering issues in 1D CFT, we must further split each region into two. Overall, we thus have six regions of AdS. Without loss of generality, we focus on region $R=R_1\cup R_2$, where $R_1$ and $R_2$ are shown in figure \ref{fig:pizza slices}. The other five regions can be obtained by permuting the $i,j,k$ scaling dimensions.

Let us start from the conformal block decomposition of the $B\partial\partial\partial$ correlator. After expanding the bulk operator, we do the OPE twice, first for $\OO_j\times \OO_k$ and then for $\OO_i\times \OO_m$ (without loss of generality we assume $\t_1<\t_2<\t_3$). This yields
\begin{equation}
\braket{\hat{\Phi}(\tau,z)\OO_i(\tau_1)\OO_j(\tau_2)\OO_k(\tau_3)} = \sum_{l,m} \bb{\hat{\Phi}}{l}C_{lim} C_{jkm}K^{\D_i \D_j \D_k}_{\D_l \D_m}(\t,z,\t_1,\t_2,\t_3)\,.
\end{equation}
We call this OPE channel the $s$-channel. Using the integral repackaging of the OPE (\ref{OPE:integralrepr}) twice, the s-channel bulk-boundary-boundary-boundary block admits the integral representation
\begin{align}\label{eq:int rep of bk-bd^3 block}
    &K^{\D_i \D_j \D_k}_{\D_l \D_m}(\t,z,\t_1,\t_2,\t_3)\\
    &\qquad = N \int_{\t_2}^{\t_3}d\t_4
    \int_{\t_1}^{\t_4} d\t_5
\<\!\< \mathcal{O}_j(\t_2)\mathcal{O}_k(\t_3)\tilde{\mathcal{O}}_m(\t_4)\>\!\>
\<\!\< \mathcal{O}_i(\t_1)\mathcal{O}_m(\t_4)\tilde{\mathcal{O}}_l(\t_5)\rangle
\<\!\< \mathcal{O}_l(\t_5)\hat{\Phi}(\t,z)\>\!\> \,,\nonumber
\end{align}
where the normalization constant is given by (using \eqref{eq:OPE block normalisation})
\begin{align}
    N= 
    \frac{\Gamma(2\Delta_m)}{\Gamma(\Delta_{mjk})\Gamma(\Delta_{mkj})}\frac{\Gamma(2\Delta_l)}{\Gamma(\Delta_{lim})\Gamma(\Delta_{lmi})}\, ,
\end{align}
and the double braket notation means the OPE and BOE coefficients have been removed from the correlation functions.
By conformal invariance we can define the scale-less integrated block $\KK$ as
\begin{align}
    \KK_{\D_i \D_j \D_k} (\D_l,\D_m)
    \assign
    -\int_{R} \frac{d\t dz}{z^2} \frac{K^{\D_i \D_j \D_k}_{\D_l \D_m}(\t,z,\t_1,\t_2,\t_3)}{\<\!\< \mathcal{O}_i(\t_1) \mathcal{O}_j(\t_2) \mathcal{O}_k(\t_3)\>\!\>} \,,
    \label{eq:covariantKint}
\end{align} 
The idea to integrate the block is through analytic continuation: we will first obtain a result that converges for certain values of $\D$'s and then analytically continue. Thus let us for the moment be careless and interchange the integrals (the ones appearing in \eqref{eq:int rep of bk-bd^3 block} and the integral of the block over AdS$_2$) freely.  Furthermore, let us focus separately on $R_1$ and $R_2$.
\subsubsection{Integral over $R_1$}
We focus on the integral over $R_1$ first. 
\begin{equation}
    I_{R_1}(\tau_5)\equiv\int_{R_1}\frac{dz}{z^2}\<\!\< O_l(\t_5)\hat{\Phi}(\t,z)\>\!\> 
\end{equation}
For this integral, it is useful to go in the frame
\begin{equation}
    \tau_1=0\,,\qquad \tau_2=1\,,\qquad \tau_3=\infty\,,
    \label{eq:frameR1}
\end{equation}
where the regime of integration, as shown in figure \ref{fig:pizza slices}, is $z>0$ with $z^2+\tau^2<1$ and $\tau<0$.

We begin by noticing that the bulk-to-boundary propagator admits the following exponentially convergent Mellin representation:
\begin{equation}
    \tau_5^{\Delta_l}\<\!\< O_l(\t_5)\hat{\Phi}(\t,z)\>\!\> =\left(\frac{z\tau_5}{z^2+(\tau-\tau_5)^2}\right)^{\Delta_l}=\int\frac{dcdu}{(2\pi i)^2}T(c,u)\left(\frac{\tau_5}{z}\right)^c\left(-\frac{\tau}{z}\right)^u
\end{equation}
with
\begin{equation}
    T(c,u)=\frac{\Gamma(-u)\Gamma(\Delta_l-s)\Gamma(\frac{\Delta_l-c-u}{2})\Gamma(\frac{\Delta_l+c+u}{2})}{2\Gamma(\Delta_l)\Gamma(\Delta_l-c-u)}\,,\quad |\text{Re}c|<\Delta_l\,, 
    \label{eq:integrandT}
\end{equation}
with the contour restrictions
\begin{equation}
-\text{Re}c-\Delta_l<\text{Re}u<0\,.
\end{equation}
This Mellin integral converges in the region of integration $R_1$.
We derived it by applying twice the basic identity
\begin{align}
    (x+y)^{-a} =\int \frac{dc}{2 \pi i}\frac{
    \Gamma(\frac{a}{2} -c) \Gamma(\frac{a}{2} +c) }{ \Gamma(a)}x^{-\frac{a}{2}+c}
    y^{-\frac{a}{2}-c}\,,\qquad -\frac{a}{2}<\text{Re}c<\frac{a}{2}\,.
\end{align}
The integrand (\ref{eq:integrandT}) decays exponentially in the Mellin variables, making the integral particularly amenable to numerical evaluation. Namely, it decays as $e^{-\frac{\pi}{2}\text{Im}c}(\text{Im}c)^{\text{Re}u+\Delta_l-1}$ and $e^{-\frac{\pi}{2}\text{Im}u}(\text{Im}u)^{\text{Re}c-1}$ in the $c$ and $u$ directions respectively.

Now we can carry out the integral over AdS explicitly.
\begin{align}
 I_{R_1}(\tau_5)&=\tau_5^{-\Delta_l}
\int \frac{dc du}{(2 \pi i)^2}
T(c,u)  \int_0^1 d\t \int_0^{\sqrt{1-\t^2}} \frac{dz}{z^2}
    \left(\frac{\tau_5}{z}\right)^{c}  \left(\frac{\t}{z}\right)^{u}\\
    &=-
\int \frac{dc du}{(2 \pi i)^2}
T(c,u)  \tau_5^{c-\Delta_l} 
\int_0^1 d\t  
\frac{
(1-\tau^2)^{-\frac{c+u+1}{2}}}
{1+c+u}
      \t^u \\  
      &=-
\int \frac{dc du}{(2 \pi i)^2}
T(c,u)  \tau_5^{c-\Delta_l} \frac{\Gamma \left(\frac{u+1}{2}\right) \Gamma \left(\frac{1-c-u}{2}
   \right)}{2(1+c+u) \Gamma \left(1-\frac{c}{2}\right)}
\end{align}
The integrals converge conditionally, updating the contour restrictions to
\begin{equation}
    |\text{Re}c|<\Delta_l,\quad -\text{Re}c-\Delta_l<\text{Re}u<0,\quad\text{Re}c+\text{Re}u<-1,\quad \text{Re}u>-1\,.
\end{equation}
From the definition of the integrated block (\ref{eq:covariantKint}) applied to this frame, we have
\begin{equation}
\begin{aligned}
    \mathcal{K}^{R_1}_{\Delta_i\Delta_j\Delta_k}(\Delta_l,\Delta_m)&=\int\frac{dcdu}{(2\pi i)^2}\tilde T(c,u)\int_{1}^\infty d\tau_4\ \frac{\tau_4^{1-\Delta_i-\Delta_l-\Delta_m}}{(\tau_4-1)^{1-\Delta_{kmj}}}\int_0^{\tau_4}d\tau_5\frac{\tau_5^{\Delta_{mi}+c-1}}{(\tau_4-\tau_5)^{1-\Delta_{ilm}}}\\
    &=\int\frac{dcdu}{(2\pi i)^2}\tilde T(c,u)\frac{\Gamma(\Delta_{ilm})\Gamma(c+\Delta_{mi})}{\Gamma(c+\Delta_l)}\int_{1}^\infty d\tau_4\frac{\tau_4^{c-\Delta_i-\Delta_m}}{(\tau_4-1)^{1-\Delta_{kmj}}}\\
    &=\int\frac{dcdu}{(2\pi i)^2}\tilde T(c,u)\frac{\Gamma(\Delta_{ilm})\Gamma(c+\Delta_{mi})}{\Gamma(c+\Delta_l)}\frac{\Gamma(\Delta_{ijk}-c)\Gamma(\Delta_{kmj})}{\Gamma(\Delta_i+\Delta_m-c)}
    \label{eq:KR1fin}
\end{aligned}
\end{equation}
where 
\begin{equation}
    \tilde T(c,u)\equiv N\frac{\Gamma \left(\frac{u+1}{2}\right) \Gamma \left(\frac{1-c-u}{2}
   \right)}{2(1+c+u) \Gamma \left(1-\frac{c}{2}\right)}T(c,u)
\end{equation}
The added conditions for convergence are
\begin{equation}
    -\Delta_{mi}<\text{Re}c<\Delta_{ijk}\,,\qquad \text{Re}c>\Delta_i+\Delta_m-1\,.
\end{equation}
Even though the original integral converged only if $\Delta_{ilm}>0$ and $\Delta_{kmj}>0$, the current integral representation defines an analytic continuation beyond those values.

Performing the change of variable $u\to-c-2s$, we finally match the expression presented in (\ref{eq:K}).

\subsubsection{Integral over $R_2$}
To carry out the integral over $R_2$, a convenient frame is the following
\begin{equation}
    \tau_1=\infty\,,\qquad \tau_2=0\,,\qquad \tau_3=1\,.
    \label{eq:frameR2}
\end{equation}
\begin{figure}[t!]
  \centering
    \includegraphics[width=0.85\textwidth]{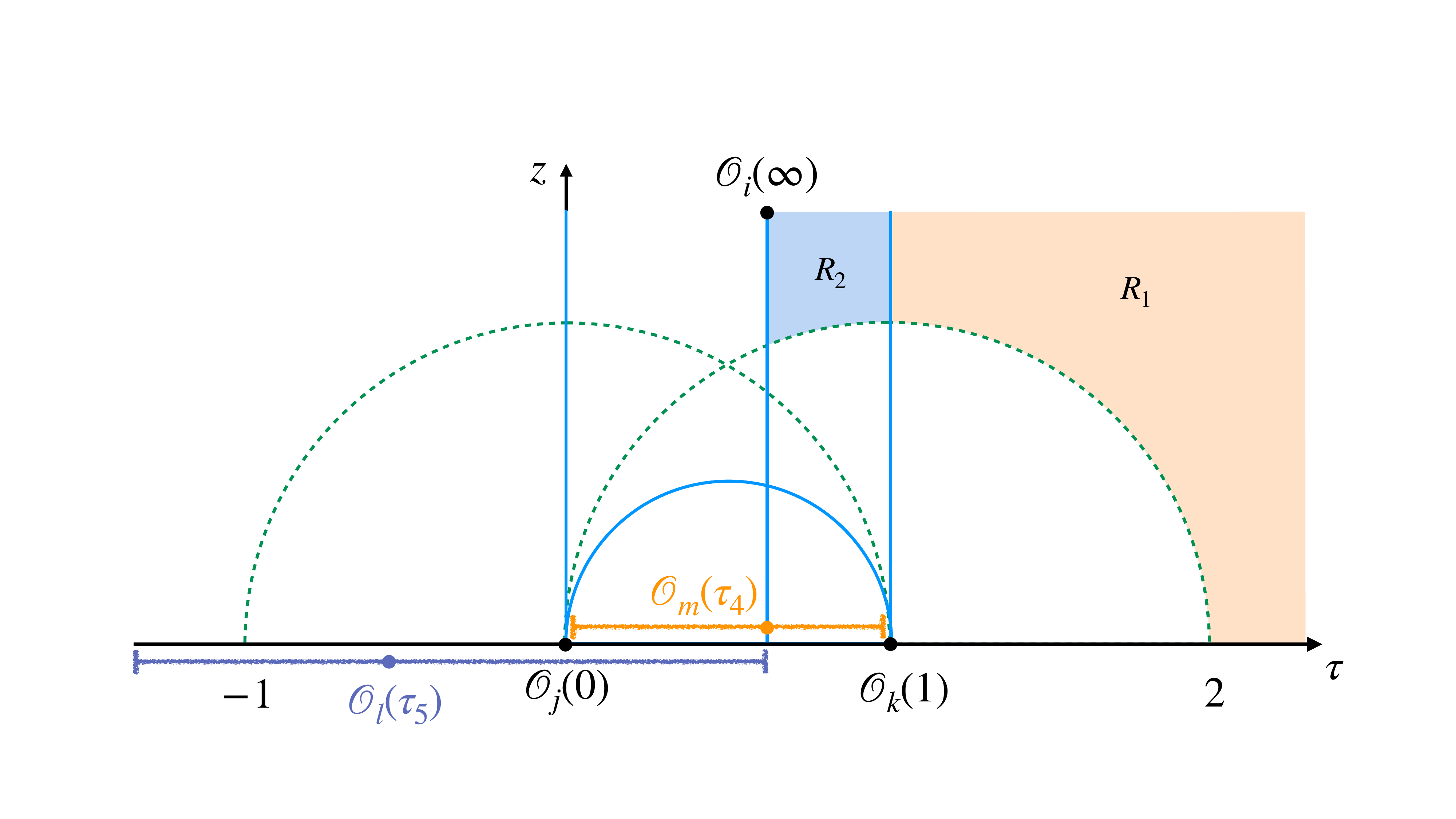}
  \caption{The regions $R_1$ and $R_2$ in the frame (\ref{eq:frameR2}).}
  \label{fig:frameR2}
\end{figure}
Figure \ref{fig:frameR2} represents the region of integration in this frame, which in equations is
\begin{equation}
    z>\max\left(\sqrt{1-\tau^2},\sqrt{\tau(2-\tau)}\right)\,,\qquad \tau_4<\tau<1
\end{equation}
The strategy for this integral will be a bit different. First of all, we use the following representation of the bulk-to-boundary propagator
\begin{equation}
    \<\!\< O_l(\t_5)\hat{\Phi}(\t,z)\>\!\> =\left(\frac{z}{z^2+(\tau-\tau_5)^2}\right)^{\Delta_l}=\int\frac{ds}{2\pi i}\frac{\Gamma(\frac{\Delta_l}{2}+s)\Gamma(\frac{\Delta_l}{2}-s)}{\Gamma(\Delta_l)}\frac{z^{2s}}{((\tau-\tau_5)^2)^{s+\frac{\Delta_l}{2}}}
\end{equation}
with the contour restriction $-\frac{\Delta_l}{2}<s<\frac{\Delta_l}{2}$.

Then, we carry out the integral over $z$, giving
\begin{equation}
    \int_{\max_\tau}^\infty\frac{dz}{z^2} \<\!\< O_l(\t_5)\hat{\Phi}(\t,z)\>\!\> =\int\frac{ds}{2\pi i}\frac{\Gamma(\frac{\Delta_l}{2}+s)\Gamma(\frac{\Delta_l}{2}-s)}{\Gamma(\Delta_l)(1-2s)}\frac{\max_\tau^{2s-1}}{((\tau-\tau_5)^2)^{s+\frac{\Delta_l}{2}}}
\end{equation}
where $\max_\tau\equiv\max\left(\sqrt{1-\tau^2},\sqrt{\tau(2-\tau)}\right)$ and the new contour condition is $\text{Re}s<\frac{1}{2}$. From the definition of the integrated block (\ref{eq:covariantKint}) applied to this frame, we have
\begin{equation}
\begin{aligned}
    &\mathcal{K}^{R_2}_{\Delta_i\Delta_j\Delta_k}(\Delta_l,\Delta_m)=\int\frac{dsC(s)}{2\pi i}\int_{0}^1 \frac{d\tau_4\ \tau_4^{\Delta_{kmj}-1}}{(1-\tau_4)^{1-\Delta_{jmk}}}\int_{\tau_4}^1 \frac{d\tau}{\max_\tau^{1-2s}}\int_{-\infty}^{\tau_4}\frac{d\tau_5(\tau_4-\tau_5)^{\Delta_{ilm}-1}}{(\tau-\tau_5)^{2s+\Delta_l}}\\
    &=\Gamma(\Delta_{ilm})\int\frac{dsC(s)}{2\pi i}\frac{\Gamma(2s+\Delta_{mi})}{\Gamma(2s+\Delta_l)}\int_{0}^1 \frac{d\tau_4\ \tau_4^{\Delta_{kmj}-1}}{(1-\tau_4)^{1-\Delta_{jmk}}}\int_{\tau_4}^1 \frac{d\tau}{\max_\tau^{1-2s}(\tau-\tau_4)^{2s+\Delta_{mi}}}\\
    &=\Gamma(\Delta_{ilm})\int\frac{dsC(s)}{2\pi i}\frac{\Gamma(2s+\Delta_{mi})}{\Gamma(2s+\Delta_l)}\int_{0}^1 \frac{d\tau}{\max_\tau^{1-2s}}\int_{0}^\tau \frac{d\tau_4\ \tau_4^{\Delta_{kmj}-1}(\tau-\tau_4)^{2s+\Delta_{mi}}}{(1-\tau_4)^{1-\Delta_{jmk}}}\\
    &=\Gamma(\Delta_{ilm})\int\frac{dsC(s)}{2\pi i}\frac{\csc(\pi(2s+\Delta_{mi}))\csc(\pi\Delta_{kmj})}{\pi^{-2}\Gamma(2s+\Delta_l)\Gamma(1-\Delta_{kmj})}\int_{0}^1 \frac{d\tau\ \tau^{\Delta_{ikj}-2s}}{\max_\tau^{1-2s}}\mathbf{F}\left(\begin{matrix}
        1-\Delta_{jmk}, & \Delta_{kmj}\\
        1-2s+\Delta_{ikj} &
    \end{matrix};\tau\right)
    \label{eq:KR2fin}
\end{aligned}
\end{equation}
where
\begin{equation}
    C(s)\equiv-N\frac{\Gamma(\frac{\Delta_l}{2}+s)\Gamma(\frac{\Delta_l}{2}-s)}{\Gamma(\Delta_l)(1-2s)}
\end{equation}
and the final contour constraints are
\begin{equation}
    -\frac{\Delta_l}{2}<\text{Re}s<\frac{\Delta_l}{2}\,,\quad \frac{\Delta_i-\Delta_m}{2}<\text{Re}s<\frac{1+\Delta_i-\Delta_m}{2}\,,\qquad \text{Re}s<\frac{1}{2}\,,
\end{equation}
with the convergence conditions
\begin{equation}
    \Delta_{ilm}>0\,,\qquad \Delta_{mkj}>0\,,\qquad \Delta_l>1\,.
\end{equation}
Two observations are in order:
\begin{itemize}
    \item It is not always possible to find a straight contour for $s$ which satisfies these constraints. Those cases can be dealt with through analytic continuation.
    \item The integral over $\tau$ does not always converge. In fact, the integrand goes as $\tau^{\Delta_{ikj}-2s}$ and $(1-\tau)^{\Delta_{ijk}-2s}$ near the extremes of integration. This issue can be dealt with through subtractions or carefully shifting the $s$ contour.
\end{itemize}
Finally, substituting $s\to u$ we match the expression presented in (\ref{eq:K}) for the total integrated block:
\begin{equation}
    \mathcal{K}_{\Delta_i\Delta_j\Delta_k}(\Delta_l,\Delta_m)=\mathcal{K}^{R_1}_{\Delta_i\Delta_j\Delta_k}(\Delta_l,\Delta_m)+\mathcal{K}^{R_2}_{\Delta_i\Delta_j\Delta_k}(\Delta_l,\Delta_m)
\end{equation}
Overall, vertical integration contours in (\ref{eq:K}) are available when
\begin{equation}
\begin{aligned}
\label{eq:Kcontours}
    -\frac{\Delta_l}{2}<\text{Re}u<\frac{\Delta_l}{2}\,,\quad \frac{\Delta_{im}}{2}<\text{Re}u<\frac{1+\Delta_{im}}{2}\,,\qquad \text{Re}u<\frac{1}{2}\,,\qquad \text{Re}s>\frac{1}{2}\,,\qquad \text{Re}s<\frac{\Delta_l}{2}\\
    \Delta_{im}<\text{Re}c<\Delta_{ijk}\,,\qquad \text{Re}c>\Delta_i+\Delta_m+1,\qquad |\text{Re}c|<\Delta_l\,,\qquad 0<\text{Re}c+2\text{Re}s<1
\end{aligned}
\end{equation}

 \section{Integrated local blocks from an integral transform }
\label{app:local block kernel}
\begin{figure}[!t]
    \centering
    \includegraphics[width=0.85\linewidth]{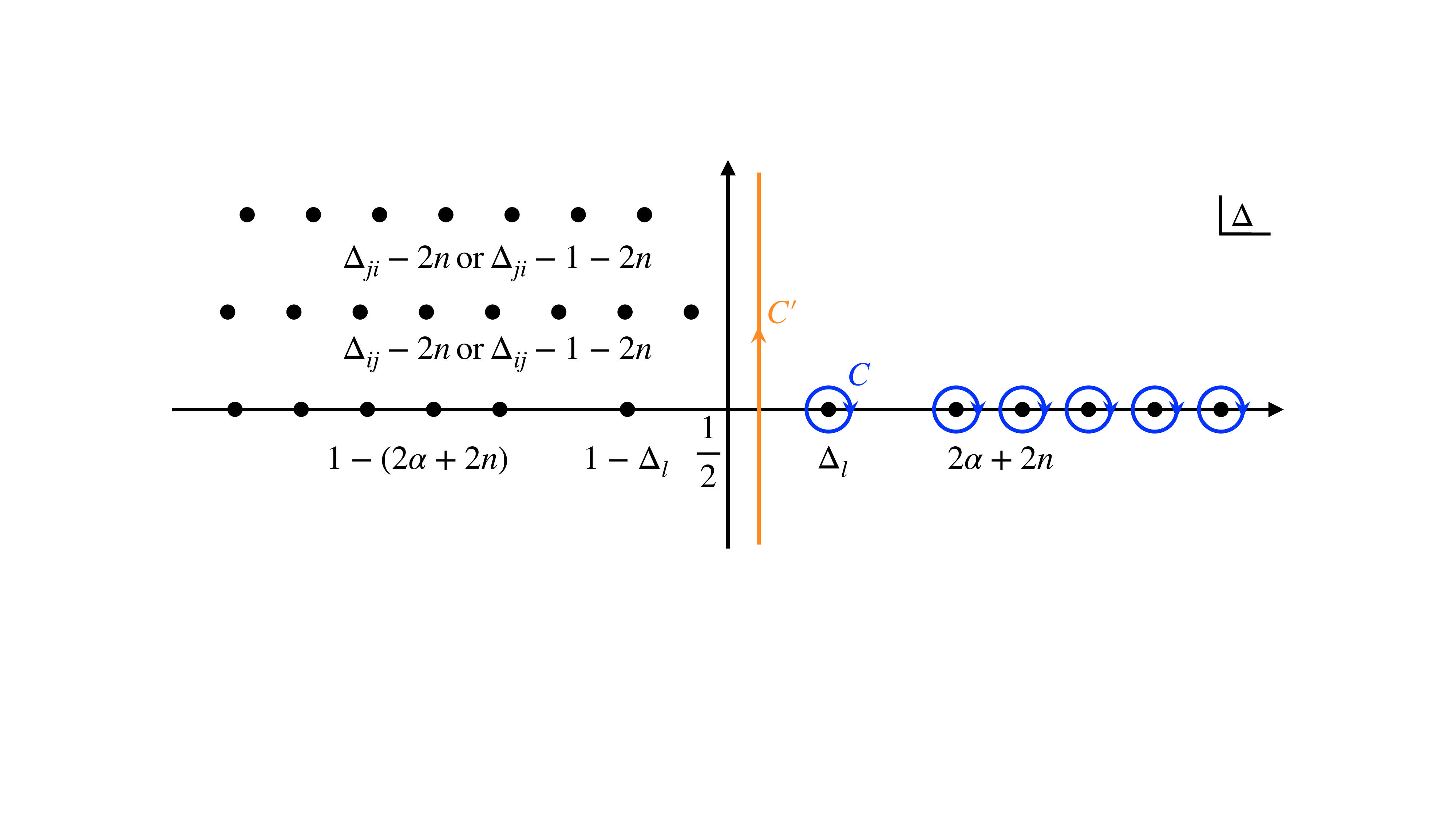}
    \caption{Integration contours $C$ (blue) and $C'$ (orange) of the integral transform \eqref{localblock:int_rep}. The poles starting from $\D_{ij}$ and $\D_{ji}$ belong to the even local block integral representation while those starting from $\D_{ij}-1$ and $\D_{ji}-1$ belong to the odd one. These poles are shifted along the imaginary axis to avoid cluttering. Notice the origin is at $\D=\tfrac{1}{2}$. This figure is taken from \cite{Loparco:2025aag}.}
    \label{fig:integral transform}
\end{figure}

As we explored in detail in \cite{Loparco:2025aag} and reviewed here in \ref{subsec:local block}, correlation functions involving bulk and boundary operators in AdS$_2$ admit decompositions in even and odd local conformal blocks. In the bulk-boundary-boundary case, the even and odd local blocks are defined as
\begin{equation}
	\begin{split}
    \label{eq:evenoddGtheta}
		G^{\Delta_i\Delta_j,(\alpha)}_{\Delta_l}(\chi)
		&=G^{\Delta_i\Delta_j}_{\Delta_l}(\chi)
		-\sum_{n=0}^{\infty}\theta^{(\alpha)}_n(\Delta_l;\Delta_{ij},\Delta_{ji})G^{\Delta_i\Delta_j}_{2\alpha+2n}(\chi)\,, \\
		\tilde{G}^{\Delta_{ij},\Delta_{ji},(\alpha)}_{\Delta_l}(\chi)
		&=\frac{1}{\sqrt{1-\chi}}\left[G^{\Delta_i\Delta_j}_{\Delta_l}(\chi)
		-\sum_{n=0}^{\infty}\theta^{(\alpha)}_n(\Delta_l;\Delta_{ij}+1,\Delta_{ji}+1)G^{\Delta_i\Delta_j}_{2\alpha+2n}(\chi)\right]\,, \\
	\end{split}
\end{equation}
where the coefficient function $\theta^{(\alpha)}_n$ is \cite{Levine:2023ywq}\footnote{We use slightly different notation than \cite[eq.(3.27)]{Levine:2023ywq}. We have set dimension $d=1$ and $\tilde\alpha_{\text{theirs}}=\alpha_{\text{ours}}-1$, $n_{\rm theirs}=n_{\rm ours}+1$.}
\begin{equation}\label{def:theta}
	\begin{split}
		\theta^{(\alpha)}_n(\Delta_l;a,b)&:=\frac{4 (-1)^n }{n!\,(2 \alpha +\Delta_l+2 n-1) (2
			\alpha +2n-\Delta_l)} \\
		&\quad\times\frac{\Gamma \left(\Delta_l+\frac{1}{2}\right) \Gamma \left(n+2 \alpha
			-\frac{1}{2}\right) \Gamma \left(\frac{a}{2}+n+\alpha \right) \Gamma
			\left(\frac{b}{2}+n+\alpha \right)}{\Gamma \left(\frac{a+\Delta_l}{2}\right)
			\Gamma \left(\frac{b+\Delta_l}{2}\right) \Gamma \left(\frac{\Delta_l-1}{2}+\alpha
			\right) \Gamma \left(\alpha -\frac{\Delta_l}{2}\right)  \Gamma \left(2 n+2 \alpha -\frac{1}{2}\right)}\,.
	\end{split}
\end{equation}
The right hand side of \ref{eq:evenoddGtheta} can be repackaged in terms of an integral transform of normal conformal blocks
\begin{equation}\label{localblock:int_rep}
	\begin{split}
		G^{\Delta_i\Delta_j,(\alpha)}_{\Delta_l}(\chi)&=
		\int_{C}\frac{d\Delta}{2\pi i}\,
		K^{(\alpha)}(\Delta,\Delta_l;\Delta_{ij},\Delta_{ji})\,
		G^{\Delta_i\Delta_j}_{\Delta}(\chi)\,, \\
		\tilde{G}^{\Delta_i\Delta_j,(\alpha)}_{\Delta_l}(\chi)
		&=\frac{1}{\sqrt{1-\chi}}
		\int_{C}\frac{d\Delta}{2\pi i}\,
		K^{(\alpha)}(\Delta,\Delta_l;\Delta_{ij}+1,\Delta_{ji}+1)\,
		G^{\Delta_i\Delta_j}_{\Delta}(\chi)\,, \\
	\end{split}
\end{equation}
where the kernel $K$ is defined by
\begin{equation}
	\begin{split}
		K^{(\alpha)}(\Delta,\Delta_l;a,b)
		:=\frac{1-2\Delta_l}{(\Delta_l-\Delta)(1-\Delta_l-\Delta)}
		\frac{\Gamma(\alpha-\tfrac{\Delta}{2})
			\Gamma(\alpha-\tfrac{1-\Delta}{2})
			\Gamma(\Delta_l-\tfrac{1}{2})
			\Gamma(\tfrac{\Delta+a}{2})
			\Gamma(\tfrac{\Delta+b}{2})}
		{\Gamma(\alpha-\tfrac{\Delta_l}{2})
			\Gamma(\alpha-\tfrac{1-\Delta_l}{2})
			\Gamma(\Delta-\tfrac{1}{2})
			\Gamma(\tfrac{\Delta_l+a}{2})
			\Gamma(\tfrac{\Delta_l+b}{2})}\,.
	\end{split}
\label{eq:local block kernel}
\end{equation}
The integration contour $C$ in \eqref{localblock:int_rep} is chosen such that the pole at $\Delta=\Delta_l$ and the poles at $\Delta=2\alpha+2n$ ($n\in\mathbb{N}_0$) are enclosed clockwise; no other poles are included (see figure \ref{fig:integral transform}). The residues of the kernel at $\Delta=2\alpha+2n$ are the coefficients $\theta^{(\alpha)}_n$ in \eqref{def:theta}:
\begin{equation}
	\underset{\Delta=2\alpha+2n}{\text{Res}}
	K^{(\alpha)}(\Delta,\Delta_l;a,b)
    =
    \theta^{(\alpha)}_n(\Delta_l;a,b)\,.
\end{equation}
In addition, the kernel has the property
\begin{equation}
\underset{\Delta=\Delta_l}{\text{Res}}
	K^{(\alpha)}(\Delta,\Delta_l;a,b)=-1\, .
\end{equation}
Throughout this appendix we will use the following shorthand notation for the even and odd kernels
\begin{equation}
    \begin{aligned}
        K_{\Delta_i,\Delta_j}^{(\alpha)}(\Delta,\Delta_l)&:=K^{(\alpha)}(\Delta,\Delta_l;\Delta_{ij},\Delta_{ji})\,,\\ 
        \tilde K_{\Delta_i,\Delta_j}^{(\alpha)}(\Delta,\Delta_l)&:=K^{(\alpha)}(\Delta,\Delta_l;\Delta_{ij}+1,\Delta_{ji}+1)
        \label{eq:defKKtilde}
    \end{aligned}
\end{equation}
In this appendix we will show that applying this integral transform to the integrated blocks (\ref{eq:I}) and (\ref{eq:J}) gives precisely the integrated local blocks (\ref{eq:I alpha}) and (\ref{eq:J alpha}). We then define the integrated local bulk-boundary-boundary-boundary block $\mathcal{K}^{(\alpha)}$ through the integral transform of the normal integrated block (\ref{eq:K}).

\subsection{Bulk-boundary-boundary ($B\partial\partial$)}
The identity we want to check is  
\begin{equation}
\II^{(\alpha)}(\D_l)\stackrel{?}{=} \int_C\frac{d\Delta}{2\pi i}K^{(\alpha)}_{\Delta_i,\Delta_i}(\Delta,\Delta_l)\mathcal{I}(\Delta)
\label{eq:IalphaK tocheck}
\end{equation}
with $\II^{(\alpha)}(\D)$ in (\ref{eq:I alpha}). In this case there is no contribution from the odd integral kernel.

The first step we take is to deform the contour from $C$ in figure \ref{fig:integral transform} to a vertical contour along the imaginary direction in the complex $\Delta$ plane. The integrated block $\mathcal{I}(\Delta)$ (see the explicit expression in (\ref{eq:I})) has poles at $\Delta=1$ and $\Delta=-\frac{1}{2}-\mathbb{N}_0$, but the latter are canceled by zeroes in the kernel $K$. We represent the final pole structure of the integrand in figure \ref{fig:polesIK}.
\begin{figure}
    \centering
    \begin{tikzpicture}[scale=2]
    \definecolor{medium-gray}{gray}{0.6}
\definecolor{light-gray}{gray}{0.92}
        \draw[-stealth,thick,medium-gray] (-2.5,0)--(3,0);
        \draw[-stealth,thick,medium-gray] (0,-1.5)--(0,1.5);
        \draw (3-3/10,1.5)--(3-3/10,1.5-3/10)--(3,1.5-3/10);
        \node[scale=1.2] at (3-3/20,1.5-3/20) {$\Delta$}; 
        \filldraw[orange] (1,0) circle (0.05);
        \node[orange] at (1,0.5) {$1$};
        \draw[orange] (1,0.05)--(1,0.4);
        \filldraw[blue] (0,0) circle (0.05);
        \node[blue] at (0,-0.5) {$0$};
        \draw[blue] (0,-0.05)--(0,-0.35);
        \filldraw[blue] (7/4,0) circle (0.05);
        \node[blue] at (7/4,0.5) {$\Delta_l$};
        \draw[blue] (7/4,0.05)--(7/4,0.4);
        \filldraw[blue] (-3/4,0) circle (0.05);
         \node[blue] at (-3/4,-0.5) {$1-\Delta_l$};
        \draw[blue] (-3/4,-0.05)--(-3/4,-0.4);
        \filldraw[blue] (2.3,0) circle (0.05);
         \node[blue] at (2.3,-0.5) {$2\alpha+2\mathbb{N}_0$};
        \draw[blue] (2.3,-0.05)--(2.3,-0.4);
        \filldraw[blue] (-1.3,0) circle (0.05);
         \node[blue] at (-1.3,0.5) {$1-(2\alpha+2\mathbb{N}_0)$};
        \draw[blue] (-1.3,0.05)--(-1.3,0.4);
        \draw[green!50!black,thick] (5/4+1/8,-1.5)--(5/4+1/8,1.5);
        \draw[-stealth,green!50!black,thick] (5/4+1/8,0.75)--(5/4+1/8,0.75+0.01);
        \draw[-stealth,green!50!black,thick] (5/4+1/8,-0.75-0.01)--(5/4+1/8,-0.75);
        \node[green!50!black] at (5/4+1/8+0.2,1.2) {$C_{\mathcal{I}}''$};
        \draw[red,thick] (7/4,0) circle (0.15);
        \draw[red,thick] (2.3,0) circle (0.15);
        \draw[-stealth,red,thick] (7/4+0.15,0)--(7/4+0.15,-0.01);
        \draw[-stealth,red,thick] (2.45,0)--(2.45,-0.01);
        \node[red] at (2.5,0.25) {$C$};
    \end{tikzpicture}
    \caption{Pole structure of the integrand in (\ref{eq:IalphaK tocheck}). In orange, poles that originate from $\mathcal{I}$, in blue, poles that originate from the kernel $K$. In red, the original contour $C$ in (\ref{eq:IalphaK tocheck}). The green vertical line is the integration contour $C_{\mathcal{I}}''$ in (\ref{eq:CppIalpha}), obtained after deforming $C$.}
    \label{fig:polesIK}
\end{figure}
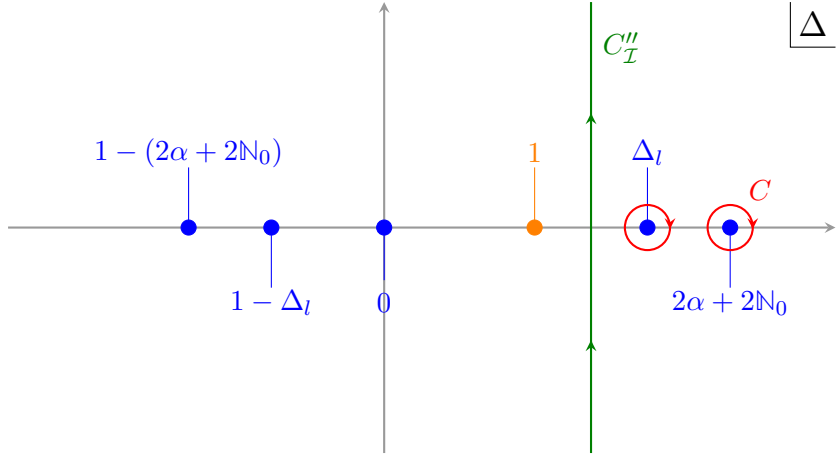
Given that in this case $\Delta_{ij}=\Delta_{ji}=0$, a vertical contour would thus have to satisfy 
\begin{equation}
    C''_\mathcal{I}:\qquad 1<\text{Re}\Delta<\min\left(\Delta_l,2\alpha\right)\,.
\end{equation}
Explicitly, the equation we want to check is thus
\begin{equation}
    \II^{(\alpha)}(\D)\stackrel{?}{=} \frac{\Gamma(\Delta_l+\frac{1}{2})}{\Gamma(\alpha-\frac{\Delta_l}{2})\Gamma(\alpha-\frac{1-\Delta_l}{2})\Gamma(\frac{\Delta_l}{2})^2}\int_{C''_\mathcal{I}}\frac{d\Delta}{2\pi i}\frac{4\sqrt{\pi}(1-2\Delta)\Gamma(\alpha-\frac{\Delta}{2})\Gamma(\alpha-\frac{1-\Delta}{2})}{\Delta(\Delta-1)(\Delta-\Delta_l)(\Delta-1+\Delta_l)}
    \label{eq:CppIalpha}
\end{equation}
The integrand is odd under $\Delta\to1-\Delta$. If we shift the vertical contour to lie at $\text{Re}\Delta=\frac{1}{2}$, the result is thus simply the residue of the integrand at $\Delta=1$:
\begin{equation}
\II^{(\alpha)}(\D)= 
  \frac{2^{3-2\alpha}\pi\Gamma(\D+\frac{1}{2})\Gamma(2\alpha-1)}{(\D-1)\Gamma(\frac{\D}{2})\Gamma(\frac{\D}{2}+1)\Gamma(\frac{\D-1}{2}+\alpha)\Gamma(\alpha-\frac{\D}{2})}\,,
\end{equation}
which is the expected result \eqref{eq:I alpha}.

\subsection{Bulk-bulk-boundary ($BB\del$)}
The identity we want to check here is
\begin{equation}
    \mathcal{J}^{(\alpha)}_{\Delta_i}(\Delta_l,\Delta_j)\stackrel{?}{=}\int_C\frac{d\Delta}{2\pi i}K^{(\alpha)}_{\Delta_i,\Delta_j}(\Delta,\Delta_l)\mathcal{J}_{\Delta_i}(\Delta_l,\Delta_j)\,,
    \label{eq:JalphaKtocheck}
\end{equation}
with $\mathcal{J}$ and $\mathcal{J}^{(\alpha)}$ respectively given in (\ref{eq:J}) and (\ref{eq:J alpha}).
Once again, there is no need for the odd kernel.

We repeat the same steps as in the previous case. The poles of $\mathcal{J}$ as a function of $\Delta$ are at $\Delta=1-2\mathbb{N}_0$ and $\Delta=-\frac{1}{2}-\mathbb{N}_0$. Moreover, zeroes of $\mathcal{J}$ at $\Delta=\Delta_{ij}-\mathbb{N}_0$ cancel a series of poles from the integral kernel.
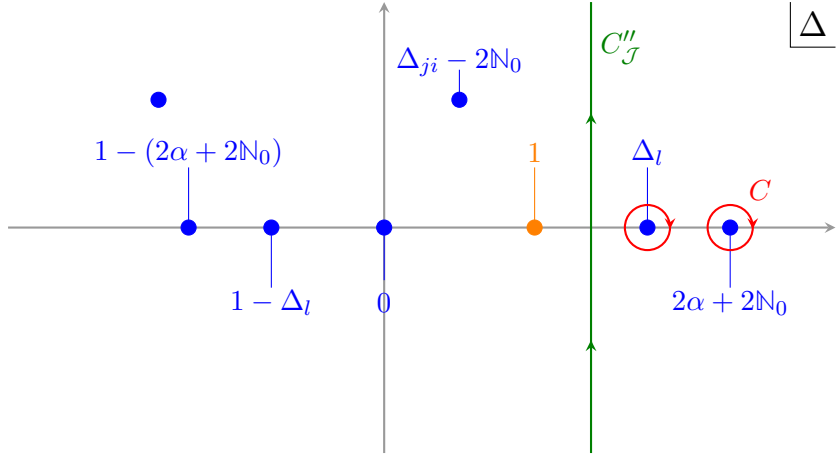
\begin{figure}
    \centering
    \begin{tikzpicture}[scale=2]
    \definecolor{medium-gray}{gray}{0.6}
\definecolor{light-gray}{gray}{0.92}
        \draw[-stealth,thick,medium-gray] (-2.5,0)--(3,0);
        \draw[-stealth,thick,medium-gray] (0,-1.5)--(0,1.5);
        \draw (3-3/10,1.5)--(3-3/10,1.5-3/10)--(3,1.5-3/10);
        \node[scale=1.2] at (3-3/20,1.5-3/20) {$\Delta$}; 
        \filldraw[orange] (1,0) circle (0.05);
        \node[orange] at (1,0.5) {$1$};
        \draw[orange] (1,0.05)--(1,0.4);
        \filldraw[blue] (0,0) circle (0.05);
        \node[blue] at (0,-0.5) {$0$};
        \draw[blue] (0,-0.05)--(0,-0.35);
        \filldraw[blue] (7/4,0) circle (0.05);
        \node[blue] at (7/4,0.5) {$\Delta_l$};
        \draw[blue] (7/4,0.05)--(7/4,0.4);
        \filldraw[blue] (-3/4,0) circle (0.05);
         \node[blue] at (-3/4,-0.5) {$1-\Delta_l$};
        \draw[blue] (-3/4,-0.05)--(-3/4,-0.4);
        \filldraw[blue] (2.3,0) circle (0.05);
         \node[blue] at (2.3,-0.5) {$2\alpha+2\mathbb{N}_0$};
        \draw[blue] (2.3,-0.05)--(2.3,-0.4);
        \filldraw[blue] (-1.3,0) circle (0.05);
         \node[blue] at (-1.3,0.5) {$1-(2\alpha+2\mathbb{N}_0)$};
        \draw[blue] (-1.3,0.05)--(-1.3,0.4);
        \draw[green!50!black,thick] (5/4+1/8,-1.5)--(5/4+1/8,1.5);
        \draw[-stealth,green!50!black,thick] (5/4+1/8,0.75)--(5/4+1/8,0.75+0.01);
        \draw[-stealth,green!50!black,thick] (5/4+1/8,-0.75-0.01)--(5/4+1/8,-0.75);
        \node[green!50!black,thick] at (5/4+1/8+0.2,1.2) {$C_{\mathcal{J}}''$};
        \filldraw[blue] (0.5,0.85) circle (0.05);
         \node[blue] at (0.5,1.1) {$\Delta_{ji}-2\mathbb{N}_0$};
        \draw[blue] (0.5,0.85)--(0.5,1.05);
        \filldraw[blue] (-3/2,0.85) circle (0.05);
        \draw[red,thick] (7/4,0) circle (0.15);
        \draw[red,thick] (2.3,0) circle (0.15);
        \draw[-stealth,red,thick] (7/4+0.15,0)--(7/4+0.15,-0.01);
        \draw[-stealth,red,thick] (2.45,0)--(2.45,-0.01);
        \node[red] at (2.5,0.25) {$C$};
    \end{tikzpicture}
    \caption{Pole structure of the integrand in (\ref{eq:IalphaK tocheck}). In orange, poles that originate from $\mathcal{J}$, in blue, poles that originate from the kernel $K$. In red, the original contour $C$ in (\ref{eq:JalphaKtocheck}). The green vertical line is the integration contour $C_{\mathcal{J}}''$ in (\ref{eq:CppIJ}), obtained after deforming $C$.}
    \label{fig:polesIJ}
\end{figure}
Overall, if $\Delta_{ji}<\Delta_l$, we can find a vertical integral contour satisfying
\begin{equation}
    C''_{\mathcal{J}}:\qquad \max\left(\Delta_{ji},1\right)<\text{Re}\Delta<\min\left(\Delta_l,2\alpha\right)\,.
    \label{eq:CppIJ}
\end{equation}
As in the previous section, we further deform the contour to lie at $\text{Re}\Delta=\frac{1}{2}$ . If we assume $\Delta_{ji}<1$\footnote{The case $\Delta_{ji}\geq1$ can be retrieved by analytic continuation of the final result.}, in this process we only pick up the pole at $\Delta=1$
\be
\underset{\D=1}{\rm Res}K^{(\a)}\JJ=
\frac{\sqrt{\pi } 4^{2-\alpha } \Gamma (2 \alpha -1) \Gamma \left(\D_l+\frac{1}{2}\right) \Gamma \left(\D_j+\frac{1}{2}\right) \Gamma \left(\frac{1+\D_{ij}}{2} \right)}
{(\D_l-1) \D_l \D_{ji}\Gamma \left(\alpha -\frac{\D_l}{2}\right) \Gamma \left(\alpha +\frac{\D_l}{2}-\frac{1}{2}\right) \Gamma \left(\frac{\D_j+\D_i+1}{2}\right) \Gamma \left(\frac{\D_{lji}}{2}\right) \Gamma \left(\frac{\D_{lij}}{2}\right)}\,.
\ee
In this case, the integrand is no longer antisymmetric under $\Delta\to1-\Delta$. Since the integral measure and the extremes of integration are symmetric under that transformation, setting $\Delta=\frac{1}{2}+i\mu$ with $\mu\in\mathbb{R}$, we can reduce the integrand to the part that is even in $\mu$.
\al{\spl{
K^{(\a)}\JJ|_{\rm even}&=\frac{4 \sqrt{\pi } \Gamma \left(\alpha -\frac{1}{2}\right) \Gamma \left(\D_l+\frac{1}{2}\right) \Gamma \left(\D_j+\frac{1}{2}\right) \Gamma \left(\alpha +\frac{\D_{ij}}{2}\right)}
{(\D_j-\D_i) \Gamma \left(\alpha -\frac{\D_l}{2}\right) \Gamma \left(\alpha +\frac{\D_l}{2}-\frac{1}{2}\right) \Gamma \left(\frac{\D_j+\D_i+1}{2}\right) \Gamma \left(\frac{\D_{lji}}{2}\right) \Gamma \left(\frac{\D_{lij}}{2}\right)}
\times
\\
&\quad\times
\frac{\Gamma \left(a_1\pm\frac{c}{2}\right) 
\Gamma \left(a_2\pm\frac{c}{2}\right)
\Gamma \left(a_3\pm\frac{c}{2}\right)}
{4 \left(b^2-c^2\right) \Gamma (\pm c) \Gamma (a_1+a_2) \Gamma (a_1+a_3) \Gamma (a_2+a_3)}
}}
with the parameters $a_1=-\frac14,\,a_2=\a-\frac14,\,a_3=\frac14(1+2\D_{ij}),\,b=\frac12(1-2\D_l)$ and the integration variable $c\equiv i\m$. Notice that $b<0$ in our case. We have also used the abbreviation $\G(x\pm y)\equiv\G(x+y)\G(x-y)$. The reason for rewriting the integrand in the above form is because now we can apply the following Mellin-Barnes type integral formula\footnote{This formula is derived by using the fact that the integral is meromorphic in $a_1$, so it can be reconstructed from the poles and residues in $a_1$. Here we briefly summarize the idea and see \cite[sec.2]{Penedones:2010ue} for more details. The poles in $a_1$ only come from the (double) pinching of the $c$-contour between poles $\pm c=2a_1+2n$ and $\pm c=b$ where the signs are correlated (recall $b<0$). The pinching between $\pm c = 2a_1+2n$ with other poles are canceled by the denominator. To extract the poles and residues from the pinching between, e.g., $c=2a_1+2n$ and $c=b$ we shift the $c$-contour to the left of $b$ and pick up the pole at $c=b$. Now the contour integral is regular and the pole in $a_1$ at $a_1=b/2-n$ can be extracted from the residue at $c=b$. Summing over all the poles gives the formula \eqref{eq:Mellin-Barnes formula}.}
\al{\spl{
&\int_{-i\infty}^{i\infty} \frac{dc}{2\pi i} 
\frac{\Gamma \left(a_1\pm\frac{c}{2}\right) 
\Gamma \left(a_2\pm\frac{c}{2}\right)
\Gamma \left(a_3\pm\frac{c}{2}\right)}
{4 \left(b^2-c^2\right) \Gamma (\pm c) \Gamma (a_1+a_2) \Gamma (a_1+a_3) \Gamma (a_2+a_3)}
=
\\
&\quad=
\frac{\Gamma \left(\frac{2 a_2-b}{2} \right) \Gamma \left(\frac{2 a_3-b}{2} \right)}{2 (2 a_1-b) \Gamma (1-b) \Gamma (a_2+a_3)}
{}_3\tilde{F}_2\left( {\begin{array}{*{20}{c}}
{a_1-\frac{b}{2},-a_2-\frac{b}{2}+1,-a_3-\frac{b}{2}+1}\\
{1-b,a_1-\frac{b}{2}+1}
\end{array};1} \right)
}\label{eq:Mellin-Barnes formula}}
for $a_1,a_2,a_3>0$ and $b<0$. (For the $b>0$ case one simply flips the signs of $b$ on the RHS.) Since in our case $a_1=-\frac14$, we need to pick up two extra poles at $c=\pm 2a_1$. The final result is
\al{\spl{
&\int_{C''_{\mathcal{J}}}
\frac{d\D}{2\pi i} K^{(\alpha)}_{\Delta_i,\Delta_j}(\Delta,\Delta_l) \JJ_{\D_i}(\D,\D_j)=
\\
&\quad=
\frac{\sqrt{\pi } \Gamma \left(\alpha -\frac{1}{2}\right) \Gamma \left(\frac{\D_l-1}{2}\right) \Gamma \left(\D_l+\frac{1}{2}\right) \Gamma \left(\D_j+\frac{1}{2}\right)}
{(\D_j-\D_i) \Gamma \left(\alpha -\frac{\D_l}{2}\right) \Gamma \left(\frac{\D_j+\D_i+1}{2} \right) \Gamma \left(\frac{\D_{lji}}{2}\right)}
\,{}_3\tilde{F}_2\left( {\begin{array}{*{20}{c}}
{\frac{\D_l-1}{2},\frac{\D_l+2-2\a}{2},\frac{\D_{lji}+1}{2}}\\
{\frac{\D_l+1}{2},\D_l+\frac{1}{2}}
\end{array};1} \right)\,,
}
\label{eq:J alpha in app}
}
which indeed gives \eqref{eq:J alpha} upon using the identity
\al{
{}_3F_2\left( {\begin{array}{*{20}{c}}
{a_1,a_2,a_3}\\
{b_1,b_2}
\end{array};1} \right)
=\frac{\Gamma \left(b_1\right) \Gamma \left(-a_1-a_2-a_3+b_1+b_2\right)}{\Gamma \left(b_1-a_1\right) \Gamma \left(-a_2-a_3+b_1+b_2\right)}
{}_3F_2\left( {\begin{array}{*{20}{c}}
{a_1,b_2-a_2,b_2-a_3}\\
{b_2,-a_2-a_3+b_1+b_2}
\end{array};1} \right)\,.
}

\subsection{Bulk-boundary-boundary-boundary ($B\del\del\del$)}
\label{subsec:bkbdbdbd integral transform}
In this case, we need to use both the even and the odd kernels, as discussed in section \ref{subsec:localblock_bkbdbdbd}. Let us focus on only one of the elements appearing in the even and odd integrated local blocks from eq. (\ref{eq:evenoddK})
\begin{equation}
    \bar{\mathcal{K}}^{(\alpha)}_{\Delta_i\Delta_j\Delta_k}(\Delta_l,\Delta_m):=\int_C\frac{d\Delta}{2\pi i}K^{(\alpha)}_{\Delta_i,\Delta_m}(\Delta,\Delta_l)\mathcal{K}_{\Delta_i\Delta_j\Delta_k}(\Delta,\Delta_m)
    \label{eq:Kbaralphadef}
\end{equation}
with $\mathcal{K}$ in (\ref{eq:K}). There is little difference between the even and odd kernels, so we will treat here the even kernel and highlight the difference with the odd case at the end.

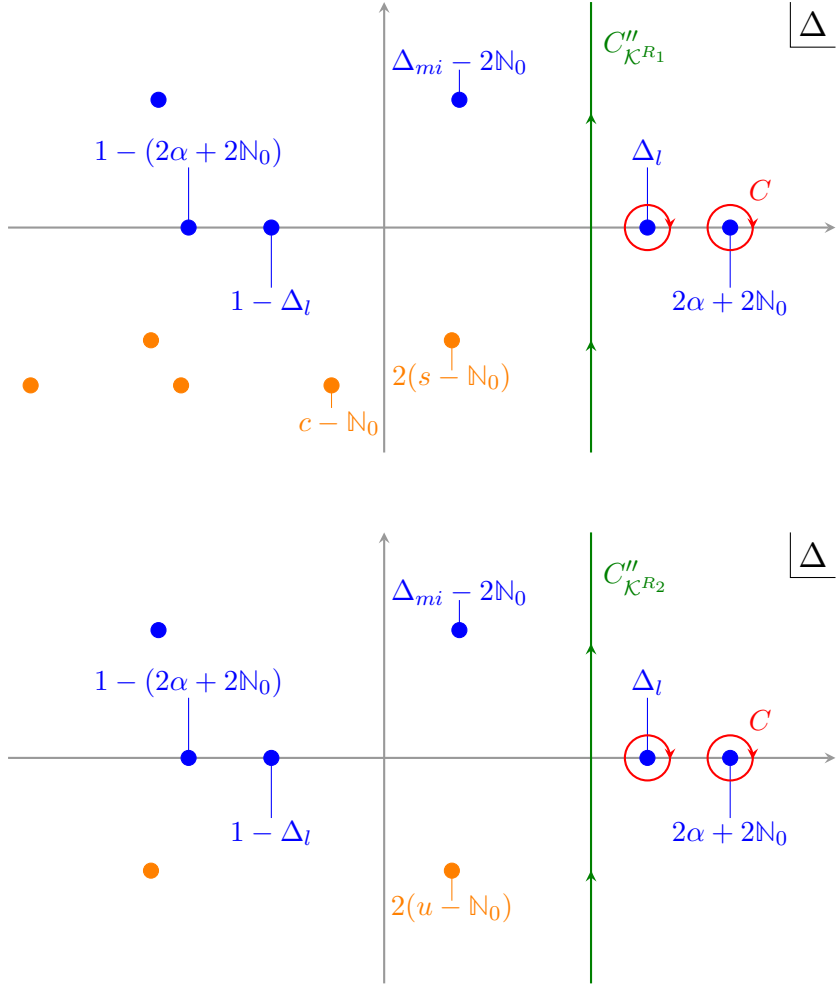
\begin{figure}
    \centering
    \begin{tikzpicture}[scale=2]
    \definecolor{medium-gray}{gray}{0.6}
\definecolor{light-gray}{gray}{0.92}
        \draw[-stealth,thick,medium-gray] (-2.5,0)--(3,0);
        \draw[-stealth,thick,medium-gray] (0,-1.5)--(0,1.5);
        \draw (3-3/10,1.5)--(3-3/10,1.5-3/10)--(3,1.5-3/10);
        \node[scale=1.2] at (3-3/20,1.5-3/20) {$\Delta$}; 
        \filldraw[blue] (7/4,0) circle (0.05);
        \node[blue] at (7/4,0.5) {$\Delta_l$};
        \draw[blue] (7/4,0.05)--(7/4,0.4);
        \filldraw[blue] (-3/4,0) circle (0.05);
         \node[blue] at (-3/4,-0.5) {$1-\Delta_l$};
        \draw[blue] (-3/4,-0.05)--(-3/4,-0.4);
        \filldraw[blue] (2.3,0) circle (0.05);
         \node[blue] at (2.3,-0.5) {$2\alpha+2\mathbb{N}_0$};
        \draw[blue] (2.3,-0.05)--(2.3,-0.4);
        \filldraw[blue] (-1.3,0) circle (0.05);
         \node[blue] at (-1.3,0.5) {$1-(2\alpha+2\mathbb{N}_0)$};
        \draw[blue] (-1.3,0.05)--(-1.3,0.4);
        \draw[green!50!black,thick] (5/4+1/8,-1.5)--(5/4+1/8,1.5);
        \draw[-stealth,green!50!black,thick] (5/4+1/8,0.75)--(5/4+1/8,0.75+0.01);
        \draw[-stealth,green!50!black,thick] (5/4+1/8,-0.75-0.01)--(5/4+1/8,-0.75);
        \node[green!50!black,thick] at (5/4+1/8+0.3,1.2) {$C_{\mathcal{K}^{R_1}}''$};
        \filldraw[blue] (0.5,0.85) circle (0.05);
         \node[blue] at (0.5,1.1) {$\Delta_{mi}-2\mathbb{N}_0$};
        \draw[blue] (0.5,0.85)--(0.5,1.05);
        \filldraw[blue] (-3/2,0.85) circle (0.05);
        \filldraw[orange] (0.45,-0.75) circle (0.05);
        \node[orange] at (0.45,-1) {$2(s-\mathbb{N}_0)$};
        \draw[orange] (0.45,-0.75)--(0.45,-0.95);
        \filldraw[orange] (0.45-2,-0.75) circle (0.05);
         \filldraw[orange] (-0.35,-1.05) circle (0.05);
        \node[orange] at (-0.3,-1.3) {$c-\mathbb{N}_0$};
        \draw[orange] (-0.35,-1.05)--(-0.35,-1.2);
        \filldraw[orange] (-1.35,-1.05) circle (0.05);
        \filldraw[orange] (-2.35,-1.05) circle (0.05);
            \draw[red,thick] (7/4,0) circle (0.15);
        \draw[red,thick] (2.3,0) circle (0.15);
        \draw[-stealth,red,thick] (7/4+0.15,0)--(7/4+0.15,-0.01);
        \draw[-stealth,red,thick] (2.45,0)--(2.45,-0.01);
        \node[red] at (2.5,0.25) {$C$};
        \filldraw[white] (-2.5,-1.5)--(3,-1.5)--(3,-2)--(-2.5,-2);
    \end{tikzpicture}
    \begin{tikzpicture}[scale=2]
    \definecolor{medium-gray}{gray}{0.6}
\definecolor{light-gray}{gray}{0.92}
        \draw[-stealth,thick,medium-gray] (-2.5,0)--(3,0);
        \draw[-stealth,thick,medium-gray] (0,-1.5)--(0,1.5);
        \draw (3-3/10,1.5)--(3-3/10,1.5-3/10)--(3,1.5-3/10);
        \node[scale=1.2] at (3-3/20,1.5-3/20) {$\Delta$}; 
        \filldraw[blue] (7/4,0) circle (0.05);
        \node[blue] at (7/4,0.5) {$\Delta_l$};
        \draw[blue] (7/4,0.05)--(7/4,0.4);
        \filldraw[blue] (-3/4,0) circle (0.05);
         \node[blue] at (-3/4,-0.5) {$1-\Delta_l$};
        \draw[blue] (-3/4,-0.05)--(-3/4,-0.4);
        \filldraw[blue] (2.3,0) circle (0.05);
         \node[blue] at (2.3,-0.5) {$2\alpha+2\mathbb{N}_0$};
        \draw[blue] (2.3,-0.05)--(2.3,-0.4);
        \filldraw[blue] (-1.3,0) circle (0.05);
         \node[blue] at (-1.3,0.5) {$1-(2\alpha+2\mathbb{N}_0)$};
        \draw[blue] (-1.3,0.05)--(-1.3,0.4);
        \draw[green!50!black,thick] (5/4+1/8,-1.5)--(5/4+1/8,1.5);
        \draw[-stealth,green!50!black,thick] (5/4+1/8,0.75)--(5/4+1/8,0.75+0.01);
        \draw[-stealth,green!50!black,thick] (5/4+1/8,-0.75-0.01)--(5/4+1/8,-0.75);
        \node[green!50!black,thick] at (5/4+1/8+0.3,1.2) {$C_{\mathcal{K}^{R_2}}''$};
        \filldraw[blue] (0.5,0.85) circle (0.05);
         \node[blue] at (0.5,1.1) {$\Delta_{mi}-2\mathbb{N}_0$};
        \draw[blue] (0.5,0.85)--(0.5,1.05);
        \filldraw[blue] (-3/2,0.85) circle (0.05);
        \filldraw[orange] (0.45,-0.75) circle (0.05);
        \node[orange] at (0.45,-1) {$2(u-\mathbb{N}_0)$};
        \draw[orange] (0.45,-0.75)--(0.45,-0.95);
        \filldraw[orange] (0.45-2,-0.75) circle (0.05);
            \draw[red,thick] (7/4,0) circle (0.15);
        \draw[red,thick] (2.3,0) circle (0.15);
        \draw[-stealth,red,thick] (7/4+0.15,0)--(7/4+0.15,-0.01);
        \draw[-stealth,red,thick] (2.45,0)--(2.45,-0.01);
        \node[red] at (2.5,0.25) {$C$};
    \end{tikzpicture}
    \caption{Pole structure of the integrands in (\ref{eq:KR1R2alpha}). On the top (bottom), the pole structure for $R_1$ ($R_2$). In orange, poles originating from $\mathcal{K}$. In blue, poles that originate from the kernel $K$. In red, the original contour $C$ in (\ref{eq:Kbaralphadef}).
    The green vertical line is the integration contour $C_{\mathcal{K}^{R_{1,2}}}''$ in (\ref{eq:CppKR1R2}). This is the structure for the even coefficient. The odd coefficient has $\Delta_{mi}\to\Delta_{mi}-1$.}
    \label{fig:polesIK}
\end{figure}
The pole structure in the $\Delta$ plane is a bit more complicated for this case, due to the fact that we know $\mathcal{K}$ only through an integral representation. Moreover, as discussed in \ref{subsec:intbkbdbdbd}, this integral representation naturally splits in two terms associated to the integrals over regions $R_1$ and $R_2$ from figure \ref{fig:pizza slices}. In the current section it will be useful to consider the integral transform of these two terms separately, defining 
\begin{equation}
\bar{\mathcal{K}}^{R_{1,2}(\alpha)}_{\Delta_i\Delta_j\Delta_k}(\Delta_l,\Delta_m):=\int_C\frac{d\Delta}{2\pi i}K^{(\alpha)}_{\Delta_i,\Delta_m}(\Delta,\Delta_l)\mathcal{K}^{R_{1,2}}_{\Delta_i\Delta_j\Delta_k}(\Delta,\Delta_m)
\label{eq:KR1R2alpha}
\end{equation}
with $\mathcal{K}^{R_1}$ and $\mathcal{K}^{R_2}$  given in (\ref{eq:KR1fin}) and (\ref{eq:KR2fin}) respectively. We will consider the full integrand, in other words we will study a function of $\Delta$ but also of $s$, $c$, $t$, $u$, the integration variables in the definition of $\mathcal{K}^{R_{1,2}}$.
We will assume that the scaling dimensions of all operators are in a regime that allows for the conditions in (\ref{eq:Kcontours}) to be satisfied by some vertical contours for the Mellin variables $s$, $c$ and $u$. Taking all of this into account, the vertical contours for the integration variable $\Delta$ should then satisfy
\begin{equation}
    \begin{aligned}
        C''_{\mathcal{K}^{R_1}}:&\qquad \max\left(2\text{Re}s,\Delta_{mi},\text{Re}c\right)<\text{Re}\Delta<\min\left(\Delta_l,2\alpha\right)\,,\\
        C''_{\mathcal{K}^{R_2}}:&\qquad \max\left(2\text{Re}u,\Delta_{mi},1-\Delta_l\right)<\text{Re}\Delta<\min\left(\Delta_l,2\alpha\right)\,.
        \label{eq:CppKR1R2}
    \end{aligned}
\end{equation}
We represent in figure \ref{fig:polesIK} the full pole structure of the integrands in eq. (\ref{eq:KR1R2alpha}) as a function of $\Delta$. 

For the odd kernel, the contour restrictions are the same up to a shift $\Delta_{mi}\to\Delta_{mi}-1$.

\section{Convergence of the flow equations and a lower bound on $\alpha$}
\label{sec:convergence of flow eqn}
In section \ref{sec:integrals of blocks}, each of the local blocks has a parameter $\alpha$ that can be tuned as the infinite sums in the flow equations do not depend on its value. In fact, there is a lower bound on $\a$ for the local block expansions of correlation functions to be valid. For the case of $B\del\del$ this is explained in \cite{Meineri:2023mps,Levine:2023ywq} and the lower bound is
\be
\a>\D_i + \frac{{\D}_{\hat\Phi}}{2}\,,
\ee
where $\D_i$ is the scaling dimension of the two identical boundary operators and ${\Delta}_{\hat\Phi}$ is the scaling dimension of the \emph{bulk} UV CFT. 

In this section we derive the lower bound of $\alpha$ for the flow equation \eqref{eq:dimension flow}, reproducing the above result, and also determine lower bounds on the value of $\alpha$ for the flow equation \eqref{eq:BOE flow}. 
This is achieved by analyzing the asymptotic behavior of the QFT data and the integrated blocks, in the limit where the boundary operators appearing in the BOE of the deformation bulk operator $\hat\Phi$ become heavy.

\subsection{Asymptotics of $\bb{\hat{\mathcal{O}}}{i}$}
The asymptotic behavior of the BOE coefficient $\bb{\hat{\OO}}{i}$ can be determined by studying the bulk two-point function. In particular, we consider a bulk operator $\hat{\mathcal{O}}$ with (bulk) UV dimension  $0<\Delta_{\hat{\mathcal{O}}}<2$ whose two-point function is,  close to the coincident point singularity,
\begin{equation}
    \braket{\hat{\mathcal{O}}(x)\hat{\mathcal{O}}(y)}\sim \frac{1}{(1-\eta)^{2\Delta_{\hat{\mathcal{O}}}}}\, ,\qquad(\eta\rightarrow1^-)\, ,
\end{equation}
with $\eta$ as defined in \eqref{def:eta}. This singularity has to be reproduced by the decomposition in conformal blocks \eqref{def:bkbk block}:
\begin{equation}
\sum_{i}\left(\bb{\hat{\mathcal{\OO}}}{i}\right)^2\eta^{\Delta_i}\,\hyperF{\Delta_i}{\frac{1}{2}}{\Delta_i+\frac{1}{2}}{\eta^2}\sim \frac{1}{(1-\eta)^{2\Delta_{\hat{\mathcal{O}}}}}.
\end{equation}
If we define $z=\frac{4\eta}{(1+\eta)^2}$, we have the identity
\begin{equation}
    (4\eta)^{\Delta_i}\hyperF{\Delta_i}{\frac{1}{2}}{\Delta_i+\frac{1}{2}}{\eta^2}=z^{\Delta_i} \hyperF{\Delta_i}{\Delta_i}{2\Delta_i}{z}.
\end{equation}
This implies 
\begin{equation}
    \sum_{i}\left(\bb{\hat{\mathcal{\OO}}}{i}\right)^2\left(\frac{z}{4}\right)^{\Delta_i} \hyperF{\Delta_i}{\Delta_i}{2\Delta_i}{z}\sim (2\sqrt{1-z})^{-2\Delta_{\hat{\mathcal{O}}}} \, ,\qquad(z\rightarrow1^-)\, .\label{eq:E5}
\end{equation}
The singularity is controlled by large $\Delta_i$, and when $\Delta_i$ is large, we have the approximation (see appendix A of \cite{Fitzpatrick:2012yx} for more details)
\begin{equation}
    z^{\Delta_i} \,\hyperF{\Delta_i}{\Delta_i}{2\Delta_i}{z}\approx4^\Delta_i\sqrt{\frac{\Delta_i}{\pi}}K_0(2\Delta_i\sqrt{1-z})\, ,\qquad \Delta_i\to\infty
\end{equation}
Comparing with \eqref{eq:E5}, we obtain
\begin{equation}
    \sum_{i}\left(\bb{\hat{\mathcal{\OO}}}{i}\right)^2\sqrt{\frac{\Delta_i}{\pi}}K_0\left(2\Delta_i\sqrt{1-z}\right)\sim (2\sqrt{1-z})^{-2\Delta_{\hat{\mathcal{O}}}}\, ,\qquad (z\rightarrow1^-)\, .\label{eq:E7}
\end{equation}
To be able to use Tauberian theorems, we change variables as $Y= (2\sqrt{1-z})^{-1}$ in \eqref{eq:E7} such that 
\begin{equation}
    \sum_i\left(\bb{\hat{\mathcal{\OO}}}{i}\right)^2\sqrt{\frac{\Delta_i}{\pi}}K_0\left(\frac{\Delta_i}{Y}\right)\sim Y^{2\Delta_{\hat{\mathcal{O}}}}\, ,\qquad (Y\rightarrow\infty)\, .\label{eq:E8}
\end{equation} We then want to use the Tauberian theorem presented in \cite{Qiao:2017xif} in equation (4.7). We define $w_1(x)=K_0(x)$ and $w_2(x)= \Theta(0\leqslant x\leqslant1)$. We moreover define 
\begin{equation}
    q(\Delta) = \sum_{i}\left(\bb{\hat{\mathcal{\OO}}}{i}\right)^2\sqrt{\frac{\D_i}{\pi}}\delta(\D-\D_i)\, .
\end{equation}
We can then rewrite \eqref{eq:E8} as
\begin{equation}
   \int_0^\infty d\Delta\, q(\Delta)w_1\left(\frac{\Delta}{Y}\right)\sim Y^{2\Delta_{\hat{\OO}}}\, ,\qquad (Y\rightarrow \infty)\, .
\end{equation}
Now using the actual (Wiener-type) Tauberian theorem, we get 
\begin{equation}
   \int_0^\infty d\Delta\, q(\Delta)w_2\left(\frac{\Delta}{Y}\right)\sim \frac{I_2}{I_1}Y^{2\Delta_{\hat{\OO}}}\, ,\qquad (Y\rightarrow \infty)\, ,
\end{equation}
with $I_i = \int_0^\infty dx x^{2\Delta_{\hat{\OO}}-1}w_i(x)$. Setting $Y=\Delta^*$, we obtain 

\begin{equation}
    \sum_{\Delta_i\leqslant\Delta_*}\left(\bb{\hat{\mathcal{\OO}}}{i}\right)^2\sqrt{\frac{\Delta_i}{\pi}}\sim \frac{2^{1-2\Delta_{\hat{\mathcal{O}}}}}{\Gamma(\Delta_{\hat{\mathcal{O}}})\Gamma(\Delta_{\hat{\mathcal{O}}}+1)}\Delta_*^{2\Delta_{\hat{\mathcal{O}}}}\,,\qquad(\Delta_*\rightarrow\infty)\, .\label{eq:E13}
\end{equation}
If we assume that $\bb{\hat{\OO}}{i}$ has power law behavior at large $\Delta_i$, we can obtain its asymptotics using \eqref{eq:E13}. This yields 
\begin{equation}
    \bb{\hat{\mathcal{O}}}{i}\sim\#\Delta_i^{\Delta_{\hat{\mathcal{O}}}-3/4}\, ,\qquad(\Delta_i\rightarrow\infty).
\end{equation}
When $\Delta_{\hat{\mathcal{O}}}=0$, the situation is very different because we resum logarithms. Nonetheless, the averaged power-law growth of $\bb{\hat{\mathcal{O}}}{i}$ cannot be faster than $\Delta_i^{-3/4}$ as otherwise the sum will produce power-law divergences.

\subsection{Asymptotics of $C_{ijk}$}
Consider the four-point function of two sets of identical boundary operators in the specific configuration:
\begin{equation}
    \begin{split}
\braket{\mathcal{O}_i(0)\mathcal{O}_j(z)\mathcal{O}_j(1)\mathcal{O}_i(\infty)}&=z^{-\Delta_i-\Delta_j}G(z)\, , \\
    G(z)&=\sum_{k}C_{ijk}^2 z^{\Delta_k}\hyperF{\Delta_{kji}}{\Delta_{kji}}{2\Delta_k}{z}\, . \\
    \end{split}
\end{equation}
When $z\rightarrow1^-$, the $t$-channel expansion yields 
\begin{equation}
    z^{-\Delta_i-\Delta_j}G(z)\sim(1-z)^{2\Delta_j}\, .
\end{equation}
After approximating the conformal blocks by Bessel functions (see  equation (6.18) in \cite{Qiao:2017xif}), we get
\begin{equation}
    \sum_{k}C_{ijk}^2 4^{\Delta_k}\sqrt{\frac{\Delta_k}{\pi}}K_{2\Delta_{ij}}(2\Delta_k\sqrt{1-z})\sim(1-z)^{-\Delta_i-\Delta_j}\, ,\qquad(z\rightarrow1^-)\, .
\end{equation}
Then using Tauberian theorem we get
\begin{equation}
    \sum_{\Delta_k\leqslant\Delta_*}C_{ijk}^2 4^{\Delta_k}\sqrt{\frac{\Delta_k}{\pi}}\sim\frac{2}{(\Delta_i+\Delta_j)\Gamma(2\Delta_i)\Gamma(2\Delta_j)}\Delta_*^{2\Delta_i+2\Delta_j}\, ,\qquad(\Delta_*\rightarrow\infty)\, .
\end{equation}
Assuming again a power law behavior, this suggests the asymptotics of $C_{ijk}$
\begin{equation}
    C_{ijk}\sim \#2^{-\Delta_k}\Delta_k^{\Delta_i+\Delta_j-3/4}\, ,\qquad(\Delta_k\rightarrow\infty)\, .
\end{equation}

\subsection{Asymptotics of scaling dimension flow equation}
\label{subsec:asymptoticsdeltaflow}
\subsubsection{With normal blocks}
The starting point is the integral over conformal blocks derived in \eqref{eq:I}, which is 
\begin{equation}
\II(\D_l)
=\frac{2 \sqrt{\pi } \Gamma \left(\Delta_l +\frac{1}{2}\right)}{(\Delta_l -1) \Gamma \left(\frac{\Delta_l}{2}\right) \Gamma \left(\frac{\Delta_l}{2}+1\right)}\,.
\end{equation}
Using Stirling's approximation, putting it together with the asymptotics of $b$ and $C$, we get
\begin{equation}
    \bb{\hat\Phi}{l} C_{iil}\mathcal{I}(\Delta_l)\sim \Delta_l^{2 \Delta_i+\Delta_{\hat{\Phi}} -\frac{5}{2}}\, ,\qquad (\D_l\rightarrow \infty)\, .
\end{equation}
The error when truncating at $\Delta_{\text{max}}$ is thus
\begin{equation}
    \text{error}=\int_{\Delta_{\text{max}}}^\infty d\Delta_l\ \Delta_l^{2 \Delta_i+\Delta_{\hat{\Phi}} -\frac{5}{2}}\sim \Delta_{\text{max}}^{2\Delta_i+\Delta_{\hat\Phi}-\frac{3}{2}}\,.
\end{equation}
For a free theory where $\hat\Phi=\phi^2$, given that the UV dimensions are $\Delta_{\hat{\Phi}}=0$, we get that the flow equation with normal blocks converges only if $\Delta_{i}<\frac{3}{4}$.

\subsubsection{With local blocks}
We can do the same exercise but with local blocks instead of normal blocks. In this case, we want to understand the large $\Delta_l$ and $\alpha$ behavior of the integral over the local $\mathcal{I}^{(\alpha)}(\Delta_l)$ that was derived in \eqref{eq:I alpha}. The original formula is 
\begin{equation}
    \II^{(\a)}(\Delta_l)=\frac{\pi\Gamma(\Delta_l+\frac{1}{2})\Gamma(2\alpha-1)}{2^{2\alpha-3}(\Delta_l-1)\Gamma(\frac{\Delta_l}{2})\Gamma(\frac{\Delta_l}{2}+1)\Gamma(\frac{\Delta_l-1}{2}+\alpha)\Gamma(\alpha-\frac{\Delta_l}{2})}\,,
\end{equation}
We first need to determine for what value of $\alpha$ does the sum rule converge. To understand this, we fix $\alpha$ and take $\Delta_l$ to infinity using the Stirling approximation. We obtain
\begin{equation}
    \bb{\hat{\Phi}}{l}C_{iil}\II^{(\a)}(\Delta_l)\underset{\text{fixed }\alpha}{\sim} \Delta_l^{2(\Delta_i-\alpha)-1+\Delta_{\hat\Phi}}\,,\qquad (\Delta_l\rightarrow\infty)\, .
\end{equation}
In this case, the error that we make with a given truncation $\Delta_{\text{max}}$ is 
\begin{equation}
    \text{error}\sim\int_{\Delta_{\text{max}}}^\infty d\Delta_l\ \Delta_l^{2(\Delta_i-\alpha)+\Delta_{\hat\Phi}-1}\sim \Delta_{\text{max}}^{2(\Delta_i-\alpha)+\Delta_{\hat\Phi}}\, ,
\end{equation}
such that the sum has power-law convergence for $\alpha>\Delta_i+\frac{1}{2}\Delta_{\hat{\Phi}}$.

We would like to understand the optimal value of $\alpha$ for a given truncation $\Delta_{\text{max}}$ and a given fixed $\Delta_i$. To do that, we take $\alpha$ to be of the same order of $\Delta_l$ and take both large. This is harder because of the factor $\Gamma(\alpha-\frac{\Delta_l}{2})$. The large $\alpha$ and $\Delta_l$ expression is indeed
\begin{equation}
    \bb{\hat{\Phi}}{l}C_{iil} \II^{(\a)}(\Delta_l)\sim \frac{\alpha ^{2 \alpha -\frac{3}{2}} e^{\frac{\Delta_l}{2}-\alpha } \left(\alpha +\frac{\Delta_l}{2}\right)^{-\alpha -\frac{\Delta_l}{2}+1} \Delta_l^{2 \Delta_i+\Delta_{\hat{\Phi}} -\frac{5}{2}}}{\Gamma \left(\alpha -\frac{\Delta_l}{2}\right)}
\, ,\qquad (\Delta_l\rightarrow\infty,\alpha\rightarrow\infty)\, .\end{equation}
To proceed we take $\Delta_l=\Delta_{\text{max}}+\beta$ and $\alpha=\gamma \Delta_{\text{max}}$ and send $\Delta_{\text{max}}$ to infinity. We need to distinguish two cases:
\begin{itemize}
    \item $\gamma>\frac{1}{2}$. This is the region of integration before the zeroes. We get
    \begin{equation}
   \left|\bb{\hat{\Phi}}{l}C_{iil} \II^{(\a)}(\Delta_l)\right|\sim \frac{(2\gamma)^{2\gamma\Delta_{\text{max}}}}{2\gamma^{\frac{3}{2}}}\frac{(2\gamma-1)^{\frac{1+(1-2\gamma)\Delta_{\text{max}}}{2}}}{(2\gamma+1)^{\frac{(1+2\gamma)\Delta_{\text{max}}}{2}-1}}\Delta_{\text{max}}^{-\frac{5}{2}+2\Delta_i+\Delta_{\hat\Phi}}\left(\frac{e(2\gamma-1)}{2\gamma+1}\right)^\frac{\beta}{2}\,.
    \end{equation}
    We integrate $\beta\in(0,\infty)$ to get the tail. This converges if $\gamma\in(\frac{1}{2},\frac{1}{2}\frac{e+1}{e-1})\sim(\frac{1}{2},1.08)\,.$ We get
    \begin{equation}
        \text{error}\sim\frac{(2\gamma)^{2\gamma\Delta_{\text{max}}}}{\gamma^{\frac{3}{2}}}\frac{(2\gamma-1)^{\frac{1+(1-2\gamma)\Delta_{\text{max}}}{2}}}{(2\gamma+1)^{\frac{(1+2\gamma)\Delta_{\text{max}}}{2}-1}}\Delta_{\text{max}}^{-\frac{5}{2}+2\Delta_i+\Delta_{\hat\Phi}}\frac{1}{\log\left(\frac{2\gamma+1}{2\gamma-1}\right)-1}
\, .    \end{equation}
    This is a monotonic function of $\gamma$, with minimum at $\gamma=\frac{1}{2}$. At the same time, the derivative there diverges.
    \item $0<\gamma<\frac{1}{2}$. This is the region of integration with lots of zeroes from the integrated block.  We get
    \begin{equation}
    \begin{aligned}
       \left| \bb{\hat{\Phi}}{l}C_{iil} \II^{(\a)}(\Delta_l)\right|\sim &-\frac{(2\gamma)^{2\gamma\Delta_{\text{max}}}}{\gamma^{\frac{3}{2}}}\frac{(1-2\gamma)^{\frac{1+(1-2\gamma)\Delta_{\text{max}}}{2}}}{(1+2\gamma)^{\frac{(1+2\gamma)\Delta_{\text{max}}}{2}-1}}\Delta_{\text{max}}^{-\frac{5}{2}+2\Delta_i+\Delta_{\hat\Phi}}\\
        &\times\left(\frac{e(1-2\gamma)}{1+2\gamma}\right)^{\frac{\beta}{2}}\left|\sin\left[\frac{\pi}{2}(\beta+\Delta_{\text{max}}(1-2\gamma))\right]\right|\, .
    \end{aligned}
    \end{equation}
   The integral over $\beta\in(0,\infty)$ converges if $\gamma\in(\frac{e-1}{2(e+1)},\frac{1}{2})\sim(0.23,\frac{1}{2})$. We estimate the sine by 1, and we get
    \begin{equation}
    \begin{aligned}
        \text{error}\sim &\frac{(2\gamma)^{2\gamma\Delta_{\text{max}}}}{\gamma^{\frac{3}{2}}}\frac{(1-2\gamma)^{\frac{1+(1-2\gamma)\Delta_{\text{max}}}{2}}}{(1+2\gamma)^{\frac{(1+2\gamma)\Delta_{\text{max}}}{2}-1}}\Delta_{\text{max}}^{-\frac{5}{2}+2\Delta_i+\Delta_{\hat\Phi}}\frac{1}{1+\log\left(\frac{1-2\gamma}{1+2\gamma}\right)}\, .
        \end{aligned}
    \end{equation}
    This expression is zero when $\gamma=\frac{1}{2}$ and monotonically decreases for lower $\gamma$.
\end{itemize}
The optimal choice is then to take $\alpha_{\text{optimal}}=\frac{\Delta_{\text{max}}}{2}\,.$
\subsection{Asymptotics of BOE coefficient flow equation}

\subsubsection{With normal blocks}
The second flow equation, \eqref{eq:BOE flow app},
\begin{align}
    \frac{db^{\hat{\Phi}}_i}{d\lambda}
    &=\sum_{j,l} b^{\hat\Phi}_l b^{\hat\Phi}_j \frac{C_{l ji}+C_{lij}}{2} \JJ_{\D_i}(\D_l,\D_j)
    \label{eq:BOE flow app}
\end{align} 
presents a sum over $l$ and $j$, where $l$ is associated to the deforming operator while $j$ is associated to the bulk operator we are flowing. Let us consider their asymptotics separately 
\begin{equation}
\begin{aligned}
    \bb{\hat\Phi}{l}\bb{\hat\Phi}{j}\frac{C_{lji}+ C_{lij}}{2}\mathcal{J}_{\D_i}(\D_l,\D_j)&\underset{\text{fixed } \Delta_i,\Delta_j}{\sim} \Delta_l^{2\Delta_i+\Delta_{\hat\Phi}-\frac{5}{2}}\\
    \bb{\hat\Phi}{l}\bb{\hat\Phi}{i}\frac{C_{lji}+C_{lij}}{2}\mathcal{J}_{\D_i}(\D_l,\D_j)&\underset{\text{fixed } \Delta_l,\Delta_i}{\sim} \Delta_j^{\Delta_i+\Delta_{\hat{\Phi}}-\frac{3}{2}}\, .
\end{aligned}
\end{equation}
The sums in (\ref{eq:BOE flow app}) with normal blocks thus diverge if $\Delta_i>\frac{1-2\Delta_{\hat{\Phi}}}{2}$. 


\subsubsection{With local blocks}
Recall the integrated $BB\del$ local block is
\begin{equation}
\begin{aligned}
    \JJ_{\D_i}^{(\a)}(\D_l,\D_j)
    =
    &\frac{2 \sqrt{\pi } \Gamma \left(\alpha -\frac{1}{2}\right) \Gamma \left(\D_j+\frac{1}{2}\right) \Gamma \left(\frac{\D_l+1}{2}\right) \Gamma \left(\alpha +\frac{\D_{ij}}{2}\right)}
    {(\D_l-1) (\D_{j}-\D_{i}) \Gamma \left(\alpha -\frac{\D_l}{2}\right) \Gamma \left(\frac{\D_i+\D_j+1}{2}\right) \Gamma \left(\frac{\D_{lji}}{2}\right) \Gamma \left(\alpha +\frac{\D_{lij}-1}{2}\right)}\times
    \\
    &\times
    \,\pFq{3}{2}{ \frac{\Delta_{lij}}{2} , \frac{\Delta_l-1}{2} , \alpha+\frac{\Delta_{l}-1}{2}}{\Delta_l+\frac{1}{2} , \alpha+\frac{\Delta_{lij}-1}{2}}{1}\, .
\end{aligned}
\end{equation}
To analyze the large $\D_l$ asymptotics with $\D_i$, $\D_j$, $\a$ fixed, we use the following integral representation of the ${}_3F_2$
\al{
\pFq{3}{2}{a_1, a_2, a_3}{b_1,b_2}{z}=
\frac{\G(b_2)}{\G(a_3)\G(b_2-a_3)}\int_0^1 dt\,
t^{a_3-1}(1-t)^{-a_3+b_2-1} \hyperF{a_1}{a_2}{b_1}{tz}\, ,
\label{eq:3F2 int rep}
}
with ${\rm Re}(b_2)>{\rm Re}(a_3)>0$. In the rest of this argument, we choose 


 \begin{equation}
    a_1 = \frac{\Delta_{lij}}{2}\, ,\quad a_2 =\alpha + \frac{\Delta_l-1}{2}\, ,\quad a_3 = \frac{\Delta_l-1}{2}\, ,\quad b_1 = \alpha + \frac{\Delta_{lij}-1}{2}\, ,\quad b_2 = \Delta_l+\frac{1}{2}\, .
    \label{eq:a1a2a3b1b2}
\end{equation}
Then
\begin{equation}
\pFq{3}{2}{a_1, a_2, a_3}{b_1,b_2}{1}=
\frac{\G(b_2)}{\G(a_3)\G(b_2-a_3)}\int_0^1 dt\,
t^{\frac{\Delta_l-3}{2}}(1-t)^{\frac{\Delta_l}{2}} \hyperF{a_1}{a_2}{b_1}{t}\, .
\label{eq:3F2 int rep 2}
\end{equation}
We then change variables as $s = 1-t$, such that we get 
\begin{equation}
\pFq{3}{2}{a_1, a_2, a_3}{b_1,b_2}{1}=
\frac{\G(b_2)}{\G(a_3)\G(b_2-a_3)}\int_0^1 ds\,
(1-s)^{\frac{\Delta_l-3}{2}}s^{\frac{\Delta_l}{2}} \hyperF{a_1}{a_2}{b_1}{1-s}\, .
\label{eq:3F2 int rep 3}
\end{equation}
This integral is dominated by $t\rightarrow 1$, which corresponds to $s\rightarrow 0$.
Then we use the standard result\footnote{One can verify that $s$ scales as $O(\D_l^{-1})$, and that \eqref{eq:2F1 small s} applies in the limit $\D_l\to\infty,\,s=O(\D_l^{-1})$ with the choice \eqref{eq:a1a2a3b1b2}.} 
\begin{equation}
   {}_2F_1(a_1,a_2;b_1;1-s) = \frac{\Gamma(b_1)\Gamma(a_1+a_2-b_1)}{\Gamma(a_1)\Gamma(a_2)}s^{b_1-a_1-a_2}\, ,\qquad (s\rightarrow 0)\, . 
   \label{eq:2F1 small s}
\end{equation}
such that 
\al{
\pFq{3}{2}{a_1, a_2, a_3}{b_1,b_2}{1}=
\frac{\G(b_2)}{\G(a_3)\G(b_2-a_3)}\frac{\Gamma(b_1)\Gamma(a_1+a_2-b_1)}{\Gamma(a_1)\Gamma(a_2)}\int_0^1 ds\,
(1-s)^{\frac{\Delta_l-3}{2}}s^{\frac{\Delta_l}{2}} s^{-\frac{\D_l}{2}}\, .
\label{eq:3F2 int rep}
}
Explicitly performing the $s$ integral (and taking $\Delta_l$ large), we obtain 
\al{
\pFq{3}{2}{a_1, a_2, a_3}{b_1,b_2}{1}
&=
\frac{\Gamma
   \left(\frac{\Delta_l}{2}\right) \Gamma
   \left(\D_l+\frac{1}{2}\right) \Gamma
   \left(\alpha +\frac{\Delta_{lij}-1}{2}
   \right)}{\Gamma
   \left(\frac{\Delta_l}{2}-\frac{1}{2}\right) \Gamma
   \left(\frac{\Delta_l}{2}+1\right) \Gamma
   \left(\alpha +\frac{\Delta_l-1}{2}\right) \Gamma
   \left(\frac{\Delta_{lij}}{2} \right)}\frac{2}{\Delta_l}\, .
\label{eq:3F2 int rep}
}Now, we can use the Stirling's approximation of the gamma function $ 
\G(x)= \sqrt{2\pi}x^{x-\frac{1}{2}}e^{-x} $, such that we get  
\begin{equation}
\pFq{3}{2}{a_1, a_2, a_3}{b_1,b_2}{1}\approx\frac{1}{\sqrt{\pi }} 2^{\Delta_l} \D_l^{-\frac12}
\, .
\end{equation}

Including the prefactor and the BOE and OPE coefficients we have
\be
\JJ_{\D_i}^{(\a)}(\D_l,\D_j)
\overset{\text{large }\D_l,\,\text{fixed }\a}{\sim}
\# 2^{\D_l} \D_l^{\frac12-2\a}\,,
\ee
and
\be
\bb{\hat\Phi}{l}\bb{\hat\OO}{j}C_{lji}\JJ_{\D_i}^{(\a)}(\D_l,\D_j)
\overset{\text{large }\D_l,\,\text{fixed }\a}{\sim}
\# \Delta_l^{\Delta_i+\Delta_j+\Delta_{\hat{\Phi}}-2\a-1}\,.
\ee
Therefore, for fixed $\D_j$, we need
\be
\a>\frac{\Delta_i+\Delta_j+\Delta_{\hat{\Phi}}}{2}\,.
\label{eq:loweralphaboe}
\ee
for the sum over $\D_l$ in the BOE coefficient flow equation to converge.

\section{Details of free scalar theory}
\label{app:free theory data}
In section \ref{sec:test of flow eqn} of the main text, we have tested our flow equations in the free scalar theory. Here we report the associated QFT data, which is known exactly along the flow. The notation used here is explained in section \ref{sec:test of flow eqn}.
\subsection{Construction of boundary primary operators}
\subsubsection{BOE of $\hat\phi$}
The bulk field $\hat{\phi}(\tau,z)$ can be expanded in terms of boundary operators $\f(\tau)$ using the boundary operator expansion, which we discussed in  \eqref{eq:BOEdef}. In this case, the expansion contains a single primary operator $\phi$. The remaining sum is over its descendants:
\begin{equation}
    \hat{\phi}(\tau,z) = b^{\hat{\phi}}_{\f}\sum_n \frac{(-1)^n(\Delta_\f)_n}{n!(2\Delta_\f)_{2n}}z^{\Delta_\f + 2n}\partial_\tau^{2n}\f(\tau)\, ,
    \end{equation}
with $\Delta_\f$ defined as in \eqref{eq:DefDelta}. The boundary expansion coefficient $b^{\hat{\phi}}_{\f}$ is given in \eqref{eq:BOEfreescalar} and can be trivially read off from (\ref{eq:bulktwopt:free}).

\subsubsection{BOE of $\hat{\phi}^2(\tau,z)$}

The boundary operator expansion of $\hat{\phi}^2(\tau,z)$ contains more primary operators. In particular, by Wick theorem, all double trace primaries of the form \cite{Mikhailov:2002bp,Penedones:2010ue}
\begin{equation}
    [\f^2]_n\propto \sum_{k=0}^n\frac{(-1)^k(2n)!(2\Delta_\f)_{2n}}{k!(2n-k)!(2\Delta_\f)_k(2\Delta_\f)_{2n-k}}:(\partial^k\f(\tau))(\partial^{2n-k}\f(\tau)):\, ,
\end{equation}
contribute, together with their descendants. The scaling dimensions of such operators are $\Delta_{[\f^2]_n} = 2\Delta_\f+2n$, and their normalization is fixed as 
\begin{align}
		(a)&:\quad\braket{[\f^2]_n(\tau_1)[\f^2]_n(\tau_2)}=\frac{1}{\abs{\tau_1-\tau_2}^{4\Delta_\f+4n}}\, ,\\
		(b)&:\quad\braket{\hat\phi^2(\tau_1,z_2)[\f^2]_n(\tau_2)}>0\, .
\end{align}
This implies that the bulk field $\hat{\phi}^2(\tau,z)$ can be expanded as 
\begin{equation}\label{eq:BOEf5}
\hat{\phi}^2(\tau,z) = \sum_n b^{\hat{\phi}^2}_{[\f^2]_n}\sum_p \frac{(-1)^p(2\Delta_\f + 2n)_p}{p!(4\Delta_\f + 4n)_{2p}}z^{2(\Delta_\f+n+p)}\partial_\tau^{2p}[\f^2]_n(\tau)\, ,
\end{equation} 
where the boundary expansion coefficients $b^{\hat{\phi}^2}_{[\f^2]_n}$ are derived in equation \eqref{eq:BOEfreescalarO2}.


\subsection{Scaling dimensions} 
We will be interested in the primaries $\phi$, $[\phi^2]_n$ and $[\phi^3]_q$, which have respectively scaling dimensions $\Delta_\phi$, $2\Delta_\phi+2n$ and $3\Delta_\phi+q$, where the relation between $\Delta_\phi$ and the mass coupling $m^2$ is
\begin{equation}
    \Delta_\phi(\Delta_\phi-1)=m^2\,.
\end{equation}

\subsection{BOE coefficients} 

The BOE coefficients can be determined by decomposing two-point functions close to the boundary. In the limit $z_1,z_2\rightarrow 0$ with $\tau_{1,2}$ fixed, the cross ratio $\eta$ behaves as 
\begin{equation}
    \eta \sim \frac{z_1z_2}{(\tau_1-\tau_2)^2}\, ,
\end{equation}
and the asymptotic behavior of the bulk two-point function \eqref{eq:bulktwopt:free} is 
\begin{equation}
    \langle\hat\phi(x_1)\hat\phi(x_2)\rangle= \frac{\Gamma(\Delta_\phi)}{2\sqrt{{\pi}}\Gamma(\Delta_\phi+\frac{1}{2})}\left[\frac{z_1z_2}{(\tau_1-\tau_2)^2}\right]^{\Delta_\phi}\, .
\end{equation}
It follows that
\begin{equation}\label{eq:BOEfreescalar}
b^{\hat\phi }_\phi=\sqrt{\frac{\Gamma(\Delta_\f)
}{2\sqrt{\pi}\Gamma\left(\Delta_\f+\frac{1}{2}\right)}}\, .
\end{equation}
Note that we thus define the boundary operator $\f(\tau)$ for a free massive scalar as 
\begin{equation}
    \f(\tau)\equiv \frac{\sqrt{2}\pi^{1/4}\Gamma\left(\Delta_\f + \frac{1}{2}\right)^{1/2}}{\Gamma\left(\Delta_\f \right)^{1/2}}\lim_{z\rightarrow 0}\left[z^{-\Delta_\f}\hat{\phi}(\tau,z)\right]\, .
\end{equation}
To obtain the BOE coefficients of $\hat{\phi}^2$, consider its two-point function:
\begin{align}
\spl{
\braket{\hat{\phi}^2(x_1)\hat{\phi}^2(x_2)} &= 2\langle\hat\phi(x_1)\hat\phi(x_2)\rangle^2\\ 
&= \frac{\Gamma(\Delta_\f)^2}{2\pi \Gamma\left(\Delta_\f+\frac{1}{2}\right)^2}\eta^{2\Delta_\f}\,_{2}F_1\left(\Delta_\f,\frac{1}{2};\Delta_\f+\frac{1}{2};\eta^2\right)^2\,.
}
\end{align}
We then use the following identity, which is valid for $0\leqslant z<1$:
\begin{equation}
    _{2}F_1\left(\Delta,\frac{1}{2};\Delta+\frac{1}{2};z\right)^2= \sum_{n=0}^\infty c_nz^n\,_{2}F_1\left(2\Delta+2n,\frac{1}{2};2\Delta+2n+\frac{1}{2};z\right)\, ,
\end{equation}
where the coefficients $c_n$ are given by 
\begin{equation}
    c_n = \frac{\left(\frac{1}{2}\right)_n(\Delta)_n^3(2\Delta-\frac{1}{2})_n}{n!(2\Delta)_n(\Delta+\frac{1}{2})_n(\Delta+\frac{1}{4})_n(\Delta-\frac{1}{4})_n}\, .
\end{equation}
This implies that the BOE coefficients of $\hat{\phi}^2$ are 
\begin{equation}\label{eq:BOEfreescalarO2}
b^{\hat\phi^2}_{[\f^2]_n}
    =
    \frac{\Gamma(\Delta_\f)}{\Gamma\left(\Delta_\f+\frac{1}{2}\right)}\sqrt{\frac{\left(\frac{1}{2}\right)_n\left(\Delta_\f\right)_n^3\left(2\Delta_\f-\frac{1}{2}\right)_n}{2\pi\,n!\,\left(2\Delta_\f\right)_n\,\left(\Delta_\phi+\frac{1}{2}\right)_n\,\left(\Delta_\f+\frac{1}{4}\right)_n\,\left(\Delta_\f-\frac{1}{4}\right)_n}}\,.
\end{equation}

\subsection{OPE coefficients}\label{app:subsec:OPECoeff}

\paragraph{OPE coefficient $C_{\f \f [\f^2]_n}$}
The OPE coefficient between two generalized free scalars $\f$ and a double-twist operator $[\f^2]_n$ is $C_{\f\f[\f^2]_n}$.
There are multiple ways to derive this quantity, but one of the easiest is to use the $B\partial\partial$ three-point function $\braket{:\phi^2(\tau,z):\f(\tau_1)\f(\tau_2)}$. Using Wick's theorem, we get 
\begin{equation}
\braket{:\hat{\phi}^2(\tau,z):\phi(\tau_1)\phi(\tau_2)} = 2 (\bb{\hat{\phi}}{\f})^2\Big[\frac{z}{(\tau-\tau_1)^2+z^2}\Big]^{\D_\f}\Big[\frac{z}{(\tau-\tau_2)^2+z^2}\Big]^{\D_\f}\, ,
\end{equation}
but this can also be obtained using the BOE \eqref{eq:BOEf5}  
\begin{equation}
\braket{:\hat{\phi}^2(\tau,z):\f(\tau_1)\f(\tau_2)} =\sum_{n,q}\bb{\hat{\phi}^2}{[\f^2]_n}\frac{(-1)^q(2\D_\f+2n)_q}{q!(4\D_\f+4n)_{2q}}z^{2\D_\phi+2n+2q}\partial_{\tau}^{2q}\braket{[\phi^2]_n(\tau)\phi(\tau_1)\phi(\tau_2)}\, .
\end{equation}
By choosing a specific configuration (such as $\tau_1=0,\, \tau_2=1$) and matching the expansion order-by-order in $z$, we obtain \cite{Fitzpatrick:2011dm} 
\begin{align}
C_{\f\f[\f^2]_n}
  &= (-1)^n\sqrt{\frac{(2\Delta_\f)_n(2\Delta_\f)_{2n}}{2^{2n-1}(2n)!(2\Delta_\f+n-1/2)_{n}}}\, ,
\end{align}
where $[\f^2]_n$ has scaling dimension $2\D_\f+2n$.

\paragraph{OPE coefficient $C_{[\f^2]_n [\f^2]_m [\f^2]_l}$}
The OPE coefficient of three double-twist operators is
\al{\spl{
&C_{[\f^2]_n [\f^2]_m [\f^2]_l}=\\
&=\left(2^{2n}\Gamma (-4 \Delta_\f -4 n+2)
\sqrt{\frac{\Gamma \left(2 \Delta_\f +2 n-\frac{1}{2}\right)}{(2 n)! \Gamma (2 (\Delta_\f +l)) \Gamma (4 \Delta_\f +2 n-1)}}\right)
\times(n\leftrightarrow m)\times (n\leftrightarrow l)
\\[6pt]
&\quad\times
\frac{2^{6 \Delta_\f -1/2}}{\pi^{3/4}}\cos (2 \pi  \Delta_\f ) (-1)^{l+m+n+1}
\Gamma (2 \Delta_\f +2 l) \Gamma (4 \Delta_\f +2 l-1)  \Gamma (2 (\Delta_\f +l-m+n))
\\[6pt]
&\quad\times
{}_4\tilde{F}_3\left( {\begin{array}{*{20}{c}}
{-2m,  -2 (3 \Delta_\f +l+m+n-1),  -2 \Delta_\f -2 m+1,  -2 \Delta_\f +2 l-2 m-2 n+1}\\
{-4 \Delta_\f -4 m+2,  -2 (2 \Delta_\f +m+n-1),  -2 \Delta_\f -2 m-2 n+1}
\end{array};1} \right)
\,.
}
\label{eq:C_O2O2O2}}
This is computed by applying harmonic analysis (or the Euclidean inversion formula) to the six-point GFF correlator in the snowflake channel. The computation details are nicely explained in \cite[App.C]{Antunes:2021kmm} and here we only make some brief comments. There are eight Wick contractions contributing to the final result, but one can verify that they are all equal. In the convention of \cite{Antunes:2021kmm} the OPE density is
\al{\spl{
&c(\D_A,\D_B,\D_C) =\\
&\quad
8\times
S([\tilde{A}]\tilde{B}\tilde{C})
S([\tilde{B}]A\tilde{C})
S([\tilde{C}]AB)\times
\frac{S([\tilde{\f}]\tilde{\f}\tilde{C})
S([\tilde{\f}]\f\tilde{C})
S([\tilde{\f}]\tilde{\f}\tilde{B})
B(\D_A)}{B(\D_A)B(\D_B)B(\D_C)}
\times 
\left\{ 
\begin{matrix}
\D & \tilde{\D}_B & \D_A
\\
\tilde{\D} & \tilde{\D} & \D_C
\end{matrix}
\right\}\,,
}}
where $S$ is the shadow coefficient \cite{Karateev:2018oml} satisfying
\begin{equation}
\int d^d x\<\!\<\widetilde{\mathcal{O}}(y) \widetilde{\mathcal{O}}^{\dagger}(x)\>\!\>\<\!\<\OO_1 \OO_2 \mathcal{O}(x)\>\!\>=S\left(\OO_1 \OO_2[\mathcal{O}]\right)\<\!\<\OO_1 \OO_2 \widetilde{\mathcal{O}}(y)\>\!\>\,,
\end{equation}
and $B$ is given by
\begin{equation}
B(\D_\OO) = \frac{\left(\<\!\<\OO_1 \OO_2 \OO\>\!\>,\<\!\<\tilde{\OO}_1 \tilde{\OO}_2 \tilde{\OO}\>\!\>\right)}{\m(\D_\OO)}\,,
\end{equation}
and the last factor is the one-dimensional $6j$ symbol (see e.g. \cite[App.B]{Liu:2018jhs}, but note that \cite{Antunes:2021kmm} and \cite{Liu:2018jhs} use different normalization convention). Taking (minus) the residues at $\D_A=2\D_\f+2n,\,\D_B=2\D_\f+2m,\,\D_C=2\D_\f+2l$ and diving out $C_{\f\f[\f^2]_n},\,C_{\f\f[\f^2]_m},\,C_{\f\f[\f^2]_l}$, one obtains \eqref{eq:C_O2O2O2}.

\paragraph{OPE coefficient $C_{\f [\f^2]_m [\f^3]_q} C_{[\f^3]_q [\f^2]_n \f}$}
Notice that here $[\f^3]_q$ denotes a collection of operators with degenerate scaling dimension $3\D_\f+q$ when $q$ is large enough.\footnote{More precisely, the generating function that counts triple-twist operators comes from
\begin{equation}
\sum_{0\leqslant n_1\leqslant n_2\leqslant n_3}t^{n_1+n_2+n_3}
=\frac{1}{(1-t)(1-t^2)(1-t^3)}\,,
\end{equation}
where $t$ is an auxiliary variable. 
The factor $\frac{1}{1-t}$ counts descendants and so
\begin{equation}
\frac{1}{(1-t^2)(1-t^3)}=
1+t^2+t^3+t^4+t^5+2 t^6+t^7+2 t^8+O\left(t^9\right)
\end{equation}
counts the degeneracy of primaries. For example, at $q=6$ and $q=8$ there are two triple-twist operators with the same scaling dimension.} This coefficient can be calculated in a similar way using harmonic analysis. The main difference is that one needs to use conformal partial waves in the comb channel instead of the snowflake channel. Additional subtleties appear because the triple-twist operator can be either parity-even ($q$ even) or parity-odd ($q$ odd), and one can obtain a closed form result for each sector. The full expression of $C_{\f [\f^2]_m [\f^3]_q} C_{[\f^3]_q [\f^2]_n \f}$ can be found in the ancillary notebook \texttt{testflows.nb}.\footnote{We thank Sebastian Harris for sharing with us the unpublished result of $C_{\f [\f^2]_m [\f^3]_q} C_{[\f^3]_q [\f^2]_n \f}$.}\ Importantly, they vanish when $n>\lfloor\frac{q}{2}\rfloor$.  For a simpler expression valid when $m=n$, see eq. (D.12) in \cite{Antunes:2023kyz}

\paragraph{OPE coefficient $C_{[\phi^2]_n\phi^4\phi^4}$}
This OPE coefficient can be obtained in a way similar to that for $C_{\f\f[\f^2]_n}$ above. We consider the $B\del\del$ three-point function
\be
\left\langle \hat{\phi }^2(\tau=0,z=\xi)\phi ^4 \left(\tau_1=1\right)\phi ^4 \left(\tau_2=\infty \right)\right\rangle =
\frac{192}{4!} \left(b_{\phi }^{\hat{\phi }}\right)^2 \chi ^{\Delta _{\phi }}
=\sum _{n\geqslant 0}
b_{\left[\phi ^2\right]_n}^{\hat{\phi }^2}
C_{\phi^4 \phi^4 [\phi^2]_n}  
G_{2 \Delta _{\phi }+2 n}(\chi)\,,
\ee
where in the first equality the numerator 192 comes from all possible Wick contractions and the denominator comes from the normalisation of the operator $\f^4$ such that it has unit two-point function. Recall the cross ratios are related through $\chi=\frac{\xi^2}{1+\xi^2}$ (see section \ref{subsec:Confblock}).

Since both $b_{\phi }^{\hat{\phi}}$ and $b_{[\phi ^2]_n}^{\hat{\phi}^2}$ are already given in \eqref{eq:BOEfreescalar} and \eqref{eq:BOEfreescalarO2}, respectively, we can solve for $C_{[\phi^2]_n\phi^4\phi^4}$ by expanding in $\chi$. The result is
\be
C_{[\phi^2]_n\phi^4\phi^4}
=
(-1)^n 2^{2 n+\frac{5}{2}} (\Delta_\phi) _n \left(\Delta_\phi +\frac{1}{2}\right)_n \sqrt{\frac{\Gamma (4 \Delta_\phi +2 n-1)}{\Gamma (2 n+1) \Gamma (4 \Delta_\phi +4 n-1)}}\,.
\label{eq:C_phi2phi4phi4}
\ee
In section \ref{subsec:weakly relevant boundary op} we need to consider the case $\D_\f=\frac14$. Here the $n=0$ term is a little subtle and we need to set $n=0$ before sending $\D_\f\to\frac14$.

\bibliography{bibliography}
\bibliographystyle{utphys}
\end{document}

%% file: topmatter.tex
\let\a=\alpha 
\let\b=\beta 
\let\g=\gamma 
\let\d=\delta 
\let\e=\epsilon

\let\z=\zeta 
\let\h=\eta 
\let\q=\theta

\let\l=\lambda 
\let\m=\mu 
 
\let\x=\xi 
\let\r=\rho
\let\s=\sigma 
\let\t=\tau  
\let\f=\phi

\let\w=\omega  

\let\D=\Delta

\let\F=\Phi

\let\W=\Omega   
\let\G=\Gamma

\let\del=\partial
\let\<=\langle
\let\>=\rangle
\newcommand{\BB}{\mathcal{B}}
\newcommand{\CC}{\mathcal{C}}
\newcommand{\DD}{\mathcal{D}}

\newcommand{\II}{\mathcal{I}}
\newcommand{\JJ}{\mathcal{J}}
\newcommand{\KK}{\mathcal{K}}

\newcommand{\OO}{\mathcal{O}}

\newcommand{\QQ}{\mathcal{Q}}
\newcommand{\RR}{\mathcal{R}}

\newcommand{\al}[1]{\begin{align}#1\end{align}}
\newcommand{\spl}[1]{\begin{split}#1\end{split}}

\def \bmat {\begin{pmatrix}}
\def \emat {\end{pmatrix}}

\newcommand{\be}{\begin{equation}}
\newcommand{\ee}{\end{equation}}